



\font\smb=cmr8


\font\smbfb=cmbx8


\def\menouno{-{\underline 1}}
\def\piuuno{+{\underline 1}}
\def\zero{{\underline 0}}

\def\grle{{\scriptscriptstyle {>\atop <}}}

\def\rect#1#2{{\vcenter{\vbox{\hrule height.3pt
	    \hbox{\vrule width.3pt height#2truecm \kern#1truecm
	    \vrule width.3pt}
	    \hrule height.3pt}}}}


\def\birect#1#2#3#4#5#6#7{{\vcenter{\vbox{\hrule height.6pt
	                 \hbox{\vrule width.6pt height#2truecm \kern#5truecm
		            {\raise#7truecm\vbox{\hrule height.6pt
	    		    \hbox{\vrule width.6pt height#4truecm \kern#3truecm
	    		    \vrule width.6pt}
	    		    \hrule height.6pt}}
		         \kern#6truecm\vrule width.6pt}
		         \hrule height.6pt}}}}

\def\square{\rect{0.2}{0.2}}
\magnification=\magstep1\hoffset=0.cm
\voffset=1truecm\hsize=16.5truecm \vsize=21.truecm
\baselineskip=14pt plus0.1pt minus0.1pt \parindent=12pt
\lineskip=4pt\lineskiplimit=0.1pt      \parskip=0.1pt plus1pt
\def\st{\scriptstyle}

 \let\b=\beta   \let\d=\delta  \let\e=\varepsilon
 \let\g=\gamma \let\h=\eta      \let\l=\lambda
\let\m=\mu      \let\o=\omega      
  \let\s=\sigma \let\t=\tau   
 \let\x=\xi \let\z=\zeta
\let\D=\Delta   \let\G=\Gamma  \let\L=\Lambda 
\let\O=\Omega      

 
 \def\\{\noindent}

\def\tende#1{\vtop{\ialign{##\crcr\rightarrowfill\crcr
              \noalign{\kern-1pt\nointerlineskip}
              \hskip3.pt${\scriptstyle #1}$\hskip3.pt\crcr}}}
\def\otto{{\kern-1.truept\leftarrow\kern-5.truept\to\kern-1.truept}}

\def\data{\number\day/\ifcase\month\or gennaio \or febbraio \or marzo \or
aprile \or maggio \or giugno \or luglio \or agosto \or settembre
\or ottobre \or novembre \or dicembre \fi/\number\year}
\setbox200\hbox{$\scriptscriptstyle \data $}
\newcount\pgn \pgn=1
\def\foglio{\number\numsec:\number\pgn
\global\advance\pgn by 1}
\def\foglioa{A\number\numsec:\number\pgn
\global\advance\pgn by 1}
%
\global\newcount\numsec\global\newcount\numfor
\global\newcount\numfig
\gdef\profonditastruttura{\dp\strutbox}
\def\senondefinito#1{\expandafter\ifx\csname#1\endcsname\relax}
\def\SIA #1,#2,#3 {\senondefinito{#1#2}
\expandafter\xdef\csname #1#2\endcsname{#3} \else
\write16{???? ma #1,#2 e' gia' stato definito !!!!} \fi}
\def\etichetta(#1){(\veroparagrafo.\veraformula)
\SIA e,#1,(\veroparagrafo.\veraformula)
 \global\advance\numfor by 1
 \write15{\string\FU (#1){\equ(#1)}}
 \write16{ EQ \equ(#1) == #1  }}
\def \FU(#1)#2{\SIA fu,#1,#2 }
\def\etichettaa(#1){(A\veroparagrafo.\veraformula)
 \SIA e,#1,(A\veroparagrafo.\veraformula)
 \global\advance\numfor by 1
 \write15{\string\FU (#1){\equ(#1)}}
 \write16{ EQ \equ(#1) == #1  }}
\def\getichetta(#1){Fig. \verafigura
 \SIA e,#1,{\verafigura}
 \global\advance\numfig by 1
 \write15{\string\FU (#1){\equ(#1)}}
 \write16{ Fig. \equ(#1) ha simbolo  #1  }}
\newdimen\gwidth
\def\BOZZA{
\def\alato(##1){
 {\vtop to \profonditastruttura{\baselineskip
 \profonditastruttura\vss
 \rlap{\kern-\hsize\kern-1.2truecm{$\scriptstyle##1$}}}}}
\def\galato(##1){ \gwidth=\hsize \divide\gwidth by 2
 {\vtop to \profonditastruttura{\baselineskip
 \profonditastruttura\vss
 \rlap{\kern-\gwidth\kern-1.2truecm{$\scriptstyle##1$}}}}}
}
\def\alato(#1){}
\def\galato(#1){}
\def\veroparagrafo{\number\numsec}\def\veraformula{\number\numfor}
\def\verafigura{\number\numfig}
\def\Eq(#1){\eqno{\etichetta(#1)\alato(#1)}}
\def\eq(#1){\etichetta(#1)\alato(#1)}
\def\Eqa(#1){\eqno{\etichettaa(#1)\alato(#1)}}
\def\eqa(#1){\etichettaa(#1)\alato(#1)}
\def\eqv(#1){\senondefinito{fu#1}$\clubsuit$#1\else\csname fu#1\endcsname\fi}
\def\equ(#1){\senondefinito{e#1}eqv(#1)\else\csname
e#1\endcsname\fi}

\def\include#1{
\openin13=#1.aux \ifeof13 \relax \else
\input #1.aux \closein13 \fi}
\openin14=\jobname.aux \ifeof14 \relax \else
\input \jobname.aux \closein14 \fi
\openout15=\jobname.aux
\footline={\rlap{\hbox{\copy200}\ $\st[\number\pageno]$}\hss\tenrm
\foglio\hss}

\tolerance=10000

\numsec=0
\centerline {\bf METASTABILITY AND NUCLEATION}
\centerline {\bf FOR THE BLUME-CAPEL MODEL.}
\centerline {\bf  DIFFERENT MECHANISMS OF TRANSITION.}
\vskip 2 truecm
\centerline {Emilio N. M. Cirillo}\par\noindent
\vskip 0.5 truecm
\centerline {\it Dipartimento di Fisica dell'Universit\`a di Bari
and}\par\noindent
\centerline {\it Istituto Nazionale di Fisica Nucleare, Sezione di
Bari.}\par\noindent
\centerline {\it V. Amendola 173, I-70126 Bari, Italy.}\par\noindent
\centerline {\rm E-mail: cirillo@axpba0.ba.infn.it}
\vskip 1 truecm
\centerline {Enzo Olivieri}\par\noindent
\vskip 0.5 truecm
\centerline {\it Dipartimento di Matematica - II Universit\`a di
Roma}\par\noindent
\centerline {\it Tor Vergata - Via della Ricerca Scientifica - 00173 ROMA -
Italy}\par\noindent
\centerline {\rm E-mail: olivieri@mat.utovrm.it}
\vskip 1.5 truecm
\centerline {\bf Abstract}
\vskip 0.5 truecm
\centerline {
\vbox
{
\hsize=13truecm
\baselineskip 0.35cm
We study metastability and nucleation for the Blume-Capel model: a
ferromagnetic
nearest neighbour two-dimensional lattice system with spin variables taking
values in  $\{-1,0,+1\}$.
We consider large but finite volume, small fixed magnetic field $h$ and
chemical
potential $\lambda$ in the limit of zero temperature;
we analyze the first excursion
from the metastable $-1$ configuration to the stable $+1$ configuration.
We compute the asymptotic
behaviour of the transition time and describe the typical
tube of trajectories during the transition. We show that, unexpectedly,
the mechanism of transition changes abruptly when the line $ h = 2 \lambda$ is
crossed.
}
}
\vfill\eject

\numsec=1\numfor=1

{\bf Section 1. Introduction.}
\par
Metastability  is a relevant phenomenon
for   thermodynamic systems  close to a first order phase transition.\par
Let us start from a given pure equilibrium phase in a suitable region of
the phase diagram and  change the thermodynamic
parameters to values corresponding to a different equilibrium phase; then,
in particular experimental situations, the
system, instead of undergoing a phase transition, can still remain in a
``wrong" equilibrium, far from the ``true" one but actually close to what the
equilibrium would be at the other side of the transition.
This apparent equilibrium, often called ``metastable state", persists untill an
external perturbation or some spontaneous fluctuation leads the system to the
stable
equilibrium.\par
 For a general revue on metastability with particular attention to rigorous
results
see [PL1],[PL2].\par
There are strong arguments leading to the conclusion that neither metastability
can be included in the scheme of Gibbsian formalism, which is confined to the
description of the genuine stable equilibrium states (see [LR]); nor it can be
directly described using extrapolation beyond the condensation point, because
of the presence
of an essential singularity of the free energy (see the fundamental result due
to Isakov [I]).\par
Metastability is a genuine dynamical phenomenon. Its description on one side
has a basic importance from the point of view of fundamental physics; on the
other
side it poses interesting new mathematical problems ([CGOV], [OS1],
[OS2]).\par
Since a general approach to non-equilibrium statistical mechanics is
still missing, a
crucial role is played by particular models of microscopic dynamics.
It is remarkable that, quite recently, rigorous results have been deduced in
this field by analyzing, in particular, the geometry of the condensation nuclei
as
well as the possible coalescence between droplets. Notice that these aspects
were totally absent in  previous theories like the so called classical
theory of nucleation (see [PL1]).
\par
In the recent years many progresses have been made in the understanding of the
phenomenon of metastability in the framework of Glauber dynamics.
By Glauber dynamics we mean a stochastic time evolution of a lattice spin
system
(in continuous or discrete time) whose elementary process is a single spin
change and which is {\it reversible} (namely it satisfies the detailed balance
condition) with respect to the Gibbs measure corresponding to a given
hamiltonian.
There is a certain freedom in chosing a particular dynamics satisfying the
above mentioned requirements. A typical choice, that actually we will make in
the present paper, is called ``Metropolis dynamics" (see Eq. 2.6 below).\par
The case of standard Ising model (spin $+1$ or $-1$, ferromagnetic nearest
neighbour interaction), often referred to as {\it Stochastic Ising model}
or {\it Kinetic Ising model}, has been analyzed,
in two dimensions, in [MOS] in connection to
relaxation to equilibrium for arbitrary large (and even infinite) systems
close to the first order phase transition.\par
A  quite complete treatment appeared in [NS1] and [NS2] where
J. Neves and R. Schonmann analyzed, in
the framework of the ``pathwise approach to metastability" introduced
some years ago in [CGOV], the phenomenon of nucleation for large but finite
volume and small magnetic field in the zero temperature limit.\par
In [S1] R. Schonmann, using an argument based on reversibility,
described in detail the typical escape paths.\par
Other asymptotic regimes, very interesting from a physical point of view and
mathematically much more complicated, are considered in [S2], [SS].\par
In the same asymptotic regime as in [NS1],
different hamiltonians have been considered  in [KO1], [KO2] where it has
been shown that the typical path followed during the growth of the stable phase
in general are
not of Wulff type.
Here by Wulff (shape) we mean equilibrium shape of a droplet at zero
temperature
namely the shape minimizing the surface energy for fixed volume. This non-Wulff
growth seems to be an interesting phenomenon in the description of crystal
growth.\par
Let us now try to explain the nature of the mathematical difficulties related
to
our problem.
We notice that in the above mentioned asymptotic regime the behaviour is
similar to the one described by Freidlin and Wentzell in their analysis of
small
random perturbations of dynamical systems: the system typically performs random
oscillations around the local minima of the energy and sometimes it goes
against
the drift following some preferential ways. In particular it is interesting to
characterize the typical tube of trajectories during the first excursion from
the metastable to the stable equilibrium. This first excursion can be seen as
an
escape from a sort of generalized basin ${\cal G}$ of attraction of the
metastable equilibrium. It turns out that many local equilibria are contained
in ${\cal G}$ and
this more general situation goes beyond the approach developed in [FW]
by Freidlin and Wentzell who were able to give a full description of the
typical
tube of escape only for the case of a domain $D$ completely attracted by a
unique stable point.\par
In our more general case, as we will see, new interesting phenomena take place
involving a sort of ``temporal entropy". These phenomena are taken into account
in [OS1], [OS2], where a complete description of the typical tube of escape is
given in general.\par
For attractive short range systems the main feature of the transition appears
to be the
formation of a critical nucleus with suitable shape and size. This critical
droplet results from a competition between the bulk energy favouring the growth
and the surface energy favouring the contraction. Only for large sizes and for
particular shapes will the droplet tend to grow.\par\bigskip
The present paper is devoted to the study of metastability and nucleation in
the
framework of a dynamical Blume-Capel model.
It is a two-dimensional spin system where the single spin variable can take
three
possible values: $-1,0,+1$. It was
 originally introduced to study the $He^3- He^4 $ phase transition.
\par
One can think of it as a system of particles with spin. The value $\s_x = 0$
of the spin at the lattice site $x$ will
correspond to absence of particles (a {\it vacancy}), whereas the values
$\s_x =  +1, -1$
will correspond to the presence, at $x$, of a particle with spin $+1, -1$,
respectively.\par
The formal hamiltonian is given by:
$$
H(\sigma)=J\sum_{<x,y>}(\sigma_{x}-\sigma_{y})^{2}-\lambda\sum_{x}
\sigma_{x}^{2}-h\sum_{x}\sigma_{x}\;\; ,
\Eq (1.1)
$$
where $\lambda$ and $h$ are two  real parameters, having the meaning of the
chemical
potential and the external magnetic field, respectively;  $J$ is a real
positive
constant (ferromagnetic interaction)
 and $<x,y>$ denotes a generic pair of nearest neighbour sites in ${\bf Z}^2$.
\par
In the following we will consider the system enclosed in
 a two-dimensional torus $\L$.
Let $ \menouno,\;\zero\;$  and $
				    \;\piuuno $ denote the configurations with all the spins in $\L$ equal
to$
\;-1,0,+1$, respectively.
The structure of the set of ground states corresponding to different values of
$\l$ and $h$ will be discussed in Section 2. Now we only note that it is
immediate  to
see that for $\l = h = 0$ the configurations $ \menouno,\;\zero\;$  and $
				    \;\piuuno $ are the only ground states.
It has been shown, using Pirogov-Sinai theory, that this phase transition
persists at
positive temperature $T=1/\beta$ in the thermodinamic limit
(see [B], [C], [BS] and [DM]).\par
We will use as dynamics the Metropolis algorithm, in which the typical time
needed to overcome an energy barrier $\Delta H$ is of order $\exp (\beta
\Delta H)$. It will be defined in detail in the next section.
\par
We are interested to the case in which $ \l$ and $h$ are very small but fixed,
the
volume is large and fixed and $T$ is very small; namely we move in the
vicinities of
the triple point $h=\l=T=0$.
In particular we will consider the region $h >\l >0$ where the the most
interesting phenomena take place. The stable equilibrium, namely the absolute
minimum
of the energy, in this case, is $\piuuno$ and we suppose to start with the
system in
the configuration $\menouno$.
We want to describe the first excursion between $ \menouno$ and $\piuuno$.
It turns out that in the above region a direct interface between pluses and
minuses
is unstable and  its appearence and resistance are very unlikely.\par
The main effect which surprisingly and unexpectedly shows up is that two
different mechanisms of transition between $\menouno$ and $\piuuno$ take place
for
different values of the parameters $\l, h$.
More precisely for $ 0 < 2\l <h $ the transition takes place via the formation
of
a suitable critical droplet of zeroes keeping growing untill it covers the
whole volume. Subsequently, from the intermediate zero phase we have the
nucleation
of a critical droplet of plus spins driving eventually the system to the
stable $\piuuno$ phase.\par
Conversely, for $0<\l <h <2\l$, the plus phase is created directly from the
minus phase via
the formation of a suitable critical nucleus, a sort of ``picture frame"
(see Fig.3.10), containing in its bulk the plus spins with a thin layer of
zeroes
separating the interior pluses from the sea of minuses.
We want to stress that the line $h = 2\l$, where this abrupt variation of the
mechanism of nucleation takes place, has no meaning from the ``static" point of
view of the Gibbs states. The reason is that we are analyzing a region
of the configuration space very unlikely at the equilibrium; but, on the
other hand, this region and the form of the ``energy landscape" (see Fig.1.1)
on it plays an important role in the relaxation from metastability.
\midinsert
\vskip 9.5 truecm\noindent
\includegraphics{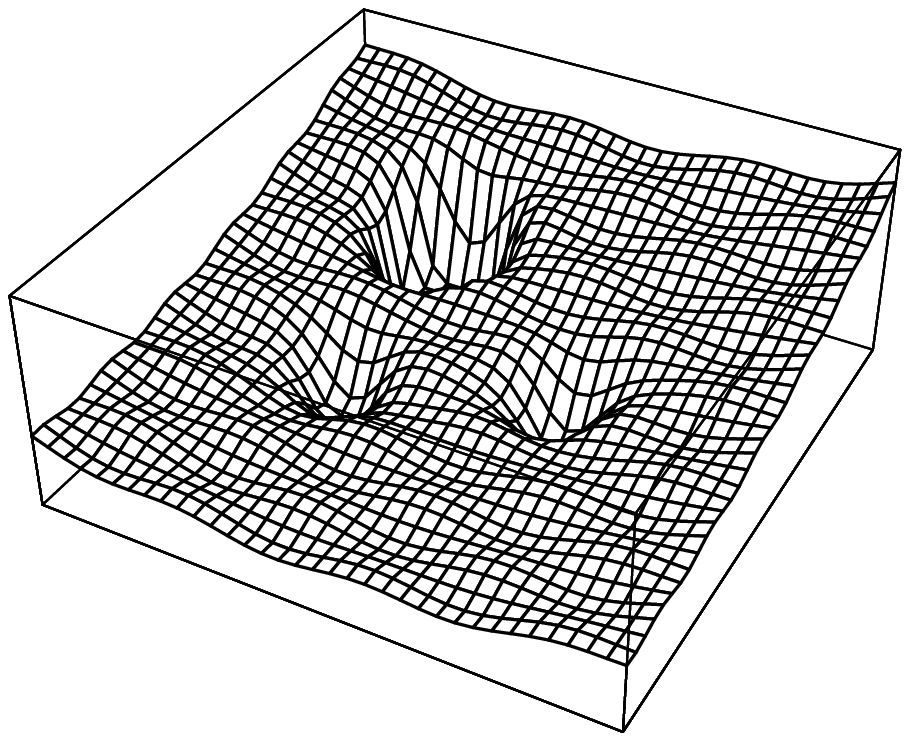}
\centerline{
{\smbfb Fig.1.1 }{\smb Energy landscape of the Blume-Capel model.}
}
\endinsert
\par
A first result that we obtain in the present paper refer to the computation of
the
asymptotic behaviour, for small temperatures, of the transition time (the
life-time
of the metastable state). Then we pass to the characterization of  the typical
trajectories during the transition; we specify  the geometrical sequence of
droplets as well as the order of magnitude of the necessary time fluctuations
during
the growth of the critical nucleus both for $ h< 2\l$ and for $h > 2\l$.
To do this we exploit some general results contained in [OS1].\par
The model-dependent part of the work consists in the solution of a well
specified
sequence of variational problems. The main difficulty is the determination of
the
``minimal global saddle" between $\menouno$ and $\piuuno$ and of the set
${\cal G}$, the
generalized basin of attraction of $\menouno$. From this we will single out an
optimal nucleation mechanism. We will
analyze the energy landscape so precisely to exclude all the other possible
mechanisms of transition. In particular we will show that any form of
coalescence will
be depressed in probability with respect to the optimal nucleation
mechanism.\par
The paper is organized in the following way:
Section 2 contains definitions and results. In particular we state Theorem 1
concerning the asymptotics of the escape time.
In Section 3 we describe the local minima of the energy. In Section 4 we
discuss
supercriticality or subcriticality of droplets namely we determine their
tendency to
grow or shrink. In Section 5 we prove a basic result on the height of different
global saddles. In Section 6 we define the set ${\cal G}$ and find the minima
of the energy
in its boundary $ \partial {\cal G}$.
In Section 7 we describe the typical tube of trajectories during the first
excursion;
then, using as preliminary results the propositions contained in the previous
section
we conclude the proof of Theorem 1; finally we state and prove Theorem 2 which
refers
to the typical tube. Section 8 contains the conclusions.
Appendix 1 contains an explicit proof of a useful result about recurrence
properties
of a general class of Markov chains.

\vfill\eject
\numsec=2\numfor=1

{\bf Section 2. Definitions and results.}
\par
We start by describing the model that we want to study. The configuration
space is
$$\O_{\L} = \{-1,0,+1\}^{\L}\;\; ,\Eq (2.1)$$
where $\L = \L_L $ is a two-dimensional torus (a square with periodic
boundary conditions) of side $L$.
\par
A configuration $\s$ is a function:
$$\s : \L \rightarrow  \{ -1,0,+1\}\;\; .$$
The energy associated to the configuration $\s$ is given by:
$$
H(\sigma)=J\sum_{<x,y>\subset
\L}(\sigma_{x}-\sigma_{y})^{2}-\lambda\sum_{x\in\L}
\sigma_{x}^{2}-h\sum_{x\in\L}\sigma_{x}\;\; ,
\Eq (2.2)
$$
where $<x,y>$ denotes a generic pair of nearest
neighbours sites in $\L$
and we suppose $0 < \lambda <h < J$. We also introduce the following
restriction
$$\l < {2J\over 2a^2+a-1}\;\; ,$$
where $a:={h\over\l}$; the meaning of this condition will be clear later on
(see (3.17)).
\par
The {\it Gibbs
measure} in the torus $\L$ is given by: $$
\m_{\L} \; = \; {\exp ( - \b H(\s)) \over Z_{\L} } \;\; ,\Eq (2.2')
$$
where $\b $ represents the inverse temperature and $Z_{\L}$ is the
normalization
factor called {\it partition function}.\par
We describe now the structure of the ground states corresponding to the
different values of our parameters $\l$ and $h$.\par
Let $\menouno,\;\zero$  and $\piuuno$ denote the configurations with all
the spins in $\L$ equal to $-1,0,+1$, respectively. We have:
$$\eqalign {
 {\rm for } \;\;\lambda=h=0\;\;\;\;\;\;      &{\rm the \; ground \; state\;
is\;
			      three \; times \;}\cr
{\phantom{\lambda=h=0\;\;\;\;\;\;}}&{\rm degenerate,\; the \; configurations\;
				     minimizing \; the\;}\cr
{\phantom{\lambda=h=0\;\;\;\;\;\;}}&{\rm  energy\;are\;\menouno,\;\zero\; and
				    \;\piuuno\; ;}\cr
{\rm for }\;\; h>0\; {\rm and} \; h>-\lambda\;\;\;\;\;\;
&{\rm the \; ground \; state\; is
				  \;\piuuno\; ;}\cr
 {\rm for }\;\; h<0\; {\rm and }\; h<\lambda\;\;\;\;\;\;  &{\rm the \; ground
\;
state\; is
				  \; \menouno\; ;}\cr
 {\rm for }\;\; \lambda<0\; {\rm and }\;
\lambda<h<-\lambda\;\;\;\;\;\; &{\rm the \;
ground \; state
						\;is \; \zero\; ;}\cr}$$
for $h=0,\l > 0:\; \piuuno,\menouno$ coexist.
For $h=\l <0:\; \menouno,\zero$ coexist.
For $h=-\l >0:\; \piuuno,\zero$ coexist.
These results are summarized  in Fig.2.1 where the coexistence
lines are shown.
\par\noindent
\midinsert
\vskip 7 truecm
\par\noindent
\includegraphics{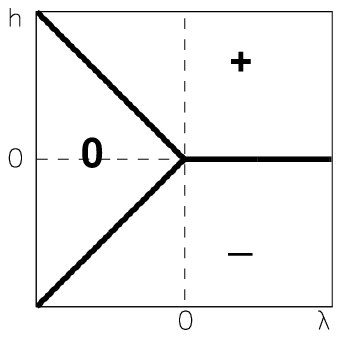}
\par\noindent
\vskip 1 truecm
\centerline {{\smbfb Fig.2.1 }{\smb Ground states for the Blume-Capel model.}}
\endinsert
\par
We want now to introduce a dynamics in our model. It will be a discrete time
Glauber dynamics namely a Markov chain with state space given by
$\O_{\L}$, where \par
1) the
allowed transitions are between {\it nearest neighbour configurations } namely
pairs $\x$ and $\eta$ of
configurations differing only in one spin: $\x = \eta^{x, b}$, with
$$\eta^{x,b}(y):=\left\{ \eqalign {\eta(y&)\;\;\;\;\;\;\;\forall y\in\L\;,\;
                                                                 y\not=x\cr
			               b\;&\;\;\;\;\;\;\;\; {\rm for }\;\; y=x}
\right. \;\; ,\Eq (2.3)$$
with $b\in \{ -1,0,+1\}$.
\par
2) It is {\it reversible} w.r.t. the Gibbs measure $\m_{\L}$ for the
Blume-Capel
model; namely the transition probabilities $P(\s, \s')$ of the chain satisfy:
$$
\m_{\L} (\s) P(\s, \s')\; = \; \m_{\L} (\s') P(\s', \s)\;\; . \Eq (2.4)
$$
Our choice is the so called {Metropolis algorithm} where the transition
probabilities, for pairs of different configurations $\s, \eta$, are defined as
$$P(\sigma ,\eta):=\left\{ \matrix {
{1\over 2|\L|} e^{-\beta [H(\eta)-H(\sigma)]^{+}}&\sigma ,\eta\;
						{\rm nearest \; neighbours}\cr
0&{\rm otherwise}\cr}\right.\;\; ,\Eq (2.5)$$
where:
$$a^{+}:=\left\{ \matrix {
a&{\rm if}\;a\geq 0\cr
0&{\rm if}\;a<0\cr}\right. \;\;\forall a\in {\bf R}\;\; .\Eq (2.6)$$
\par
The {\it space of trajectories} of the process is
$$\Xi :=\bigl(\O_{\L} \bigr)^{{\bf N}}\;\; .$$
An element in $\Xi$ is denoted by $\omega$; it is a function
$$\omega:{\bf N} \to \O_{\L}.$$
We often write $\omega=\sigma_0,\sigma_1,\dots,\sigma_t,\dots$.
\par
We will call {\it path} an {\it allowed} trajectory namely:
$\omega=\sigma_0,\sigma_1,\dots,\sigma_t,\dots$ is a path iff
$\sigma_j$ and $\sigma_{j+1}\; \forall j$ are {\it connected} in the sense that
$\sigma_{j+1} = \sigma_j^{ x,b}$ for some $x\in\L$ and $b\in\{ -1,0,+1\}$.
We use the notation  $\o :\s \rightarrow \h$ to denote a path $\o$ joining
$\s$ to $\h$.
\par
A path $\o=\s_0,\s_1,\dots ,\s_n$ is called {\it downhill}
({\it uphill}) iff $H(\s_{j+1})\le H(\s_j)$ ($H(\s_{j+1})\ge H(\s_j)$)
$\forall j=0,1,\dots ,n-1$. We will use the convention that a
downhill path can (and will) end only in a local minimum.
\par
A set $Q$ of configurations, $Q \in\O_{\L}$, is said to be
{\it connected} iff for every pair of configurations $\s,\h\in Q$, $\exists$ a
path $\o :\s \rightarrow\h$ such that $\o\subset Q$.
\par
We say that a configuration $\s$ is {\it downhill connected} to $\h$ iff there
exists a downhill path $\o :\s\rightarrow\h$.
\par
We will denote by $M$ the set of all the locally stable configurations namely
the
set of all the local minima of the energy. More precisely:  $\s\in M$ iff for
every  $x\in\L ,\; b\in \{ -1,0,+1\}$ the corresponding increment in energy,
given by
$$\D _{x,b} H(\s):= H(\s^{ x,b}) - H(\s) \Eq (2.6')$$
is positive.
\par
It is easy to see that in our model with the choice of the paremeters $J,h,\l$
that we have made, the quantity $\D _{x,b} H(\s)$ will be always non-zero and
this somehow simplifies some arguments.
\par
Given $Q\subset \O_{\L}$
we define the (outer) boundary $\partial Q$ of $Q$ as the set:
$$
\partial Q := \{\s \not\in Q : \exists  \s' \in Q : P(\s',\s) >0\}\;\; ,
$$
namely
$$  \partial Q := \{\s \not\in Q :\exists x\in \L , \; b \in \{ -1, 0, +1 \}
\;\hbox {  such that}\; \s' = \s^{x,b}\in Q\}\;\; .\Eq (2.6'')
$$
We denote by $U=U(Q)$ the set of all the minima of the energy
on the boundary $\partial Q $ of $Q$:
$$
U(Q) := \{ \z \in \partial Q : \min _{ \s \in \partial Q } H(\s) = H(\z) \}
\Eq(2.6''') $$
and we define $H(U(Q)):=H(\xi)$ with $\xi\in U(Q)$.
\par
We denote by $F=F(Q)$ the set of all minima of the energy
on $ Q $:
$$
F(Q) := \{ \z\in  Q : \min _{ \s\in Q } H(\s) = H(\z) \}\;\; . \Eq(2.6 '''')
$$
Given a stable state $\s\in M$ i.e. a  local minimum for the energy,
 we define the following {\it basins} for $\s$:\par
\noindent
i) the {\it wide basin of attraction of } $\s$  :
$$
\hat B (\s) := \{ \z : \exists \; \hbox {downhill path } \o : \z \to \s\}\;\; ;
\Eq(2.6a)
$$
\par\noindent
ii) the   {\it basin of attraction
of } $\s$  given by:
$$
 B(\s) := \{ \z :  \hbox {every downhill path starting from }\z \;  \hbox {ends
 in} \; \s\}\;\; , \Eq(2.6b) $$
$B(\s) $ can be seen as the usual basin of attraction of
$\s$
with respect  to the $\b = \infty$ dynamics.
\par
\noindent
iii)  $\bar B(\s) =$ the {\it strict basin or attraction of } $\s$ given by
:\par
$$
\bar B(\s) := \{ \z \in B(\s) \; : \; H(\z) < H(U( B(\s)))\}\;\; . \Eq(2.6c)
$$
\par
We introduce now the useful the notion of cycle. A connected set $A$ which
satisfies:
$$
\max _ {\s\in A} H(\s) = \bar H < \min _ {\z \in \partial A } H(\z) = H(U(A))
$$
is called {\it cycle}. Notice that every local minimum for the energy is a
(trivial) cycle.
\par
The following simple properties of the cycles are true. Their proof
is immediate  (see, for instance [OS1]).
\par\noindent
\item{1.} Given a state $\bar \s \in \O_{\L} $
and a real number $c$ the set of all $\s$'s
connected to $\bar\s $ by paths with energy always below $c$ either
coincides with $\O_{\L}$ or it is a cycle $A$ with
$$
H(U(A)) \geq c\;\; .
$$
\item{2.} Given two cycles $A_1,\; A_2$, either
i) $ A_1 \cap A_2 = \emptyset$
or ii) $A_1 \subset A_2 $ or, viceversa, $A_2 \subset A_1 $ .
\par
We give now some more definitions: a cycle $A$ for which there exists
$\h^*\in U(A)$  downhill connected to some
point $\s$ in $A^c$, is called {\it transient}; given a transient cycle $A$
the points $\h^*$ downhill connected
to  $A^c$ are called {\it minimal saddles}. The set of all the minimal saddles
of a transient cycle A is denoted by ${\cal S}(A)$.
\par
A transient cycle $A$ such that $ \exists \; \bar \s \not \in A $ with
$H(\bar \s) < H(F(A))$, there exists $\h^* \in {\cal S}(A)$ and a path
$ \o : \h^* \to \bar \s$ {\it below } $\h^*$ (namely $\forall\s \in \o :
H(\s) < H(\h^*)$), is called {\it metastable}.
\par
For each pair of states $\s,\h\in \O_{\L}$  we define their minimal saddle
$ {\cal S}(x,y)$ as the set of states corresponding to the
solution of the following minimax problem: let, for any path $\o$
$$
\hat H(\o)\;  := \max _{ \z\in \o} H(\z), \;\; \;\;\;\;\;\;\bar H_{\s,\h}
\;:=\;\min_{\o : \s \to \h} \hat H (\o)\;\; ,
$$
find
$$
{\cal S}(\s,\h):= \{ \z:\;  H(\z) = \bar H_{\s,\h} ; \;
\exists  \;\o : \;\s \to \h ,\; \o
\ni \z,\;\hat H(\o) = \bar H_{\s,\h} \}\;\; .
$$
One immediately verifies that a strict basin of attraction of a local minimum
is a
transient cycle. This case corresponds to a ``one well" structure.
More general cases involve the presence of ``internal saddles" and correspond
to a
``several wells" situation.
\par
Given any set of configurations $A\subset \O_{\L}$, we use  $\tau_A$ to denote
the {\it first hitting time} to $A$:
$$
\tau_A :=\inf\{t\geq 0 \;: \sigma_t\in A\}\;\; .
\Eq (2.7)
$$
We use $P_\eta(\cdot)$ to denote the probability distribution
over the process starting at $t=0$ from the configuration $\eta$.\par
We are interested in dynamics at very low temperatures. Namely,
we will discuss the asymptotic behaviour, in the limit $\beta\to\infty$,
of typical paths of the first escape from $-\underline 1$ to
$+\underline 1$ for fixed $h, \; \l $ and $\L$.
\par
Let us now better clarify the asymptotic regime in which we will operate:
the volume $|\L|$, the magnetic field $h$ and the chemical
potential $\l$ are fixed  and we
consider  asymptotic estimates for $\b$ very large. This regime was studied in
the
case of standard Ising model in 2D by J. Neves and R. Schonmann in [NS1] where
the
point of view of the {\it pathwise approach to metastability}, introduced in
[CGOV], was assumed.
\par
One can think, for instance, to take $\l$ very small,
$h=a\l$, $a$ fixed positive number,  $|\L|$ of order,
say, of ${1\over  h^2}$ and $\b$ of order $ 1\over  h^5$; physically this
corresponds to a regime in which, at the equilibrium, the energy dominates
w.r.t. the
entropy.\par
In the above described situation the qualitative behaviour of our stochastic
time
evolution can be described as follows: the system will spend the majority of
the
time in the local minima of the energy. Sometimes it escapes from them but
there is
always a natural  tendency to follow a downhill path and an occasional, random
and rather unprobable, uphill move.\par
An important role will be played by the  saddles separating different ``basins
of attraction'' (or generalized basins of attraction, see below) w.r.t.
the $\b =\infty$ dynamics.
\par
We will see that the local minima will correspond to  particular geometric
shapes
that will be called {\it plurirectangles} (see Fig.3.8); we will analyze, in
particular, the special saddles between ``contiguous" local minima
(see Lemma 5.1).
\par
A {\it global saddle point} is any configuration
$$\bar\sigma \; \in {\cal S}(- \underline 1,+ \underline 1)\;\; .$$
In Section 6 we will see that the set of  these global minima ${\cal P}$ are
substantially different according to the values of the parameters $\l$ and
$h$.
\par\noindent
1) For $ h < 2\l$ we distinguish two cases:\par
a) if $\d := l^* - { 2J -(h-\l)\over h} < { h +\l \over 2h}$,
${\cal P}$ is of the form ${\cal P}_{1,a}$
given in Fig.5.1 (the two critical lengths $l^*$ and $M^*$ are defined in
$(3.12)$ and $(3.15)$); namely it contains a ``droplet"
with external rectangle given by a square of side $l^*+2$; the
internal shape given by a rectangle with sides $ l^*,\; l^* -1$, at a distance
one from the external rectangle and with a unit square protuberance attached
to the longest ``free" side; the internal shape is full of
pluses, the spins lying outside to the exterior rectangle are minuses; finally
between the interior shape and the external rectangle there are zeroes.\par
b) If $\d  > { h +\l \over 2h}$,
${\cal P}$ is of the form ${\cal P}_{1,b }$ depicted in Fig.5.1.
${\cal P}_{1,b }$  is similar to ${\cal P}_{1,a }$ but now the external
rectangle
has sides $l^*+1,\; l^* +2$ and internally we have a square with sides
$ l^*-1$ with a unit square protuberance attached to the
shortest ``free" side.
\par \noindent
2) For $ h > 2\l$, ${\cal P}$ is  of the
form ${\cal P}_2$
given in Fig.5.1; namely it is given by a rectangle of sides $M^*$ and
$M^*-1$, with a unit square protuberance attached to one of its longest sides,
full of zeroes in a ``sea" of minuses.
\par
We set:
$$\G \; := \; H ( {\cal P}) - H ( \menouno)\;\; . \Eq (2.10)$$
Let us now summarize our main results.\par
We shall prove that the first excursion from $-\underline 1$
to $+\underline 1$ typically passes through a configuration from ${\cal P}$
and the time needed for this to happen is of the order $\exp(\beta\Gamma)$;
this
is the content of Theorem 1  that we are now going to state.
\par
Theorem 2  will characterize the typical trajectories of the first excursion.
The proof of Theorems 1,2 and even the statement of Theorem 2 will
need many more definitions and propositions. For this reasons they will be
postponed to Section 7.
\par
Theorem 1 is based, in particular,  on Propositions 4.1, 4.2 and 4.3 given in
Section 4. These propositions refer to the tendency of a given
minimum $\eta$ of the energy to
evolve towards $ \piuuno$ or to $\menouno$ namely they establish under which
conditions a {\it droplet} is {\it supercritical} or not.\par
It will be crucial to introduce a sort of generalized basin of
attraction of $\menouno$. Indeed we will
reduce the proof of Theorem 1  to finding a certain set ${\cal G}$ of
configurations satisfying suitable properties. In order to explicitly construct
this
set ${\cal G}$ we will need the results contained in Propositions 4.1, 4.2 and
4.3. This construction will be achieved in Section 6.
\vskip 0.35 truecm
\noindent
{\bf Theorem 1.}\par\noindent
Let $\bar\tau_{-\underline 1}$ be the last instant in which $\sigma_t=
-\underline 1$ before $\tau_{+\underline 1}$:
$$
\bar\tau _{-\underline 1} :=\max \{t<\tau_{+\underline 1}:
\sigma_t=-\underline 1\}\;\; .\Eq (2.11)
$$
Let
$$ \bar\tau_{\cal P} :=\min\{t> \bar\tau_{-\underline 1}
: \sigma_t = \cal P\}\;\; ; \Eq (2.12)
$$
for every  $\varepsilon > 0$:
\par\noindent
\vskip 0.35 truecm
i)
$$
\lim_{\beta\to\infty}P_{-\underline 1}(\bar\tau_{\cal P}
<\tau_{+\underline 1})=1\;\; ; \Eq (2.13)
$$
\par
ii)
$$
\lim_{\beta\to\infty}P_{-\underline 1}(\exp[\beta(\Gamma-\varepsilon)]
<\tau_{+\underline 1}<\exp[\beta(\Gamma+\varepsilon)])=1\;\; .\Eq (2.14)
$$

\vfill\eject
\numsec=3\numfor=1

{\bf Section 3. Local minima of the hamiltonian $H(\s)$.}
\par
In this section we want to analyze the geometrical structure of the local
minima of the energy.
\par
For any configuration $\s\in\O_\L$ we denote by $c^{+}(\s)$, $c^{-}(\s)$ and
$c^{0}(\s)$ the union of all closed unit squares centered
at sites $x\in\L$ with $\s(x)$ respectivly equal to $+1,-1$ and $0$.
$c^{+}(\s)$, $c^{-}(\s)$ and
$c^{0}(\s)$ decompose into maximal connected components
$c^{+,0,-}_j, \; j=1, \dots , k^{+,0,-}$.\par
The centers of $c^{+,0,-}_j$ form a $\star$--cluster in the sense of sites
percolation, namely they are maximally connected components in the sense of the
next  nearest neighbours. The $c^{+,0,-}_j$ will be simply called {\it
clusters}.
\par
To any such $c^{+,0,-}_j$ we assign its {\it rectangular envelope} defined
as the minimal closed rectangle  $R(c^{\pm ,0}_{j})$ containing it; if none of
the rectangles $R(c^{+,0}_{j})$ is winding around the torus, we call the
corresponding configuration {\it acceptable}.
\par
Let $\s$ be an acceptable configuration, we denote by $\g^{+,0}_{j}$
the boundary of $c^{+,0}_{j}\;\;\forall j\in\{1,...,k^{+,0}\}$; the
internal component ${\check\g^{+,0}_{j}}$ of the boundary is defined as
follows: let {\it s} be a unit segment of the dual lattice
${\bf Z}^{2}+({1\over 2},{1\over 2})$ belonging to $\g^{+,0}_{j}$, we say that
$s\in{\check\g^{+,0}_{j}}$ if and only if all the paths joining nearest
neighbour sites of $\L$ and starting from the site adjacent to {\it s} and
not in $c^{+,0}_{j}$, necessarily reach a site in $c^{+,0}_{j}$ before
touching the cluster $c^{-}_{j}$ winding around the torus.
The external component ${\hat\g^{+,0}_{j}}$ of the boundary of
$c^{+,0}_{j}$ is defined as $\g^{+,0}_{j}\setminus{\check\g^{+,0}_{j}}$.
Of course ${\check\g^{+,0}_{j}}$ can be empty.
\midinsert
\vskip 0.5 truecm
\vbox{\font\amgr=cmr10 at
10truept\baselineskip0.1466667truein\lineskiplimit-\maxdimen
\catcode`\-=\active\catcode`\~=\active\def~{{\char32}}\def-{{\char1}}%
\hbox{\amgr ~~~~~~~~~~~~~0~~~$\phantom {\left\{\matrix {0&
{\char45}6J{\char45}(h{\char45}\l)\cr +&
{\char45}4J{\char45}2h\cr}\right.}$~~~~~~~~~~~~~~~~0}
\hbox{\amgr ~~~~~~~~~~~+~{\char45}~0~~~$\left\{\matrix {0&
{\char45}6J{\char45}(h{\char45}\l)\cr
+&
{\char45}4J{\char45}2h\cr}\right.$~~~~~~~~~~~~+~{\char45}~0~~~$\left\{\matrix
{0& {\char45}8J{\char45}(h{\char45}\l)\cr
+& {\char45}8J{\char45}2h\cr}\right.$}
\hbox{\amgr ~~~~~~~~~~~~~0~~~$\phantom {\left\{\matrix {0&
{\char45}6J{\char45}(h{\char45}\l)\cr +&
{\char45}4J{\char45}2h\cr}\right.}$~~~~~~~~~~~~~~~~+}
\hbox{\amgr }
\hbox{\amgr }
\hbox{\amgr ~~~~~~~~~~~~~0~~~$\phantom {\left\{\matrix {0&
{\char45}4J{\char45}(h{\char45}\l)\cr +&
{\char45}3J{\char45}2h\cr}\right.}$~~~~~~~~~~~~~~~~0}
\hbox{\amgr ~~~~~~~~~~~+~{\char45}~0~~~$\left\{\matrix {0&
{\char45}4J{\char45}(h{\char45}\l)\cr
+&
{\char45}3J{\char45}2h\cr}\right.$~~~~~~~~~~~~+~{\char45}~+~~~$\left\{\matrix
{0& {\char45}10J{\char45}(h{\char45}\l)\cr
+& {\char45}12J{\char45}2h\cr}\right.$}
\hbox{\amgr ~~~~~~~~~~~~~{\char45}~~~$\phantom {\left\{\matrix {0&
{\char45}4J{\char45}(h{\char45}\l)\cr +&
{\char45}3J{\char45}2h\cr}\right.}$~~~~~~~~~~~~~~~~+}
\hbox{\amgr }
\hbox{\amgr }
\hbox{\amgr ~~~~~~~~~~~~~0~~~$\phantom {\left\{\matrix {0&
{\char45}6J{\char45}(h{\char45}\l)\cr +&
{\char45}4J{\char45}2h\cr}\right.}$~~~~~~~~~~~~~~~~0}
\hbox{\amgr ~~~~~~~~~~~+~{\char45}~+~~~$\left\{\matrix {0&
{\char45}6J{\char45}(h{\char45}\l)\cr
+&
{\char45}4J{\char45}2h\cr}\right.$~~~~~~~~~~~~+~{\char45}~{\char45}~~~$\left\{\matrix {0& {\char45}2J{\char45}(h{\char45}\l)\cr
+& +4J{\char45}2h\cr}\right.$}
\hbox{\amgr ~~~~~~~~~~~~~{\char45}~~~$\phantom {\left\{\matrix {0&
{\char45}6J{\char45}(h{\char45}\l)\cr +&
{\char45}4J{\char45}2h\cr}\right.}$~~~~~~~~~~~~~~~~{\char45}}
\hbox{\amgr }
\hbox{\amgr }
\hbox{\amgr ~~~~~~~~~~~~~+~~~$\phantom {\left\{\matrix {0&
{\char45}12J{\char45}(h{\char45}\l)\cr +&
{\char45}16J{\char45}2h\cr}\right.}$~~~~~~~~~~~~~~~+}
\hbox{\amgr ~~~~~~~~~~~+~{\char45}~+~~~$\left\{\matrix {0&
{\char45}12J{\char45}(h{\char45}\l)\cr
+&
{\char45}16J{\char45}2h\cr}\right.$~~~~~~~~~~~+~{\char45}~+~~~$\left\{\matrix
{0& {\char45}8J{\char45}(h{\char45}\l)\cr
+& {\char45}8J{\char45}2h\cr}\right.$}
\hbox{\amgr ~~~~~~~~~~~~~+~~~$\phantom {\left\{\matrix {0&
{\char45}12J{\char45}(h{\char45}\l)\cr +&
{\char45}16J{\char45}2h\cr}\right.}$~~~~~~~~~~~~~~~{\char45}}
\hbox{\amgr }
\hbox{\amgr }
\hbox{\amgr ~~~~~~~~~~~~~+~~~$\phantom {\left\{\matrix {0&
{\char45}4J{\char45}(h{\char45}\l)\cr +&
{\char45}2h\cr}\right.}$~~~~~~~~~~~~~~~~{\char45}}
\hbox{\amgr ~~~~~~~~~~~+~{\char45}~{\char45}~~~$\left\{\matrix {0&
{\char45}4J{\char45}(h{\char45}\l)\cr
+& {\char45}2h\cr}\right.$~~~~~~~~~~~~+~{\char45}~{\char45}~~~$\left\{\matrix
{0& {\char45}(h{\char45}\l)\cr
+& +8J{\char45}2h\cr}\right.$}
\hbox{\amgr ~~~~~~~~~~~~~{\char45}~~~$\phantom {\left\{\matrix {0&
{\char45}4J{\char45}(h{\char45}\l)\cr +&
{\char45}2h\cr}\right.}$~~~~~~~~~~~~~~~~{\char45}}}
\vskip 0.35 truecm
\par\noindent
\centerline {\smbfb Fig.3.1}
\endinsert
\par
In order to construct the local minima of the hamiltonian we first prove
that direct $+-$ interfaces cannot exist in such configurations; in Fig.3.1
we analyze the interaction of a minus spin with its neighbouring sites. We
examine all the possible cases and we show that it is always possible to
construct a lower energy configuration by changing the minus spin adjacent
to the interface.
\par
Let $\s$ be an acceptable configuration such that there exists only one
cluster of $0$ spins $c^0$ and no plus spins; it can be proved that
$$\sigma\;{\rm is\; a\; local\; minimum\; of\;}H(\s)\;\Longleftrightarrow\;
\left\{\eqalign {
\g^{0}&={\hat\g^{0}}\;{\rm is\; a\; rectangle\; whose}\cr
{\rm si}&{\rm des\; are\; longer\; than\; two}\cr}\right.\;\; .\Eq (3.1)$$
Indeed, if $\s$ is a local minimum and there exists a minus spin inside
the cluster $c^0$, then, as a consequence of the fact that $c^0$ does not wind
around the torus, one has that necessarily it must exist at least one minus
spin with at least two nearest neighbour sites occupied by $0$ spins
(see Fig.3.2).
\vskip 0.5 truecm
\vbox{\font\amgr=cmr10 at
10truept\baselineskip0.1466667truein\lineskiplimit-\maxdimen
\catcode`\-=\active\catcode`\~=\active\def~{{\char32}}\def-{{\char1}}%
\hbox{\amgr ~~~~~~~~~~~~~~~~~~~~~~~~~~~~~~~~~~~{\char2}-----{\char3}}
\hbox{\amgr
{}~~~~~~~~~~~~~~~~~~~~~~~~~~~~~~~~~~~{\char0}~~0~~{\char0}~~~{\char2}--{\char3}}
\hbox{\amgr
{}~~~~~~~~~~~~~~~~~~~~~~~~~~~~~~~~~--{\char4}~~~~~{\char0}~~~{\char0}~~{\char5}--{\char3}}
\hbox{\amgr
{}~~~~~~~~~~~~~~~~~~~~~~~~~~~~~~~~~~~~~~~~~{\char5}---{\char4}~~0~~{\char0}}
\hbox{\amgr
{}~~~~~~~~~~~~~~~~~~~~~~~~~~~~~~~{\char45}~{\char45}~{\char45}~{\char45}~---------{\char3}~~{\char5}-{\char3}}
\hbox{\amgr
{}~~~~~~~~~~~~~~~~~~~~~~~~~~~~~~~~~~~~~~~~~~~~~~-~{\char0}~0~~{\char0}}}
\vskip 0.35 truecm
\par\noindent
\centerline {\smbfb Fig.3.2}
\par\noindent
This minus spin can be changed into $+$ or $0$ by obtaining, in this way,
a lower energy configuration, as shown in Fig.3.3; this is an absurd.
\vskip 0.5 truecm
\vbox{\font\amgr=cmr10 at
10truept\baselineskip0.1466667truein\lineskiplimit-\maxdimen
\catcode`\-=\active\catcode`\~=\active\def~{{\char32}}\def-{{\char1}}%
\hbox{\amgr ~~~~~~~~~~~~~0~~~$\phantom {\left\{\matrix {0&
{\char45}4J{\char45}(h{\char45}\l)\cr +&
{\char45}2h\cr}\right.}$~~~~~~~~~~~~~~~~0}
\hbox{\amgr ~~~~~~~~~~~0~{\char45}~0~~~$\left\{\matrix {0&
{\char45}4J{\char45}(h{\char45}\l)\cr +&
{\char45}2h\cr}\right.$~~~~~~~~~~~~{\char45}~{\char45}~0~~~$\left\{\matrix {0&
{\char45}(h{\char45}\l)\cr +&  +8J{\char45}2h\cr}\right.$}
\hbox{\amgr ~~~~~~~~~~~~~0~~~$\phantom {\left\{\matrix {0&
{\char45}4J{\char45}(h{\char45}\l)\cr +&
{\char45}2h\cr}\right.}$~~~~~~~~~~~~~~~~+}
\hbox{\amgr }
\hbox{\amgr }
\hbox{\amgr ~~~~~~~~~~~~~~~~~~~~~~~~~~~~0}
\hbox{\amgr ~~~~~~~~~~~~~~~~~~~~~~~~~~{\char45}~{\char45}~0~~~$\left\{\matrix
{0& {\char45}2J{\char45}(h{\char45}\l)\cr +& +4J{\char45}2h\cr}\right.$}
\hbox{\amgr ~~~~~~~~~~~~~~~~~~~~~~~~~~~~0}}
\vskip 0.35 truecm
\par\noindent
\centerline {\smbfb Fig.3.3}
\par\noindent
We can conclude that no minus spins can be inside $c^0$, that is
${\check\g^0}=\{\emptyset\}$. In a similar way it can be proved that
${\hat\g^0}$ is a rectangle and its sides are longer than two.
\par
The proof of the implication $\Leftarrow$ is in Fig.3.4, where it is shown
that all the possible nearest neighbour configurations of $\s$ are at
higher energy; in Fig.3.4 the modified spin is represented by a unit
empty square.
\midinsert
\vskip 0.5 truecm
\vbox{\font\amgr=cmr10 at
10truept\baselineskip0.1466667truein\lineskiplimit-\maxdimen
\catcode`\-=\active\catcode`\~=\active\def~{{\char32}}\def-{{\char1}}%
\hbox{\amgr ~~~~~~~~{\char2}------------------------{\char3}}
\hbox{\amgr ~~~~~~~~{\char0}~~~~~~~~~~~~~~~~~~~~~~~~{\char0}~~~-}
\hbox{\amgr
{}~~~~~~~~{\char0}~~~~~~~~~~0~~~~~~~~~~~~~{\char0}~{\char2}-{\char3}~~~~~~~~~~$\left\{ \matrix {
0& +4J{\char45}(h{\char45}\l)\cr
+& +16J{\char45}2h\cr}\right.$}
\hbox{\amgr ~~~~~~~~{\char0}~~~~~~~~~~~~~~~~~~~~~~~~{\char0}~{\char5}-{\char4}}
\hbox{\amgr ~~~~~~~~{\char0}~~~~~~~~~~~~~~~~~~~~~~~~{\char0}}
\hbox{\amgr ~~~~~~~~{\char5}------------------------{\char4}}
\hbox{\amgr }
\hbox{\amgr }
\hbox{\amgr ~~~~~~~~{\char2}------------------------{\char3}}
\hbox{\amgr ~~~~~~~~{\char0}~~~~~~~~~~~~~~~~~~~~~~~~{\char0}~}
\hbox{\amgr
{}~~~~~~~~{\char0}~~~~~~~~~~~~~~~~~~~~~~~~{\char15}-{\char3}~~~~~~~~~~~~$\left\{ \matrix {
0& +2J{\char45}(h{\char45}\l)\cr
+& +12J{\char45}2h\cr}\right.$}
\hbox{\amgr ~~~~~~~~{\char0}~~~~~~~~~~~~~~~~~~~~~~~~{\char15}-{\char4}}
\hbox{\amgr ~~~~~~~~{\char0}~~~~~~~~~~~~~~~~~~~~~~~~{\char0}}
\hbox{\amgr ~~~~~~~~{\char5}------------------------{\char4}}
\hbox{\amgr }
\hbox{\amgr }
\hbox{\amgr ~~~~~~~~{\char2}----------------------{\char18}-{\char3}}
\hbox{\amgr ~~~~~~~~{\char0}~~~~~~~~~~~~~~~~~~~~~~{\char5}-{\char16}~~~~~~~~}
\hbox{\amgr
{}~~~~~~~~{\char0}~~~~~~~~~~~~~~~~~~~~~~~~{\char0}~~~~~~~~~~~~~~~$\left\{
\matrix {
{\char45}& +(h{\char45}\l)\cr
+& +8J{\char45}(h+\l)\cr}\right.$}
\hbox{\amgr ~~~~~~~~{\char0}~~~~~~~~~~~~~~~~~~~~~~~~{\char0}}
\hbox{\amgr ~~~~~~~~{\char0}~~~~~~~~~~~~~~~~~~~~~~~~{\char0}}
\hbox{\amgr ~~~~~~~~{\char5}------------------------{\char4}}
\hbox{\amgr ~}
\hbox{\amgr }
\hbox{\amgr ~~~~~~~~{\char2}------------------------{\char3}}
\hbox{\amgr ~~~~~~~~{\char0}~~~~~~~~~~~~~~~~~~~~~~~~{\char0}~}
\hbox{\amgr
{}~~~~~~~~{\char0}~~~~~~~~~~~~~~~~~~~~~~{\char2}-{\char16}~~~~~~~~~~~~~~~$\left\{ \matrix {
{\char45}& +2J+(h{\char45}\l)\cr
+& +6J{\char45}(h+\l)\cr}\right.$}
\hbox{\amgr ~~~~~~~~{\char0}~~~~~~~~~~~~~~~~~~~~~~{\char5}-{\char16}}
\hbox{\amgr ~~~~~~~~{\char0}~~~~~~~~~~~~~~~~~~~~~~~~{\char0}}
\hbox{\amgr ~~~~~~~~{\char5}------------------------{\char4}}
\hbox{\amgr }
\hbox{\amgr }
\hbox{\amgr ~~~~~~~~{\char2}------------------------{\char3}}
\hbox{\amgr ~~~~~~~~{\char0}~~~~~~~~~~~~~~~~~~~~~~~~{\char0}~}
\hbox{\amgr
{}~~~~~~~~{\char0}~~~~~~~~~~~~~~~~~~~~{\char2}-{\char3}~{\char0}~~~~~~~~~~~~~~~$\left\{ \matrix {
{\char45}& +4J+(h{\char45}\l)\cr
+& +4J{\char45}(h+\l)\cr}\right.$}
\hbox{\amgr ~~~~~~~~{\char0}~~~~~~~~~~~~~~~~~~~~{\char5}-{\char4}~{\char0}}
\hbox{\amgr ~~~~~~~~{\char0}~~~~~~~~~~~~~~~~~~~~~~~~{\char0}}
\hbox{\amgr ~~~~~~~~{\char5}------------------------{\char4}}}
\vskip 0.35 truecm
\par\noindent
\centerline {\smbfb Fig.3.4}
\endinsert
\par
Now let $\s$ be an acceptable configuration such that the following
conditions are satisfied:
there exists just one cluster $c^0$ of $0$ spins touching $c^-$,
${\hat\g^0}$ is a rectangle, no minus spin is inside clusters of $0$ spins;
all plus spins are in the cluster $c^+$ and ${\hat\g^+}={\check\g^0}$
(see Fig.3.5).
\midinsert
\vskip 0.5 truecm
\vbox{\font\amgr=cmr10 at
10truept\baselineskip0.1466667truein\lineskiplimit-\maxdimen
\catcode`\-=\active\catcode`\~=\active\def~{{\char32}}\def-{{\char1}}%
\hbox{\amgr
{}~~~~~~~~~~~~~~~~~~~~~~{\char2}---------------------------------------{\char3}}
\hbox{\amgr
{}~~~~~~~~~~~~~~~~~~~~~~{\char0}~~~~~~~~~~~~~~~~~~~~~~~~~~~~~~~~~~0~~~~{\char0}}
\hbox{\amgr
{}~~~~~~~~~~~~~~~~~~~~~~{\char0}~~~~{\char2}-{\char3}~~~0~~{\char2}-----{\char3}~~~~~~{\char2}---{\char3}~~~~~~~~{\char0}%
}
\hbox{\amgr
{}~~~~~~~~~~~~~~~~~~~~~~{\char0}~~~~{\char0}~{\char5}------{\char4}~~+~~{\char5}-{\char3}~~~~{\char0}~~~{\char5}-----%
{\char3}~~{\char0}}
\hbox{\amgr
{}~~~~~~~~~~~~~~~~~~~~~~{\char0}~~~~{\char0}~~~~~~~~~~~~~~~~{\char5}----{\char4}~~~~~+~~~{\char0}~~{\char0}}
\hbox{\amgr
{}~~~~~~~~~~~~~~~~~~~~~~{\char0}~~~~{\char0}~~~~~{\char2}----{\char3}~~~~~~~~~~~~~{\char2}----{\char3}~{\char0}~~{\char0}%
}
\hbox{\amgr
{}~~~~~~~~~~~~~~~~~~~~~~{\char0}~~{\char2}-{\char4}~~~{\char2}-{\char4}~~~~{\char5}----{\char3}~~~{\char2}----{\char4}%
{}~~~~{\char0}~{\char0}~~{\char0}}
\hbox{\amgr
{}~~~~~~~~~~~~~~~~~~~~~~{\char0}~~{\char0}~~~~~{\char0}~~~~0~~~~~~{\char0}~~~{\char5}-{\char3}~~~~{\char45}~~{\char0}~%
{\char0}~~{\char0}}
\hbox{\amgr
{}~~~~~~~~~~~~~~~~~~~~~~{\char0}~~{\char5}---{\char3}~{\char5}---{\char3}~~~0~~~{\char0}~~~~~{\char0}~{\char45}~~~{\char2}%
-{\char4}~{\char0}~~{\char0}}
\hbox{\amgr
{}~~~~~~~~~~~~~~~~~~~~~~{\char0}~~~~~~{\char0}~~~~~{\char0}~~~~~{\char2}-{\char4}~~~~~{\char5}-----{\char4}~~~{\char0}%
{}~~{\char0}}
\hbox{\amgr
{}~~~~~~~~~~~~~~~~~~~~~~{\char0}~~~~~~{\char5}--{\char3}~~{\char5}-{\char3}~~~{\char0}~~+~~~~~~~~+~~~{\char2}-{\char4}%
{}~~{\char0}}
\hbox{\amgr
{}~~~~~~~~~~~~~~~~~~~~~~{\char0}~~~~~~~~~{\char0}~~~~{\char5}---{\char4}~~~~{\char2}----{\char3}~~~{\char2}-{\char4}~%
{}~~~{\char0}}
\hbox{\amgr
{}~~~~~~~~~~~~~~~~~~~~~~{\char0}~~~0~~~~~{\char0}~~~~~~~~~~~~~{\char0}~~~~{\char5}---{\char4}~~~~~~{\char0}}
\hbox{\amgr
{}~~~~~~~~~~~~~~~~~~~~~~{\char0}~~~~~~~~~{\char5}----{\char3}~~~~~~{\char2}-{\char4}~~~~~~~~~~0~~~~{\char0}}
\hbox{\amgr
{}~~~~~~~~~~~~~~~~~~~~~~{\char0}~~~~~~~~~~~~~~{\char5}------{\char4}~~~~~~~~~~~~~~~~~{\char0}}
\hbox{\amgr
{}~~~~~~~~~~~~~~~~~~~~~~{\char0}~~~~~~~~~~~~~~~~~~~~~~~~~~~~~~~~~~~~~~~{\char0}}
\hbox{\amgr
{}~~~~~~~~~~~~~~~~~~~~~~{\char5}---------------------------------------{\char4}}}
\vskip 0.35 truecm
\par\noindent
\centerline {\smbfb Fig.3.5}
\endinsert
\par\noindent
With arguments similar to the ones used before, it can be proved that
$$\sigma\;{\rm is\; a\; local\; minimum\; of\;}H(\s)\;\Longleftrightarrow\;
\left\{\eqalign {
\g^+&={\hat \g^{+}}\;{\rm is\; a\; rectangle\; whose}\cr
{\rm si}&{\rm des\; are\; longer\; than\; two}\cr}\right.\;\; .\Eq (3.2)$$
In the proof it is crucial that the energy of a configuration can be lowered
by properly changing a $0$ spin having at most two zero spins and no
minus spins among its nearest neighbour sites; all the possible situations
are shown in Fig.3.6.
\vskip 0.5 truecm
\vbox{\font\amgr=cmr10 at
10truept\baselineskip0.1466667truein\lineskiplimit-\maxdimen
\catcode`\-=\active\catcode`\~=\active\def~{{\char32}}\def-{{\char1}}%
\hbox{\amgr ~~~~~~~~~~~~~+~~~$\phantom {\left\{\matrix {0& 8J+(h{\char45}\l)\cr
+& {\char45}(h+\l)\cr}\right.}$~~~~~~~~~~~~~~~~+}
\hbox{\amgr ~~~~~~~~~~~0~0~+~~~$\left\{\matrix {{\char45}& 8J+(h{\char45}\l)\cr
+& {\char45}(h+\l)\cr}\right.$~~~~~~~~~~~~0~0~+~~~$\left\{\matrix
{{\char45}& 10J+(h{\char45}\l)\cr
+& {\char45}2J{\char45}(h+\l)\cr}\right.$}
\hbox{\amgr ~~~~~~~~~~~~~0~~~$\phantom {\left\{\matrix {{\char45}&
8J+(h{\char45}\l)\cr +& {\char45}(h+\l)\cr}\right.}$~~~~~~~~~~~~~~~~+}
\hbox{\amgr }
\hbox{\amgr }
\hbox{\amgr ~~~~~~~~~~~~~~~~~~~~~~~~~~~~+}
\hbox{\amgr ~~~~~~~~~~~~~~~~~~~~~~~~~~+~0~+~~~$\left\{\matrix {{\char45}&
12J+(h{\char45}\l)\cr +& {\char45}4J{\char45}(h+\l)\cr}\right.$}
\hbox{\amgr ~~~~~~~~~~~~~~~~~~~~~~~~~~~~+}}
\vskip 0.35 truecm
\par\noindent
\centerline {\smbfb Fig.3.6}
\par
Hence we have proved that configurations like the one in Fig.3.7 are
local minima of $H(\s)$; these configurations are called {\it birectangles}
and are denoted by the symbol $R(L_{1},L_{2};M_{1},M_{2})$ where
$$\left\{ \matrix {
M_{1}\geq L_{1}+2,M_{2}\geq L_{2}+2 &\; {\rm if}\;\; L_{1},L_{2}\geq 2\cr
M_{1},M_{2}\geq 2 &\; {\rm if}\;\; L_{1}=L_{2}=0\cr}\right.\;\; .\Eq (3.3)$$
\midinsert
\vskip 0.5 truecm
\vbox{\font\amgr=cmr10 at
10truept\baselineskip0.1466667truein\lineskiplimit-\maxdimen
\catcode`\-=\active\catcode`\~=\active\def~{{\char32}}\def-{{\char1}}%
\hbox{\amgr ~~~~~~~~~~~~~~~~~~~~~~~~~~~}
\hbox{\amgr
{}~~~~~~~~~~~~~~~~~~~~~~~~~~~{\char2}-------------------------------{\char3}}
\hbox{\amgr
{}~~~~~~~~~~~~~~~~~~~~~~~~~~~{\char0}~~~~~~~~~~~~~~~~~~~~~~~~~~~~~~~{\char0}}
\hbox{\amgr
{}~~~~~~~~~~~~~~~~~~~~~~~~~~~{\char0}~~~~~~~~~~~~~~~~~~~~~~~~~~~~~~~{\char0}}
\hbox{\amgr
{}~~~~~~~~~~~~~~~~~~~~~~~~~~~{\char0}~~~~~$L_{1}\;\,\,$~~~~~~~~~~~~~~~~~~~~~~{\char0}}
\hbox{\amgr
{}~~~~~~~~~~~~~~~~~~~~~~~~~~~{\char0}~{\char2}-------{\char3}~~~~~~~~~~~~~~~~~~~~~{\char0}}
\hbox{\amgr
{}~~~~~~~~~~~~~~~~~~~~~~~~~~~{\char0}~{\char0}~~~~~~~{\char0}~~~~~~~~~~~~~~~~~~~~~{\char0}~$M_{2}$}
\hbox{\amgr
{}~~~~~~~~~~~~~~~~~~~~~~~~~~~{\char0}~{\char0}~~~~~~~{\char0}~~~~~~~~~~~~~~~~~~~~~{\char0}}
\hbox{\amgr
{}~~~~~~~~~~~~~~~~~~~~~~~~~~~{\char0}~{\char0}~~~~~~~{\char0}~$L_{2}\;\,
\,$~~~~~~~~~~~~~~~~{\char0}}
\hbox{\amgr
{}~~~~~~~~~~~~~~~~~~~~~~~~~~~{\char0}~{\char0}~~~~~~~{\char0}~~~~~~~~~~~~~~~~~~~~~{\char0}}
\hbox{\amgr
{}~~~~~~~~~~~~~~~~~~~~~~~~~~~{\char0}~{\char5}-------{\char4}~~~~~~~~~~~~~~~~~~~~~{\char0}}
\hbox{\amgr
{}~~~~~~~~~~~~~~~~~~~~~~~~~~~{\char5}-------------------------------{\char4}}
\hbox{\amgr ~~~~~~~~~~~~~~~~~~~~~~~~~~~~~~~~~~~~~~~~~$M_{1}$}}
\vskip 0.35 truecm
\par\noindent
\centerline {\smbfb Fig.3.7}
\endinsert
\par
It is easy to understand that the most general local minimum of $H(\s)$ is
not a birectangle, but, rather, a more complicate configuration that we call
family of {\it
plurirectangle} (see Fig.3.8). It is an acceptable configuration satisfying
the following conditions:
\vskip 0.35 truecm
\itemitem{i)} there are $k^0$ clusters $c^0_1,...,c^0_{k_0}$ of
$0$ spin touching $c^-$;
\itemitem{ii)} ${\hat\g^0_1},...,{\hat\g^0_{k_0}}$ are non interacting
rectangles whose sides are longer than two;
\itemitem{iii)} in every cluster $c^0_j$ there are $k^+_{j}$
clusters $c^+_1,...,c^+_{k^+_j}$ of +1 spins;
\itemitem{iv)}  $\forall j\in\{1,...,k^{0}\}\;$
${\hat\g^+_{j,1}},...,{\hat\g^+_{j,k^+_j}}$ are non interacting rectangles
whose sides are longer than two.
\vskip 0.35 truecm
\par\noindent
We have a single plurirectangle when $k^0=1$.
\par
We have used, above, the geometric notion of interacting rectangles:
given two rectangles $R_1$ and $R_2$ with boundaries on the dual lattice
${\bf Z}^{2}+({1\over 2},{1\over 2})$, we say that they {\it interact} if
and only if one of the two following conditions occurs:
\vskip 0.35 truecm
\itemitem{i)} their boundaries intersect;
\itemitem{ii)} there exists a unit square centered at some lattice site
such that two of its edges are opposite and lie respectively on the boundaries
of $R_1$ and $R_2$.
\vskip 0.35 truecm
\par
\midinsert
\vskip 0.5 truecm
\vbox{\font\amgr=cmr10 at
10truept\baselineskip0.1466667truein\lineskiplimit-\maxdimen
\catcode`\-=\active\catcode`\~=\active\def~{{\char32}}\def-{{\char1}}%
\hbox{\amgr
{}~~~~~~~~~~~~~~{\char2}----------------------------------------------------------{\char3}}
\hbox{\amgr
{}~~~~~~~~~~~~~~{\char0}~~~~~~~~~~~~~~~~~~~~~~~~~~~~~~~~~~~~~~~~~~~~~~~~~~~~~~~~~~{\char0}}
\hbox{\amgr
{}~~~~~~~~~~~~~~{\char0}~~~~~~~{\char2}---------------------{\char3}~~~~~~~~~~~~~~~~~~~~~{\char45}~~~~~~{\char0}%
}
\hbox{\amgr
{}~~~~~~~~~~~~~~{\char0}~~~~~~~{\char0}~~~~~~~~~~~~~~~~~~~~~{\char0}~~~~~~~~~~~~~~~~~~~~~~~~~~~~{\char0}}
\hbox{\amgr
{}~~~~~~~~~~~~~~{\char0}~~~~~~~{\char0}~{\char2}-------{\char3}~~~~~0~~~~~{\char0}~~~{\char2}-----------{\char3}%
{}~~~~~~~~~~~~{\char0}}
\hbox{\amgr
{}~~~~~~~~~~~~~~{\char0}~~~~~~~{\char0}~{\char0}~~~~~~~{\char0}~~{\char2}----{\char3}~~~{\char0}~~~{\char0}~~%
{}~~~~~~~~~{\char0}~~~~~~~~~~~~{\char0}}
\hbox{\amgr
{}~~~~~~~~~~~~~~{\char0}~~~~~~~{\char0}~{\char0}~~~~~~~{\char0}~~{\char0}~~~~{\char0}~~~{\char0}~~~{\char0}~~%
{}~~~~~~~~~{\char0}~~~~~~~~~~~~{\char0}}
\hbox{\amgr
{}~~~~~~~~~~~~~~{\char0}~~~~~~~{\char0}~{\char0}~~~+~~~{\char0}~~{\char0}~~+~{\char0}~~~{\char0}~~~{\char0}~~%
{}~~~0~~~~~{\char0}~~~~~~~~~~~~{\char0}}
\hbox{\amgr
{}~~~~~~~~~~~~~~{\char0}~~~~~~~{\char0}~{\char0}~~~~~~~{\char0}~~{\char0}~~~~{\char0}~~~{\char0}~~~{\char0}~~%
{}~~~~~~~~~{\char0}~~~~~~~~~~~~{\char0}}
\hbox{\amgr
{}~~~~~~~~~~~~~~{\char0}~~~~~~~{\char0}~{\char0}~~~~~~~{\char0}~~{\char5}----{\char4}~~~{\char0}~~~{\char5}--%
---------{\char4}~~~~~~~~~~~~{\char0}}
\hbox{\amgr
{}~~~~~~~~~~~~~~{\char0}~~~~~~~{\char0}~{\char0}~~~~~~~{\char0}~~~~~~~~~~~{\char0}~~~~~~~~~~~~~~~~~~~~~~~~~~~%
{}~{\char0}}
\hbox{\amgr
{}~~~~~~~~~~~~~~{\char0}~~~~~~~{\char0}~{\char5}-------{\char4}~{\char2}------{\char3}~~{\char0}~~~~~~~~~{\char45}%
{}~~~~~~~~~~~~~~~~~~{\char0}}
\hbox{\amgr
{}~~~~~~~~~~~~~~{\char0}~~~~~~~{\char0}~~~~~~~~~~~{\char0}~~~~~~{\char0}~~{\char0}~~~~~~~~~~~~~~~~~~~~~~~~~~~%
{}~{\char0}}
\hbox{\amgr
{}~~~~~~~~~~~~~~{\char0}~~~~~~~{\char0}~~~~~0~~~~~{\char0}~~+~~~{\char0}~~{\char0}~~~~~~~~~{\char2}----------%
-----{\char3}~~{\char0}}
\hbox{\amgr
{}~~~~~~~~~~~~~~{\char0}~~~~~~~{\char0}~~~~~~~~~~~{\char0}~~~~~~{\char0}~~{\char0}~~~~~~~~~{\char0}~~~~~~~~~~%
{}~~~~~{\char0}~~{\char0}}
\hbox{\amgr
{}~~~~~~~~~~~~~~{\char0}~~~~~~~{\char0}~~~~~~~~~~~{\char5}------{\char4}~~{\char0}~~~~~~~~~{\char0}~{\char2}-%
----------{\char3}~{\char0}~~{\char0}}
\hbox{\amgr
{}~~~~~~~~~~~~~~{\char0}~~~~~~~{\char5}---------------------{\char4}~~~~~~~~~{\char0}~{\char0}~~~~~~~~~~~{\char0}%
{}~{\char0}~~{\char0}}
\hbox{\amgr
{}~~~~~~~~~~~~~~{\char0}~~~~~~~~~~~~~~~~~~~~~~~~~~~~~~~~~~~~~~~{\char0}~{\char0}~~~~~~~~~~~{\char0}~{\char0}~%
{}~{\char0}}
\hbox{\amgr
{}~~~~~~~~~~~~~~{\char0}~~~~~~~~~~~~~~~~~~~~~~~~~~~~~~~~~~~~~~~{\char0}~{\char0}~~~~~+~~~~~{\char0}~{\char0}~%
{}~{\char0}}
\hbox{\amgr
{}~~~~~~~~~~~~~~{\char0}~~~~~~~~~~~~~~~~~~~~~~~~~~~~~~~~~~~~~~~{\char0}~{\char0}~~~~~~~~~~~{\char0}~{\char0}~%
{}~{\char0}}
\hbox{\amgr
{}~~~~~~~~~~~~~~{\char0}~~~~~~~~~~~~~~~~~{\char45}~~~~~~~~~~~~~~~~~~~~~{\char0}~{\char5}-----------{\char4}~{\char0}%
{}~~{\char0}}
\hbox{\amgr
{}~~~~~~~~~~~~~~{\char0}~~~~~~~~~~~~~~~~~~~~~~~~~~~~~~~~~~~~~~~{\char0}~~~~~~~~~~~~~~~{\char0}~~{\char0}}
\hbox{\amgr
{}~~~~~~~~~~~~~~{\char0}~~~~~~~~~~~~~~~~~~~~~~~~~~~~~~~~~~~~~~~{\char5}---------------{\char4}~~{\char0}}
\hbox{\amgr
{}~~~~~~~~~~~~~~{\char0}~~~~~~~~~~~~~~~~~~~~~~~~~~~~~~~~~~~~~~~~~~~~~~~~~~~~~~~~~~{\char0}}
\hbox{\amgr
{}~~~~~~~~~~~~~~{\char5}----------------------------------------------------------{\char4}~~~~}}
\vskip 0.35 truecm
\par\noindent
\centerline {\smbfb Fig.3.8}
\endinsert
\par
We have to compute the energy of such local minima as a first step in the
description of their tendency to shrink or grow of the stables clusters.
\par
We say that a local minimum $\s$ is {\it subcritical} if and only if
$$\lim_{\beta\to\infty}P_{\s}(\tau_{\menouno}<\tau_{\piuuno})=1\;\; ;
\Eq (3.3.1)$$
one of the main problems that we have to solve is to understand when a
local minimum is subcritical.\par
The energy of a birectangle $R(L_{1},L_{2};M_{1},M_{2})$ is
$$H(R(L_{1},L_{2};M_{1},M_{2}))-H(\menouno)=$$
$$=(2M_{1}+2M_{2})J+(2L_{1}+2L_{2})J-
M_{1}M_{2}(h-\l)-L_{1}L_{2}(h+\l)\;\; .\Eq (3.4)$$
The above formula can be easily generalized to the case of a general
plurirectangle $\s$, characterized by the parameters
$M_{1,j},M_{2,j},L_{1,j,i}\;{\rm e}\; L_{2,j,i}\;\;\forall
j\in\{1,...,k^{0}\}\; {\rm and }\;\forall i\in\{1,...,k^{+}_{j}\}$,
with obvious meaning of the notation. One has
$$\eqalign {
H(\sigma)-H(\menouno)=\sum_{j=1}^{k^{0}}&\big\{
(2M_{1,j}+2M_{2,j})J-M_{1,j}M_{2,j}(h-\l)+\cr
+\sum_{i=1}^{k^{+}_{j}} [
(2L_{1,j,i}+&2L_{2,j,i})J-L_{1,j,i}L_{2,j,i}(h+\l)]\big\}\cr}\;\;
.\Eq (3.5)$$
\par
Now we consider a squared birectangle $Q(L,M):=R(L,L;M,M)$, whose energy
$e(M,L):=H(Q(L,M))-H(\menouno)$ is given by
$$e(M,L)=4MJ+4LJ-M^{2}(h-\l)-L^{2}(h+\l)\;\; .\Eq (3.6)$$
The graph of this function $ e: {\bf R}^2 \to {\bf R}$ is a paraboloid
with elliptical section and
downhill concavity, the coordinates of the vertex are
$$M={2J\over h-\l}\;\;\;\;\;\;L={2J\over h+\l};\Eq (3.7)$$
the level curves of $e(M,L)$ are represented in Fig.3.9.
\midinsert
\vskip 9 truecm\noindent
\includegraphics{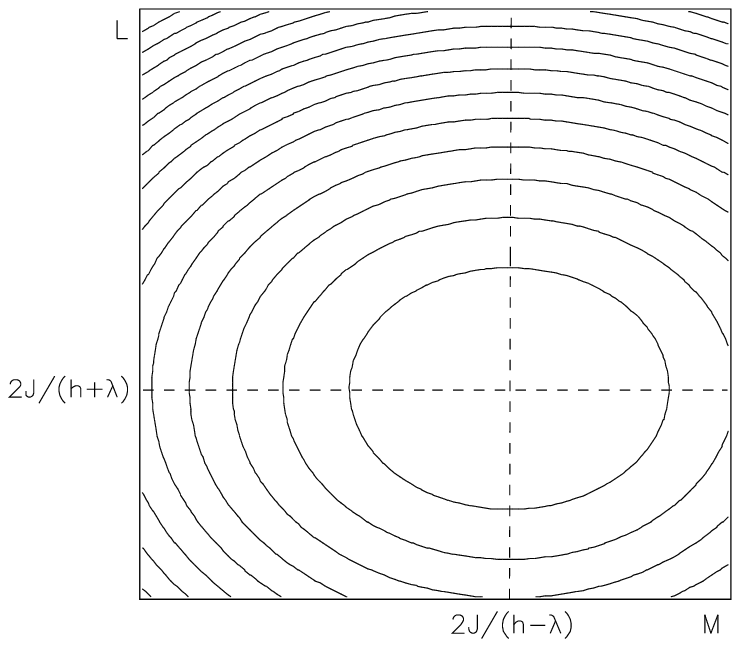}
\centerline {\smbfb Fig.3.9}
\endinsert
\par
Let us consider a droplet $Q(M,L)$ such that $M<{2J\over h-\l}$ and
$L<{2J\over h+\l}$: if these conditions are satisfied $e(M,L)$ is an
increasing function of $M$ and $L$, so we expect that this droplet will shrink.
On the other hand if $M>{2J\over h-\l}$, since $e(M,L)$ is a decreasing
function of $M$, we expect that the external
cluster of the droplet will grow; this suggests that
$M^{*}:=\left[ {2J\over h-\l}\right] +1$ is the {\it critical dimension} for
the external cluster of a local minimum.
After the growth of the external cluster, we look at what will happen
to the internal one; with similar arguments one can convince himself that
$L^{*}:=\left[ {2J\over h+\l}\right] +1$ appears to play the role of the
critical dimension. Obviously these two processes of growth cannot be
inverted, in fact a plus spin droplet can ``live" only inside a zero spin
droplet.
\par
But it can also happen that the plus spin phase is reached directly,
without passing through the zero spin phase; this happens if the droplet
$Q(M,L)$ grows moving along the line $M=L+2$. In this case one can see that
the system reaches the stable phase through a sequence of frames
(picture frames).
We call {\it squared frame} a birectangle $C(l,l):=R(l,l;l+2,l+2)$ with
$l\ge 2$. The most general {\it frame} is a rectangular one (see Fig.3.10)
$$C(l_{1},l_{2}):=R(l_{1},l_{2};l_{1}+2,l_{2}+2)\;\; ,\Eq (3.8)$$
where $l_1,l_2\ge 2$.
\midinsert
\vskip 0.5 truecm
\vbox{\font\amgr=cmr10 at
10truept\baselineskip0.1466667truein\lineskiplimit-\maxdimen
\catcode`\-=\active\catcode`\~=\active\def~{{\char32}}\def-{{\char1}}%
\hbox{\amgr
{}~~~~~~~~~~~~~~~~~~~~~~~~~~~~~~~~{\char2}---------------------{\char3}}
\hbox{\amgr
{}~~~~~~~~~~~~~~~~~~~~~~~~~~~~~~~~{\char0}~{\char2}-----------------{\char3}~{\char0}}
\hbox{\amgr
{}~~~~~~~~~~~~~~~~~~~~~~~~~~~~~~~~{\char0}~{\char0}~~~~~~~~~~~~~~~~~{\char0}~{\char0}}
\hbox{\amgr
{}~~~~~~~~~~~~~~~~~~~~~~~~~~~~~~~~{\char0}~{\char0}~~~~~~~~~~~~~~~~~{\char0}~{\char0}}
\hbox{\amgr
{}~~~~~~~~~~~~~~~~~~~~~~~~~~~~~~~~{\char0}0{\char0}~~~~~~~~~~~~~~~~~{\char0}0{\char0}}
\hbox{\amgr
{}~~~~~~~~~~~~~~~~~~~~~~~~~~~~~~~~{\char0}~{\char0}~~~~~~~~+~~~~~~~~{\char0}~{\char0}~~~$l_2+2$}
\hbox{\amgr
{}~~~~~~~~~~~~~~~~~~~~~~~~~~~~~~~~{\char0}~{\char0}~~~~~~~~~~~~~~~~~{\char0}~{\char0}}
\hbox{\amgr
{}~~~~~~~~~~~~~~~~~~~~~~~~~~~~~~~~{\char0}~{\char0}~~~~~~~~~~~~~~~~~{\char0}~{\char0}}
\hbox{\amgr
{}~~~~~~~~~~~~~~~~~~~~~~~~~~~~~~~~{\char0}~{\char5}-----------------{\char4}~{\char0}}
\hbox{\amgr
{}~~~~~~~~~~~~~~~~~~~~~~~~~~~~~~~~{\char5}---------------------{\char4}}
\hbox{\amgr }
\hbox{\amgr ~~~~~~~~~~~~~~~~~~~~~~~~~~~~~~~~~~~~~~~~~~$l_1+2$~~~~~~~~~~}}
\vskip 0.35 truecm
\par\noindent
\centerline {\smbfb Fig.3.10 }
\endinsert
\par
Now we consider the energy of a squared frame $e(l):=H(C(l,l))
-H(\menouno)$, using equality $\equ (3.6)$ we have
$$e(l)=-2hl^{2}+l[8J-4(h-\l)]+[8J-4(h-\l)]\;\; ;\Eq (3.9)$$
the graph of this function is a concave parabola, whose vertex coordinate
is
$$l={2J-(h-\l)\over h}\;\; .\Eq (3.10)$$
We expect that $C(l,l)$ will grow if $l\ge l^*$, where
$$l^*:=\left[ {2J-(h-\l)\over h}\right] +1\;\; ,\Eq (3.10.1)$$
otherwise it will shrink;
hence $l^*$ should be the critical dimension of a squared frame.
\par
In order to describe the behaviour of a general birectangle
$R=R(L_1,L_2;M_1,M_2)$, we must study
growth and contraction mechanisms of a droplet; like in the Ising model
these are mainly: growth of a (unit square) protuberance and corner erosion.
But in Blume-Capel model the relevant local minima are made of two components,
the internal and the external ones, and both of them can grow or shrink
independently. The mechanisms of growth and contraction are explained in
Fig.3.11, they corrispond to:
\vskip 0.35 truecm
\itemitem{1)} creation of a $+$ protuberance adjacent from the exterior to
the internal rectangle;
\itemitem{2)} creation of a $0$ protuberance adjacent from the exterior to
the external rectangle;
\itemitem{3)} erosion $(+\rightarrow 0)$ of all but one $+$ spin in a row
or column of the internal rectangle;
\itemitem{4)} erosion $(0\rightarrow -)$ of all but one $0$ spin in a row
or column of the external rectangle.
\vskip 0.35 truecm
\par\noindent
Their typical times are
$$\eqalign {
t_{1}=e^{\b[2J-(h+\l)]}\;\;\;&t_{2}=e^{\b[2J-(h-\l)]}\cr
t_{3}=e^{\b(h+\l)(L-1)}\;\;\;&t_{4}=e^{\b(h-\l)(M-1)}\cr}
\;\; ,\Eq (3.11)$$
where $L:=\min \{L_{1},L_{2}\}$ and $M:=\min \{M_{1},M_{2}\}$.
\midinsert
\vskip 0.35 truecm
\vbox{\font\amgr=cmr10 at
10truept\baselineskip0.1466667truein\lineskiplimit-\maxdimen
\catcode`\-=\active\catcode`\~=\active\def~{{\char32}}\def-{{\char1}}%
\hbox{\amgr
{}~~~~~~~~{\char2}-----------------------{\char3}~~~~~~~~~~~~~~~{\char2}-----------------------{\char3}%
}
\hbox{\amgr
{}~~~~~~~~{\char0}~~~~~~~~~~~~~~~~~~~~~~~{\char0}~~~~~~~~~~~~~~~{\char5}-{\char3}~~~~~~~~~~~~~~~~~~~~~{\char0}%
}
\hbox{\amgr
{}~~~~~~~~{\char0}~~~~~~~~~~~~~~~~~~~~~~~{\char0}~~~~~~~~~~~~~~~~~{\char0}~~~~~~~~~~~~~~~~~~~~~{\char0}%
}
\hbox{\amgr
{}~~~~~~~~{\char0}~~~~~~~{\char2}----------{\char3}~~~~{\char0}~~~~~~~~~~~~~~~|~{\char0}~~~~~{\char2}--%
--------{\char3}~~~~{\char0}}
\hbox{\amgr
{}~~~~~~~~{\char0}~~~~~~~{\char0}~~~~~~~~~~{\char0}~~~~{\char0}~~~~~~~~~~~~~~~~~{\char0}~~~~~{\char5}-{\char3}%
{}~~~~~~~~{\char0}~~~~{\char5}-{\char3}}
\hbox{\amgr
{}~~~~~~~~{\char0}~~~~~~~{\char0}~~~~~~~~~~{\char0}~~~~{\char0}~~~~~~~~~~~~~~~|~{\char0}~~~~~~~{\char0}%
{}~~~~~~~~{\char0}~~~~{\char2}-{\char4}}
\hbox{\amgr
{}~~~~~~~~{\char0}~~~~~~~{\char0}~~~~~~~~~~{\char0}~~~~{\char0}~~~~~~~~~~~~~~4~~{\char0}~~~~~|~{\char0}%
{}~~~~~~~~{\char0}~~~~{\char0}~2}
\hbox{\amgr
{}~~~~~~~~{\char0}~~~~~~~{\char0}~~~~~~~~~~{\char0}~~~~{\char0}~~~~~~~~~~~~~~~|~{\char0}~~~3~~~{\char0}%
{}~~~~~~~~{\char5}-{\char3}~~{\char0}}
\hbox{\amgr
{}~~~~~~~~{\char0}~~~~~~~{\char0}~~~~~~~~~~{\char0}~~~~{\char0}~~~~~~~~~~~~~~~~~{\char0}~~~~~|~{\char0}%
{}~~~~~~~~{\char2}-{\char4}~~{\char0}}
\hbox{\amgr
{}~~~~~~~~{\char0}~~~~~~~{\char0}~~~~~~~~~~{\char0}~~~~{\char0}~~~~~~~~~~~~~~~|~{\char0}~~~~~~~{\char0}%
{}~~~~~~~~{\char0}~1~~{\char0}}
\hbox{\amgr
{}~~~~~~~~{\char0}~~~~~~~{\char5}----------{\char4}~~~~{\char0}~~~~~~~~~~~~~~~~~{\char0}~~~~~{\char5}~{\char17}%
--------{\char4}~~~~{\char0}}
\hbox{\amgr
{}~~~~~~~~{\char0}~~~~~~~~~~~~~~~~~~~~~~~{\char0}~~~~~~~~~~~~~~~|~{\char0}~~~~~~~~~~~~~~~~~~~~~{\char0}%
}
\hbox{\amgr
{}~~~~~~~~{\char0}~~~~~~~~~~~~~~~~~~~~~~~{\char0}~~~~~~~~~~~~~~~~~{\char0}~~~~~~~~~~~~~~~~~~~~~{\char0}%
}
\hbox{\amgr
{}~~~~~~~~{\char5}-----------------------{\char4}~~~~~~~~~~~~~~~{\char5}~{\char17}---------------------{\char4}%
}
\hbox{\amgr }}
\vskip 0.35 truecm
\par\noindent
\centerline {\smbfb Fig.3.11}
\endinsert
\par
By comparing times $t_1,\dots ,t_4$, we observe that the growth of an internal
protuberance is always faster than the growth of an external one, indeed
$$2J-(h+\l)<2J-(h-\l)\;\Rightarrow\;t_{1}<t_{2}\;\; .\Eq (3.12)$$
Then we introduce the following critical dimensions
$$\eqalign {
L^{*}:=\left[ {2J\over h+\l} \right] +1\;\;\;&
{\widetilde L}\; :=\left[ {2J+2\l\over h+\l} \right] +1\cr
M^{*}:=\left[ {2J\over h-\l} \right] +1\;\;\;&
{\widetilde M}:=\left[ {2J-2\l\over h-\l} \right] +1\cr}\;\; ;
\Eq (3.13)$$
whose meaning is explained below
\vskip 0.35truecm
\item{} $L<L^{*}\;\Leftrightarrow\; (h+\l)(L-1)<2J-(h+\l)$:
internal contraction is faster than  growth, that is the internal
component of the local minimum is (relatively) {\it subcritical};
\item{} $L<{\widetilde L}\;\Leftrightarrow\; (h+\l)(L-1)<
2J-(h-\l)$: internal contraction is faster than external growth;
\item{} $M<{\widetilde M}\;\Leftrightarrow\; (h-\l)(M-1)<
2J-(h+\l)$: external contraction is faster than internal growth;
\item{} $M<M^{*}\;\Leftrightarrow\; (h-\l)(M-1)<
2J-(h-\l)$: external contraction is faster than growth, that is the
external component of the local minimum is (relatively) {\it subcritical}.
\vskip 0.35truecm
\par
As we will see in the next section, another interesting length will be
$l_0:=\left[ {h\over\l} \right] +1$.
\par
We choose the parameters $J,h$ and $\l$ in such a way that ${2J\over h+\l}$,
${2J+2\l\over h+\l}$, ${2J\over h-\l}$,
${2J-2\l\over h-\l}$ , ${2J-(h-\l)\over h}$ and ${h\over \l}$
are not integer, so that ambiguos situations, here and in the following, are
avoided. \par
The behaviour of our birectangle $R$ depends on its dimensions, some of the
possible cases are described below:
\vskip 0.35truecm
\item{} $L<L^{*}\;{\rm and}\;M<M^{*}$: both internal and external
component are subcritical, $R$ is subcritical;
\item{} $L<L^{*}\;{\rm and}\;M>M^{*}$: the internal component is
subcritical, but not the external one, $R$ is supercritical and the system
starting from $R$ will reach $\zero$;
\item{} $L>L^{*}\;{\rm and}\;M>M^{*}$: both internal and external component
are supercritical; $R$ is supercritical and the system starting from $R$
will reach $\piuuno$ by passing through $C(M_1-2,M_2-2)$
(internal growth is faster than external one);
\item{} $L>L^{*}\;{\rm and}\;M<M^{*}$: internal component is supercritical
while external one is subcritical, the future of the system starting from
$R$ depends on the relation $M\grle{\widetilde M}$.
\vskip0.35truecm
\par\noindent
Many different situations can take place, the last one is
surely the most difficult but also the most interesting that we have to
examine.
\par
Growth and contraction of a frame are based on the same elementary mechanisms
described before, but they take place in more than one step. The
possible contraction
of a squared frame $C(l,l)$ starts with the contraction
of its internal component: our system typically
first reaches the configuration $S(l,l)$, increasing
its energy of the quantity $H(S(l,l))-H(C(l,l))= (h+\l)(l-1)$,
and then the configuration $R(l,l):=R(l-1,l;l+2,l+2)$,
lowering its energy of the quantity $H(S(l,l))-H(R(l,l))=2J-(h+\l)$
(see Fig.3.12).
\midinsert
\vskip 0.35 truecm
\vbox{\font\amgr=cmr10 at
10truept\baselineskip0.1466667truein\lineskiplimit-\maxdimen
\catcode`\-=\active\catcode`\~=\active\def~{{\char32}}\def-{{\char1}}%
\hbox{\amgr ~~{\char2}--------------------{\char3}~~${\phantom
{\rightarrow}}$~~{\char2}--------------------{\char3}~~${\phantom
{\rightarrow}}$~~{\char2}-------------------%
-{\char3}}
\hbox{\amgr ~~{\char0}~{\char2}----------------{\char3}~{\char0}~~${\phantom
{\rightarrow}}$~~{\char0}~{\char2}--------------{\char3}~~~{\char0}%
{}~~${\phantom
{\rightarrow}}$~~{\char0}~{\char2}--------------{\char3}~~~{\char0}}
\hbox{\amgr ~~{\char0}~{\char0}~~~~~~~~~~~~~~~~{\char0}~{\char0}~~${\phantom
{\rightarrow}}$~~{\char0}~{\char0}~~~~~~~~~~~~~~{\char0}~~~{\char0}%
{}~~${\phantom
{\rightarrow}}$~~{\char0}~{\char0}~~~~~~~~~~~~~~{\char0}~~~{\char0}}
\hbox{\amgr ~~{\char0}~{\char0}~~~~~~~~~~~~~~~~{\char0}~{\char0}~~${\phantom
{\rightarrow}}$~~{\char0}~{\char0}~~~~~~~~~~~~~~{\char0}~~~{\char0}%
{}~~${\phantom
{\rightarrow}}$~~{\char0}~{\char0}~~~~~~~~~~~~~~{\char0}~~~{\char0}}
\hbox{\amgr ~~{\char0}~{\char0}~~~~~~~~~~~~~~~~{\char0}~{\char0}~~${\phantom
{\rightarrow}}$~~{\char0}~{\char0}~~~~~~~~~~~~~~{\char0}~~~{\char0}%
{}~~${\phantom
{\rightarrow}}$~~{\char0}~{\char0}~~~~~~~~~~~~~~{\char0}~~~{\char0}}
\hbox{\amgr
{}~~{\char0}~{\char0}~~~~~~~~~~~~~~~~{\char0}~{\char0}~~$\rightarrow$~~{\char0}~{\char0}~~~~~~~~~~~~~~{\char0}~~~{\char0}~~$\rightarrow$~~{\char0}~{\char0}~~~~~~~~~~~~~~{\char0}~~~{\char0}}
\hbox{\amgr ~~{\char0}~{\char0}~~~~~~~~~~~~~~~~{\char0}~{\char0}~~${\phantom
{\rightarrow}}$~~{\char0}~{\char0}~~~~~~~~~~~~~~{\char0}~~~{\char0}%
{}~~${\phantom
{\rightarrow}}$~~{\char0}~{\char0}~~~~~~~~~~~~~~{\char0}~~~{\char0}}
\hbox{\amgr ~~{\char0}~{\char0}~~~~~~~~~~~~~~~~{\char0}~{\char0}~~${\phantom
{\rightarrow}}$~~{\char0}~{\char0}~~~~~~~~~~~~~~{\char0}~~~{\char0}%
{}~~${\phantom
{\rightarrow}}$~~{\char0}~{\char0}~~~~~~~~~~~~~~{\char0}~~~{\char0}}
\hbox{\amgr ~~{\char0}~{\char0}~~~~~~~~~~~~~~~~{\char0}~{\char0}~~${\phantom
{\rightarrow}}$~~{\char0}~{\char0}~~~~~~~~~~~~~~{\char5}-{\char3}~{\char0}%
{}~~${\phantom
{\rightarrow}}$~~{\char0}~{\char0}~~~~~~~~~~~~~~{\char0}~~~{\char0}}
\hbox{\amgr ~~{\char0}~{\char5}----------------{\char4}~{\char0}~~${\phantom
{\rightarrow}}$~~{\char0}~{\char5}----------------{\char4}~{\char0}%
{}~~${\phantom
{\rightarrow}}$~~{\char0}~{\char5}--------------{\char4}~~~{\char0}}
\hbox{\amgr ~~{\char5}--------------------{\char4}~~${\phantom
{\rightarrow}}$~~{\char5}--------------------{\char4}~~${\phantom
{\rightarrow}}$~~{\char5}-------------------%
-{\char4}}
\hbox{\amgr ~~~~~~~~~~~$C(l,l)$~~~~~~~~~~~${\phantom
{\rightarrow}}$~~~~~~~~~$S(l,l)$~~~~~${\phantom
{\rightarrow}}$~~~~~~~~~~~~~~~$R(l,l)$~~~~~~~~}}
\vskip 0.35 truecm
\par\noindent
\centerline {\smbfb \quad Fig.3.12}
\endinsert
\par\noindent
At this level it is not easy to describe the future evolution of the system:
the internal component could continue to shrink or the external component
could start its contraction; but we remark that the first step in
the contraction of $C(l,l)$  always involves  bypassing of an energetical
barrier whose height is $(h+\l)(l-1)$.
\par
On the other hand
the possible expansion of $C(l,l)$ starts with the growth of an external
protuberance: the system typically reaches  the configuration $G(l,l)$
by overcoming the
energetical barrier $H(G(l,l))-H(C(l,l))=2J-(h-\lambda)$ and then it
goes down to $R(l+1,l):=R(l,l;l+3,l+2)$ lowering its energy of the quantity
$H(G(l,l))-H(R(l+1,l))=(h-\lambda)(l+1)$ (see Fig.3.13). We have supposed,
without loss of generality, that the growth is horizontal.
\midinsert
\vskip 0.35 truecm
\vbox{\font\amgr=cmr10 at
10truept\baselineskip0.1466667truein\lineskiplimit-\maxdimen
\catcode`\-=\active\catcode`\~=\active\def~{{\char32}}\def-{{\char1}}%
\hbox{\amgr
{}~~{\char2}--------------------{\char3}~~~~{\char2}--------------------{\char3}~~~~~~{\char2}-----------%
------------{\char3}}
\hbox{\amgr
{}~~{\char0}~{\char2}----------------{\char3}~{\char0}~~~~{\char0}~{\char2}----------------{\char3}~{\char0}%
{}~~~~~~{\char0}~{\char2}----------------{\char3}~~~~{\char0}}
\hbox{\amgr
{}~~{\char0}~{\char0}~~~~~~~~~~~~~~~~{\char0}~{\char0}~~~~{\char0}~{\char0}~~~~~~~~~~~~~~~~{\char0}~{\char0}%
{}~~~~~~{\char0}~{\char0}~~~~~~~~~~~~~~~~{\char0}~~~~{\char0}}
\hbox{\amgr
{}~~{\char0}~{\char0}~~~~~~~~~~~~~~~~{\char0}~{\char0}~~~~{\char0}~{\char0}~~~~~~~~~~~~~~~~{\char0}~{\char0}%
{}~~~~~~{\char0}~{\char0}~~~~~~~~~~~~~~~~{\char0}~~~~{\char0}}
\hbox{\amgr
{}~~{\char0}~{\char0}~~~~~~~~~~~~~~~~{\char0}~{\char0}~~~~{\char0}~{\char0}~~~~~~~~~~~~~~~~{\char0}~{\char0}%
{}~~~~~~{\char0}~{\char0}~~~~~~~~~~~~~~~~{\char0}~~~~{\char0}}
\hbox{\amgr
{}~~{\char0}~{\char0}~~~~~~~~~~~~~~~~{\char0}~{\char0}~->~{\char0}~{\char0}~~~~~~~~~~~~~~~~{\char0}~{\char5}%
-{\char3}~->~{\char0}~{\char0}~~~~~~~~~~~~~~~~{\char0}~~~~{\char0}}
\hbox{\amgr
{}~~{\char0}~{\char0}~~~~~~~~~~~~~~~~{\char0}~{\char0}~~~~{\char0}~{\char0}~~~~~~~~~~~~~~~~{\char0}~{\char2}%
-{\char4}~~~~{\char0}~{\char0}~~~~~~~~~~~~~~~~{\char0}~~~~{\char0}}
\hbox{\amgr
{}~~{\char0}~{\char0}~~~~~~~~~~~~~~~~{\char0}~{\char0}~~~~{\char0}~{\char0}~~~~~~~~~~~~~~~~{\char0}~{\char0}%
{}~~~~~~{\char0}~{\char0}~~~~~~~~~~~~~~~~{\char0}~~~~{\char0}}
\hbox{\amgr
{}~~{\char0}~{\char0}~~~~~~~~~~~~~~~~{\char0}~{\char0}~~~~{\char0}~{\char0}~~~~~~~~~~~~~~~~{\char0}~{\char0}%
{}~~~~~~{\char0}~{\char0}~~~~~~~~~~~~~~~~{\char0}~~~~{\char0}}
\hbox{\amgr
{}~~{\char0}~{\char5}----------------{\char4}~{\char0}~~~~{\char0}~{\char5}----------------{\char4}~{\char0}%
{}~~~~~~{\char0}~{\char5}----------------{\char4}~~~~{\char0}~}
\hbox{\amgr
{}~~{\char5}--------------------{\char4}~~~~{\char5}--------------------{\char4}~~~~~~{\char5}-----------%
------------{\char4}}
\hbox{\amgr
{}~~~~~~~~~~~$C(l,l)$~~~~~~~~~~~~~~~~~~~~$G(l,l)$~~~~~~~~~~~~~~~~~~~~~$R(l+1,l)$~~~~~~~~}}
\vskip 0.35 truecm
\par\noindent
\centerline {\smbfb Fig.3.13}
\endinsert
\par\noindent
As a consequence of the fact that it is always $t_1<t_2$,
the second step in the expansion of the droplet will be the growth of an
internal protuberance: the system reaches the configuration $S(l+1,l)$
by overcoming the energetical barrier $H(S(l+1,l))-H(R(l+1,l))=2J-(h+\l)$ and
then goes down to the frame $C(l+1,l)$ lowering its energy of the
quantity $H(S(l+1,l))-H(C(l+1,l))=(h+\l)(l-1)$.
\par
In order to describe the future probable evolution of the system, starting from
$C(l,l)$, and establish its tendency to shrink or grow we have to
distinguish the following four cases:
$$\eqalign {
l<{\widetilde L}&\Rightarrow H(S(l,l))
				      <H(G(l,l))\cr
{\widetilde L}<l<l^{*}&\Rightarrow \left\{ \eqalign {
H(G(l,l))&<H(S(l,l))\cr
H(S(l,l))&<H(S(l+1,l))\cr}\right.\cr
{l}^{*}<l,\; l+2<{\widetilde M}&\Rightarrow \left\{ \eqalign {
H(G(l,l))&<H(S(l+1,l))\cr
H(S(l+1,l))&<H(S(l,l))\cr}\right.\cr
l^{*}<l,\;{\widetilde M}<l+2&\Rightarrow \left\{ \eqalign {
H(G(l,l))&<H(S(l,l))\cr
H(S(l+1,l))-H(R(l+1,l))&<H(G(l,l))-H(R(l+1,l))\cr}\right.\cr}\;\; ;
$$
these four cases are illustrated in Fig.3.14.
\par\noindent
Even the analysis of growth and contraction mechanisms leads to conclude that
$l^*$ is the critical dimension of a square frame.
\par
\par
We close this section remarking that parameter $\l$ may be choosen
sufficiently small, that is
$$\l<{2J\over 2a^2+a-1}\Eq (3.14')$$
where $a={h\over\l}$,
so that the following inequalities are satisfied:
$$\eqalign{
1)&\; L^*+1\leq l^*\cr
2)&\; L^{*}\leq {\widetilde L}< l^{*}<l^{*}+3\leq {\widetilde M}
\leq M^{*}\cr}\;\; .\Eq (3.14)$$
\par
\vfill\eject
\midinsert
{\centerline {$l<{\widetilde L}$\hskip 5.5 truecm${\widetilde L}<l<l^{*}$}}
\vskip 12 truecm\noindent
\includegraphics{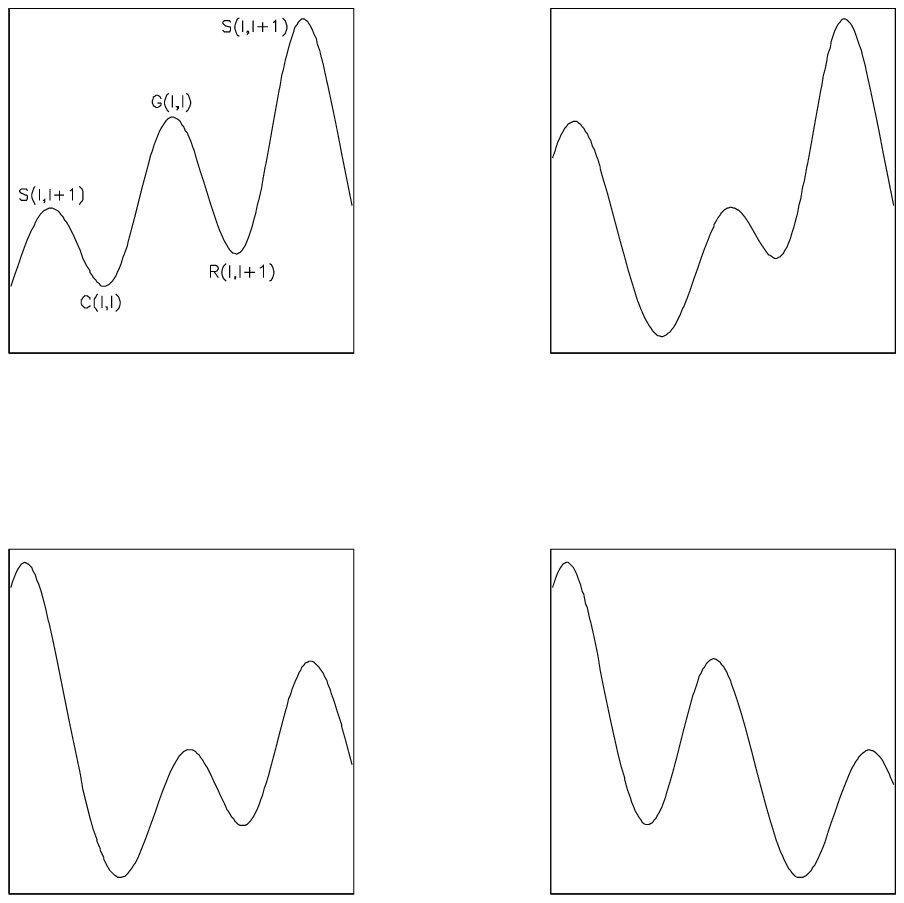}
\par\noindent
{\centerline {$l^{*}<l,\;l+2<
{\widetilde M}$\hskip 3.5 truecm$l^{*}<l,\;{\widetilde M}<l+2$}}
\par\noindent
\vskip 1 truecm
\centerline {\smbfb Fig.3.14}
\endinsert

\vfill\eject
\numsec=4\numfor=1

{\bf Section 4. Supercriticality and subcriticality of local minima.}
\par
In this section we want to prove rigorous results about supercriticality
and subcriticality of local minima. Namely we want to give criteria to
estabilish the natural tendency of the geometrical structures representing
the minima for H to shrink or grow. We will first analyze the ``frames",
then the generic birectangles and finally the plurirectangles.
\par
First of all we state the following proposition:
\vskip 0.35 truecm
\noindent
{\bf Proposition 4.1.}\par\noindent
Let us consider the configuration $C(l_{1},l_{2})$; we set
$l:=\min\{l_{1},l_{2}\}$ and $m:=\max\{l_{1},l_{2}\}$. Let $\e>0$, we have:
$$l<l^{*}\; {\rm and}\; m<m^{*}(l)\Rightarrow \left\{ \eqalign {
\lim_{\beta\to\infty}P_{C(l_{1},l_{2})}&(\tau_{\menouno}<\tau_{\piuuno})
						=1\cr
\lim_{\beta\to\infty}P_{C(l_{1},l_{2})}&(T_{-}^{s}(\e)<
					\tau_{\menouno}<T_{+}^{s}(\e))
					=1\cr}\right.\;\; ,$$
where
$$T_{\pm}^{s}(\e):=\left\{ \matrix {
e^{\beta(h-\lambda)(l+1)\pm\beta\e}&l<[{h\over\lambda}]+1\cr
e^{\beta(h+\lambda)(l-1)\pm\beta\e}&l\geq [{h\over\lambda}]+1\cr}\right. $$
and
$$m^{*}(l):=\left[ {2h\over h-\lambda} {2J-(h-\lambda)\over h}- {h+\lambda\over
h-\lambda}l\right]+1\;\; .$$
Moreover
$${\widetilde L}\le l<l^{*}\; {\rm and}\; m\geq m^{*}(l)\Rightarrow
\left\{ \eqalign {
\lim_{\beta\to\infty}P_{C(l_{1},l_{2})}&(\tau_{\piuuno}<\tau_{\menouno})
                                                =1\cr
\lim_{\beta\to\infty}P_{C(l_{1},l_{2})}&(T_{-}^{g,1}(\e)<
                                        \tau_{\piuuno}<T_{+}^{g,1}(\e))
                                        =1\cr}\right.$$
where
$$T_{\pm}^{g,1}(\e):=\left\{ \matrix {
e^{\beta\{[2J-(h-\lambda)]-(h-\lambda)(m+1)+[2J-(h+\lambda)]\}\pm\beta\e}
		&m<{\widetilde M}-2\cr
e^{\beta [2J-(h-\lambda)]\pm\beta\e}
		&m\geq {\widetilde M}-2\cr}\right. $$
Finally
$$l\geq l^{*}\Rightarrow \left\{ \eqalign {
\lim_{\beta\to\infty}P_{C(l_{1},l_{2})}&(\tau_{\piuuno}<\tau_{\menouno})
						=1\cr
\lim_{\beta\to\infty}P_{C(l_{1},l_{2})}&(T_{-}^{g,2}(\e)<
                                        \tau_{\piuuno}<T_{+}^{g,2}(\e))
                                        =1\cr}\right.\;\; ,$$
where
$$T_{\pm}^{g,2}(\e):=\left\{ \matrix {
e^{\beta\{[2J-(h-\lambda)]-(h-\lambda)(m+1)+[2J-(h+\lambda)]\}\pm\beta\e}
		&l,m<{\widetilde M}-2\cr
e^{\beta [2J-(h-\lambda)]\pm\beta\e}
		&{\rm otherwise}\cr}\right.\;\; . $$
\par\noindent
{\it Proof.}
\vskip 0.5 truecm
\par
Let us consider the frame $C:=C(l_1,l_2)$ with
$l:=\min\{l_1,l_2\}<{\widetilde L}$, its basin of attraction
$B:=B(C(l_1,l_2))$ and the relative boundary $\partial B$.
Let us denote by $S_1\in\partial B$ the set of configurations obtained by
changing
into zero $l-1$ plus spin adjacent to one of the shortest sides of the internal
rectangle of $C$ (see Fig.4.1); we claim that
$$\min_{\sigma\in\partial B}H(\sigma)=H(S_{1})\;\; .\Eq (4.3)$$
\par\bigskip
{\bf Remark.}\par
In the following we will consider: \par
1. configurations $\s$ containing a unique droplet $\g$ with a given particular
shape, size and location; for example a rectangle of zeroes (with given
location and
horizontal and vertical sizes) or a birectangle with given location and
external,
internal horizontal and vertical sizes.\par
2. The equivalence class of all the configurations  $\s'$ with a unique droplet
$\g'$ obtained from $\g$ by symmetries like rotations, translations, inversions
w.r.t lattice axes and even displacements along sides of unit square
protuberances.
\par
In the following, to avoid lengthy specification and to accelerate the
exposition we
often interchange the two above objects and we even use the same symbols to
denote them. The reader will easily deduce the meaning of our statements from
the
context.\par
For example sometimes we will denote  by $S_1$ also a particular droplet
obtained from a particular configuration in $C$ by substituting one
particular smaller internal side with a particular unit square protuberance.
\par\bigskip
\midinsert
\vskip 0.5 truecm
\vbox{\font\amgr=cmr10 at
10truept\baselineskip0.1466667truein\lineskiplimit-\maxdimen
\catcode`\-=\active\catcode`\~=\active\def~{{\char32}}\def-{{\char1}}%
\hbox{\amgr ~~~~~~~~~~~~~~~~~~~~~~~~$\phantom
{l+2}$~{\char2}----------------------------{\char3}}
\hbox{\amgr ~~~~~~~~~~~~~~~~~~~~~~~~$\phantom
{l+2}$~{\char0}~{\char2}----------------------{\char3}~~~{\char0}}
\hbox{\amgr ~~~~~~~~~~~~~~~~~~~~~~~~$\phantom
{l+2}$~{\char0}~{\char0}~~~~~~~~~~~~~~~~~~~~~~{\char0}~~~{\char0}}
\hbox{\amgr
{}~~~~~~~~~~~~~~~~~~~~~~~~${l+2}$~{\char0}~{\char0}~~~~~~~~~~~+~~~~~~~~~~{\char5}-{\char3}~{\char0}~~$S_{1}$}
\hbox{\amgr ~~~~~~~~~~~~~~~~~~~~~~~~$\phantom
{l+2}$~{\char0}~{\char0}~~~~~~~~~~~~~~~~~~~~~~{\char2}-{\char4}~{\char0}}
\hbox{\amgr ~~~~~~~~~~~~~~~~~~~~~~~~$\phantom
{l+2}$~{\char0}~{\char0}~~~~~~~~~~~~~~~~~~~~~~{\char0}~~~{\char0}}
\hbox{\amgr ~~~~~~~~~~~~~~~~~~~~~~~~$\phantom
{l+2}$~{\char0}~{\char5}----------------------{\char4}~0~{\char0}~~-}
\hbox{\amgr ~~~~~~~~~~~~~~~~~~~~~~~~$\phantom
{l+2}$~{\char5}----------------------------{\char4}}
\hbox{\amgr ~~~~~~~~~~~~~~~~~~~~~~~~$\phantom
{l+2}$~~~~~~~~~~~~~$m+2$~~~~~~~~~~~~~~~}}
\vskip 0.5 truecm
\par\noindent
\centerline {\smbfb Fig.4.1}
\endinsert
\par
\midinsert
\vskip -0.5 truecm
\vbox{\font\amgr=cmr10 at
10truept\baselineskip0.1466667truein\lineskiplimit-\maxdimen
\catcode`\-=\active\catcode`\~=\active\def~{{\char32}}\def-{{\char1}}%
\hbox{\amgr ~~~~~~~~{\char2}------------------------{\char3}}
\hbox{\amgr ~~~~~~~~{\char0}~{\char2}--------------------{\char3}~{\char0}~}
\hbox{\amgr
{}~~~~~~~~{\char0}~{\char0}~~~~~~~~~~~~~~~~~~~~{\char0}~{\char0}~{\char2}-{\char3}~}
\hbox{\amgr
{}~~~~~~~~{\char0}~{\char0}~~~~~~~~~~~~~~~~~~~~{\char0}~{\char0}~{\char5}-{\char4}~~~~~~~~~~~~$\left\{ \eqalign {
+&\;\;\;\;\Delta H_{11}=16J{\char45}2h\cr
0&\;\;\;\;\Delta H_{12}=4J{\char45}(h{\char45}\lambda)\cr}\right.$}
\hbox{\amgr ~~~~~~~~{\char0}~{\char0}~~~~~~~~~~~~~~~~~~~~{\char0}~{\char0}}
\hbox{\amgr ~~~~~~~~{\char0}~{\char5}--------------------{\char4}~{\char0}}
\hbox{\amgr ~~~~~~~~{\char5}------------------------{\char4}}
\hbox{\amgr }
\hbox{\amgr ~~~~~~~~{\char2}------------------------{\char3}}
\hbox{\amgr ~~~~~~~~{\char0}~{\char2}--------------------{\char3}~{\char0}~}
\hbox{\amgr
{}~~~~~~~~{\char0}~{\char0}~~~~~~~~~~~~~~~~~~~~{\char0}~{\char15}-{\char3}~}
\hbox{\amgr
{}~~~~~~~~{\char0}~{\char0}~~~~~~~~~~~~~~~~~~~~{\char0}~{\char15}-{\char4}~~~~~~~~~~~~~~$\left\{ \eqalign {
+&\;\;\;\;\Delta H_{21}=12J{\char45}2h\cr
0&\;\;\;\;\Delta H_{22}=2J{\char45}(h{\char45}\lambda)\cr}\right.$}
\hbox{\amgr ~~~~~~~~{\char0}~{\char0}~~~~~~~~~~~~~~~~~~~~{\char0}~{\char0}}
\hbox{\amgr ~~~~~~~~{\char0}~{\char5}--------------------{\char4}~{\char0}}
\hbox{\amgr ~~~~~~~~{\char5}------------------------{\char4}}
\hbox{\amgr }
\hbox{\amgr ~~~~~~~~{\char2}----------------------{\char18}-{\char3}}
\hbox{\amgr ~~~~~~~~{\char0}~{\char2}--------------------{\char14}-{\char16}~}
\hbox{\amgr ~~~~~~~~{\char0}~{\char0}~~~~~~~~~~~~~~~~~~~~{\char0}~{\char0}}
\hbox{\amgr
{}~~~~~~~~{\char0}~{\char0}~~~~~~~~~~~~~~~~~~~~{\char0}~{\char0}~~~~~~~~~~~~~~~~$\left\{ \eqalign {
+&\;\;\;\;\Delta H_{31}=8J{\char45}(h+\lambda)\cr
{\char45}&\;\;\;\;\Delta H_{32}=(h{\char45}\lambda)\cr}\right.$}
\hbox{\amgr ~~~~~~~~{\char0}~{\char0}~~~~~~~~~~~~~~~~~~~~{\char0}~{\char0}}
\hbox{\amgr ~~~~~~~~{\char0}~{\char5}--------------------{\char4}~{\char0}}
\hbox{\amgr ~~~~~~~~{\char5}------------------------{\char4}}
\hbox{\amgr ~}
\hbox{\amgr ~~~~~~~~{\char2}------------------------{\char3}}
\hbox{\amgr ~~~~~~~~{\char0}~{\char2}--------------------{\char3}~{\char0}}
\hbox{\amgr ~~~~~~~~{\char0}~{\char0}~~~~~~~~~~~~~~~~~~~~{\char15}-{\char16}~}
\hbox{\amgr
{}~~~~~~~~{\char0}~{\char0}~~~~~~~~~~~~~~~~~~~~{\char15}-{\char16}~~~~~~~~~~~~~~~~$\left\{ \eqalign {
+&\;\;\;\;\Delta H_{41}=4J{\char45}(h+\lambda)\cr
{\char45}&\;\;\;\;\Delta H_{42}=4J+(h{\char45}\lambda)\cr}\right.$}
\hbox{\amgr ~~~~~~~~{\char0}~{\char0}~~~~~~~~~~~~~~~~~~~~{\char0}~{\char0}}
\hbox{\amgr ~~~~~~~~{\char0}~{\char5}--------------------{\char4}~{\char0}}
\hbox{\amgr ~~~~~~~~{\char5}------------------------{\char4}}
\hbox{\amgr }
\hbox{\amgr ~~~~~~~~{\char2}------------------------{\char3}}
\hbox{\amgr
{}~~~~~~~~{\char0}~{\char2}------------------{\char18}-{\char3}~{\char0}}
\hbox{\amgr
{}~~~~~~~~{\char0}~{\char0}~~~~~~~~~~~~~~~~~~{\char5}-{\char16}~{\char0}}
\hbox{\amgr
{}~~~~~~~~{\char0}~{\char0}~~~~~~~~~~~~~~~~~~~~{\char0}~{\char0}~~~~~~~~~~~~~~~~$\left\{ \eqalign {
0&\;\;\;\;\Delta H_{51}=+(h+\lambda)\cr
{\char45}&\;\;\;\;\Delta H_{52}=8J+2h\cr}\right.$}
\hbox{\amgr ~~~~~~~~{\char0}~{\char0}~~~~~~~~~~~~~~~~~~~~{\char0}~{\char0}}
\hbox{\amgr ~~~~~~~~{\char0}~{\char5}--------------------{\char4}~{\char0}}
\hbox{\amgr ~~~~~~~~{\char5}------------------------{\char4}}
\hbox{\amgr }
\hbox{\amgr ~~~~~~~~{\char2}------------------------{\char3}}
\hbox{\amgr ~~~~~~~~{\char0}~{\char2}--------------------{\char3}~{\char0}}
\hbox{\amgr ~~~~~~~~{\char0}~{\char0}~~~~~~~~~~~~~~~~~~~~{\char0}~{\char0}}
\hbox{\amgr
{}~~~~~~~~{\char0}~{\char0}~~~~~~~~~~~~~~~~~~{\char2}-{\char16}~{\char0}~~~~~~~~~~~~~~~~$\left\{ \eqalign {
0&\;\;\;\;\Delta H_{61}=2J+(h+\lambda)\cr
{\char45}&\;\;\;\;\Delta H_{62}=12J+2h\cr}\right.$}
\hbox{\amgr
{}~~~~~~~~{\char0}~{\char0}~~~~~~~~~~~~~~~~~~{\char5}-{\char16}~{\char0}}
\hbox{\amgr ~~~~~~~~{\char0}~{\char5}--------------------{\char4}~{\char0}}
\hbox{\amgr ~~~~~~~~{\char5}------------------------{\char4}}
\hbox{\amgr }
\hbox{\amgr ~~~~~~~~{\char2}------------------------{\char3}}
\hbox{\amgr ~~~~~~~~{\char0}~{\char2}--------------------{\char3}~{\char0}}
\hbox{\amgr ~~~~~~~~{\char0}~{\char0}~~~~~~~~~~~~~~~~~~~~{\char0}~{\char0}}
\hbox{\amgr
{}~~~~~~~~{\char0}~{\char0}~~~~~~~~~~~~~~~~{\char2}-{\char3}~{\char0}~{\char0}~~~~~~~~~~~~~~~~$\left\{ \eqalign {
0&\;\;\;\;\Delta H_{71}=4J+(h+\lambda)\cr
{\char45}&\;\;\;\;\Delta H_{72}=16J+2h\cr}\right.$}
\hbox{\amgr
{}~~~~~~~~{\char0}~{\char0}~~~~~~~~~~~~~~~~{\char5}-{\char4}~{\char0}~{\char0}}
\hbox{\amgr ~~~~~~~~{\char0}~{\char5}--------------------{\char4}~{\char0}}
\hbox{\amgr ~~~~~~~~{\char5}------------------------{\char4}}}
\vskip 0.5 truecm
\par\noindent
\centerline {\smbfb Fig.4.2}
\endinsert
\par
Let us now continue the proof of Proposition 4.1.
\par
In order to prove $\equ (4.3)$ we observe that, starting from $C$ and
considering all the possible uphill path, one is able to examine
all the configurations in $\partial B$. The energy cost of all the possible
first steps of the above mentioned paths are given in Fig.4.2; here we mark
by a unitary square the site whose spin is changed and we denote by a couple
of positive integer numbers $(i,j)$ the generic first step
of our uphill path. We denote by
$C_{i,j}$ the configuration reached after the step $(i,j)$. We observe
that $C_{2,2}\in\partial B$ and that
$\Delta H_{ij}>\Delta H_{22}\;\;\forall (i,j){\not\in}\{ (5,1),(3,2),(2,2)\}$.
Hence, all the paths whose first step is different from (5,1) and (3,2) lead
to a boundary configuration whose energy is greater than $H(C_{2,2})$.
\par
Starting from $C_{3,2}$ or $C_{5,1}$ an uphill path can
continue by following one of the ways shown in Fig.4.2 and in Fig.4.3. It can
be easily shown that the steps $(8,j)$ can be neglected as well.
\par
\midinsert
\vskip 0.5 truecm
\vbox{\font\amgr=cmr10 at
10truept\baselineskip0.1466667truein\lineskiplimit-\maxdimen
\catcode`\-=\active\catcode`\~=\active\def~{{\char32}}\def-{{\char1}}%
\hbox{\amgr ~~~~~~~~{\char2}----------------------{\char3}}
\hbox{\amgr ~~~~~~~~{\char0}~{\char2}--------------------{\char14}-{\char3}~}
\hbox{\amgr ~~~~~~~~{\char0}~{\char0}~~~~~~~~~~~~~~~~~~~~{\char15}-{\char16}}
\hbox{\amgr
{}~~~~~~~~{\char0}~{\char0}~~~~~~~~~~~~~~~~~~~~{\char0}~{\char0}~~~~~~~~~~~~~~~~$\left\{ \eqalign {
+&\;\;\;\;\Delta H_{81}=6J{\char45}(h+\lambda)\cr
{\char45}&\;\;\;\;\Delta H_{82}=2J+(h{\char45}\lambda)\cr}\right.$}
\hbox{\amgr ~~~~~~~~{\char0}~{\char0}~~~~~~~~~~~~~~~~~~~~{\char0}~{\char0}}
\hbox{\amgr ~~~~~~~~{\char0}~{\char5}--------------------{\char4}~{\char0}}
\hbox{\amgr ~~~~~~~~{\char5}------------------------{\char4}}}
\vskip 0.5 truecm
\par\noindent
\centerline {\smbfb Fig.4.3}
\endinsert
\par
In conclusion, only the paths made by steps (3,2) and (5,1) can lead to a
configuration whose energy is lower than $H(C_{2,2})$.
\par
Now, let $\s$ be an acceptable configuration such that the following
conditions are satisfied:
there exists just one cluster $c^0$ of $0$ spins
which touches the sea of minuses namely the cluster $c^-$
winding around the torus, no minus spins are inside $c^0$;
all plus spins are in a unique cluster $c^+$
included in $c^0$ and ${\hat\g^+}={\check\g^0}$.
If $\s\in B$ then the following propositions are true:
\itemitem{$i)$} $R(c^0)\equiv$ the external rectangle $(l_1+2)\times (l_2+2)$
of
the frame $C$;
\itemitem{$ii)$} $R(c^+)\equiv$ the internal rectangle $l_1\times l_2$ of the
frame $C$;
\itemitem{$iii)$}
the intersection of each one of the four sides of $R(c^0)$ with
$\hat\g^0$ contains at least a segment of length greater or equal to 2;
\itemitem{$iv$)}the intersection of each one of the four sides of $R(c^+)$ with
$\hat\g^+$ contains at least a segment of length greater or equal to 2.
\par\noindent
We prove $(i)$ by absurd: let us suppose that $R(c^0)$ is different
from  $(l_1+2)\times (l_2+2)$ and that ${\hat\g^+}$ is a rectangle. We can
construct a downhill path which leads to a local minimum different from
$C$ by filling with $0$ spins the region $R(c^0)\setminus c^+$. Thus
$\s\not\in B$, and this is an absurd. $(ii)$ can be proved in a
similar way. $(iii)$ is proved by absurd as well: suppose that the intersection
between $\hat\g^0$ and one of the sides of $R(c^0)$
contains only isolated intervals of length
1, namely there
is a certain number of spins $0$ with three minus spins among their nearest
neighbour sites. By
changing this $0$ spins into $-1$ we construct a configuration at a lower
energy
level and characterized by a cluster of $0$ spins $c'^0$ such that $R(c'^0)$
is different from the rectangle $(l_1+2)\times (l_2+2)$; then there
exists a downhill path which connects $\s$ to a local minimum different from
$C$. Hence the absurd $\s\not\in B$ is obtained. $(iv)$ is
proved in a similar way.
\par
But, as we noticed before, all the uphill paths starting from $C$ and
leading to configurations in $\partial B$  with energy smaller than
$H(C_{2,2})$ necessarily can only be made by steps
$(5,1)$ and $(3,2)$.
\par
It is clear that, by virtue of the necessary conditions stated above,
we cannot reach $\partial B$ starting from $C$ with less than $l-1$ steps
(5.1). On the other hand, since $S_1\in\partial B$, with more that $l-1$
steps $(5.1)$ we certainly get an energy larger that $H(S_1)$ and so a
configuration which cannot be of minimal energy in $\partial B$.
\par
In this way we can only reach configurations with a unique cluster of
pluses, so any boundary configuration with minimal energy
is characterized by an external cluster $c^0$, such that the
intersection between ${\hat\g^0}$ and all the sides of $R(c^0)$ is at least
of length 2, and an internal cluster $c^+$, such that the intersection between
${\hat\g^+}$ and one of the sides of $R(c^+)$ has length 1 (see Fig.4.4). Among
all these configurations it is easily seen that the one with lowest energy is
$S_1$.
\par
\midinsert
\vskip 0.5 truecm
\vbox{\font\amgr=cmr10 at
10truept\baselineskip0.1466667truein\lineskiplimit-\maxdimen
\catcode`\-=\active\catcode`\~=\active\def~{{\char32}}\def-{{\char1}}%
\hbox{\amgr
{}~~~~~~~~~~~~~~~~~~~~~~~~~~{\char2}---------{\char18}---{\char18}----------------{\char3}}
\hbox{\amgr
{}~~~~~~~~~~~~~~~~~~~~~~~~~~{\char0}~~~~~~~~~{\char0}~~~{\char0}~~~~~~~~~~~~~{\char45}~~{\char0}}
\hbox{\amgr
{}~~~~~~~~~~~~~~~~~~~~~~~~~~{\char0}~~{\char45}~~{\char2}---{\char4}~~~{\char5}------------{\char3}~~~{\char0}}
\hbox{\amgr
{}~~~~~~~~~~~~~~~~~~~~~~~~~~{\char0}~~~~~{\char0}~~~~~{\char2}-----{\char3}~~~~0~~~{\char0}~~~{\char0}}
\hbox{\amgr
{}~~~~~~~~~~~~~~~~~~~~~~~~~~{\char15}-----{\char4}~~0~~{\char0}~~+~~{\char0}~~~~~~~~{\char0}~{\char45}~{\char0}}
\hbox{\amgr
{}~~~~~~~~~~~~~~~~~~~~~~~~~~{\char0}~~~~~~~{\char2}---{\char4}~~~~~{\char5}----{\char3}~~~{\char5}---{\char16}}
\hbox{\amgr
{}~~~~~~~~~~~~~~~~~~~~~~~~~~{\char15}---{\char3}~~~{\char5}-----{\char3}~~~~~~~~{\char0}~~~~~~~{\char0}}
\hbox{\amgr
{}~~~~~~~~~~~~~~~~~~~~~~~~~~{\char0}~~~{\char0}~~~~~~~~~{\char0}~~~+~~~~{\char0}~~~{\char2}---{\char16}}
\hbox{\amgr
{}~~~~~~~~~~~~~~~~~~~~~~~~~~{\char0}~~~{\char5}-----{\char3}~~~{\char5}--------{\char4}~0~{\char0}~~~{\char0}}
\hbox{\amgr
{}~~~~~~~~~~~~~~~~~~~~~~~~~~{\char0}~~~~~~~~~{\char0}~~~~~~~~~~~~~~~~{\char0}~~~{\char0}}
\hbox{\amgr
{}~~~~~~~~~~~~~~~~~~~~~~~~~~{\char0}~~~{\char45}~~~~~{\char0}~0~{\char2}------------{\char4}~~~{\char0}}
\hbox{\amgr
{}~~~~~~~~~~~~~~~~~~~~~~~~~~{\char5}---------{\char17}---{\char17}----------------{\char4}}}
\vskip 0.5 truecm
\par\noindent
\centerline {\smbfb Fig.4.4}
\endinsert
\par
In conclusion we have to compare $H(S_1)$ with $H(C_{2,2})$.
Equality $\equ (4.3)$ follows from $l<{\widetilde L}$, $H(S_1)-H(C)=
(h+\l)(l-1)$ and $H(C_{2,2})-H(C)=2J-(h-\l)$.
\par
Now we want to apply to the description of the first escape from $B$
the approach developed in [OS1], which is based on the properties of
the above defined sets called cycles.
\par
It is easy to see that the basin of attraction $B:=B(C(l_1,l_2))$
defined in $\equ (2.6b)$ satisfies the following properties:
\itemitem{$i)$} $B$ is connected;
\itemitem{$ii)$} $S_1\subset\partial B$, and
$$\min_{\s\in\partial B} H(\s)=H(S_1), \;
  \min_{\s\in\partial B\setminus S_1} H(\s)>H(S_1)$$
\itemitem{$iii)$} $\forall \h\in S_1$ there exists a path
$\o :\h\rightarrow C$ such that $\forall\s\in\o\setminus \{\h\}$ one
has $\s\in B$ and $H(\s)<H(S_1)$.
\par
As it was noticed in [OS1] (see Proposition 3.4 therein), properties $i),\;
ii)$ and $iii)$ imply that the set ${\bar B}$ defined as the maximal
connected set containing $C$ and with energy less then
$H(S_1)$ is a cycle with $S_1$ belonging to its boundary
$\partial {\bar B}$. Moreover we notice, here, the following obvious
properties:
\itemitem{$\bullet$} any point $\h\in\partial {\bar B}$ necessarily is such
that
$H(\h)\geq H(S_1)$;
\itemitem{$\bullet$} if $H(\h)=H(S_1)$ and $\h\not\in\partial B$
necessarily any downhill path starting from $\h$ ends in $C$.
\par
We recall that, given any set $A\subset \O_{\L}$, we have denoted
by ${\cal S}(A)$ the possibly empty subset of $U(A)$ (see $\equ (2.6''')$),
which is downhill connected to $A^c$; ${\cal S}(A)$ was called the set of
minimal saddles of $A$. We can write
$$S({\bar B})=S_1\;\; .\Eq (4.3.3)$$
\par
{}From Proposition 3.7 in [OS1], from reversibility of the dynamics (see Lemma
1 in [KO1]) and from $\equ (4.3.3)$ we easily get that
$\forall\s\in {\bar B}$
$$\lim_{\b\to\infty} P_{\s}(\s_{\t_{(B\cup\partial B)^c}-1}\in
                            S_1)=1\;\; .\Eq (4.3.4)$$
\par
Since $H(S_1)-H(C)=(h+\l)(l-1)$ we deduce that for every
$\e>0$
$$\lim_{\b\to\infty} P_C(
  e^{\b (h+\l)(l-1)-\b\e}<\t_{\partial B}<e^{\b (h+\l)(l-1)+\b\e})=1
\;\; .\Eq (4.3.6)$$
\par
Up to now we have described how the system reaches $\partial B$ starting from
$C$; now we want to describe its further evolution.
\par
Two things can happen: the system gets back to $B$ or it goes to the
birectangle $R_1:=R(l_1-1,l;l_1+2,l+2)$ (see Fig.4.5); we have supposed,
without loss of generality, that $l=l_2$.
\par
\midinsert
\vskip 0.5 truecm
\vbox{\font\amgr=cmr10 at
10truept\baselineskip0.1466667truein\lineskiplimit-\maxdimen
\catcode`\-=\active\catcode`\~=\active\def~{{\char32}}\def-{{\char1}}%
\hbox{\amgr
{}~~~~~{\char2}-------------------{\char3}~~~~~{\char2}-------------------{\char3}~~~~~{\char2}-------------%
------{\char3}}
\hbox{\amgr
{}~~~~~{\char0}~{\char2}-------------{\char3}~~~{\char0}~~~~~{\char0}~{\char2}-------------{\char3}~~~{\char0}%
{}~~~~~{\char0}~{\char2}-------------{\char3}~~~{\char0}}
\hbox{\amgr
{}~~~~~{\char0}~{\char0}~~~~~~~~~~~~~{\char0}~~~{\char0}~~~~~{\char0}~{\char0}~~~~~~~~~~~~~{\char0}~~~{\char0}%
{}~~~~~{\char0}~{\char0}~~~~~~~~~~~~~{\char0}~~~{\char0}}
\hbox{\amgr
{}~~~~~{\char0}~{\char0}~~~~~~~~~~~~~{\char0}~~~{\char0}~~~~~{\char0}~{\char0}~~~~~~~~~~~~~{\char5}-{\char3}%
{}~{\char0}~~~~~{\char0}~{\char0}~~~~~~~~~~~~~{\char5}-{\char3}~{\char0}}
\hbox{\amgr
{}~~~~~{\char0}~{\char0}~~~~~~~~~~~~~{\char0}~~~{\char0}~<-~~{\char0}~{\char0}~~~~~~~~~~~~~{\char2}-{\char4}%
{}~{\char0}~~->~{\char0}~{\char0}~~~~~~~~~~~~~~~{\char0}~{\char0}}
\hbox{\amgr
{}~~~~~{\char0}~{\char0}~~~~~~~~~~~~~{\char0}~~~{\char0}~~~~~{\char0}~{\char0}~~~~~~~~~~~~~{\char0}~~~{\char0}%
{}~~~~~{\char0}~{\char0}~~~~~~~~~~~~~{\char2}-{\char4}~{\char0}}
\hbox{\amgr
{}~~~~~{\char0}~{\char0}~~~~~~~~~~~~~{\char0}~~~{\char0}~~~~~{\char0}~{\char0}~~~~~~~~~~~~~{\char0}~~~{\char0}%
{}~~~~~{\char0}~{\char0}~~~~~~~~~~~~~{\char0}~~~{\char0}}
\hbox{\amgr
{}~~~~~{\char0}~{\char5}-------------{\char4}~~~{\char0}~~~~~{\char0}~{\char5}-------------{\char4}~~~{\char0}%
{}~~~~~{\char0}~{\char5}-------------{\char4}~~~{\char0}}
\hbox{\amgr
{}~~~~~{\char5}-------------------{\char4}~~~~~{\char5}-------------------{\char4}~~~~~{\char5}-------------%
------{\char4}}}
\vskip 0.5 truecm
\par\noindent
\centerline {\smbfb Fig.4.5}
\endinsert
\par
In Appendix 1 we give a general argument showing that, with high
probability, our process, possibly after many attempts,
soon or later, will eventually get out of $B \cup \partial B$
through $S_1$ reaching $R_1$ before
touching any other local minimum
and:
$$\eqalign {
\lim_{\beta\to\infty}P_{C}&(\tau_{R_{1}}<\tau_{\piuuno})=1\cr
\lim_{\beta\to\infty}P_{C}&(\tau_{R_{1}}<e^{\beta(h+\lambda)(l-1)
+\beta\e})=1\cr}\;\; .\Eq (4.13)$$
\par
Now we have to describe the further evolution of our Markov chain starting from
the
birectangle $R_1$. We denote by $B_1:=B(R_1)$ the basin of attraction of
$R_1$. Let us first consider the case $\min\{l_1-1,l\}=l$ (this is equivalent
to suppose that $C$ is not a squared frame). We denote by $S_2$
the configuration obtained by changing into minus  $l+1$ of the $0$ spins on
the ``free" side of the external rectangle and
by $S_3$ the configuration obtained by changing into zero $l-1$ of the
plus spins of one of the shortest sides of the internal rectangle of $R_1$
(see Fig.4.6). The following is true:
$$\min_{\sigma\in\partial B_{1}}H(\sigma)=\left\{ \eqalign {
H(S_{3})&\;\; {\rm if\;} l<\left[{h\over\lambda}\right]+1\cr
H(S_{2})&\;\; {\rm if\;} l\geq \left[{h\over\lambda}\right]+1\cr}\right.
\;\; .\Eq (4.14)$$
\midinsert
\vskip 0.5 truecm
\vbox{\font\amgr=cmr10 at
10truept\baselineskip0.1466667truein\lineskiplimit-\maxdimen
\catcode`\-=\active\catcode`\~=\active\def~{{\char32}}\def-{{\char1}}%
\hbox{\amgr ~~~~~~~~~{\char2}-----------------{\char3}~~$\phantom
{S_{2}}$~~~~~~~~~~~~~~~{\char2}-------------------{\char3}~~}
\hbox{\amgr ~~~~~~~~~{\char0}~{\char2}-------------{\char3}~{\char0}~~$\phantom
{S_{2}}$~~~~~~~~~~~~~~~{\char0}~{\char2}-----------{\char3}~%
{}~~~~{\char0}~~}
\hbox{\amgr ~~~~~~~~~{\char0}~{\char0}~~~~~~~~~~~~~{\char0}~{\char0}~~$\phantom
{S_{2}}$~~~~~~~~~~~~~~~{\char0}~{\char0}~~~~~~~~~~~{\char0}~%
{}~~~~{\char0}~~}
\hbox{\amgr ~~~~~~~~~{\char0}~{\char0}~~~~~~~~~~~~~{\char0}~{\char0}~~$\phantom
{S_{2}}$~~~~~~~~~~~~~~~{\char0}~{\char0}~~~~~~~~~~~{\char0}~%
{}~~~~{\char0}~~}
\hbox{\amgr
{}~~~~~~~~~{\char0}~{\char0}~~~~~~~~~~~~~{\char0}~{\char0}~~~~$S_{2}$~~~~~~~~~~~~~{\char0}~{\char0}~~~~~~~~~~~{\char0}~~~~~{\char0}~~$S_{3}$}
\hbox{\amgr ~~~~~~~~~{\char0}~{\char0}~~~~~~~~~~~~~{\char0}~{\char0}~~$\phantom
{S_{2}}$~~~~~~~~~~~~~~~{\char0}~{\char0}~~~~~~~~~~~{\char0}~%
{}~~~~{\char0}~~}
\hbox{\amgr ~~~~~~~~~{\char0}~{\char0}~~~~~~~~~~~~~{\char0}~{\char0}~~$\phantom
{S_{2}}$~~~~~~~~~~~~~~~{\char0}~{\char0}~~~~~~~~~~~{\char5}-%
{\char3}~~~{\char0}}
\hbox{\amgr
{}~~~~~~~~~{\char0}~{\char5}-------------{\char4}~{\char5}-{\char3}~~$\phantom
{S_{2}}$~~~~~~~~~~~~~{\char0}~{\char5}-------------%
{\char4}~~~{\char0}~~}
\hbox{\amgr ~~~~~~~~~{\char5}-------------------{\char4}~~$\phantom
{S_{2}}$~~~~~~~~~~~~~{\char5}-------------------{\char4}~~}}
\vskip 0.5 truecm
\par\noindent
\centerline {\smbfb Fig.4.6}
\endinsert
\par
Equality $\equ (4.14)$ can be proved with arguments similar to those used in
the case of the local minimum $C(l_1,l_2)$ and observing that
$H(S_{2})-H(R_{1})=(h-\lambda)(l+1)$ and
$H(S_{3})-H(R_{1})=(h+\lambda)(l-1)$ even though, in this case, there are
other possible first steps (with
high increment in energy).
They are shown in Fig.4.7.
\midinsert
\vskip 0.5 truecm
\vbox{\font\amgr=cmr10 at
10truept\baselineskip0.1466667truein\lineskiplimit-\maxdimen
\catcode`\-=\active\catcode`\~=\active\def~{{\char32}}\def-{{\char1}}%
\hbox{\amgr ~~~~~~~~{\char2}--------------------{\char3}}
\hbox{\amgr ~~~~~~~~{\char0}~{\char2}--------------{\char3}~~~{\char0}}
\hbox{\amgr
{}~~~~~~~~{\char0}~{\char0}~~~~~~~~~~~~~~{\char0}~{\char2}-{\char16}~}
\hbox{\amgr
{}~~~~~~~~{\char0}~{\char0}~~~~~~~~~~~~~~{\char0}~{\char5}-{\char16}~~~~~~~~~~~~~~~~$\left\{ \eqalign {
+&\;\;\;\;\Delta H_{91}=6J{\char45}(h+\lambda)\cr
{\char45}&\;\;\;\;\Delta H_{82}=2J+(h{\char45}\lambda)\cr}\right.$}
\hbox{\amgr ~~~~~~~~{\char0}~{\char0}~~~~~~~~~~~~~~{\char0}~~~{\char0}}
\hbox{\amgr ~~~~~~~~{\char0}~{\char5}--------------{\char4}~~~{\char0}}
\hbox{\amgr ~~~~~~~~{\char5}--------------------{\char4}}
\hbox{\amgr }
\hbox{\amgr }
\hbox{\amgr ~~~~~~~~{\char2}--------------------{\char3}}
\hbox{\amgr ~~~~~~~~{\char0}~{\char2}--------------{\char18}-{\char3}~{\char0}}
\hbox{\amgr
{}~~~~~~~~{\char0}~{\char0}~~~~~~~~~~~~~~{\char15}-{\char4}~{\char0}~}
\hbox{\amgr
{}~~~~~~~~{\char0}~{\char0}~~~~~~~~~~~~~~{\char0}~~~{\char0}~~~~~~~~~~~~~~~~$\left\{ \eqalign {
+&\;\;\;\;\Delta H_{10\; 1}=2J{\char45}(h+\lambda)\cr
{\char45}&\;\;\;\;\Delta H_{10\; 2}=6J+(h{\char45}\lambda)\cr}\right.$}
\hbox{\amgr ~~~~~~~~{\char0}~{\char0}~~~~~~~~~~~~~~{\char0}~~~{\char0}}
\hbox{\amgr ~~~~~~~~{\char0}~{\char5}--------------{\char4}~~~{\char0}}
\hbox{\amgr ~~~~~~~~{\char5}--------------------{\char4}}}
\vskip 0.5 truecm
\par\noindent
\centerline {\smbfb Fig.4.7}
\endinsert
\par
With arguments similar to those used before we get that
the typical time of first escape from $\partial B_1$ is of the order of
$e^{\beta[\min_{\sigma\in\partial B_{1}}H(\sigma)-H(R_{1})]}$ and that
the system hits for the first time the boundary $\partial B_1$ in
$S_3$ if $l<\left[{h\over \l}\right]+1$ and in
$S_2$ if $l\geq\left[{h\over \l}\right]+1$.
Notice that if $1 < {h\over \l} < 2$ the integer
$\left[{h\over \l}\right]+1$ equals $2$ so that $S_2$
is  preferred.
\par
We have that, with probability tending to 1 as $\b\rightarrow\infty$, our
droplet continues its contraction:
the system reaches another local minimum $R_2$ strictly contained in $R_1$,
that is
$$R_{2}\prec R_{1}\prec C\;\; ;\Eq (4.15)$$
where we have introduced the following {\it partial order relation} in
$\O_{\L}$
$$\sigma\prec\eta\Leftrightarrow \sigma (x)\leq \eta (x)\; \forall x\in\L\;\;
.\Eq (4.16)$$
We also have that, given $\e>0$,
$e^{\beta[\min_{\sigma\in\partial B_{1}}H(\sigma)-H(R_{1})]+\beta\e}$
is an upper bound, in the limit $\b\to\infty$ to the first hitting
time to $R_2$ of the Markov chain starting from $R_1$.
\par
In conclusion we can say that the Markov chain starting from $C$ visits
smaller and smaller local minima untill it reaches the configuration
$\menouno$; this completes the proof of the statement
$P_C(\t_{\menouno}<\t_{\piuuno})
{\buildrel\beta\rightarrow\infty\over\longrightarrow}\; 1$.
\par
Each step of the shrinking process is characterized by a typical time
$t_{\beta}$ whose asymptotic behaviour, exponentially in $\beta$, is known
in the sense that we control
$$\lim_{\beta\to\infty} {1\over \beta} \log t_{\beta}\;\; ;$$
we say that:
$$t^1_{\beta}, t^2_{\beta}\; {\rm are} \; logarithmically\; equivalent\;
\Leftrightarrow \; \lim_{\b\to\infty} {1\over\b} t^1_{\beta}=
\lim_{\b\to\infty} {1\over\b} t^2_{\beta}\;\; .$$
\par
By using Markov property we can say that the typical time of the whole
shrinking event is given by the largest time among all the {\it partial
shrinking times}. Then the proof of Proposition 4.1 is completed in the case
$l<{\widetilde L}$ when $C$ is a rectangular frame.
\par
Next, we consider the case when $C$ is a squared frame: the boundary
configuration $S_3$ is now the one represented in Fig.4.8,
$H(S_3)-H(R_1)=(h+\l)(l-2)$ and $\min_{\s\in\partial B_1} H(\s)
=H({\cal S}_3)$ if $l<\left[ {3\over 2}{h\over\l}+{1\over 2}\right] +1$.
We obtain results similar to those obtained in the previous case of a
rectangular frame.
\midinsert
\vskip 0.5 truecm
\vbox{\font\amgr=cmr10 at
10truept\baselineskip0.1466667truein\lineskiplimit-\maxdimen
\catcode`\-=\active\catcode`\~=\active\def~{{\char32}}\def-{{\char1}}%
\hbox{\amgr
{}~~~~~~~~~~~~~~~~~~~~~~~~~~~~~~~~~~~{\char2}---------------{\char3}~~}
\hbox{\amgr
{}~~~~~~~~~~~~~~~~~~~~~~~~~~~~~~~~~~~{\char0}~{\char2}-{\char3}~~~~~~~~~~~{\char0}~~}
\hbox{\amgr
{}~~~~~~~~~~~~~~~~~~~~~~~~~~~~~~~~~~~{\char0}~{\char0}~{\char5}-------{\char3}~~~{\char0}~~~~}
\hbox{\amgr
{}~~~~~~~~~~~~~~~~~~~~~~~~~~~~~~~~~~~{\char0}~{\char0}~~~~~~~~~{\char0}~~~{\char0}~~}
\hbox{\amgr
{}~~~~~~~~~~~~~~~~~~~~~~~~~~~~~~~~~~~{\char0}~{\char0}~~~~~~~~~{\char0}~~~{\char0}~~$l+2$}
\hbox{\amgr
{}~~~~~~~~~~~~~~~~~~~~~~~~~~~~~~~~~~~{\char0}~{\char0}~~~~~~~~~{\char0}~~~{\char0}~~}
\hbox{\amgr
{}~~~~~~~~~~~~~~~~~~~~~~~~~~~~~~~~~~~{\char0}~{\char0}~~~~~~~~~{\char0}~~~{\char0}~}
\hbox{\amgr
{}~~~~~~~~~~~~~~~~~~~~~~~~~~~~~~~~~~~{\char0}~{\char5}---------{\char4}~~~{\char0}~~}
\hbox{\amgr
{}~~~~~~~~~~~~~~~~~~~~~~~~~~~~~~~~~~~{\char5}---------------{\char4}~~}
\hbox{\amgr ~~~~~~~~~~~~~~~~~~~~~~~~~~~~~~~~~~~~~~~~~$l+2$}
\hbox{\amgr }}
\vskip 0.5 truecm
\par\noindent
\centerline {\smbfb Fig.4.8}
\endinsert
\par
Now we consider the frame $C:=C(l_1,l_2)$; we suppose that
${\widetilde L}\leq l:=\min \{l_1,l_2\}<l^*$ and
$m:=\max \{l_1,l_2\}<m^*(l)$; we denote by $B$ the basin of attraction of
the frame $C$ and by ${\partial B}$ its boundary.
We denote by $S_4$ the set of configurations obtained by attaching
a unit square protuberance (with a zero spin) to
one of the four sides of the external rectangle of
$C$ (see Fig.4.9). By considering all the uphill paths starting from
$C$ it can be proved that
$$\min_{\s\in\partial B} H(\s)=H(S_4),\;\;
  \min_{\s\in\partial B\setminus S_4} H(\s)>H(S_4)
\;\; ,\Eq (4.17)$$
namely
$$U(B)=S_4\;\; ;\Eq (4.17.1)$$
we remark that $H(S_4)-H(C)=2J-(h-\l)$.
\par
\bigskip
\midinsert
\vskip 0.5 truecm
\vbox{\font\amgr=cmr10 at
10truept\baselineskip0.1466667truein\lineskiplimit-\maxdimen
\catcode`\-=\active\catcode`\~=\active\def~{{\char32}}\def-{{\char1}}%
\hbox{\amgr ~~~~~~~~$\phantom
{S_{4,\perp}}$~~~~~{\char2}---------------------{\char3}~~~~~~~~~{\char2}---------------------{\char3}}
\hbox{\amgr ~~~~~~~~$\phantom
{S_{4,\perp}}$~~~~~{\char0}~{\char2}-----------------{\char3}~{\char0}~~~~~~~~~{\char0}~{\char2}-----------------{\char3}~{\char0}}
\hbox{\amgr ~~~~~~~~$\phantom
{S_{4,\perp}}$~~~~~{\char0}~{\char0}~~~~~~~~~~~~~~~~~{\char0}~{\char5}-{\char3}~~~~~{\char2}-{\char4}~{\char0}~~~~~~~~~~~~~~~~~{\char0}~{\char0}}
\hbox{\amgr ~~~~~~~~$
{S_{4,\perp}}$~~~~~{\char0}~{\char0}~~~~~~~~~~~~~~~~~{\char0}~{\char2}-{\char4}~~~~~{\char5}-{\char3}~{\char0}~~~~~~~~~~~~~~~~~{\char0}~{\char0}~$l+2$}
\hbox{\amgr ~~~~~~~~$\phantom
{S_{4,\perp}}$~~~~~{\char0}~{\char0}~~~~~~~~~~~~~~~~~{\char0}~{\char0}~~~~~~~~~{\char0}~{\char0}~~~~~~~~~~~~~~~~~{\char0}~{\char0}}
\hbox{\amgr ~~~~~~~~$\phantom
{S_{4,\perp}}$~~~~~{\char0}~{\char5}-----------------{\char4}~{\char0}~~~~~~~~~{\char0}~{\char5}-----------------{\char4}~{\char0}}
\hbox{\amgr ~~~~~~~~$\phantom
{S_{4,\perp}}$~~~~~{\char5}---------------------{\char4}~~~~~~~~~{\char5}---------------------{\char4}}
\hbox{\amgr ~~~~~~~~$\phantom
{S_{4,\perp}}$~~~~~~~~~~~~~$m+2$~~~~~~~~~~~~~~~~~~~~~~~~~~~~~$m+2$~~~~~~~~~~~~}
\hbox{\amgr }
\hbox{\amgr }
\hbox{\amgr ~~~~~~~~$\phantom
{S_{4,\parallel}}$~~~~~~~~~~~~~~{\char2}-{\char3}}
\hbox{\amgr ~~~~~~~~$\phantom
{S_{4,\parallel}}$~~~~~{\char2}--------{\char4}~{\char5}----------{\char3}~~~~~~~~~{\char2}---------------------{\char3}}
\hbox{\amgr ~~~~~~~~$\phantom
{S_{4,\parallel}}$~~~~~{\char0}~{\char2}-----------------{\char3}~{\char0}~~~~~~~~~{\char0}~{\char2}-----------------{\char3}~{\char0}}
\hbox{\amgr ~~~~~~~~$\phantom
{S_{4,\parallel}}$~~~~~{\char0}~{\char0}~~~~~~~~~~~~~~~~~{\char0}~{\char0}~~~~~~~~~{\char0}~{\char0}~~~~~~~~~~~~~~~~~{\char0}~{\char0}}
\hbox{\amgr ~~~~~~~~$
{S_{4,\parallel}}$~~~~~{\char0}~{\char0}~~~~~~~~~~~~~~~~~{\char0}~{\char0}~~~~~~~~~{\char0}~{\char0}~~~~~~~~~~~~~~~~~{\char0}~{\char0}~$l+2$}
\hbox{\amgr ~~~~~~~~$\phantom
{S_{4,\parallel}}$~~~~~{\char0}~{\char0}~~~~~~~~~~~~~~~~~{\char0}~{\char0}~~~~~~~~~{\char0}~{\char0}~~~~~~~~~~~~~~~~~{\char0}~{\char0}}
\hbox{\amgr ~~~~~~~~$\phantom
{S_{4,\parallel}}$~~~~~{\char0}~{\char5}-----------------{\char4}~{\char0}~~~~~~~~~{\char0}~{\char5}-----------------{\char4}~{\char0}}
\hbox{\amgr ~~~~~~~~$\phantom
{S_{4,\parallel}}$~~~~~{\char5}---------------------{\char4}~~~~~~~~~{\char5}--------{\char3}~{\char2}----------{\char4}}
\hbox{\amgr ~~~~~~~~$\phantom
{S_{4,\parallel}}$~~~~~~~~~~~~~~~~~~~~~~~~~~~~~~~~~~~~~~~~~~~~~~{\char5}-{\char4}~~}
\hbox{\amgr ~~~~~~~~$\phantom
{S_{4,\parallel}}$~~~~~~~~~~~~~$m+2$~~~~~~~~~~~~~~~~~~~~~~~~~~~~~$m+2$~~~~~~~~~~~~}}
\vskip 0.5 truecm
\par\noindent
\centerline {\smbfb Fig.4.9}
\endinsert
\par
Without loss of generality we suppose that $l=l_2$ and $m=l_1$. By
arguments similar to those used before it can be proved that in a typical
time $e^{[2J-(h-\l)]}$ the Markov chain starting from $C$,
with high probability, will visit
$R_{2,\perp}:=R(l_1,l_2;l_1+3,l_2+2)$ or
$R_{2,\parallel}:=R(l_1,l_2;l_1+2,l_2+3)$. The symbol $\perp$ denotes the
fact that the frame is growing in a direction perpendicular to its shortest
side (see Fig.4.10).
\midinsert
\vskip 0.5 truecm
\vbox{\font\amgr=cmr10 at
10truept\baselineskip0.1466667truein\lineskiplimit-\maxdimen
\catcode`\-=\active\catcode`\~=\active\def~{{\char32}}\def-{{\char1}}%
\hbox{\amgr ~~~~~~~~$\phantom
{R_{2,\perp}}$~~~~~{\char2}-----------------------{\char3}~~~~~{\char2}-----------------------{\char3}}
\hbox{\amgr ~~~~~~~~$\phantom
{R_{2,\perp}}$~~~~~{\char0}~{\char2}-----------------{\char3}~~~{\char0}~~~~~{\char0}~~~{\char2}-----------------{\char3}~{\char0}}
\hbox{\amgr ~~~~~~~~$\phantom
{R_{2,\perp}}$~~~~~{\char0}~{\char0}~~~~~~~~~~~~~~~~~{\char0}~~~{\char0}~~~~~{\char0}~~~{\char0}~~~~~~~~~~~~~~~~~{\char0}~{\char0}}
\hbox{\amgr ~~~~~~~~$
{R_{2,\perp}}$~~~~~{\char0}~{\char0}~~~~~~~~~~~~~~~~~{\char0}~~~{\char0}~~~~~{\char0}~~~{\char0}~~~~~~~~~~~~~~~~~{\char0}~{\char0}~$l+2$}
\hbox{\amgr ~~~~~~~~$\phantom
{R_{2,\perp}}$~~~~~{\char0}~{\char0}~~~~~~~~~~~~~~~~~{\char0}~~~{\char0}~~~~~{\char0}~~~{\char0}~~~~~~~~~~~~~~~~~{\char0}~{\char0}}
\hbox{\amgr ~~~~~~~~$\phantom
{R_{2,\perp}}$~~~~~{\char0}~{\char5}-----------------{\char4}~~~{\char0}~~~~~{\char0}~~~{\char5}-----------------{\char4}~{\char0}}
\hbox{\amgr ~~~~~~~~$\phantom
{R_{2,\perp}}$~~~~~{\char5}-----------------------{\char4}~~~~~{\char5}-----------------------{\char4}}
\hbox{\amgr ~~~~~~~~$\phantom
{R_{2,\perp}}$~~~~~~~~~~~~~$m+3$~~~~~~~~~~~~~~~~~~~~~~~~~~$m+3$~~~~~~~~~~~~~~~}
\hbox{\amgr }
\hbox{\amgr }
\hbox{\amgr ~~~~~~~~$\phantom
{R_{2,\parallel}}$~~~~~{\char2}---------------------{\char3}~~~~~~~~~}
\hbox{\amgr ~~~~~~~~$\phantom
{R_{2,\parallel}}$~~~~~{\char0}~~~~~~~~~~~~~~~~~~~~~{\char0}~~~~~~~~~{\char2}---------------------{\char3}}
\hbox{\amgr ~~~~~~~~$\phantom
{R_{2,\parallel}}$~~~~~{\char0}~{\char2}-----------------{\char3}~{\char0}~~~~~~~~~{\char0}~{\char2}-----------------{\char3}~{\char0}}
\hbox{\amgr ~~~~~~~~$\phantom
{R_{2,\parallel}}$~~~~~{\char0}~{\char0}~~~~~~~~~~~~~~~~~{\char0}~{\char0}~~~~~~~~~{\char0}~{\char0}~~~~~~~~~~~~~~~~~{\char0}~{\char0}}
\hbox{\amgr ~~~~~~~~$
{R_{2,\parallel}}$~~~~~{\char0}~{\char0}~~~~~~~~~~~~~~~~~{\char0}~{\char0}~~~~~~~~~{\char0}~{\char0}~~~~~~~~~~~~~~~~~{\char0}~{\char0}~$l+3$}
\hbox{\amgr ~~~~~~~~$\phantom
{R_{2,\parallel}}$~~~~~{\char0}~{\char0}~~~~~~~~~~~~~~~~~{\char0}~{\char0}~~~~~~~~~{\char0}~{\char0}~~~~~~~~~~~~~~~~~{\char0}~{\char0}}
\hbox{\amgr ~~~~~~~~$\phantom
{R_{2,\parallel}}$~~~~~{\char0}~{\char5}-----------------{\char4}~{\char0}~~~~~~~~~{\char0}~{\char5}-----------------{\char4}~{\char0}}
\hbox{\amgr ~~~~~~~~$\phantom
{R_{2,\parallel}}$~~~~~{\char5}---------------------{\char4}~~~~~~~~~{\char0}~~~~~~~~~~~~~~~~~~~~~{\char0}}
\hbox{\amgr ~~~~~~~~$\phantom
{R_{2,\parallel}}$~~~~~~~~~~~~~~~~~~~~~~~~~~~~~~~~~~~~~{\char5}---------------------{\char4}}
\hbox{\amgr ~~~~~~~~$\phantom
{R_{2,\parallel}}$~~~~~~~~~~~~~$m+2$~~~~~~~~~~~~~~~~~~~~~~~~~~~~~$m+2$~~~~~~~~~~~~}}
\vskip 0.5 truecm
\par\noindent
\centerline {\smbfb Fig.4.10}
\endinsert
\par
We denote by $B_{2,\perp}$ and $B_{2,\parallel}$ the basins of attraction
of $R_{2,\perp}$ and $R_{2,\parallel}$; their boundaries are
respectively denoted by $\partial B_{2,\perp}$ and $\partial B_{2,\parallel}$.
By considering all the uphill paths starting from $R_{2,\perp}$ and
$R_{2,\parallel}$ we get that
$$\eqalign {
\min_{\sigma\in\partial B_{2,\perp}}H(\sigma)=H(S_{4})\cr
\min_{\sigma\in\partial B_{2,\parallel}}H(\sigma)=H(S_{4})\cr}
\;\; ,\Eq (4.18)$$
more precisely
$$U(B_{2,\perp})=S_4,\;\; U(B_{2,\parallel})=S_4
\;\; .\Eq (4.18.1)$$
Indeed the most relevant inequalities in the proof of $\equ (4.18)$ are the
following ones
$$\eqalign{
(h+\lambda)(l-1)>&2J-(h-\lambda)>2J-(h+\lambda)>(h-\lambda)(l+1)\cr
(h+\lambda)(l-1)>&2J-(h-\lambda)>2J-(h+\lambda)>(h-\lambda)(m+1)\cr}\;\; .
\Eq (4.19)$$
We remark that $H(S_{4})-H(R_{2,\perp})=(h-\lambda)(l+1)$ and
$H(S_{4})-H(R_{2,\parallel})=(h-\lambda)(m+1)$.
In order to prove the first one of the equalities $\equ (4.19)$
we notice that
$$\eqalign {
l\geq {\widetilde L}&\Rightarrow (h+\lambda)(l-1)>2J-(h-\lambda)\cr
l+2< {\widetilde M}&\Rightarrow 2J-(h+\lambda)>(h-\lambda)(l+1)\cr}\;\; .$$
In order to prove the second one we notice that
$$l\geq {\widetilde L}\Rightarrow m^{*}(l)+2\leq {\widetilde M}$$
and that
$$m< m^{*}(l)\Rightarrow m+2<{\widetilde M}\Rightarrow
(h-\lambda)(m+1)<2J-(h+\lambda)\;\; .$$
\par
Starting from $R_{2,\perp}$ or from $R_{2,\parallel}$ the system will typically
go back
to $C$ before visiting other frames;
these phenomena take place, respectively, in the two
typical times $e^{\beta(h-\lambda)(l+1)}$ and $e^{\beta(h-\lambda)(m+1)}$.
It appears clear that the system, before eventually leaving $C$ to
reach another frame,
will wander, performing random oscillations, in the union of the basins $B$,
$B_{2,\perp}$ and $B_{2,\parallel}$. Then, in order to understand whether the
frame will shrink or grow we have to describe its behaviour in a
larger basin, containing $B\cup B_{2,\perp} \cup B_{2,\parallel}$.
This basin is denoted by ${\cal D}$ and it is defined as follows
$${\cal D}:= \{\h :{\rm every\; downhill\; path\; starting\; from\;} \h\;
{\rm ends\; in\;} C\; {\rm or}\; R_{2,\perp}\; {\rm or}\; R_{2,\parallel}
\}\;\; .\Eq (4.20)$$
\par
We denote by $S_{5,\perp}$ and $S_{5,\parallel}$ the
configurations obtained by attaching a unit square protuberance to the free
side of the internal rectangle of $R_{2,\perp}$ and $R_{2,\parallel}$
(see Fig.4.11).
By considering all the uphill paths starting from $C$, $R_{2,\perp}$
and $R_{2,\parallel}$, we are able to examine all the configurations in
$\partial {\cal D}$. We get:
$$\min_{\sigma\in\partial {\cal D}}H(\sigma)=H(S_{1}),\;\;\;\;
U({\cal D})=S_1\;\; .\Eq (4.21)$$
The most relevant inequalities in the proof of equation $\equ (4.21)$ are
$$\eqalign{
l<l^{*}&\Rightarrow (h+\lambda)(l-1)<[2J-(h-\lambda)]-(h-\lambda)(l+1)+
		[2J-(h+\lambda)]\cr
m<m^{*}(l)&\Rightarrow (h+\lambda)(l-1)<[2J-(h-\lambda)]-(h-\lambda)(m+1)+
		[2J-(h+\lambda)]\cr}
\;\; ,\Eq (4.22)$$
they mean, respectively, $H(S_{5,\perp})>H(S_1)$ and
$H(S_{5,\parallel})>H(S_1)$. Of course it is always
$H(S_{5,\parallel})<H(S_{5,\perp})$.
\midinsert
\vskip 0.5 truecm
\vbox{\font\amgr=cmr10 at
10truept\baselineskip0.1466667truein\lineskiplimit-\maxdimen
\catcode`\-=\active\catcode`\~=\active\def~{{\char32}}\def-{{\char1}}%
\hbox{\amgr ~~~~~~~~~~$\phantom
{S_{5,\perp}}$~~~~~{\char2}-----------------------{\char3}~~~~~{\char2}-----------------------{\char3}}
\hbox{\amgr ~~~~~~~~~~$\phantom
{S_{5,\perp}}$~~~~~{\char0}~{\char2}-----------------{\char3}~~~{\char0}~~~~~{\char0}~~~{\char2}-----------------{\char3}~{\char0}}
\hbox{\amgr ~~~~~~~~~~$\phantom
{S_{5,\perp}}$~~~~~{\char0}~{\char0}~~~~~~~~~~~~~~~~~{\char0}~~~{\char0}~~~~~{\char0}~~~{\char0}~~~~~~~~~~~~~~~~~{\char0}~{\char0}}
\hbox{\amgr ~~~~~~~~~~$
{S_{5,\perp}}$~~~~~{\char0}~{\char0}~~~~~~~~~~~~~~~~~{\char5}-{\char3}~{\char0}~~~~~{\char0}~{\char2}-{\char4}~~~~~~~~~~~~~~~~~{\char0}~{\char0}~$l+2$}
\hbox{\amgr ~~~~~~~~~~$\phantom
{S_{5,\perp}}$~~~~~{\char0}~{\char0}~~~~~~~~~~~~~~~~~{\char2}-{\char4}~{\char0}~~~~~{\char0}~{\char5}-{\char3}~~~~~~~~~~~~~~~~~{\char0}~{\char0}~}
\hbox{\amgr ~~~~~~~~~~$\phantom
{S_{5,\perp}}$~~~~~{\char0}~{\char5}-----------------{\char4}~~~{\char0}~~~~~{\char0}~~~{\char5}-----------------{\char4}~{\char0}}
\hbox{\amgr ~~~~~~~~~~$\phantom
{S_{5,\perp}}$~~~~~{\char5}-----------------------{\char4}~~~~~{\char5}-----------------------{\char4}}
\hbox{\amgr ~~~~~~~~~~$\phantom
{S_{5,\perp}}$~~~~~~~~~~~~~$m+3$~~~~~~~~~~~~~~~~~~~~~~~~~~$m+3$~~~~~~~~~~~~~~~}
\hbox{\amgr }
\hbox{\amgr }
\hbox{\amgr ~~~~~~~~~~$\phantom
{S_{5,\parallel}}$~~~~~{\char2}---------------------{\char3}~~~~~~~~~}
\hbox{\amgr ~~~~~~~~~~$\phantom
{S_{5,\parallel}}$~~~~~{\char0}~~~~~~~~{\char2}-{\char3}~~~~~~~~~~{\char0}~~~~~~~~~{\char2}---------------------{\char3}}
\hbox{\amgr ~~~~~~~~~~$\phantom
{S_{5,\parallel}}$~~~~~{\char0}~{\char2}------{\char4}~{\char5}--------{\char3}~{\char0}~~~~~~~~~{\char0}~{\char2}-----------------{\char3}~{\char0}}
\hbox{\amgr ~~~~~~~~~~$\phantom
{S_{5,\parallel}}$~~~~~{\char0}~{\char0}~~~~~~~~~~~~~~~~~{\char0}~{\char0}~~~~~~~~~{\char0}~{\char0}~~~~~~~~~~~~~~~~~{\char0}~{\char0}}
\hbox{\amgr ~~~~~~~~~~$
{S_{5,\parallel}}$~~~~~{\char0}~{\char0}~~~~~~~~~~~~~~~~~{\char0}~{\char0}~~~~~~~~~{\char0}~{\char0}~~~~~~~~~~~~~~~~~{\char0}~{\char0}}
\hbox{\amgr ~~~~~~~~~~$\phantom
{S_{5,\parallel}}$~~~~~{\char0}~{\char0}~~~~~~~~~~~~~~~~~{\char0}~{\char0}~~~~~~~~~{\char0}~{\char0}~~~~~~~~~~~~~~~~~{\char0}~{\char0}~$l+3$}
\hbox{\amgr ~~~~~~~~~~$\phantom
{S_{5,\parallel}}$~~~~~{\char0}~{\char5}-----------------{\char4}~{\char0}~~~~~~~~~{\char0}~{\char5}-----{\char3}~{\char2}---------{\char4}~{\char0}}
\hbox{\amgr ~~~~~~~~~~$\phantom
{S_{5,\parallel}}$~~~~~{\char5}---------------------{\char4}~~~~~~~~~{\char0}~~~~~~~{\char5}-{\char4}~~~~~~~~~~~{\char0}}
\hbox{\amgr ~~~~~~~~~~$\phantom
{S_{5,\parallel}}$~~~~~~~~~~~~~~~~~~~~~~~~~~~~~~~~~~~~~{\char5}---------------------{\char4}}
\hbox{\amgr ~~~~~~~~~~$\phantom
{S_{5,\parallel}}$~~~~~~~~~~~~~$m+2$~~~~~~~~~~~~~~~~~~~~~~~~~~~~~$m+2$~~~~~~~~~~~~}}
\vskip 0.5 truecm
\par\noindent
\centerline {\smbfb Fig.4.11}
\endinsert
\par
We notice that ${\cal D}$ is a sort of generalized basin of attraction of
$ C$; indeed it is easy to see that as a consequence of $m < M^*$ the ``bottom"
of
 ${\cal D}$ reduces to $ C$  in the sense that $ C$ are  the
only absolute
minima of the energy in ${\cal D}$ and, as it is easy to see, starting from any
initial configuration $\s \in {\cal D}$ our process,  with high
probability for large $\b$,  will visit $C$ before exiting from ${\cal D}$.
{}From ${\cal D}$ one can easily obtain, by suitably cutting in energy, a cycle
having
the same minimal saddles in its boundary:\par
take the maximal connected set  $\bar {\cal D}$ of
configurations containing $ C $ with energy less than $H(S_1)$.
Since it is easy to see that properties $i)$, $ii)$, $iii)$ of
$B=B(C(l_1,l_2))$
are still verified with ${\cal D}$ in place of $B$ for
$\widetilde L \leq l < l^*, \;\; m < m^* (l)$
 one
immediately gets: ${\cal S}(\bar {\cal D}) \ni S_1$; moreover, $\forall\s\in
{\cal D}$,
$$\lim_{\b\to\infty} P_{\s}(\s_{\t_{({\cal D}\cup\partial {\cal D})^c}-1}\in
                            S_1)=1 \;\; $$
and for every
$\e>0$
$$\lim_{\b\to\infty} P_C(
  e^{\b (h+\l)(l-1)-\b\e}<\t_{\partial {\cal D}} <e^{\b (h+\l)(l-1)+\b\e})=1
\;\; .$$
\par
We want to stress that the cycle $\bar {\cal D}$ is not the strict basin of
attraction
of any stable equilibium point but, rather, it has a several-well structure:
it contains
in its interior, beyond $ C$, the equilibria
$R_{2,\perp}$ and
$R_{2,\parallel}$; moreover it contains the internal saddles $S_4$.
The difference w.r.t the previous case of $l<\widetilde L$ (where
 we had to consider the cycle
$\bar B( C)$  in place of $\bar {\cal D}$) is that now,  not all
the downhill paths emerging from $\s \in \bar {\cal D}$ end in
$C$ and the system, before escaping from $\bar {\cal D}$
 will typically make many transitions, back and forth, between
$\bar B(C)$ and $\bar B(R_{2,\perp})$
$\bar B(R_{2,\parallel})$ through $S_4$.
\par
We consider, now, the frame $C$ and suppose that
${\widetilde L}\leq l<l^*$ and $m^*(l)\leq m<{\widetilde M}-2$.
We have $H(S_1)> H(S_{5,\parallel})$, $H(S_4)< H(S_{5,\parallel})$.
With the usual arguments one can prove that
$$\min_{\s\in\partial {\cal D}} H(\s)=H(S_{5,\parallel}),\;\;\;
  U({\cal D})=S_{5,\parallel}\;\; ;\Eq (4.23)$$
hence the frame $C$ is supercritical and the system starting from
$C$ will hit ${\piuuno}$ in a typical time
$e^{\beta\{[2J-(h-\lambda)]-(h-\lambda)(m+1)+[2J-(h+\lambda)]\}}$.
\par
In the case ${\widetilde L}\leq l<l^*$ and
$m^*(l)\leq {\widetilde M}-2\leq m$ it can be proved that
$\min_{\s\in\partial {\cal D}} H(\s)=H(S_{5,\parallel})$,
$U({\cal D})=S_{5,\parallel}$ and
$H(S_{5,\parallel})<H(S_4)$. Hence the frame is
supercritical, but the typical escape time is $e^{\b [2J-(h-\l)]}$.
We remark that in this case $H(S_4)<H(S_1)<H(S_{5,\perp})$ and
$H(S_{5,\parallel})<H(S_4)$ (see Fig.4.12). Hence the set
$$
\bar {\cal D} := \{ \s \in {\cal D} ; \; H(\s) \leq H(S_4)\}
$$
is a {\it generalized} cycle like the set $Q_1$ defined in [OS1].
$\bar {\cal D}$ is a set of cycles  communicating through the minimal
saddles in $S_4$. Starting from ${\cal D}$ the system, before leaving
${\cal D}$ will not necessarily visit all the cycles contained in
$\bar {\cal D}$ with energy less than $H(S_4)$.
\par
The system can leave $B$ either through $S_{4,\perp}$ or through
$S_{4,\parallel}$. In the first case it will enter into $ B_{2,\perp}$
visiting all the configurations of the cycle:
$$
\bar B_{2,\perp} := \{ \s \in B_{2,\perp} : \;
H(\s) < H(S_4)\}
$$
before leaving $B_{2,\perp}$ and passing again through
$S_{4,\perp}$. In the second case it will directly get out of ${\cal D}$.
\midinsert
\vskip 13truecm\noindent
\includegraphics{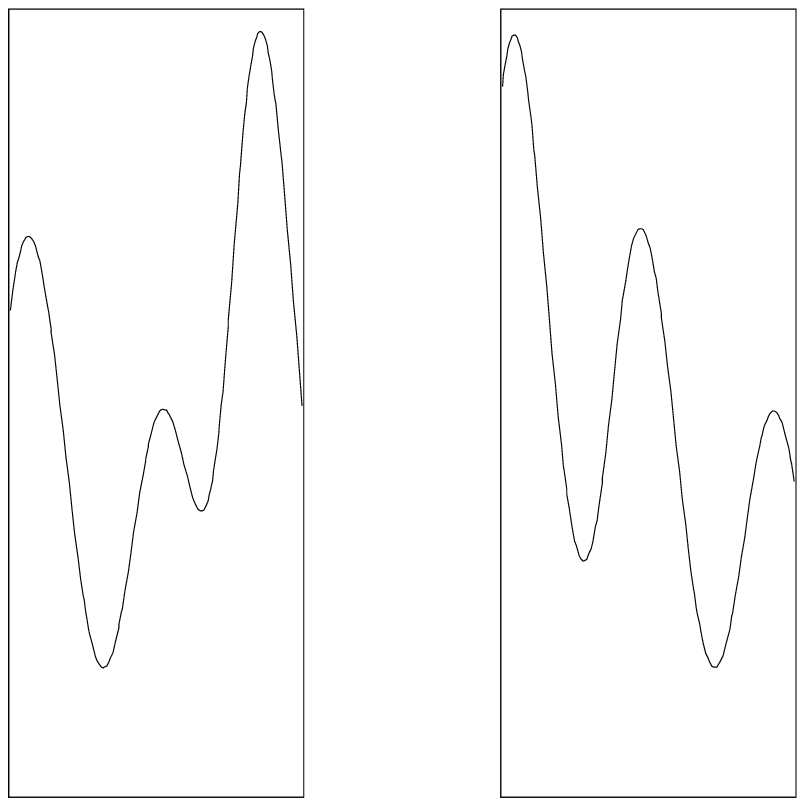}
\par\noindent
\vskip 1.7 truecm
{\vskip -5.1 truecm \hskip 4.1 truecm{$C$}}
{\vskip -2.6 truecm \hskip 5.2 truecm{$R_{2,\perp}$}}
{\vskip  0.5 truecm \hskip 9.9 truecm{$C$}}
{\vskip  0.8 truecm \hskip 11.5 truecm{$R_{2,\parallel}$}}
\par\noindent
\vskip 1.5 truecm
\centerline {\smbfb Fig.4.12}
\endinsert
\par
In the case $l^*\leq l<{\widetilde M}-2$ and $m<{\widetilde M}-2$
we have that $\min_{\s\in\partial {\cal D}} H(\s)=H({\cal
P}_{5,\parallel})$, $U({\cal D})=S_{5,\parallel}$ and
$H(S_{5,\parallel})>H(S_4)$. Hence the frame is
supercritical and the typical escape time is
$e^{\beta\{[2J-(h-\lambda)]-(h-\lambda)(m+1)+[2J-(h+\lambda)]\}}$. In this
case the most important inequalities are
$$\eqalign {
(h+\lambda)(l-1)>&[2J-(h-\lambda)]-(h-\lambda)(l+1)+[2J-(h+\lambda)]>\cr
	        >&[2J-(h-\lambda)]-(h-\lambda)(m+1)+[2J-(h+\lambda)]\cr}
\;\; ,\Eq (4.24)$$
we remark that $H(S_1)-H(C)=(h+\lambda)(l-1)$,
$H(S_{5,\perp})-H(C)=[2J-(h-\lambda)]-(h-\lambda)(l+1)
+[2J-(h+\lambda)]$
and $H(S_{5,\parallel})-H(C)=
[2J-(h-\lambda)]-(h-\lambda)(m+1)+[2J-(h+\lambda)]$.
\par
In the case $l^*\leq l<{\widetilde M}-2$ and ${\widetilde M}-2\leq m$
we have that $\min_{\s\in\partial {\cal D}} H(\s)=H(S_{5,\parallel})$ and
$H(S_{5,\parallel})<H(S_4)$. Hence the frame is
supercritical and the typical escape time is $e^{\b [2J-(h-\l)]}$.
In this case we have that ${\cal D}$ contains again a generalized cycle.
\par
We remark that in the supercritical cases discussed above, namely for
$l<{\widetilde M}-2$, the growth of a rectangular frame is asymmetric. The
frame grows in a direction parallel to its shortest side towards a squared
frame. Notice that the same tendency to be attracted by a squared shape is
present also in the contraction of a subcritical frame which, as we have
seen above, prefers to shrink in the direction orthogonal to its smallest
side.
\par
Finally we consider the case $l\geq {\widetilde M}-2$. It can be proved that
$H(S_{5,\parallel})<H(S_{5,\perp})<H(S_4)
<H(S_1)$, hence the frame is supercritical and the typical time is
$e^{\b [2J-(h-\l)]}$. In this case the growth process is symmetric,
similary to what happens in the stochastic Ising model for {\it any}
supercritical rectangle.
This concludes the proof of Proposition 4.1. $\square$
\par
\vskip.5cm
{\bf Remark.}\par
In the following to avoid lengthy and repetitions we will often use
short expressions like: {\it the external rectangle shrinks in a direction
perpendicular to its shortest sides} instead of: by a comparative analysis
of the possible barriers of energy, namely looking at the set of minimal
saddles of a suitable (possibly generalized) basin of attraction, we know
that with a probability tending to one as $\beta$ tends to infinity
{\it the external rectangle shrinks in a direction
perpendicular to its shortest sides}.
\vskip.5cm
\bigskip
In Proposition 4.1 we have stated conditions of subcriticality and
supercriticality for frames, now we state similar results for
birectangles
\vskip 0.35 truecm
\noindent
{\bf Proposition 4.2.}\par\noindent
Let us consider a birectangle $R:=R(L_1,L_2;M_1,M_2)$, let
$L:=\min \{L_1,L_2\}$, ${\hat L}:=\max \{L_1,L_2\}$,
$M:=\min \{M_1,M_2\}$ and ${\hat M}:=\max \{M_1,M_2\}$.
If one of the conditions 1-5 is satisfied
$$\eqalign {
1)\; &L<L^{*},\; M<M^{*}\cr
2)\; &L\geq L^{*},\; M<{\widetilde M},\; {\hat L}+2<{\widetilde M},\;
	L<{\widetilde L}\cr
3)\; &L\geq L^{*},\; M<{\widetilde M},\; {\hat L}+2<{\widetilde M},\;
	{\widetilde L}\leq L<l^{*},\; {\hat L}<m^{*}(L)\cr
4)\; &L\geq L^{*},\; M<{\widetilde M},\; {\hat L}+2\geq {\widetilde M},\;
	M-2<{\widetilde L}\cr
5)\; &L\geq L^{*},\; M<{\widetilde M},\; {\hat L}+2\geq {\widetilde M},\;
	{\widetilde L}\leq M-2<l^{*},\; {\hat L}<m^{*}(M-2)\cr}$$
then
$$\lim_{\beta\to\infty}P_{R}(\tau_{\menouno}<\tau_{\piuuno})=1\;\; .$$
If one of the conditions 6-11 is satisfied
$$\eqalign {
6)\; &L\geq L^{*},\; {\widetilde M}\leq M<M^{*}\cr
7)\; &M\geq M^*\cr
8)\; &L\geq L^{*},\; M<{\widetilde M},\; {\hat L}+2<{\widetilde M},\;
	L\geq l^*\cr
9)\; &L\geq L^{*},\; M<{\widetilde M},\; {\hat L}+2<{\widetilde M},\;
	{\widetilde L}\leq L<l^{*},\; {\hat L}\geq m^{*}(L)\cr
10)\; &L\geq L^{*},\; M<{\widetilde M},\; {\hat L}+2\geq {\widetilde M},\;
	M-2\geq l^*\cr
11)\; &L\geq L^{*},\; M<{\widetilde M},\; {\hat L}+2\geq {\widetilde M},\;
	{\widetilde L}\leq M-2<l^{*},\; {\hat L}\geq m^{*}(M-2)\cr}$$
then
$$\lim_{\beta\to\infty}P_{R}(\tau_{\piuuno}<\tau_{\menouno})=1\;\; .$$
\par\noindent
{\it Proof.}
\vskip 0.5 truecm
\par
Without loss of generality we can assume $ M= M_1$. Let us denote by $B:=B(R)$
the basin of attraction of $R$ and by
$\partial B$ its boundary; first af all we have to find the minimum of the
energy on the boundary $\partial B$. We examine all the uphill paths
starting from $R$, but the relevant ones are those made of steps of the kinds
$(2,2)$, $(3,2)$, $(5,1)$ and $(10,1)$ (see Fig.4.2 and Fig.4.7).
The boundary configurations $S_1$, $S_2$, $S_3$ and
$S_4$ reached by the uphill paths described above are represented in
Fig.4.13, togheter with the energy differences
$\D H_i=H(S_i)-H(R)\;\forall i=1,\dots ,4$. We remark that certainly if
both the shortest sides of the external rectangle are not ``free", then at
least one of the longest sides will be free; in this case
$\D H_2=({\hat M}-1)(h-\l)$.
\midinsert
\vskip 0.5 truecm
\vbox{\font\amgr=cmr10 at
10truept\baselineskip0.1466667truein\lineskiplimit-\maxdimen
\catcode`\-=\active\catcode`\~=\active\def~{{\char32}}\def-{{\char1}}%
\hbox{\amgr ~~~~~~~~~~~~{\char45}~{\char45}~----------{\char3}}
\hbox{\amgr ~~~~~~~~~~~~{\char45}~{\char45}~----{\char3}~~~~~{\char0}}
\hbox{\amgr ~~~~~~~~~~~~~~~~~~~~{\char0}~~~~~{\char0}}
\hbox{\amgr ~~~~~~~~~~~~~~~~~~~~{\char0}~~~~~{\char0}}
\hbox{\amgr
{}~~~~~~~~~~~~~~~~~~~~{\char0}~~~~~{\char5}-{\char3}~~$S_1$~~~~~~~$\Delta
H_{1}=2J{\char45}(h{\char45}\lambda)$}
\hbox{\amgr ~~~~~~~~~~~~~~~~~~~~{\char0}~~~~~{\char2}-{\char4}}
\hbox{\amgr ~~~~~~~~~~~~~~~~~~~~{\char0}~~~~~{\char0}}
\hbox{\amgr ~~~~~~~~~~~~{\char45}~{\char45}~----{\char4}~~~~~{\char0}}
\hbox{\amgr ~~~~~~~~~~~~{\char45}~{\char45}~----------{\char4}}
\hbox{\amgr ~~~~}
\hbox{\amgr ~~~~}
\hbox{\amgr ~~~~~~~~~~~~{\char45}~{\char45}~--------{\char3}}
\hbox{\amgr ~~~~~~~~~~~~{\char45}~{\char45}~----{\char3}~~~{\char0}}
\hbox{\amgr ~~~~~~~~~~~~~~~~~~~~{\char0}~~~{\char0}}
\hbox{\amgr ~~~~~~~~~~~~~~~~~~~~{\char0}~~~{\char0}}
\hbox{\amgr ~~~~~~~~~~~~~~~~~~~~{\char0}~~~{\char0}~~~~~~$S_2$~~~~~~~$\Delta
H_{2}=(M{\char45}1)(h{\char45}\lambda)$}
\hbox{\amgr ~~~~~~~~~~~~~~~~~~~~{\char0}~~~{\char0}}
\hbox{\amgr ~~~~~~~~~~~~~~~~~~~~{\char0}~~~{\char0}}
\hbox{\amgr ~~~~~~~~~~~~{\char45}~{\char45}~----{\char4}~~~{\char5}-{\char3}}
\hbox{\amgr ~~~~~~~~~~~~{\char45}~{\char45}~----------{\char4}}
\hbox{\amgr ~~~~}
\hbox{\amgr ~~~~}
\hbox{\amgr ~~~~~~~~~~~~{\char45}~{\char45}~----------{\char3}}
\hbox{\amgr ~~~~~~~~~~~~{\char45}~{\char45}~--{\char3}~~~~~~~{\char0}}
\hbox{\amgr ~~~~~~~~~~~~~~~~~~{\char0}~~~~~~~{\char0}}
\hbox{\amgr ~~~~~~~~~~~~~~~~~~{\char0}~~~~~~~{\char0}}
\hbox{\amgr ~~~~~~~~~~~~~~~~~~{\char0}~~~~~~~{\char0}~~~~$S_3$~~~~~~~$\Delta
H_{3}=(L{\char45}1)(h+\lambda)$}
\hbox{\amgr ~~~~~~~~~~~~~~~~~~{\char0}~~~~~~~{\char0}}
\hbox{\amgr ~~~~~~~~~~~~~~~~~~{\char5}-{\char3}~~~~~{\char0}}
\hbox{\amgr ~~~~~~~~~~~~{\char45}~{\char45}~----{\char4}~~~~~{\char0}}
\hbox{\amgr ~~~~~~~~~~~~{\char45}~{\char45}~----------{\char4}}
\hbox{\amgr ~~~~}
\hbox{\amgr ~~~~}
\hbox{\amgr ~~~~~~~~~~~~{\char45}~{\char45}~----------{\char3}}
\hbox{\amgr ~~~~~~~~~~~~{\char45}~{\char45}~----{\char3}~~~~~{\char0}}
\hbox{\amgr ~~~~~~~~~~~~~~~~~~~~{\char0}~~~~~{\char0}}
\hbox{\amgr ~~~~~~~~~~~~~~~~~~~~{\char0}~~~~~{\char0}}
\hbox{\amgr
{}~~~~~~~~~~~~~~~~~~~~{\char5}-{\char3}~~~{\char0}~~~~$S_4$~~~~~~~$\Delta
H_{4}=2J{\char45}(h+\lambda)$}
\hbox{\amgr ~~~~~~~~~~~~~~~~~~~~{\char2}-{\char4}~~~{\char0}}
\hbox{\amgr ~~~~~~~~~~~~~~~~~~~~{\char0}~~~~~{\char0}}
\hbox{\amgr ~~~~~~~~~~~~{\char45}~{\char45}~----{\char4}~~~~~{\char0}}
\hbox{\amgr ~~~~~~~~~~~~{\char45}~{\char45}~----------{\char4}}}
\vskip 0.5 truecm
\par\noindent
\centerline {\smbfb Fig.4.13}
\endinsert
\par
Now, we consider the case $(1)$: as a consequence of the subcriticality of
the internal and the external rectangle one has
$$\eqalign{
L<L^*&\Rightarrow (L-1)(h+\lambda)<2J-(h+\lambda)<2J-(h-\lambda)\cr
M<M^*&\Rightarrow (M-1)(h-\lambda)<2J-(h-\lambda)\cr}\;\; .\Eq (4.25)$$
But we cannot say anything about the inequality
$(L-1)(h+\lambda)\grle (M-1)(h-\lambda)$, without specifying more
conditions on $L$ and $M$; therefore the minimum of the energy on $\partial B$
is given either by $S_2$ or $S_3$, depending on the values of
$L$ and $M$. Thus, the birectangle $R$ is always subcritical and we can
express the typical time
needed by the system starting from $R$ to hit $\menouno$ as
$\max\{ e^{\beta (L-1)(h+\lambda)},e^{\beta (M-1)(h-\lambda)}\}$.
Similar results are obtained if one supposes that the shortest side of the
external rectangle is not ``free".
\par
Now, we suppose $L\geq L^*$, $M<{\widetilde M}$ and ${\hat L}+2\leq M$
(see Fig.4.14):  the internal
rectangle is supercritical, namely $2J-(h+\l)<(L-1)(h+\l)$, and the external
one is subcritical, namely $(M-1)(h-\l)<2J-(h-\l)$; moreover
$$M<{\widetilde M}\Rightarrow (M-1)(h-\l)<2J-(h+\l)\;\; .\Eq (4.26)$$
Then the minimum of the energy on the boundary $\partial B$ is
$S_2$. The external rectangle shrinks in a direction perpendicular
to its shortest sides untill it becomes a squared rectangle, then the
shrinking process goes on in both directions untill the frame
$C(L_1,L_2)$ is reached in a typical time $e^{\b (M-1)(h-\l)}$.
\midinsert
\vskip 0.5 truecm
\vbox{\font\amgr=cmr10 at
10truept\baselineskip0.1466667truein\lineskiplimit-\maxdimen
\catcode`\-=\active\catcode`\~=\active\def~{{\char32}}\def-{{\char1}}%
\hbox{\amgr
{}~~~~~~~~~~~~~~{\char2}-----------------------{\char3}~~~~~~~{\char2}-----------------------{\char3}}
\hbox{\amgr
{}~~~~~~~~~~~~~~{\char0}~~~~~~~~~~~~~~~~~~~~~~~{\char0}~~~~~~~{\char0}~~~~~~~~~~~~~~~~~~~~~~~{\char0}}
\hbox{\amgr
{}~~~~~~~~~~~~~~{\char0}~~~~~~~~~~~~~~~~~~~~~~~{\char0}~~~~~~~{\char0}~~~~~~~~~~~~~~~~~~~~~~~{\char0}}
\hbox{\amgr
{}~~~~~~~~~~~~~~{\char0}~~~~{\char2}--------------{\char3}~~~{\char0}~~~~~~~{\char0}~~~~~~~~~~~~~~~~~~~~~~~{\char0}%
}
\hbox{\amgr
{}~~~~~~~~~~~~~~{\char0}~~~~{\char0}~~~~~~~~~~~~~~{\char0}~~~{\char0}~~~~~~~{\char0}~~~~{\char2}-------{\char3}%
{}~~~~~~~~~~{\char0}}
\hbox{\amgr
{}~~~~~~~~~~~~~~{\char0}~~~~{\char0}~~~~~~~~~~~~~~{\char0}~~~{\char0}~~~~~~~{\char0}~~~~{\char0}~~~~~~~{\char0}%
{}~~~~~~~~~~{\char0}}
\hbox{\amgr
{}~~~~~~~~~~~~~~{\char0}~~~~{\char0}~~~~~~~~~~~~~~{\char0}~~~{\char0}~~~~~~~{\char0}~~~~{\char0}~~~~~~~{\char0}%
{}~~~~~~~~~~{\char0}}
\hbox{\amgr
{}~~~~~~~~~~~~~~{\char0}~~~~{\char5}--------------{\char4}~~~{\char0}~~~~~~~{\char0}~~~~{\char0}~~~~~~~{\char0}%
{}~~~~~~~~~~{\char0}}
\hbox{\amgr
{}~~~~~~~~~~~~~~{\char0}~~~~~~~~~~~~~~~~~~~~~~~{\char0}~~~~~~~{\char0}~~~~{\char0}~~~~~~~{\char0}~~~~~~~~~~{\char0}%
}
\hbox{\amgr
{}~~~~~~~~~~~~~~{\char0}~~~~~~~~~~~~~~~~~~~~~~~{\char0}~~~~~~~{\char0}~~~~{\char0}~~~~~~~{\char0}~~~~~~~~~~{\char0}%
}
\hbox{\amgr
{}~~~~~~~~~~~~~~{\char0}~~~~~~~~~~~~~~~~~~~~~~~{\char0}~~~~~~~{\char0}~~~~{\char0}~~~~~~~{\char0}~~~~~~~~~~{\char0}%
}
\hbox{\amgr
{}~~~~~~~~~~~~~~{\char0}~~~~~~~~~~~~~~~~~~~~~~~{\char0}~~~~~~~{\char0}~~~~{\char5}-------{\char4}~~~~~~~~~~{\char0}%
}
\hbox{\amgr
{}~~~~~~~~~~~~~~{\char0}~~~~~~~~~~~~~~~~~~~~~~~{\char0}~~~~~~~{\char0}~~~~~~~~~~~~~~~~~~~~~~~{\char0}}
\hbox{\amgr
{}~~~~~~~~~~~~~~{\char0}~~~~~~~~~~~~~~~~~~~~~~~{\char0}~~~~~~~{\char0}~~~~~~~~~~~~~~~~~~~~~~~{\char0}}
\hbox{\amgr
{}~~~~~~~~~~~~~~{\char0}~~~~~~~~~~~~~~~~~~~~~~~{\char0}~~~~~~~{\char0}~~~~~~~~~~~~~~~~~~~~~~~{\char0}}
\hbox{\amgr
{}~~~~~~~~~~~~~~{\char5}-----------------------{\char4}~~~~~~~{\char5}-----------------------{\char4}}}
\vskip 0.5 truecm
\par\noindent
\centerline {\smbfb Fig.4.14}
\endinsert
\par
Even in the case $L\geq L^*$, $M<{\widetilde M}$ and
$M<{\hat L}+2<{\widetilde M}$ (the longest sides of the internal and the
external rectangle are necessary parallel) the system, starting from $R$,
reaches the frame $C(L_1,L_2)$. Indeed the external rectangle shrinks
along the direction perpendicular to its shortest sides untill this process is
stopped by the internal rectangle (see Fig.4.15). In other words this
appears when the configuration
$R(L,{\hat L};M,{\hat L}+2)$ is reached (we have supposed, without
loss of generality, that $L_1=L$). At this point the external
rectangle will begin to shrink along the direction perpendicular to its
longest sides, because
${\hat L}+2<{\widetilde M}$ and then $({\hat L}+1)(h-\l)<2J-(h+\l)$.
Hence, the system, starting from $R$, reaches the frame $C(L_1,L_2)$
in a typical time $\max \{ e^{\b ({\hat L}+1)(h-\l)},e^{\b (M-1)(h-\l)}\}=
e^{\b ({\hat L}+1)(h-\l)}$.
\midinsert
\vskip 0.5 truecm
\vbox{\font\amgr=cmr10 at
10truept\baselineskip0.1466667truein\lineskiplimit-\maxdimen
\catcode`\-=\active\catcode`\~=\active\def~{{\char32}}\def-{{\char1}}%
\hbox{\amgr ~~~~~~~~~~~~~~{\char2}-----------------------{\char3}}
\hbox{\amgr ~~~~~~~~~~~~~~{\char0}~~~~~~~~~~~~~~~~~~~~~~~{\char0}~~~~~~~}
\hbox{\amgr
{}~~~~~~~~~~~~~~{\char0}~~~~~~~~~~~~~~~~~~~~~~~{\char0}~~~~~~~{\char2}-----------------------{\char3}}
\hbox{\amgr
{}~~~~~~~~~~~~~~{\char0}~~~~{\char2}-----------{\char3}~~~~~~{\char0}~~~~~~~{\char0}~~~~{\char2}-----------{\char3}%
{}~~~~~~{\char0}}
\hbox{\amgr
{}~~~~~~~~~~~~~~{\char0}~~~~{\char0}~~~~~~~~~~~{\char0}~~~~~~{\char0}~~~~~~~{\char0}~~~~{\char0}~~~~~~~~~~~{\char0}%
{}~~~~~~{\char0}~~~~~~~}
\hbox{\amgr
{}~~~~~~~~~~~~~~{\char0}~~~~{\char0}~~~~~~~~~~~{\char0}~~~~~~{\char0}~~~~~~~{\char0}~~~~{\char0}~~~~~~~~~~~{\char0}%
{}~~~~~~{\char0}}
\hbox{\amgr
{}~~~~~~~~~~~~~~{\char0}~~~~{\char0}~~~~~~~~~~~{\char0}~~~~~~{\char0}~~~~~~~{\char0}~~~~{\char0}~~~~~~~~~~~{\char0}%
{}~~~~~~{\char0}}
\hbox{\amgr
{}~~~~~~~~~~~~~~{\char0}~~~~{\char0}~~~~~~~~~~~{\char0}~~~~~~{\char0}~~~~~~~{\char0}~~~~{\char0}~~~~~~~~~~~{\char0}%
{}~~~~~~{\char0}}
\hbox{\amgr
{}~~~~~~~~~~~~~~{\char0}~~~~{\char0}~~~~~~~~~~~{\char0}~~~~~~{\char0}~~~->~~{\char0}~~~~{\char0}~~~~~~~~~~~{\char0}%
{}~~~~~~{\char0}~${\hat L}+2$}
\hbox{\amgr
{}~~~~~~~~~~~~~~{\char0}~~~~{\char0}~~~~~~~~~~~{\char0}~~~~~~{\char0}~~~~~~~{\char0}~~~~{\char0}~~~~~~~~~~~{\char0}%
{}~~~~~~{\char0}}
\hbox{\amgr
{}~~~~~~~~~~~~~~{\char0}~~~~{\char0}~~~~~~~~~~~{\char0}~~~~~~{\char0}~~~~~~~{\char0}~~~~{\char0}~~~~~~~~~~~{\char0}%
{}~~~~~~{\char0}}
\hbox{\amgr
{}~~~~~~~~~~~~~~{\char0}~~~~{\char0}~~~~~~~~~~~{\char0}~~~~~~{\char0}~~~~~~~{\char0}~~~~{\char0}~~~~~~~~~~~{\char0}%
{}~~~~~~{\char0}}
\hbox{\amgr
{}~~~~~~~~~~~~~~{\char0}~~~~{\char0}~~~~~~~~~~~{\char0}~~~~~~{\char0}~~~~~~~{\char0}~~~~{\char0}~~~~~~~~~~~{\char0}%
{}~~~~~~{\char0}}
\hbox{\amgr
{}~~~~~~~~~~~~~~{\char0}~~~~{\char0}~~~~~~~~~~~{\char0}~~~~~~{\char0}~~~~~~~{\char0}~~~~{\char0}~~~~~~~~~~~{\char0}%
{}~~~~~~{\char0}}
\hbox{\amgr
{}~~~~~~~~~~~~~~{\char0}~~~~{\char5}-----------{\char4}~~~~~~{\char0}~~~~~~~{\char0}~~~~{\char5}-----------{\char4}%
{}~~~~~~{\char0}}
\hbox{\amgr
{}~~~~~~~~~~~~~~{\char0}~~~~~~~~~~~~~~~~~~~~~~~{\char0}~~~~~~~{\char5}-----------------------{\char4}}
\hbox{\amgr ~~~~~~~~~~~~~~{\char5}-----------------------{\char4}~~~~~~~}
\hbox{\amgr
{}~~~~~~~~~~~~~~~~~~~~~~~~~~$M$~~~~~~~~~~~~~~~~~~~~~~~~~~~~~$M$~~~~~~~~~~~}}
\vskip 0.5 truecm
\par\noindent
\centerline {\smbfb Fig.4.15}
\endinsert
\par
Then, we can conclude that in the cases $(2)$ and $(3)$ the frame
$C(L_1,L_2)$ is eventually reached, but $C(L_1,L_2)$
is subcritical, hence $R$ is subcritcal, as well. For similar reasons in
the cases $(8)$ and $(9)$ the birectangle $R$ is supercritical.
\par
The typical shrinking time is given by
$$\eqalign{
\max\{e^{\b (M-1)(h-\l)},e^{\b (L-1)(h+\l)}\}&
                      \; {\rm if}\; {\hat L}+2\leq M\cr
\max\{e^{\b ({\hat L}+1)(h-\l)},e^{\b (L-1)(h+\l)}\}&
                      \; {\rm if}\; {\hat L}+2> M\cr}\;\; .$$
\par
With similar arguments it can be shown that in cases $(4)$ and $(5)$
the system, starting from $R$, hits $C(M-2,{\hat L})$ in a typical
time $e^{2J-(h+\l)}$. Hence, the birectangle $R$ is subcritical and the
typical shrinking time is $e^{2J-(h+\l)}$. In the cases $(10)$ and
$(11)$ the birectangle $R$ is supercritical, as a consequence of the
supercriticality of the frame $C(M-2,{\hat L})$.
\par
With arguments similar to those used before it can also be seen that in the
case $(6)$ the birectangle is supercritical
since it first evolves towards the frame $C (M - 2, \hat M - 2)$
which is a supercritical frame since $ \hat M - 2 > l^* $
with our choice of the paremeters.\par
Finally, in the case $(7)$, the birectangle is easily seen to be
supercritical. Indeed it follows
from an argument similar to the corresponding one valid for the
standard Ising model that starting from a configuration with $M\ge M^*$, we
get $\zero$ before $\piuuno$ in a time of order $e^{\beta [2J-(h-\l)]}$
with high probability for large $\beta$. Then, starting from $\zero$ we
tipically follow an Ising--like nucleation path (see [NS1], [S1]) leading
to $\piuuno$ through the saddles ${\cal S}(\zero ,\piuuno )$. These saddles
are given by configurations with precisely one cluster of pluses (in the
sea of zeroes), this cluster is given by a rectangle $L^*\times (L^*-1)$
with a unit square protuberance attached to one of its longest sides. It is
immediate to verify that
$$H({\cal S}(\zero , \piuuno)) < H({\cal P})\;\; .$$
\par
The proof of Proposition 4.2 is complete. $\square$
\par
\bigskip
We consider, now, a plurirectangle $R$. We denote by $M_1$ and $M_2$ the
lengths of the sides
of the external rectangle, by $L_{1,i}$ and $L_{2,i}$ $\forall
i=1,...,k^+$ the lengths of the sides of the $k^+$ internal rectangles
$R^+_i$ and we
define $M:=\min \{M_1,M_2\}$ and $L_i:=\min \{L_{1,i},L_{2,i}\}\; \forall
i=1,...,k^+$. In order to state conditions of subcriticality and
supercriticality for such configurations, we must introduce the rectangle
$R^+$ defined as the rectangular envelpe of the union of all the internal
supercritical rectangles. We denote by $L_{1,R^+}$ and $L_{2,R^+}$ the
lenghts of its sides and we define $L_{R^+}:=\min \{L_{1,R^+},L_{2,R^+}\}$ and
${\hat L}_{R^+}:=\max \{L_{1,R^+},L_{2,R^+}\}$. Suppose that
$\exists i\in\{1,2,...,k^+\}$  such that $L_i\geq L^*$,
we denote by ${\bar R}$ the birectangle obtained by removing all the
internal rectangles and by filling up with plus spin the rectangle $R^+$.
Finally we state the following
proposition
\vskip 0.35 truecm
\noindent
{\bf Proposition 4.3.}\par\noindent
If one of the two following conditiones is satisfied
\itemitem{1)} $L_i<l^*\forall i=1,...,k^+\; {\rm and}\; M<M^*$;
\itemitem{2)} $\exists i\in\{1,2,...,k^+\}$  such that $L_i\geq L^*$ and
${\bar R}$ is subcritical;
\par
then
$$\lim_{\b\to\infty} P_R(\t_{\menouno}<\t_{\piuuno})=1\;\; .$$
\par\noindent
{\it Proof.}
\vskip 0.5 truecm
\par
Let us consider the case $(1)$: we prove Proposition 4.3 by describing the
shrinking process.
\par
First af all the internal rectangles whose sides are such that
$(L_i-1)(h+\l)<(M-1)(h-\l)$ shrink in a typical time
$e^{\b (L_i-1)(h+\l)}$. We denote by $R^{(1)}$ the rectangular envelope of
the union of all the ``surviving" rectangles $R^{(1)}_i\forall i\in I^{(1)}
\subset \{1,...,k^+\}$ and by ${\hat L}^{(1)}$ its longest side.
\par
At this point the external rectangle starts shrinking (if it can). If
${\hat L}^{(1)}\leq M-2$ this contraction ends when the external rectangle
reaches $R^{(1)}$ (see Fig.4.16).
\midinsert
\vskip 0.5 truecm
\vbox{\font\amgr=cmr10 at
10truept\baselineskip0.1466667truein\lineskiplimit-\maxdimen
\catcode`\-=\active\catcode`\~=\active\def~{{\char32}}\def-{{\char1}}%
\hbox{\amgr
{}~~~~~~~~~~~~~~~~~~~~~~~~~~~~~~~{\char2}------------------------{\char3}}
\hbox{\amgr
{}~~~~~~~~~~~~~~~~~~~~~~~~~~~~~~~{\char0}~{\char2}------{\char3}~{\char45}~{\char45}~{\char45}~{\char45}~{\char45}~{\char45}~{\char3}~{\char0}}
\hbox{\amgr
{}~~~~~~~~~~~~~~~~~~~~~~~~~~~~~~~{\char0}~{\char0}~~~~~~{\char0}~~~~{\char2}---{\char3}~~~~~~{\char0}}
\hbox{\amgr
{}~~~~~~~~~~~~~~~~~~~~~~~~~~~~~~~{\char0}~{\char0}~~~~~~{\char0}~~~~{\char0}~~~{\char0}~~~~|~{\char0}}
\hbox{\amgr
{}~~~~~~~~~~~~~~~~~~~~~~~~~~~~~~~{\char0}~{\char5}------{\char4}~~~~{\char0}~~~{\char0}~~~~~~{\char0}}
\hbox{\amgr
{}~~~~~~~~~~~~~~~~~~~~~~~~~~~~~~~{\char0}~~~~~~~~~~~~~{\char5}---{\char4}~~~~|~{\char0}}
\hbox{\amgr
{}~~~~~~~~~~~~~~~~~~~~~~~~~~~~~~~{\char0}~|~{\char2}------{\char3}~~~~~~~~~~~~~{\char0}}
\hbox{\amgr
{}~~~~~~~~~~~~~~~~~~~~~~~~~~~~~~~{\char0}~~~{\char0}~~~~~~{\char0}~~~~~{\char2}-----{\char3}~{\char0}}
\hbox{\amgr
{}~~~~~~~~~~~~~~~~~~~~~~~~~~~~~~~{\char0}~|~{\char5}------{\char4}~~~~~{\char0}~~~~~{\char0}~{\char0}}
\hbox{\amgr
{}~~~~~~~~~~~~~~~~~~~~~~~~~~~~~~~{\char0}~~~~~~~~~~~~~~~~{\char0}~~~~~{\char0}~{\char0}}
\hbox{\amgr
{}~~~~~~~~~~~~~~~~~~~~~~~~~~~~~~~{\char0}~|~~~~~~~~~~~~~~{\char0}~~~~~{\char0}~{\char0}}
\hbox{\amgr
{}~~~~~~~~~~~~~~~~~~~~~~~~~~~~~~~{\char0}~~~~~~~~~~~~~~~~{\char0}~~~~~{\char0}~{\char0}}
\hbox{\amgr
{}~~~~~~~~~~~~~~~~~~~~~~~~~~~~~~~{\char0}~{\char5}~{\char45}~{\char45}~{\char45}~{\char45}~{\char45}~{\char45}~~{\char5}---%
--{\char4}~{\char0}}
\hbox{\amgr
{}~~~~~~~~~~~~~~~~~~~~~~~~~~~~~~~{\char5}------------------------{\char4}}}
\vskip 0.5 truecm
\par\noindent
\centerline {\smbfb Fig.4.16}
\endinsert
\par
Let us define $L_{\rm min}:=\min_{i\in I^{(1)}} \{L_i\}$: the internal
rectangle $R^+_i$ such that $L_i=L_{\rm min}$ starts shrinking and loses a
slide of lenght $L_i=L_{\rm min}$. There are two possible situations
(see Fig.4.17): after this contraction the external rectangle has a ``free"
side or not. In the first case the external rectangle loses another slice
and a configuration of the type described in Fig.4.16 is reached. In the
second case the internal rectangle goes on shrinking untill it disappears,
and a configuration like the one in Fig.4.16 is reached, as well. In both cases
the plurirectangle goes on shrinking by the mechanism described before
untill it disappears, hence in the case ${\hat L}^{(1)}\leq M-2$ the
plurirectangle $R$ is subcritical.
\midinsert
\vskip 0.5 truecm
\vbox{\font\amgr=cmr10 at
10truept\baselineskip0.1466667truein\lineskiplimit-\maxdimen
\catcode`\-=\active\catcode`\~=\active\def~{{\char32}}\def-{{\char1}}%
\hbox{\amgr
{}~~~~~~~~~~~{\char2}-----------------------------{\char3}~~~~{\char2}-----------------------------{\char3}%
}
\hbox{\amgr
{}~~~~~~~~~~~{\char0}~{\char2}-----------{\char3}~~~~~~~~~~~~~~~{\char0}~~~~{\char0}~{\char2}-----------{\char3}%
{}~~~~~~~~~~~~~~~{\char0}}
\hbox{\amgr
{}~~~~~~~~~~~{\char0}~{\char5}-{\char3}~~~~~~~~~{\char0}~~~~~~~~~~~~~~~{\char0}~~~~{\char0}~{\char0}~~~~~~%
{}~~~~~{\char0}~~~~~~~~~~~~~~~{\char0}}
\hbox{\amgr
{}~~~~~~~~~~~{\char0}~~~{\char0}~~~~~~~~~{\char0}~~~~~~~~~~~~~~~{\char0}~~~~{\char0}~{\char0}~~~~~~~~~~~{\char0}%
{}~~~~~~~~~~~~~~~{\char0}}
\hbox{\amgr
{}~~~~~~~~~~~{\char0}~~~{\char5}---------{\char4}~~~~~~~~~~~~~~~{\char0}~~~~{\char0}~{\char5}-----------{\char4}%
{}~~~~~~~~~~~~~~~{\char0}}
\hbox{\amgr
{}~~~~~~~~~~~{\char0}~~~~~~~~~~~~~~~~~~~~~~~~~~~~~{\char0}~~~~{\char0}~~~~~~~~~~~~~~~~~~~~~~~~~~~~~{\char0}%
}
\hbox{\amgr
{}~~~~~~~~~~~{\char0}~~~{\char2}-----------{\char3}~~~~~~~~~~~~~{\char0}~~~~{\char0}~~~{\char2}-----------%
{\char3}~~~~~~~~~~~~~{\char0}}
\hbox{\amgr
{}~~~~~~~~~~~{\char0}~~~{\char0}~~~~~~~~~~~{\char0}~~~~~~~~~~~~~{\char0}~~~~{\char0}~~~{\char0}~~~~~~~~~{\char2}%
-{\char4}~~~~~~~~~~~~~{\char0}}
\hbox{\amgr
{}~~~~~~~~~~~{\char0}~~~{\char0}~~~~~~~~~~~{\char0}~~~~~~~~~~~~~{\char0}~~~~{\char0}~~~{\char0}~~~~~~~~~{\char0}%
{}~~~~~~~~~~~~~~~{\char0}}
\hbox{\amgr
{}~~~~~~~~~~~{\char0}~~~{\char5}-----------{\char4}~~{\char2}--------{\char3}~{\char0}~~~~{\char0}~~~{\char5}%
---------{\char4}~~~~{\char2}--------{\char3}~{\char0}}
\hbox{\amgr
{}~~~~~~~~~~~{\char0}~~~~~~~~~~~~~~~~~~{\char0}~~~~~~~~{\char0}~{\char0}~~~~{\char0}~~~~~~~~~~~~~~~~~~{\char0}%
{}~~~~~~~~{\char0}~{\char0}}
\hbox{\amgr
{}~~~~~~~~~~~{\char0}~~~~~~~~~~~~~~~~~~{\char0}~~~~~~~~{\char0}~{\char0}~~~~{\char0}~~~~~~~~~~~~~~~~~~{\char0}%
{}~~~~~~~~{\char0}~{\char0}}
\hbox{\amgr
{}~~~~~~~~~~~{\char0}~~~~~~~~~~~~~~~~~~{\char0}~~~~~~~~{\char0}~{\char0}~~~~{\char0}~~~~~~~~~~~~~~~~~~{\char0}%
{}~~~~~~~~{\char0}~{\char0}}
\hbox{\amgr
{}~~~~~~~~~~~{\char0}~~~~~~~~~~~~~~~~~~{\char5}--------{\char4}~{\char0}~~~~{\char0}~~~~~~~~~~~~~~~~~~{\char5}%
--------{\char4}~{\char0}}
\hbox{\amgr
{}~~~~~~~~~~~{\char5}-----------------------------{\char4}~~~~{\char5}-----------------------------{\char4}%
}}
\vskip 0.5 truecm
\par\noindent
\centerline {\smbfb Fig.4.17}
\endinsert
\par
Now, we consider the case ${\hat L}^{(1)}> M-2$. During the second phase of
the contraction the system reaches a configuration characterized by an
external rectangle whose sides are $M$ and ${\hat L}^{(1)}+2$. The ``free"
side of the external rectangle is eventually ${\hat L}^{(1)}+2$. If
$({\hat L}^{(1)}+1)(h-\lambda)<(L_{i}-1)(h+\lambda)\; \forall i\in I^{(1)}$
the external rectangle shrinks in a direction perpendicular to its ``free"
side untill it reaches $R^{(1)}$; and then the shrinkig goes on as we have
described before. If there exists an internal rectangle
$R^+_i$ such that $({\hat L}^{(1)}+1)(h-\lambda)>(L_{i}-1)(h+\lambda)$
it disappers before anything else can happen. Then the contraction goes on as
described before.
In conclusion we have proved that in the case $(1)$ the plurirectangle $R$
is subcritical.
\par
In the case $(2)$ the proof of Proposition 4.3 can be achieved with
arguments similar to those used in the case $(1)$. $\square$
\par \bigskip
\vfill\eject
\numsec=5\numfor=1

{\bf Section 5. Comparison between special saddles.}
\par
Let us consider a subcritical frame or birectangle; we say that such a
configuration is {\it almost--supercritical} iff it can be
transformed into a supercritical minimum by attaching to one of its internal or
external sides a whole slice. By attaching a slice to an internal or external
side of a birectangle (or, in particular, of a frame) we mean transforming
from $-1$ to $0$ the value of the spins in the row or column adjacent
externally to this side. ``Removing a slice" is the inverse operation of
``attaching a slice".
\par
Let us consider, now, a supercritical frame or
birectangle; we say that such a configuration is {\it just--supercritical} iff
it can be transformed into a subcritical minimum by removing a whole slice from
one of its internal or external sides.
\par
Let us consider an almost supercritical frame or birectangle, we denote
by $u$ the internal or external side such that by attaching to it a whole
slice we obtain a supercritical configuration. We call {\it special saddle} the
configuration obtained by attaching to $u$ a plus unit protuberance, if $u$ is
an internal side, or a zero unit protuberance, if $u$ is an external one.
\par
\vskip 0.5 truecm
\vbox{\font\amgr=cmr10 at
10truept\baselineskip0.1466667truein\lineskiplimit-\maxdimen
\catcode`\-=\active\catcode`\~=\active\def~{{\char32}}\def-{{\char1}}%
\hbox{\amgr ~$\phantom {{\cal P}_1={\cal P}_{1,a}\;\; {\rm if}\;{\it \delta }
<{h+{\it \lambda }\over 2h}}$~{\char2}----------------{\char3}~~$\phantom
{l^*+2}$~~{\char2}--------------{\char3}~~~~~~~~~%
{}~~~~~~~}
\hbox{\amgr ~$\phantom {{\cal P}_1={\cal P}_{1,a}\;\; {\rm if}\;{\it \delta }
<{h+{\it \lambda }\over
2h}}$~{\char0}~{\char2}----------{\char3}~~~{\char0}~~$\phantom
{l^*+2}$~~{\char0}~~~~~{\char2}-{\char3}%
{}~~~~~~{\char0}}
\hbox{\amgr ~$\phantom {{\cal P}_1={\cal P}_{1,a}\;\; {\rm if}\;{\it \delta }
<{h+{\it \lambda }\over
2h}}$~{\char0}~{\char0}~~~~~~~~~~{\char0}~~~{\char0}~~$\phantom
{l^*+2}$~~{\char0}~{\char2}---{\char4}%
{}~{\char5}----{\char3}~{\char0}}
\hbox{\amgr ~$          {\cal P}_1={\cal P}_{1,a}\;\; {\rm if}\;{\it \delta }
<{h+{\it \lambda }\over
2h}$~{\char0}~{\char0}~~~~~~~~~~{\char0}~~~{\char0}~~$\phantom
{l^*+2}$~~{\char0}~{\char0}~~~~~~~~%
{}~~{\char0}~{\char0}~${\cal P}_1={\cal P}_{1,b}\;\; {\rm if}\;{\it \delta
}>{h+{\it \lambda }\over 2h}$}
\hbox{\amgr ~$\phantom {{\cal P}_1={\cal P}_{1,a}\;\; {\rm if}\;{\it \delta }
<{h+{\it \lambda }\over
2h}}$~{\char0}~{\char0}~~~~~~~~~~{\char5}-{\char3}~{\char0}~~$\phantom
{l^*+2}$~~{\char0}~{\char0}~%
{}~~~~~~~~~{\char0}~{\char0}}
\hbox{\amgr ~$\phantom {{\cal P}_1={\cal P}_{1,a}\;\; {\rm if}\;{\it \delta }
<{h+{\it \lambda }\over
2h}}$~{\char0}~{\char0}~~~~~~~~~~{\char2}-{\char4}~{\char0}~~${l^*+2}$~~{\char0}~{\char0}~%
{}~~~~~~~~~{\char0}~{\char0}~~}
\hbox{\amgr ~$\phantom {{\cal P}_1={\cal P}_{1,a}\;\; {\rm if}\;{\it \delta }
<{h+{\it \lambda }\over
2h}}$~{\char0}~{\char0}~~~~~~~~~~{\char0}~~~{\char0}~~$\phantom
{l^*+2}$~~{\char0}~{\char0}~~~~~~~~%
{}~~{\char0}~{\char0}}
\hbox{\amgr ~$\phantom {{\cal P}_1={\cal P}_{1,a}\;\; {\rm if}\;{\it \delta }
<{h+{\it \lambda }\over
2h}}$~{\char0}~{\char0}~~~~~~~~~~{\char0}~~~{\char0}~~$\phantom
{l^*+2}$~~{\char0}~{\char0}~~~~~~~~%
{}~~{\char0}~{\char0}}
\hbox{\amgr ~$\phantom {{\cal P}_1={\cal P}_{1,a}\;\; {\rm if}\;{\it \delta }
<{h+{\it \lambda }\over
2h}}$~{\char0}~{\char5}----------{\char4}~~~{\char0}~~$\phantom
{l^*+2}$~~{\char0}~{\char5}--------%
--{\char4}~{\char0}}
\hbox{\amgr ~$\phantom {{\cal P}_1={\cal P}_{1,a}\;\; {\rm if}\;{\it \delta }
<{h+{\it \lambda }\over 2h}}$~{\char5}----------------{\char4}~~$\phantom
{l^*+2}$~~{\char5}--------------{\char4}}
\hbox{\amgr ~$\phantom {{\cal P}_1={\cal P}_{1,a}\;\; {\rm if}\;{\it \delta }
<{h+{\it \lambda }\over 2h}}$~~~~~~${l^*+2}$~~~~~~~~~~~$\phantom
{l^*+2}$~~~~~~~$l^*+1$}
\hbox{\amgr }
\hbox{\amgr }
\hbox{\amgr ~~~~~~~~~~~~~~~~~~~~~~~~~~~~~~$\phantom {{\cal
P}_2}$~~~~~~{\char2}--------------{\char3}~~~~~~~~~~~~~~~~}
\hbox{\amgr ~~~~~~~~~~~~~~~~~~~~~~~~~~~~~~$\phantom {{\cal
P}_2}$~~~~~~{\char0}~~~~~~~~~~~~~~{\char0}~~}
\hbox{\amgr ~~~~~~~~~~~~~~~~~~~~~~~~~~~~~~$\phantom {{\cal
P}_2}$~~~~~~{\char0}~~~~~~~~~~~~~~{\char0}}
\hbox{\amgr ~~~~~~~~~~~~~~~~~~~~~~~~~~~~~~$\phantom {{\cal
P}_2}$~~~~~~{\char0}~~~~~~~~~~~~~~{\char0}}
\hbox{\amgr ~~~~~~~~~~~~~~~~~~~~~~~~~~~~~~$\phantom {{\cal
P}_2}$~~~~~~{\char0}~~~~~~~~~~~~~~{\char5}-{\char3}}
\hbox{\amgr ~~~~~~~~~~~~~~~~~~~~~~~~~~~~~~$          {\cal
P}_2$~~~~~~{\char0}~~~~~~~~~~~~~~{\char2}-{\char4}~~$M^*$}
\hbox{\amgr ~~~~~~~~~~~~~~~~~~~~~~~~~~~~~~$\phantom {{\cal
P}_2}$~~~~~~{\char0}~~~~~~~~~~~~~~{\char0}}
\hbox{\amgr ~~~~~~~~~~~~~~~~~~~~~~~~~~~~~~$\phantom {{\cal
P}_2}$~~~~~~{\char0}~~~~~~~~~~~~~~{\char0}}
\hbox{\amgr ~~~~~~~~~~~~~~~~~~~~~~~~~~~~~~$\phantom {{\cal
P}_2}$~~~~~~{\char0}~~~~~~~~~~~~~~{\char0}}
\hbox{\amgr ~~~~~~~~~~~~~~~~~~~~~~~~~~~~~~$\phantom {{\cal
P}_2}$~~~~~~{\char5}--------------{\char4}}
\hbox{\amgr ~~~~~~~~~~~~~~~~~~~~~~~~~~~~$\phantom {{\cal
P}_2}$~~~~~~~~~~~~$M^*{\char 45}1$~~~~~~~~~~~~~~~~~~~~~~~}}
\vskip 0.5 truecm
\par\noindent
\centerline {\smbfb Fig.5.1}
\par
Let us consider the set ${\hat {\cal P}}:=({\cal P}_1\cup {\cal P}_2)\subset
\O_{\L}$ with ${\cal P}_1$ and ${\cal P}_2$ the set of special saddles
shown in Fig.5.1, where we have used the following definition
$$\d:=l^*-{2J-(h-\l)\over h}\in\rbrack 0,1\lbrack\;\; .\Eq (5.1)$$
We state the following lemma:
\vskip 0.5 truecm
\noindent
{\bf Lemma 5.1.}\par\noindent
For any special saddle $S\not\in {\hat {\cal P}}$ it there exists
$S^*\in {\hat {\cal P}}$ such that
$$H(S)>H(S^*)\;\; .$$
\par\noindent
\vskip 0.5 truecm
\par
Before starting the proof, we observe that the frame $C(l^*,l^*)$ is
supercritical and $C(l^*-1,l^*-1)$ is subcritical for any choice of the
parameters $\l$ and $h$; indeed it can be proved that
$$m^*(l^*-1)\ge l^*\;{\rm for\; any\; value\; of\;} h\; {\rm and}\; \l ,
\Eq (5.1.1)$$
(see $(5.5)$). On the other hand we remark that
the criticality of the frame $C(l^*-1,l^*)$ depends on the value of the
real number $\d$ defined in $\equ (5.1)$.
By comparing the energies of the two configurations shown in Fig.5.2
one can easily convince himself that
$$\eqalign{
C(l^*-1,l^*) &\;\; {\rm subcritical\; iff}\;\d< {h+\l\over 2h}\cr
C(l^*-1,l^*) &\;\; {\rm supercritical\; iff}\;\d> {h+\l\over 2h}\cr
}\;\; ,$$
we observe that ${h+\l\over 2h}\in\rbrack 0,1\lbrack$ if ${h\over\l}>1$.
This explains the reason of the twofold definition of the configuration
${\cal P}_1$.
\vskip 0.5 truecm
\vbox{\font\amgr=cmr10 at
10truept\baselineskip0.1466667truein\lineskiplimit-\maxdimen
\catcode`\-=\active\catcode`\~=\active\def~{{\char32}}\def-{{\char1}}%
\hbox{\amgr ~~~~~~~~~~~$\phantom
{l^*+2}$~~{\char2}----------------{\char3}~~~~~~~~~~~~~~{\char2}------------------{\char3}~~~~~%
{}~~~~~~~}
\hbox{\amgr ~~~~~~~~~~~$\phantom
{l^*+2}$~~{\char0}~{\char2}-{\char3}~~~~~~~~~~~~{\char0}~~~~~~~~~~~~~~{\char0}~{\char2}--------%
----{\char3}~~~{\char0}}
\hbox{\amgr ~~~~~~~~~~~$\phantom
{l^*+2}$~~{\char0}~{\char0}~{\char5}----------{\char3}~{\char0}~~~~~~~~~~~~~~{\char0}~{\char0}~%
{}~~~~~~~~~~~{\char0}~~~{\char0}}
\hbox{\amgr ~~~~~~~~~~~$\phantom
{l^*+2}$~~{\char0}~{\char0}~~~~~~~~~~~~{\char0}~{\char0}~~~~~~~~~~~~~~{\char0}~{\char0}~~~~~~~~%
{}~~~~{\char0}~~~{\char0}}
\hbox{\amgr ~~~~~~~~~~~$\phantom
{l^*+2}$~~{\char0}~{\char0}~~~~~~~~~~~~{\char0}~{\char0}~~~~~~~~~~~~~~{\char0}~{\char0}~~~~~~~~%
{}~~~~{\char5}-{\char3}~{\char0}}
\hbox{\amgr
{}~~~~~~~~~~~$l^*+2$~~{\char0}~{\char0}~~~~~~~~~~~~{\char0}~{\char0}~~~~~~~~~~~~~~{\char0}~{\char0}~~~~~~~~%
{}~~~~{\char2}-{\char4}~{\char0}~~$l^*+2$}
\hbox{\amgr ~~~~~~~~~~~$\phantom
{l^*+2}$~~{\char0}~{\char0}~~~~~~~~~~~~{\char0}~{\char0}~~~~~~~~~~~~~~{\char0}~{\char0}~~~~~~~~%
{}~~~~{\char0}~~~{\char0}~~~~}
\hbox{\amgr ~~~~~~~~~~~$\phantom
{l^*+2}$~~{\char0}~{\char0}~~~~~~~~~~~~{\char0}~{\char0}~~~~~~~~~~~~~~{\char0}~{\char0}~~~~~~~~%
{}~~~~{\char0}~~~{\char0}}
\hbox{\amgr ~~~~~~~~~~~$\phantom
{l^*+2}$~~{\char0}~{\char5}------------{\char4}~{\char0}~~~~~~~~~~~~~~{\char0}~{\char5}--------%
----{\char4}~~~{\char0}}
\hbox{\amgr ~~~~~~~~~~~$\phantom
{l^*+2}$~~{\char5}----------------{\char4}~~~~~~~~~~~~~~{\char5}------------------{\char4}}
\hbox{\amgr ~~~~$\phantom
{l^*+2}$~~~~~~~~~~~~~~$l^*+1$~~~~~~~~~~~~~~~~~~~~~~~~~~$l^*+2$}}
\vskip 0.5 truecm
\par\noindent
\centerline {\smbfb Fig.5.2}
\par
\vskip 0.5 truecm
\par\noindent
{\it Proof of lemma} 5.1.
\vskip 0.35 truecm
\par\noindent
Let us suppose that $\d <{h+\l\over 2h}$. One can prove that for any $l$ such
that ${\widetilde L}\le l\le l^*-1$
$$m^*(l)\ge l^*+1\;\; .\Eq (5.2)$$
First of all we observe that $m^*(l)$ is a decreasing function of $l$, more
precisily one can easily prove that
$$m^*(l-1)\ge m^*(l)+1\;\;\forall l\in [{\widetilde L},l^*-1]\;\; .\Eq (5.3)$$
Therefore in order to get a lower bound to $m^*(l)$ it is sufficient to
evaluate $m^*(l^*-1)$; with some algebra one can easily obtain
$$m^*(l^*-1)=l^*+\left[ (1-\d) {2h\over h-\l}\right]\;\; .\Eq (5.4)$$
Then,
$$\d <{h+\l\over 2h}\Rightarrow (1-\d){2h\over h-\l}>1\Rightarrow
  m^*(l^*-1)\ge l^*+1\;\; ;$$
this completes the proof of inequality $\equ (5.2)$. We remark that the
validity of the equations $\equ (5.3)$ and $\equ (5.4)$ does not depend on the
value of the real number $\d$.
\par
Now, in order to prove Lemma 5.1 we have to examine all the possible special
saddles.
\vskip 0.35 truecm
\par\noindent
{\it Case C1}.
\par
We consider the special saddle $C_{1}(m)$ in Fig.5.3.
\midinsert
\vskip 0.5 truecm
\vbox{\font\amgr=cmr10 at
10truept\baselineskip0.1466667truein\lineskiplimit-\maxdimen
\catcode`\-=\active\catcode`\~=\active\def~{{\char32}}\def-{{\char1}}%
\hbox{\amgr ~~~~~~~$\phantom
{C_1(m)}$~~~~{\char2}------------------{\char3}~~~~~~~~~~~~}
\hbox{\amgr ~~~~~~~$\phantom
{C_1(m)}$~~~~{\char0}~{\char2}------------{\char3}~~~{\char0}}
\hbox{\amgr ~~~~~~~$\phantom
{C_1(m)}$~~~~{\char0}~{\char0}~~~~~~~~~~~~{\char0}~~~{\char0}}
\hbox{\amgr ~~~~~~~$\phantom
{C_1(m)}$~~~~{\char0}~{\char0}~~~~~~~~~~~~{\char0}~~~{\char0}~~$m+2$}
\hbox{\amgr ~~~~~~~$\phantom
{C_1(m)}$~~~~{\char0}~{\char0}~~~~~~~~~~~~{\char5}-{\char3}~{\char0}}
\hbox{\amgr ~~~~~~~$\phantom
{C_1(m)}$~~~~{\char0}~{\char0}~~~~~~~~~~~~{\char2}-{\char4}~{\char0}~~~}
\hbox{\amgr ~~~~~~~$
{C_1(m)}$~~~~{\char0}~{\char0}~~~~~~~~~~~~{\char0}~~~{\char0}~~~~}
\hbox{\amgr ~~~~~~~$\phantom
{C_1(m)}$~~~~{\char0}~{\char0}~~~~~~~~~~~~{\char0}~~~{\char0}~~~~~~~~${\rm
with}$~$l^*\le m\le m^*(l^*{\char45}1){\char45}1$}
\hbox{\amgr ~~~~~~~$\phantom
{C_1(m)}$~~~~{\char0}~{\char0}~~~~~~~~~~~~{\char0}~~~{\char0}~~~~}
\hbox{\amgr ~~~~~~~$\phantom
{C_1(m)}$~~~~{\char0}~{\char0}~~~~~~~~~~~~{\char0}~~~{\char0}}
\hbox{\amgr ~~~~~~~$\phantom
{C_1(m)}$~~~~{\char0}~{\char5}------------{\char4}~~~{\char0}}
\hbox{\amgr ~~~~~~~$\phantom {C_1(m)}$~~~~{\char5}------------------{\char4}}
\hbox{\amgr ~~~~~~~$\phantom {C_1(m)}$~~~~~~~~~$l^*+2$~~~~}}
\vskip 0.5 truecm
\par\noindent
\centerline {\smbfb Fig.5.3}
\endinsert
\par\noindent
It can be easily shown that $H(C_1(m))$ is an increasing function of
$m\in [l^*,m^*(l^*-1)-1]$, indeed $H(C_1(m+1))-H(C_1(m))=
(h+\l)-2h\d>0$ by virtue of the hypothesis $\d<{h+\l\over 2h}$. Hence,
$$H(C_{1}(m))\ge {\cal P}_1\;\; \forall m\in [l^*,m^*(l^*-1)-1]
\;\; ;\Eq (5.4.1)$$
we observe that the equality is verified in $\equ (5.4.1)$ iff $m=l^*$, that
is $C_1(m)\equiv {\cal P}_1$.
\vskip 0.35 truecm
\par\noindent
{\it Case C2}.
\par
We consider the special saddles
$C_{2,a}(l)$ and $C_{2,b}(l)$ in Fig.5.4.
We remark that the configuration obtained from $C_{2,b}(l)$ by removing
the protuberance is subcritical because $m^*(l-1)\ge m^*(l)+1$.
\midinsert
\vskip 0.5 truecm
\vbox{\font\amgr=cmr10 at
10truept\baselineskip0.1466667truein\lineskiplimit-\maxdimen
\catcode`\-=\active\catcode`\~=\active\def~{{\char32}}\def-{{\char1}}%
\hbox{\amgr ~~~~~~~~~~~~$\phantom
{C_{2,a}(l)}$~~~~{\char2}----------------{\char3}~~~~~~~~~~~~~~}
\hbox{\amgr ~~~~~~~~~~~~$\phantom
{C_{2,a}(l)}$~~~~{\char0}~~~~{\char2}-{\char3}~~~~~~~~~{\char0}}
\hbox{\amgr ~~~~~~~~~~~~$\phantom
{C_{2,a}(l)}$~~~~{\char0}~{\char2}--{\char4}~{\char5}-------{\char3}~{\char0}}
\hbox{\amgr ~~~~~~~~~~~~$\phantom
{C_{2,a}(l)}$~~~~{\char0}~{\char0}~~~~~~~~~~~~{\char0}~{\char0}~~$m+3$}
\hbox{\amgr ~~~~~~~~~~~~$\phantom
{C_{2,a}(l)}$~~~~{\char0}~{\char0}~~~~~~~~~~~~{\char0}~{\char0}}
\hbox{\amgr ~~~~~~~~~~~~$\phantom
{C_{2,a}(l)}$~~~~{\char0}~{\char0}~~~~~~~~~~~~{\char0}~{\char0}~~~}
\hbox{\amgr ~~~~~~~~~~~~$
{C_{2,a}(l)}$~~~~{\char0}~{\char0}~~~~~~~~~~~~{\char0}~{\char0}~~~~~~~~~~~~~~~${\widetilde L}\le l\le l^*{\char 45}1$}
\hbox{\amgr ~~~~~~~~~~~~$\phantom
{C_{2,a}(l)}$~~~~{\char0}~{\char0}~~~~~~~~~~~~{\char0}~{\char0}~~~~~~~~~~~~~~~$m=m^*(l){\char 45}1$}
\hbox{\amgr ~~~~~~~~~~~~$\phantom
{C_{2,a}(l)}$~~~~{\char0}~{\char0}~~~~~~~~~~~~{\char0}~{\char0}~~~~~~}
\hbox{\amgr ~~~~~~~~~~~~$\phantom
{C_{2,a}(l)}$~~~~{\char0}~{\char0}~~~~~~~~~~~~{\char0}~{\char0}}
\hbox{\amgr ~~~~~~~~~~~~$\phantom
{C_{2,a}(l)}$~~~~{\char0}~{\char5}------------{\char4}~{\char0}}
\hbox{\amgr ~~~~~~~~~~~~$\phantom
{C_{2,a}(l)}$~~~~{\char5}----------------{\char4}}
\hbox{\amgr ~~~~~~~~~~~~$\phantom {C_{2,a}(l)}$~~~~~~~~~$l+2$~~~~~}
\hbox{\amgr }
\hbox{\amgr }
\hbox{\amgr ~~~~~~~~~~~~$\phantom
{C_{2,b}(l)}$~~~~{\char2}----------------{\char3}~~~~~~~~~~~~~~}
\hbox{\amgr ~~~~~~~~~~~~$\phantom
{C_{2,b}(l)}$~~~~{\char0}~{\char2}----------{\char3}~~~{\char0}}
\hbox{\amgr ~~~~~~~~~~~~$\phantom
{C_{2,b}(l)}$~~~~{\char0}~{\char0}~~~~~~~~~~{\char0}~~~{\char0}~~$m+2$}
\hbox{\amgr ~~~~~~~~~~~~$\phantom
{C_{2,b}(l)}$~~~~{\char0}~{\char0}~~~~~~~~~~{\char0}~~~{\char0}}
\hbox{\amgr ~~~~~~~~~~~~$\phantom
{C_{2,b}(l)}$~~~~{\char0}~{\char0}~~~~~~~~~~{\char0}~~~{\char0}~~~}
\hbox{\amgr ~~~~~~~~~~~~$
{C_{2,b}(l)}$~~~~{\char0}~{\char0}~~~~~~~~~~{\char5}-{\char3}~{\char0}~~~~~~~~~~~~~~~${\widetilde L}\le l\le l^*{\char 45}1$}
\hbox{\amgr ~~~~~~~~~~~~$\phantom
{C_{2,b}(l)}$~~~~{\char0}~{\char0}~~~~~~~~~~{\char2}-{\char4}~{\char0}~~~~~~~~~~~~~~~$m=m^*(l)$}
\hbox{\amgr ~~~~~~~~~~~~$\phantom
{C_{2,b}(l)}$~~~~{\char0}~{\char0}~~~~~~~~~~{\char0}~~~{\char0}~~~~~~}
\hbox{\amgr ~~~~~~~~~~~~$\phantom
{C_{2,b}(l)}$~~~~{\char0}~{\char0}~~~~~~~~~~{\char0}~~~{\char0}}
\hbox{\amgr ~~~~~~~~~~~~$\phantom
{C_{2,b}(l)}$~~~~{\char0}~{\char5}----------{\char4}~~~{\char0}}
\hbox{\amgr ~~~~~~~~~~~~$\phantom
{C_{2,b}(l)}$~~~~{\char5}----------------{\char4}}
\hbox{\amgr ~~~~~~~~~~~~$\phantom {C_{2,b}(l)}$~~~~~~~~~$l+2$~~~~~}}
\vskip 0.5 truecm
\par\noindent
\centerline {\smbfb Fig.5.4}
\endinsert
\par
We have that $H(C_{2,a}(l))<H(C_{2,b}(l))$; indeed
$H(C_{2,a}(l))-H(C_{2,b}(l))=(h+\l)(l-m^*(l))$ and $l-m^*(l)<0$; the last
inequality is a consequence of the fact that $l<l^*$ and of equation
$\equ (5.1.1)$: $m^*(l)\ge m^*(l^*-1)\ge l^*>l$.
\par
We observe that $H(C_{2,a}(l))$ is a decreasing function of
$l$:
$$H(C_{2,a}(l+1))<H(C_{2,a}(l))\;\;\forall l\in [{\widetilde L}, l^*-2]
  \;\; .\Eq (5.5)$$
Indeed, it is not difficult to show that
$H(C_{2,a}(l+1))-H(C_{2,a}(l))=(h+\l)+2h\left(l^*-l-\d\right)
(m^*(l+1)-m^*(l))-2h\left( m^*(l+1)-l^*+\d\right)$ and by observing that
$l^*-l-\d>+1$, $m^*(l+1)-m^*(l)<-1$ and $m^*(l+1)-l^*+\d<0$ we obtain
$H(C_{2,a}(l+1))-H(C_{2,a}(l))<(h+\l)-2h=\l-h<0$. This completes the
proof of the inequality $\equ (5.5)$.
\par
Since $H(C_{2,a}(l))$ is a decreasing function, we have to compare the
energy of the two configurations $C_{2,a}(l^*-1)$ and ${\cal P}_1$; by
a direct calculation one obtains $H(C_{2,a}(l^*-1))>H({\cal P}_1)$.
\vskip 0.35 truecm
\par\noindent
{\it Case B1}.
\par
We consider the special saddles ${\cal B}_{1,a}(M,{\hat M};{\hat L})$ and
${\cal B}_{1,b}({\hat M};L,{\hat L})$ in Fig.5.5 (here and in the following we
use the notation introduced in Proposition 4.2 to label the internal and
external sides; we use $L$, ${\hat L}$, $M$ and ${\hat M}$ to denote the
dimensions of the birectangle obtained by removing the unit protuberance of the
special saddle).
\midinsert
\vskip 0.5 truecm
\vbox{\font\amgr=cmr10 at
10truept\baselineskip0.1466667truein\lineskiplimit-\maxdimen
\catcode`\-=\active\catcode`\~=\active\def~{{\char32}}\def-{{\char1}}%
\hbox{\amgr ~~~~~$\phantom {{\cal B}_{1,a}(M,{\hat M};{\hat
L})}$~~~~{\char2}----------------{\char3}~~~~~~~~~{\char2}--------------{\char3}}
\hbox{\amgr ~~~~~$\phantom {{\cal B}_{1,a}(M,{\hat M};{\hat
L})}$~~~~{\char0}~~~~~~~~~~~~~~~~{\char0}~~~~~~~~~{\char0}~~~~~~~~~~~~~~{\char0}}
\hbox{\amgr ~~~~~$\phantom {{\cal B}_{1,a}(M,{\hat M};{\hat
L})}$~~~~{\char0}~~~{\char2}------{\char3}~~~~~{\char0}~~~~~~~~~{\char0}~~{\char2}------{\char3}%
{}~~~~{\char0}}
\hbox{\amgr ~~~~~$\phantom {{\cal B}_{1,a}(M,{\hat M};{\hat
L})}$~~~~{\char0}~~~{\char0}~~~~~~{\char0}~~~~~{\char0}~~~~~~~~~{\char0}~~{\char0}~~~~~~{\char0}%
{}~~~~{\char0}}
\hbox{\amgr ~~~~~$\phantom {{\cal B}_{1,a}(M,{\hat M};{\hat
L})}$~~~~{\char0}~~~{\char0}~~~~~~{\char0}~~~~~{\char0}~~~~~~~~~{\char0}~~{\char0}~~~~~~{\char0}%
{}~~~~{\char0}}
\hbox{\amgr ~~~~~$\phantom {{\cal B}_{1,a}(M,{\hat M};{\hat
L})}$~~~~{\char0}~~~{\char0}~~~~~~{\char5}-{\char3}~~~{\char0}~~~~~~~~~{\char0}~~{\char0}%
{}~~~~~~{\char0}~~~~{\char5}-{\char3}}
\hbox{\amgr ~~~~~$\phantom {{\cal B}_{1,a}(M,{\hat M};{\hat
L})}$~~~~{\char0}~~~{\char0}~~~~~~{\char2}-{\char4}~~~{\char0}~~~~~~~~~{\char0}~~{\char0}%
{}~~~~~~{\char0}~~~~{\char2}-{\char4}}
\hbox{\amgr ~~~~~$         {{\cal B}_{1,a}(M,{\hat M};{\hat
L})}$~~~~{\char0}~~~{\char0}~~~~~~{\char0}~~~~~{\char0}~~~~~~~~~{\char0}~~{\char0}~~~~~~{\char0}%
{}~~~~{\char0}~~~~${\cal B}_{1,b}({\hat M};L,{\hat L})$}
\hbox{\amgr ~~~~~$\phantom {{\cal B}_{1,a}(M,{\hat M};{\hat
L})}$~~~~{\char0}~~~{\char0}~~~~~~{\char0}~~~~~{\char0}~~~~~~~~~{\char0}~~{\char0}~~~~~~{\char0}%
{}~~~~{\char0}}
\hbox{\amgr ~~~~~$\phantom {{\cal B}_{1,a}(M,{\hat M};{\hat
L})}$~~~~{\char0}~~~{\char0}~~~~~~{\char0}~~~~~{\char0}~~~~~~~~~{\char0}~~{\char0}~~~~~~{\char0}%
{}~~~~{\char0}}
\hbox{\amgr ~~~~~$\phantom {{\cal B}_{1,a}(M,{\hat M};{\hat
L})}$~~~~{\char0}~~~{\char5}------{\char4}~~~~~{\char0}~~~~~~~~~{\char0}~~{\char5}------{\char4}%
{}~~~~{\char0}}
\hbox{\amgr ~~~~~$\phantom {{\cal B}_{1,a}(M,{\hat M};{\hat
L})}$~~~~{\char5}----------------{\char4}~~~~~~~~~{\char5}--------------{\char4}}
\hbox{\amgr ~~~~~~~~$\phantom {{\cal B}_{1,a}(M,{\hat M};{\hat
L})}$~~~~~~~~~$M$~~~~~~~~~~~~~~~~~~~~~~~${\widetilde M}{\char 45}1$}
\hbox{\amgr }
\hbox{\amgr }
\hbox{\amgr ~~~~~~~~$\phantom {{\cal B}_{1,a}(M,{\hat M};{\hat
L})}$~~~~~~~~~$({\rm a})$~~~~~~~~~~~~~~~~~~~~~~~$({\rm b})$}}
\vskip 0.5 truecm
\par\noindent
\centerline{
\vbox
{
\hsize=15truecm
\baselineskip 0.35cm
\noindent
{\smbfb Fig.5.5}{\smb \quad
(a) The internal horizontal dimension is $\scriptstyle {L^*-1}$ and the
vertical one, $\scriptstyle {{\hat L}}$, is such that
$\scriptstyle {{\hat L}\ge L^*}$. The external vertical dimension is
$\scriptstyle {{\hat M}}$ and the horizontal one $\scriptstyle {M}$ is such
that $\scriptstyle {{\widetilde M}\le M<M^*}$. If we choose the parameters
$\scriptstyle {h}$ and $\scriptstyle {\l}$ such that
$\scriptstyle {{\widetilde M}=M^*}$, then the special saddle (a) does not
exist. (b) The external horizontal
dimension is $\scriptstyle {{\widetilde M}-1}$ and the vertical one is
$\scriptstyle {{\hat M}}$. The internal dimensions
$\scriptstyle {L}$ and $\scriptstyle {{\hat L}}$ are such that $L\ge L^*$ and
by removing the external unit protuberance one obtains a subcritical
birectangle.}
}
}
\endinsert
\par
We observe that
$$H({\cal B}_{1,a}(M,{\hat M};{\hat L}))\ge
H({\cal B}_{1,a}({\widetilde M},{\widetilde M};L^*))\Eq (5.7)$$
for every possible choice of the positive integer numbers ${\hat M}$, $M$ and
${\hat L}$. This is an obvious consequence of the fact that $L=L^*-1<L^*$
and $M<M^*$.
\par
Now, we transform ${\cal B}_{1,a}({\widetilde M},{\widetilde M};L^*)$ into
${\cal P}_1$ in several steps and we evaluate the energy cost $\D H_i$ of each
step.
\itemitem{$\bullet$} ${\cal B}_{1,a}({\widetilde M},{\widetilde M};L^*)
\rightarrow R(L^*-1,L^*;{\widetilde M},{\widetilde M})$,
$\D H_1:=H(R(L^*-1,L^*;{\widetilde M},{\widetilde M}))-
H({\cal B}_{1,a}({\widetilde M},{\widetilde M};L^*))=-[2J-(h+\l)]$;
\itemitem{$\bullet$} $R(L^*-1,L^*;{\widetilde M},{\widetilde M})
\rightarrow R(L^*-1,L^*;l^*+2,l^*+2)$, $\D H_2<0$ because the external
rectangle is subcritical and ${\widetilde M}>l^*+2$
(see inequalities $\equ (3.14)$);
\itemitem{$\bullet$} $R(L^*-1,L^*;l^*+2,l^*+2)\rightarrow
R(L^*,L^*;l^*+2,l^*+2)$, $\D H_3<0$ because a whole internal slice of
lenght $L^*$ has been attached to the internal (relatively) supercritical
rectangle;
\itemitem{$\bullet$} $R(L^*,L^*;l^*+2,l^*+2)\rightarrow
R(l^*-1,l^*;l^*+2,l^*+2)$, $\D H_4<0$ because the internal rectangle
is supercritical and $L^*<l^*-1$;
\itemitem{$\bullet$} $R(l^*-1,l^*;l^*+2,l^*+2)\rightarrow {\cal P}_1$,
$\D H_5=2J-(h+\l)$.
\par\noindent
One has that $\sum_{i=1}^5 \D H_i<0$, hence
$H({\cal B}_{1,a}({\widetilde M},{\widetilde M};L^*))>H({\cal P}_1)$. This
inequality and $\equ (5.7)$ lead us to the conclusion that
$$H({\cal B}_{1,a}(M,{\hat M};{\hat L}))> {\cal P}_1\Eq (5.8)$$
for every possible choice of ${\hat M}$, $M$ and
${\hat L}$.
\par
In order to carachterize the special saddle
${\cal B}_{1,b}({\hat M};L,{\hat L})$, we have to distinguish two possible
cases.
\par\noindent
Case (i) ${\hat L}+2\ge {\widetilde M}$: the birectangle
$R(L,{\hat L};{\widetilde M}-1,{\hat M})$, obtained from
${\cal B}_{1,b}({\hat M};L,{\hat L})$ by removing the external unit
protuberance, must be subcritical. Then, by virtue of Proposition 4.2, one can
say that it must necessarily be $({\widetilde M}-1)-2\le l^*-1$, that is
${\widetilde M}\le l^*+2$. This is an absurd
(see inequalities $\equ (3.14)$).
Then we can conclude that it does not exist a special saddle
${\cal B}_{1,b}({\hat M};L,{\hat L})$ such that ${\hat L}+2\ge {\widetilde M}$.
\par\noindent
Case (ii) ${\hat L}+2< {\widetilde M}$: the internal rectangle
$L\times {\hat L}$ must be contained in the rectangle $L\times (m^*(L)-1)$,
otherwise the birectangle $R(L,{\hat L};{\widetilde M}-1,{\hat M})$ would be
supercritical. Now we transform the special saddle
${\cal B}_{1,b}({\hat M};L,{\hat L})$ into $C_{2,a}(L+1)$ (notice that
$L\ge L^*\Rightarrow L+1\ge {\widetilde L}$) and we show
that the energy lowers.
\itemitem{$\bullet$} ${\cal B}_{1,b}({\hat M};L,{\hat L})\rightarrow
R(L,{\hat L};{\widetilde M}-1,{\hat M})$, $\D H_1=-[2J-(h-\l)]$;
\itemitem{$\bullet$} $R(L,{\hat L};{\widetilde M}-1,{\hat M})\rightarrow
R(L+1,{\hat L};{\widetilde M}-1,{\hat M})$, $\D H_2\le 0$ because
${\hat L}\ge L\ge L^*$;
\itemitem{$\bullet$} $R(L+1,{\hat L};{\widetilde M}-1,{\hat M})\rightarrow
R(L+1,m^*(L+1)-1;{\widetilde M}-1,{\hat M})$, $\D H_3\le 0$ because $L\ge L^*$
and ${\hat L}\le m^*(L+1)-1$;
\itemitem{$\bullet$} $R(L+1,m^*(L+1)-1;{\widetilde M}-1,{\hat M})\rightarrow
R(L+1,m^*(L+1)-1;L+2,m^*(L+1)+2)$, $\D H_4<0$ because the external rectangle is
subcritical and $L+2<{\widetilde M}-1$;
\itemitem{$\bullet$} $R(L+1,m^*(L+1)-1;L+2,m^*(L+1)+2)\rightarrow
C_{2,a}(L+1)$, $\D H_5=2J-(h+\l)$.
\par\noindent
Hence,
$$H({\cal B}_{1,b}({\hat M};L,{\hat L}))>H(C_{2,a}(L+1))>H({\cal P}_1)
\Eq (5.9)$$
for every possible choice of the dimensions ${\hat M}$, $L$ and ${\hat L}$.
\vskip 0.35 truecm
\par\noindent
{\it Case B2}.
\par
We consider the special saddle ${\cal B}_2({\hat M};L,{\hat L})$ in Fig.5.6.
Two possible cases must be considered.
\midinsert
\vskip 0.5 truecm
\vbox{\font\amgr=cmr10 at
10truept\baselineskip0.1466667truein\lineskiplimit-\maxdimen
\catcode`\-=\active\catcode`\~=\active\def~{{\char32}}\def-{{\char1}}%
\hbox{\amgr ~~~~~~~~~~~~~~~~~~~~~$\phantom {{\cal B}_2({\hat M};L,{\hat
L})}$~~~~{\char2}----------------{\char3}~~~~~~~~~~~~~~}
\hbox{\amgr ~~~~~~~~~~~~~~~~~~~~~$\phantom {{\cal B}_2({\hat M};L,{\hat
L})}$~~~~{\char0}~~~~~~~~~~~~~~~~{\char0}}
\hbox{\amgr ~~~~~~~~~~~~~~~~~~~~~$\phantom {{\cal B}_2({\hat M};L,{\hat
L})}$~~~~{\char0}~{\char2}-------{\char3}~~~~~~{\char0}}
\hbox{\amgr ~~~~~~~~~~~~~~~~~~~~~$\phantom {{\cal B}_2({\hat M};L,{\hat
L})}$~~~~{\char0}~{\char0}~~~~~~~{\char0}~~~~~~{\char0}}
\hbox{\amgr ~~~~~~~~~~~~~~~~~~~~~$\phantom {{\cal B}_2({\hat M};L,{\hat
L})}$~~~~{\char0}~{\char0}~~~~~~~{\char0}~~~~~~{\char0}}
\hbox{\amgr ~~~~~~~~~~~~~~~~~~~~~$\phantom {{\cal B}_2({\hat M};L,{\hat
L})}$~~~~{\char0}~{\char0}~~~~~~~{\char0}~~~~~~{\char5}-{\char3}}
\hbox{\amgr ~~~~~~~~~~~~~~~~~~~~~$         {{\cal B}_2({\hat M};L,{\hat
L})}$~~~~{\char0}~{\char0}~~~~~~~{\char0}~~~~~~{\char2}-{\char4}}
\hbox{\amgr ~~~~~~~~~~~~~~~~~~~~~$\phantom {{\cal B}_2({\hat M};L,{\hat
L})}$~~~~{\char0}~{\char0}~~~~~~~{\char0}~~~~~~{\char0}~~~~${\hat
M}$~~~~~~~~~~~}
\hbox{\amgr ~~~~~~~~~~~~~~~~~~~~~$\phantom {{\cal B}_2({\hat M};L,{\hat
L})}$~~~~{\char0}~{\char0}~~~~~~~{\char0}~~~~~~{\char0}~~~~~~}
\hbox{\amgr ~~~~~~~~~~~~~~~~~~~~~$\phantom {{\cal B}_2({\hat M};L,{\hat
L})}$~~~~{\char0}~{\char0}~~~~~~~{\char0}~~~~~~{\char0}}
\hbox{\amgr ~~~~~~~~~~~~~~~~~~~~~$\phantom {{\cal B}_2({\hat M};L,{\hat
L})}$~~~~{\char0}~{\char5}-------{\char4}~~~~~~{\char0}}
\hbox{\amgr ~~~~~~~~~~~~~~~~~~~~~$\phantom {{\cal B}_2({\hat M};L,{\hat
L})}$~~~~{\char5}----------------{\char4}}
\hbox{\amgr ~~~~~~~~~~~~~~~~~~~~~$\phantom {{\cal B}_2({\hat M};L,{\hat
L})}$~~~~~~~~~$M^*{\char 45}1$~~~~~}}
\vskip 0.5 truecm
\par\noindent
\centerline{
\vbox
{
\hsize=15truecm
\baselineskip 0.35cm
\noindent
{\smbfb Fig.5.6}{\smb \quad The internal dimensions $\scriptstyle {L}$ and
$\scriptstyle {{\hat L}}$ are such that the birectangle obtained by removing
the zero unit protuberance is subcritical. The
external dimensions $\scriptstyle {M^*-1}$ and $\scriptstyle {M}$ are such
that $\scriptstyle {{\hat M}\ge M^*}$.}
}
}
\endinsert
\par\noindent
Case (i) $M^*>{\widetilde M}$: the internal rectangle is subcritical, hence by
removing it we obtain a configuration at lower energy. Then by means of
arguments similar to those used in the case of standard Ising model
(see e.g. [NS1]), one can prove that
$$H({\cal B}_2({\hat M};L,{\hat L}))\ge H({\cal P}_2)\;\; ,\Eq (5.6)$$
where the equality stands iff ${\cal B}_2({\hat M};L,{\hat L})\equiv {\cal
P}_2$.
\par\noindent
Case (ii) $M^*={\widetilde M}$: see the discussion about the special saddle
${\cal B}_{1,b}({\hat M};L,{\hat L})$.
\midinsert
\vskip 0.5 truecm
\vbox{\font\amgr=cmr10 at
10truept\baselineskip0.1466667truein\lineskiplimit-\maxdimen
\catcode`\-=\active\catcode`\~=\active\def~{{\char32}}\def-{{\char1}}%
\hbox{\amgr ~~~~~~~~~~~~~~~~~~~~~$\phantom {{\cal B}_{3}(M,{\hat M};{\hat
L})}$~~~~{\char2}----------------{\char3}}
\hbox{\amgr ~~~~~~~~~~~~~~~~~~~~~$\phantom {{\cal B}_{3}(M,{\hat M};{\hat
L})}$~~~~{\char0}~~~~~~~~~~~~~~~~{\char0}}
\hbox{\amgr ~~~~~~~~~~~~~~~~~~~~~$\phantom {{\cal B}_{3}(M,{\hat M};{\hat
L})}$~~~~{\char0}~~~{\char2}------{\char3}~~~~~{\char0}}
\hbox{\amgr ~~~~~~~~~~~~~~~~~~~~~$\phantom {{\cal B}_{3}(M,{\hat M};{\hat
L})}$~~~~{\char0}~~~{\char0}~~~~~~{\char0}~~~~~{\char0}}
\hbox{\amgr ~~~~~~~~~~~~~~~~~~~~~$\phantom {{\cal B}_{3}(M,{\hat M};{\hat
L})}$~~~~{\char0}~~~{\char0}~~~~~~{\char0}~~~~~{\char0}}
\hbox{\amgr ~~~~~~~~~~~~~~~~~~~~~$\phantom {{\cal B}_{3}(M,{\hat M};{\hat
L})}$~~~~{\char0}~~~{\char0}~~~~~~{\char5}-{\char3}~~~{\char0}}
\hbox{\amgr ~~~~~~~~~~~~~~~~~~~~~$         {{\cal B}_{3}(M,{\hat M};{\hat
L})}$~~~~{\char0}~~~{\char0}~~~~~~{\char2}-{\char4}~~~{\char0}~~~${\hat M}$}
\hbox{\amgr ~~~~~~~~~~~~~~~~~~~~~$\phantom {{\cal B}_{3}(M,{\hat M};{\hat
L})}$~~~~{\char0}~~~{\char0}~~~~~~{\char0}~~~~~{\char0}}
\hbox{\amgr ~~~~~~~~~~~~~~~~~~~~~$\phantom {{\cal B}_{3}(M,{\hat M};{\hat
L})}$~~~~{\char0}~~~{\char0}~~~~~~{\char0}~~~~~{\char0}}
\hbox{\amgr ~~~~~~~~~~~~~~~~~~~~~$\phantom {{\cal B}_{3}(M,{\hat M};{\hat
L})}$~~~~{\char0}~~~{\char0}~~~~~~{\char0}~~~~~{\char0}}
\hbox{\amgr ~~~~~~~~~~~~~~~~~~~~~$\phantom {{\cal B}_{3}(M,{\hat M};{\hat
L})}$~~~~{\char0}~~~{\char5}------{\char4}~~~~~{\char0}}
\hbox{\amgr ~~~~~~~~~~~~~~~~~~~~~$\phantom {{\cal B}_{3}(M,{\hat M};{\hat
L})}$~~~~{\char5}----------------{\char4}}
\hbox{\amgr ~~~~~~~~~~~~~~~~~~~~~$\phantom {{\cal B}_{3}(M,{\hat M};{\hat
L})}$~~~~~~~~~~$M$~~~~}}
\vskip 0.5 truecm
\par\noindent
\centerline{
\vbox
{
\hsize=15truecm
\baselineskip 0.35cm
\noindent
{\smbfb Fig.5.7}{\smb \quad
The internal horizontal dimension is $\scriptstyle {l^*-1}$ and the
vertical one $\scriptstyle {{\hat L}}$ is such that
$\scriptstyle {l^*\le {\hat L}< {\widetilde M}-2}$ and
$\scriptstyle {{\hat L}< m^*(l^*-1)}$. The external vertical dimension is
$\scriptstyle {{\hat M}}$ and the horizontal one  is
$\scriptstyle {M<{\widetilde M}}$.}
}
}
\endinsert
\par
\vskip 0.35 truecm
\par\noindent
{\it Case B3}.
\par
We consider, now, the special saddle ${\cal B}_{3}(M,{\hat M};{\hat L})$ in
Fig.5.7. One can easily prove that $H({\cal B}_{3}(M,{\hat M};{\hat L}))\ge
H(C_1({\hat L}))$ by virtue of the inequalities $M<M^*$ and
${\hat L}+2<{\widetilde M}\le M^*$.
\par
Hence, we conclude that
$$H({\cal B}_{3}(M,{\hat M};{\hat L}))\ge H(C_1({\hat L}))\ge H({\cal P}_1)
\Eq (5.10)$$
for every possible choice of the dimensions $M$, ${\hat M}$ and ${\hat L}$.
We observe that in $\equ (5.10)$ the equality holds iff
${\cal B}_{3}(M,{\hat M};{\hat L})\equiv {\cal P}_1$.
\midinsert
\vskip 0.5 truecm
\vbox{\font\amgr=cmr10 at
10truept\baselineskip0.1466667truein\lineskiplimit-\maxdimen
\catcode`\-=\active\catcode`\~=\active\def~{{\char32}}\def-{{\char1}}%
\hbox{\amgr ~~$\phantom {{\cal B}_{4,a}(M,{\hat
M};L)}$~~~~{\char2}----------------{\char3}~~~~~~~~~~~~~~{\char2}----------------{\char3}}
\hbox{\amgr ~~$\phantom {{\cal B}_{4,a}(M,{\hat
M};L)}$~~~~{\char0}~~~~~~~~~~~~~~~~{\char0}~~~~~~~~~~~~~~{\char0}~~~~~~~~~~~~~~~~{\char0}}
\hbox{\amgr ~~$\phantom {{\cal B}_{4,a}(M,{\hat
M};L)}$~~~~{\char0}~~~~~{\char2}-{\char3}~~~~~~~~{\char0}~~~~~~~~~~~~~~{\char0}~~~~~~~~~~~~~~~~{\char0}%
}
\hbox{\amgr ~~$\phantom {{\cal B}_{4,a}(M,{\hat
M};L)}$~~~~{\char0}~~~{\char2}-{\char4}~{\char5}--{\char3}~~~~~{\char0}~~~~~~~~~~~~~~{\char0}~~{\char2}%
------{\char3}~~~~~~{\char0}}
\hbox{\amgr ~~$\phantom {{\cal B}_{4,a}(M,{\hat
M};L)}$~~~~{\char0}~~~{\char0}~~~~~~{\char0}~~~~~{\char0}~~~~~~~~~~~~~~{\char0}~~{\char0}~~~~~~{\char0}%
{}~~~~~~{\char0}}
\hbox{\amgr ~~$\phantom {{\cal B}_{4,a}(M,{\hat
M};L)}$~~~~{\char0}~~~{\char0}~~~~~~{\char0}~~~~~{\char0}~~~~~~~~~~~~~~{\char0}~~{\char0}~~~~~~{\char0}%
{}~~~~~~{\char0}}
\hbox{\amgr ~~$\phantom {{\cal B}_{4,a}(M,{\hat
M};L)}$~~~~{\char0}~~~{\char0}~~~~~~{\char0}~~~~~{\char0}~~~~~~~~~~~~~~{\char0}~~{\char0}~~~~~~{\char5}%
-{\char3}~~~~{\char0}}
\hbox{\amgr ~~$         {{\cal B}_{4,a}(M,{\hat
M};L)}$~~~~{\char0}~~~{\char0}~~~~~~{\char0}~~~~~{\char0}~~~~~~~~~~~~~~{\char0}~~{\char0}~~~~~~{\char2}%
-{\char4}~~~~{\char0}~~~~${\cal B}_{4,b}(M,{\hat M};L)$}
\hbox{\amgr ~~$\phantom {{\cal B}_{4,a}(M,{\hat
M};L)}$~~~~{\char0}~~~{\char0}~~~~~~{\char0}~~~~~{\char0}~~~~~~~~~~~~~~{\char0}~~{\char0}~~~~~~{\char0}%
{}~~~~~~{\char0}}
\hbox{\amgr ~~$\phantom {{\cal B}_{4,a}(M,{\hat
M};L)}$~~~~{\char0}~~~{\char0}~~~~~~{\char0}~~~~~{\char0}~~~~~~~~~~~~~~{\char0}~~{\char0}~~~~~~{\char0}%
{}~~~~~~{\char0}}
\hbox{\amgr ~~$\phantom {{\cal B}_{4,a}(M,{\hat
M};L)}$~~~~{\char0}~~~{\char0}~~~~~~{\char0}~~~~~{\char0}~~~~~~~~~~~~~~{\char0}~~{\char0}~~~~~~{\char0}%
{}~~~~~~{\char0}}
\hbox{\amgr ~~$\phantom {{\cal B}_{4,a}(M,{\hat
M};L)}$~~~~{\char0}~~~{\char5}------{\char4}~~~~~{\char0}~~~~~~~~~~~~~~{\char0}~~{\char5}------{\char4}%
{}~~~~~~{\char0}}
\hbox{\amgr ~~$\phantom {{\cal B}_{4,a}(M,{\hat
M};L)}$~~~~{\char0}~~~~~~~~~~~~~~~~{\char0}~~~~~~~~~~~~~~{\char0}~~~~~~~~~~~~~~~~{\char0}}
\hbox{\amgr ~~$\phantom {{\cal B}_{4,a}(M,{\hat
M};L)}$~~~~{\char0}~~~~~~~~~~~~~~~~{\char0}~~~~~~~~~~~~~~{\char0}~~~~~~~~~~~~~~~~{\char0}}
\hbox{\amgr ~~$\phantom {{\cal B}_{4,a}(M,{\hat
M};L)}$~~~~{\char5}----------------{\char4}~~~~~~~~~~~~~~{\char5}----------------{\char4}}
\hbox{\amgr ~~$\phantom {{\cal B}_{4,a}(M,{\hat
M};L)}$~~~~~~~~~~~$M$~~~~~~~~~~~~~~~~~~~~~~~~~~~~~~~$M$}}
\vskip 0.5 truecm
\par\noindent
\centerline{
\vbox
{
\hsize=15truecm
\baselineskip 0.35cm
\noindent
{\smbfb Fig.5.8}{\smb \quad
(a) The internal horizontal dimension $\scriptstyle {L}$ is such that
$\scriptstyle {{\widetilde L}\le L\le l^*-1}$. The vertical one
is $\scriptstyle {{\hat L}=m^*(L)-1}$ and it is such that
$\scriptstyle {{\hat L}+3<{\widetilde M}}$. The external vertical dimension is
$\scriptstyle {{\hat M}}$ and the horizontal one $\scriptstyle {M}$ is such
that $\scriptstyle {M<{\widetilde M}}$.
(b) The external  dimensions are like those in (a). The internal
horizontal dimension $\scriptstyle {L-1}$ is such that
$\scriptstyle {{\widetilde L}\le L-1\le l^*-2}$. The vertical one
is $\scriptstyle {{\hat L}=m^*(L)}$ and it is such that
$\scriptstyle {{\hat L}+2<{\widetilde M}}$. We remark that for certain choices
of the parameters $\scriptstyle {h}$ and $\scriptstyle {\l}$ the configurations
in (a) and (b) cannot be considered; it could be, indeed,
$\scriptstyle {m^*(L)\ge {\widetilde M}-2}$.}
}
}
\endinsert
\par
\vskip 0.35 truecm
\par\noindent
{\it Case B4}.
\par
We consider the special saddles ${\cal B}_{4,a}(M,{\hat M};L)$ and
${\cal B}_{4,b}(M,{\hat M};L)$ in Fig.5.8. First of all we observe that
$$H({\cal B}_{4,a}(M,{\hat M};L))\ge C_{2,a}(L)\; {\rm and}\;
  H({\cal B}_{4,b}(M,{\hat M};L))\ge C_{2,b}(L)\;\; ,\Eq (5.11)$$
for every possible choice of $M$, ${\hat M}$ and $L$. Equation $\equ (5.11)$
is a consequence of the fact that $M<{\widetilde M}\le M^*$ and
${\hat L}+3<{\widetilde M}\le M^*$ in both cases. The equalities are verified
in $\equ (5.11)$ iff ${\cal B}_{4,a}(M,{\hat M};L)\equiv C_{2,a}(L)$ or
${\cal B}_{4,b}(M,{\hat M};L)\equiv C_{2,b}(L)$.
\par
Now, by arguments similar to those used in the discussion of case {\it C2} we
can prove that $H({\cal B}_{4,a}(M,{\hat M};L))>H({\cal P}_1)$ and
$H({\cal B}_{4,b}(M,{\hat M};L))>H({\cal P}_1)$.
\vskip 0.35 truecm
\par\noindent
{\it Case B5}.
\par
We consider the special saddle ${\cal B}_{5}({\hat M};L,{\hat L})$ in Fig.5.9.
\midinsert
\vskip 0.5 truecm
\vbox{\font\amgr=cmr10 at
10truept\baselineskip0.1466667truein\lineskiplimit-\maxdimen
\catcode`\-=\active\catcode`\~=\active\def~{{\char32}}\def-{{\char1}}%
\hbox{\amgr ~~~~~~~~~~~~~~~~~~~~$\phantom {{\cal B}_{5}({\hat M};L,{\hat
L})}$~~~~{\char2}--------------{\char3}}
\hbox{\amgr ~~~~~~~~~~~~~~~~~~~~$\phantom {{\cal B}_{5}({\hat M};L,{\hat
L})}$~~~~{\char0}~~~~~~~~~~~~~~{\char0}}
\hbox{\amgr ~~~~~~~~~~~~~~~~~~~~$\phantom {{\cal B}_{5}({\hat M};L,{\hat
L})}$~~~~{\char0}~~{\char2}------{\char3}~~~~{\char0}}
\hbox{\amgr ~~~~~~~~~~~~~~~~~~~~$\phantom {{\cal B}_{5}({\hat M};L,{\hat
L})}$~~~~{\char0}~~{\char0}~~~~~~{\char0}~~~~{\char0}}
\hbox{\amgr ~~~~~~~~~~~~~~~~~~~~$\phantom {{\cal B}_{5}({\hat M};L,{\hat
L})}$~~~~{\char0}~~{\char0}~~~~~~{\char0}~~~~{\char0}}
\hbox{\amgr ~~~~~~~~~~~~~~~~~~~~$\phantom {{\cal B}_{5}({\hat M};L,{\hat
L})}$~~~~{\char0}~~{\char0}~~~~~~{\char0}~~~~{\char5}-{\char3}}
\hbox{\amgr ~~~~~~~~~~~~~~~~~~~~$\phantom {{\cal B}_{5}({\hat M};L,{\hat
L})}$~~~~{\char0}~~{\char0}~~~~~~{\char0}~~~~{\char2}-{\char4}}
\hbox{\amgr ~~~~~~~~~~~~~~~~~~~~$         {{\cal B}_{5}({\hat M};L,{\hat
L})}$~~~~{\char0}~~{\char0}~~~~~~{\char0}~~~~{\char0}}
\hbox{\amgr ~~~~~~~~~~~~~~~~~~~~$\phantom {{\cal B}_{5}({\hat M};L,{\hat
L})}$~~~~{\char0}~~{\char0}~~~~~~{\char0}~~~~{\char0}}
\hbox{\amgr ~~~~~~~~~~~~~~~~~~~~$\phantom {{\cal B}_{5}({\hat M};L,{\hat
L})}$~~~~{\char0}~~{\char0}~~~~~~{\char0}~~~~{\char0}}
\hbox{\amgr ~~~~~~~~~~~~~~~~~~~~$\phantom {{\cal B}_{5}({\hat M};L,{\hat
L})}$~~~~{\char0}~~{\char5}------{\char4}~~~~{\char0}}
\hbox{\amgr ~~~~~~~~~~~~~~~~~~~~$\phantom {{\cal B}_{5}({\hat M};L,{\hat
L})}$~~~~{\char0}~~~~~~~~~~~~~~{\char0}}
\hbox{\amgr ~~~~~~~~~~~~~~~~~~~~$\phantom {{\cal B}_{5}({\hat M};L,{\hat
L})}$~~~~{\char0}~~~~~~~~~~~~~~{\char0}}
\hbox{\amgr ~~~~~~~~~~~~~~~~~~~~$\phantom {{\cal B}_{5}({\hat M};L,{\hat
L})}$~~~~{\char5}--------------{\char4}}
\hbox{\amgr ~~~~~~~~~~~~~~~~~~~~$\phantom {{\cal B}_{5}({\hat M};L,{\hat
L})}$~~~~~~~~~$l^*+1$~~~~~}}
\vskip 0.5 truecm
\par\noindent
\centerline{
\vbox
{
\hsize=15truecm
\baselineskip 0.35cm
\noindent
{\smbfb Fig.5.9}{\smb \quad
The internal horizontal dimension $\scriptstyle {L}$ is such that
$\scriptstyle {L^*\le L\le l^*-1}$
and the vertical one $\scriptstyle {{\hat L}}$ is such that
$\scriptstyle {{\widetilde M}-2\le {\hat L}< m^*(l^*-1)}$.
The external vertical dimension is $\scriptstyle {{\hat M}}$.
We remark that this special saddle does not exist, if we choose the parameters
$\scriptstyle {h}$ and $\scriptstyle {\l}$ such that
$\scriptstyle {m^*(l^*-1)\le {\widetilde M}-2}$.}
}
}
\endinsert
\par
Now we transform the special saddle ${\cal B}_{5}({\hat M};L,{\hat L})$ into
$C_{1}({\hat L})$ and we show that the energy lowers.
\itemitem {$\bullet$} ${\cal B}_{5}({\hat M};L,{\hat L})\rightarrow
R(L,{\hat L};l^*+2,{\hat M})$, $\D H_1=-(h-\l)({\hat M}-1)$;
\itemitem {$\bullet$} $R(L,{\hat L};l^*+2,{\hat M})\rightarrow
R(l^*-1,{\hat L};l^*+2,{\hat M})$, $\D H_2\le 0$ because
$L\ge {\widetilde M}-2>L^*$ and $L\le l^*-1$;
\itemitem {$\bullet$} $R(l^*-1,{\hat L};l^*+2,{\hat M})\rightarrow
R(l^*-1,{\hat L};l^*+2,{\hat L}+2)$, $\D H_3\le 0$ since $l^*+2<M^*$;
\itemitem {$\bullet$} $R(l^*-1,{\hat L};l^*+2,{\hat L}+2)\rightarrow
C_{1}({\hat L})$, $\D H_4=2J-(h+\l)$.
\par\noindent
By a direct calculation it can be proved that $\D H_1+\D H_4\le 0$, then
$H({\cal B}_{5}({\hat M};L,{\hat L}))> H(C_{1}({\hat L}))\ge
H({\cal P}_1)$.
\vskip 0.35 truecm
\par\noindent
{\it Case B6}.
\par
We consider the special saddles ${\cal B}_{6,a}(M,{\hat M};L)$ and
${\cal B}_{6,b}(M,{\hat M};L)$ in Fig.5.10.
\midinsert
\vskip 0.5 truecm
\vbox{\font\amgr=cmr10 at
10truept\baselineskip0.1466667truein\lineskiplimit-\maxdimen
\catcode`\-=\active\catcode`\~=\active\def~{{\char32}}\def-{{\char1}}%
\hbox{\amgr ~~$\phantom {{\cal B}_{6,a}(M,{\hat
M};L)}$~~~~{\char2}----------------{\char3}~~~~~~~~~{\char2}----------------{\char3}}
\hbox{\amgr ~~$\phantom {{\cal B}_{6,a}(M,{\hat
M};L)}$~~~~{\char0}~~~~~~~~~~~~~~~~{\char0}~~~~~~~~~{\char0}~~~~~~~~~~~~~~~~{\char0}}
\hbox{\amgr ~~$\phantom {{\cal B}_{6,a}(M,{\hat
M};L)}$~~~~{\char0}~~~~~{\char2}-{\char3}~~~~~~~~{\char0}~~~~~~~~~{\char0}~~~~~~~~~~~~~~~~{\char0}%
}
\hbox{\amgr ~~$\phantom {{\cal B}_{6,a}(M,{\hat
M};L)}$~~~~{\char0}~~~{\char2}-{\char4}~{\char5}--{\char3}~~~~~{\char0}~~~~~~~~~{\char0}~~{\char2}%
------{\char3}~~~~~~{\char0}}
\hbox{\amgr ~~$\phantom {{\cal B}_{6,a}(M,{\hat
M};L)}$~~~~{\char0}~~~{\char0}~~~~~~{\char0}~~~~~{\char0}~~~~~~~~~{\char0}~~{\char0}~~~~~~{\char0}%
{}~~~~~~{\char0}}
\hbox{\amgr ~~$\phantom {{\cal B}_{6,a}(M,{\hat
M};L)}$~~~~{\char0}~~~{\char0}~~~~~~{\char0}~~~~~{\char0}~~~~~~~~~{\char0}~~{\char0}~~~~~~{\char0}%
{}~~~~~~{\char0}}
\hbox{\amgr ~~$\phantom {{\cal B}_{6,a}(M,{\hat
M};L)}$~~~~{\char0}~~~{\char0}~~~~~~{\char0}~~~~~{\char0}~~~~~~~~~{\char0}~~{\char0}~~~~~~{\char0}~~~~~~{\char5}-{\char3}}
\hbox{\amgr ~~$         {{\cal B}_{6,a}(M,{\hat
M};L)}$~~~~{\char0}~~~{\char0}~~~~~~{\char0}~~~~~{\char0}~~~~~~~~~{\char0}~~{\char0}~~~~~~{\char0}~~~~~~{\char2}-{\char4}~~~~${\cal B}_{6,b}(M,{\hat M};L)$}
\hbox{\amgr ~~$\phantom {{\cal B}_{6,a}(M,{\hat
M};L)}$~~~~{\char0}~~~{\char0}~~~~~~{\char0}~~~~~{\char0}~~~~~~~~~{\char0}~~{\char0}~~~~~~{\char0}%
{}~~~~~~{\char0}}
\hbox{\amgr ~~$\phantom {{\cal B}_{6,a}(M,{\hat
M};L)}$~~~~{\char0}~~~{\char0}~~~~~~{\char0}~~~~~{\char0}~~~~~~~~~{\char0}~~{\char0}~~~~~~{\char0}%
{}~~~~~~{\char0}}
\hbox{\amgr ~~$\phantom {{\cal B}_{6,a}(M,{\hat
M};L)}$~~~~{\char0}~~~{\char0}~~~~~~{\char0}~~~~~{\char0}~~~~~~~~~{\char0}~~{\char0}~~~~~~{\char0}%
{}~~~~~~{\char0}}
\hbox{\amgr ~~$\phantom {{\cal B}_{6,a}(M,{\hat
M};L)}$~~~~{\char0}~~~{\char5}------{\char4}~~~~~{\char0}~~~~~~~~~{\char0}~~{\char5}------{\char4}%
{}~~~~~~{\char0}}
\hbox{\amgr ~~$\phantom {{\cal B}_{6,a}(M,{\hat
M};L)}$~~~~{\char0}~~~~~~~~~~~~~~~~{\char0}~~~~~~~~~{\char0}~~~~~~~~~~~~~~~~{\char0}}
\hbox{\amgr ~~$\phantom {{\cal B}_{6,a}(M,{\hat
M};L)}$~~~~{\char0}~~~~~~~~~~~~~~~~{\char0}~~~~~~~~~{\char0}~~~~~~~~~~~~~~~~{\char0}}
\hbox{\amgr ~~$\phantom {{\cal B}_{6,a}(M,{\hat
M};L)}$~~~~{\char5}----------------{\char4}~~~~~~~~~{\char5}----------------{\char4}}
\hbox{\amgr ~~$\phantom {{\cal B}_{6,a}(M,{\hat
M};L)}$~~~~~~~~~~~$M$~~~~~~~~~~~~~~~~~~~~~~~~$M$}}
\vskip 0.5 truecm
\par\noindent
\centerline{
\vbox
{
\hsize=15truecm
\baselineskip 0.35cm
\noindent
{\smbfb Fig.5.10}{\smb \quad
(a) The internal horizontal dimension $\scriptstyle {L}$ is such that
$\scriptstyle {L\ge L^*}$. The vertical one
is $\scriptstyle {{\hat L}=m^*(M-2)-1}$ and it is such that
$\scriptstyle {{\hat L}+3\ge {\widetilde M}}$. The external vertical dimension
is $\scriptstyle {{\hat M}}$ and the horizontal one $\scriptstyle {M}$ is such
that $\scriptstyle {{\widetilde L}\le M-2\le l^*-1}$.
(b) The internal horizontal dimension $\scriptstyle {L}$ is such that
$\scriptstyle {L\ge L^*}$. The vertical one
is $\scriptstyle {{\hat L}=m^*(M-1)}$ and it is such that
$\scriptstyle {{\hat L}+2\ge {\widetilde M}}$. The external vertical dimension
is $\scriptstyle {{\hat M}}$ and the external horizontal dimension
$\scriptstyle {M}$ is such that
$\scriptstyle {{\widetilde L}\le M-1\le l^*-1}$. We remark that for certain
choices of the parameters $\scriptstyle {h}$ and $\scriptstyle {\l}$ the
configurations in (a) and (b) cannot be considered.}
}
}
\endinsert
\par
Now, we transform the special saddle ${\cal B}_{6,a}(M,{\hat M};L)$ into
$C_{2,a}(M-2)$ and show that the energy lowers.
\itemitem{$\bullet$} ${\cal B}_{6,a}(M,{\hat M};L)\rightarrow
{\cal B}_{6,a}(M,{\hat L}+3;L)$, $\D H_1\le 0$ since $M\le l^*+1<M^*$;
\itemitem{$\bullet$} ${\cal B}_{6,a}(M,{\hat L}+3;L)\rightarrow
C_{2,a}(M-2)$, $\D H_2\le 0$ because
${\hat L}\ge {\widetilde M}-3\ge l^*>L^*$.
\par\noindent
Hence, we conclude that
$H({\cal B}_{6,a}(M,{\hat M};L))\ge H(C_{2,a}(M-2))>H({\cal P}_1)$.
\par
The special saddle ${\cal B}_{6,b}(M,{\hat M};L)$ can be transformed into the
configuration $C_{2,b}(M-1)$ lowering the energy.
\itemitem{$\bullet$} ${\cal B}_{6,b}(M,{\hat M};L)\rightarrow
R(L,{\hat L};M+1,{\hat M})$, $\D H_1=-(h-\l)({\hat M}-1)$;
\itemitem{$\bullet$} $R(L,{\hat L};M+1,{\hat M})\rightarrow
R(L,{\hat L};M+1,{\hat L}+2)$, $\D H_2\le 0$ since $M+1\le l^*+1<M^*$;
\itemitem{$\bullet$} $R(L,{\hat L};M+1,{\hat L}+2)\rightarrow
R(M-2,{\hat L};M+1,{\hat L}+2)$, $\D H_3\le 0$ because
${\hat L}\ge {\widetilde M}-2>L^*$;
\itemitem{$\bullet$} $R(M-2,{\hat L};M+1,{\hat L}+2)\rightarrow
C_{2,b}(M-1)$, $\D H_4=2J-(h+\l)$.
\par\noindent
It is easily seen that $\D H_1+\D H_4<0$, hence
$H({\cal B}_{6,a}(M,{\hat M};L))>H(C_{2,b}(M-1))>H({\cal P}_1)$.
\par
This completes the proof of Lemma 5.1 in the case $\d<{h+\l\over 2h}$.
We suppose, now, $\d>{h+\l\over 2h}$ and observe that in this case
$$m^*(l^*-1)=l^*\;\; ,\Eq (5.12)$$
as it follows from equation $\equ (5.4)$. In the sequel we will analyze all the
cases that have to be discussed with arguments different from those used
before.
\vskip 0.35 truecm
\par\noindent
{\it Case C1}.
\par
The special saddle $C_1(m)$ with $l^*\le m\le m^*(l^*-1)-1$ cannot be
considered, since $m^*(l^*-1)-1=l^*-1$ (see $\equ (5.12)$).
\vskip 0.35 truecm
\par\noindent
{\it Case C2}.
\par
We proved above that $C_{2,a}(l^*-1)$ is the special saddle with lowest
energy among $C_{2,a}(L)$ and $C_{2,b}(L)$. This result is not dependent
on the value of the real number $\d$. Hence, one can say
$H(C_{2,b}(l))>H(C_{2,a}(l))\ge H(C_{2,a}(l^*-1))
=H({\cal P}_1)$ (we remark that in the case $\d>{h+\l\over 2h}$ the special
saddle $C_{2,a}(l^*-1)$ and the global saddle ${\cal P}_1$ coincide).
\vskip 0.35 truecm
\par\noindent
{\it Case B1}.
\par
In order to prove that $H({\cal B}_{1,a}(M,{\hat M};{\hat L}))>H({\cal P}_1)$
we have to consider two different cases.
\par\noindent
Case (i) ${\hat L}\ge l^*-1$: we transform the special saddle
${\cal B}_{1,a}(M,{\hat M};{\hat L})$ into ${\cal P}_1$ and we prove that the
energy lowers.
\itemitem{$\bullet$} ${\cal B}_{1,a}(M,{\hat M};{\hat L})\rightarrow
R(L^*-1,{\hat L};M,{\hat M})$, $\D H_1=-[2J-(h+\l)]$;
\itemitem{$\bullet$} $R(L^*-1,{\hat L};M,{\hat M})\rightarrow
R(L^*-1,l^*-1;M,{\hat M})$, $\D H_2\le 0$ since $L^*-1<L^*$ and
${\hat L}\ge l^*-1$;
\itemitem{$\bullet$} $R(L^*-1,l^*-1;M,{\hat M})\rightarrow
R(l^*-1,l^*-1;M,{\hat M})$, $\D H_3<0$ because $l^*-1\ge L^*$;
\itemitem{$\bullet$} $R(l^*-1,l^*-1;M,{\hat M})\rightarrow
R(l^*-1,l^*-1;l^*+2,l^*+1)$, $\D H_4<0$ since $M<M^*$ and $M>l^*+2$;
\itemitem{$\bullet$} $R(l^*-1,l^*-1;l^*+2,l^*+1)\rightarrow
{\cal P}_1$, $\D H_5=2J-(h+\l)$.
\par\noindent
We conclude that $H({\cal B}_{1,a}(M,{\hat M};{\hat L}))>H({\cal P}_1)$ since
$\sum_{1=1}^5 \D H_i<0$.
\par\noindent
Case (ii) $L^*\le {\hat L}<l^*-1$: first of all we notice that this case can be
considered only if $l^*-1>L^*$. Now we transform
${\cal B}_{1,a}(M,{\hat M};{\hat L})$ into ${\cal P}_1$.
\itemitem{$\bullet$} ${\cal B}_{1,a}(M,{\hat M};{\hat L})\rightarrow
R(L^*-1,{\hat L};M,{\hat M})$, $\D H_1=-[2J-(h+\l)]$;
\itemitem{$\bullet$} $R(L^*-1,{\hat L};M,{\hat M})\rightarrow
R(l^*-1,{\hat L};M,{\hat M})$, $\D H_2< 0$ since $L^*-1<l^*-1$ and
${\hat L}\ge L^*$;
\itemitem{$\bullet$} $R(l^*-1,{\hat L};M,{\hat M})\rightarrow
R(l^*-1,l^*-1;M,{\hat M})$, $\D H_3<0$ because $l^*-1\ge L^*$ and
${\hat L}<l^*-1$;
\itemitem{$\bullet$} $R(l^*-1,l^*-1;M,{\hat M})\rightarrow
R(l^*-1,l^*-1;l^*+2,l^*+1)$, $\D H_4<0$ since $M<M^*$ and $M>l^*+2$;
\itemitem{$\bullet$} $R(l^*-1,l^*-1;l^*+2,l^*+1)\rightarrow
{\cal P}_1$, $\D H_5=2J-(h+\l)$.
\par\noindent
Also in this case we conclude that
$H({\cal B}_{1,a}(M,{\hat M};{\hat L}))>H({\cal P}_1)$.
\par
Finally, with arguments similar to those used in the case $\d<{h+\l\over 2h}$
one can show that
$H({\cal B}_{1,b}({\hat M};L,{\hat L}))>H(C_{2,a}(L))>H({\cal P}_1)$.
\vskip 0.35 truecm
\par\noindent
{\it Case B3}.
\par
This case cannot be considered because the inequalities
$l^*\le {\hat L}<m^*(l^*-1)$ cannot be verified (see $\equ (5.12)$).
\vskip 0.35 truecm
\par\noindent
{\it Case B5}.
\par
This case cannot be considered because the inequalities
${\widetilde M}-2\le {\hat L}<m^*(l^*-1)$ cannot be verified. Indeed,
from $(3.15)$
one has $l^*+3\le {\widetilde M}$; hence $l^*<{\widetilde M}-2$. Finally,
$m^*(l^*-1)=l^*\Rightarrow m^*(l^*-1)<{\widetilde M}-2$.
\par
The proof of Lemma 5.1 is now complete. $\square$
\par \bigskip
\vfill\eject
\numsec=6\numfor=1

{\bf Section 6. The set ${\cal G}$ and the minimum of the energy on
$\partial {\cal G}$.}
\par
In this section we define a set ${\cal G}$ of configurations which will
play a basic role in the proof of our results. ${\cal G}$ will constitute
an ``upper estimate" of the generalized basin of attraction of $\menouno$,
in the sense that every subcritical configuration, that is a configuration
$\sigma$ such that
$$\lim_{\b\to\infty} P_{\s}(\t_{\menouno}<\t_{\piuuno})=1\;\; ,\Eq (6.1)$$
will belong to ${\cal G}$; moreover given any $\eta\in {\cal G}$ there
exists a downhill path leading to a configuration $\sigma$ satisfying $\equ
(6.1)$. On the other hand there are configurations $\eta\in {\cal G}$ which
are supercritical in the sense that
$$\lim_{\b\to\infty} P_{\eta}(\t_{\menouno}>\t_{\piuuno})=1\;\; .\Eq (6.1')$$
\par
The crucial property of ${\cal G}$ will be that the minimum of the energy
in its boundary $\partial {\cal G}$ will be given by ${\cal P}_1$ or
${\cal P}_2$.
\par
We will see that this implies that for every configuration $\sigma$ with
sufficiently low energy ($H(\sigma)<\min\{ H({\cal P}_1), H({\cal P}_2) \}$)
$\equ (6.1)$ is verified.
\par
Let us now give an example of a configuration belonging to ${\cal G}$ which
is potentially supercritical in the sense that $\equ (6.1)$ fails.
\par
Consider an acceptable configuration $\eta$ which is different from
$\menouno$ only in a square $\L_0:=\L_{L_0}$ ($\eta (x)=-1\; \forall x\in
\L\setminus\L_0$); the even integer $L_0$ will be chosen later on.
\par
Consider the four sublattices of spacing $2$ into which ${\bf Z}^{2}$ is
partitioned and write
$${\bf Z}^{2}\equiv {\bf Z}^{2}_1=
{\bf Z}^{2}_{2,a}\cup {\bf Z}^{2}_{2,b}\cup {\bf Z}^{2}_{2,c}\cup
{\bf Z}^{2}_{2,d}$$
where the subscript $1$ or $2$ denotes the spacing; $a,\; b,\; c,\; d$ label
the four sublattices of spacing $2$.
\par
Suppose that ${\bf Z}^{2}_{2,b}$ and ${\bf Z}^{2}_{2,d}$ belong to the same
sublattice of spacing $\sqrt 2$ and set
$$\eta (x)=\left\{ \eqalign{
+1& \;\;\; \forall x\in {\bf Z}^{2}_{2,a} \cap \L_0\cr
-1& \;\;\; \forall x\in {\bf Z}^{2}_{2,b} \cap \L_0\cr
-1& \;\;\; \forall x\in {\bf Z}^{2}_{2,d} \cap \L_0\cr
 0& \;\;\; \forall x\in {\bf Z}^{2}_{2,c} \cap \L_0\cr}\right. \;\; ;$$
(see Fig.6.1).
\par
Starting from our definition of ${\cal G}$ we have first to transform the
$-1$ with a plus spin among its nearest neighbours into $0$. In this way we
get the configuration $\eta_1$ depicted in the left hand side of Fig.6.1.
Eventually, we get a configuration $\hat\eta$ with a unique plurirectangle
with external edges with length $L_0+1$ and many non--interacting unit
squares of pluses in its interior.
\par
On the other hand, starting from $\eta$ we can change the $-1$ in $\L_0$
with two plus spins among their nearest neighbours into $+1$ by decreasing
the energy; we obtain the configuration $\eta_2$ depicted in the right hand
side of Fig.6.1. Subsequently, still decreasing the energy the configuration
$\eta_2$ can be transformed into the configuration $\eta^*=C(L_0-1,L_0-1)$.
\par
Now, if $L_0$ is chosen such that $l^*+1\le L_0<M^*-1$ we have that
$\hat\eta$ (and so $\eta$) is subcritical and then it belongs to ${\cal G}$,
but $\eta^*$ (to which we arrived starting from $\eta$ with a downhill
path) is supercritical.
\midinsert
\vskip 0.5 truecm
\vbox{\font\amgr=cmr10 at
10truept\baselineskip0.1466667truein\lineskiplimit-\maxdimen
\catcode`\-=\active\catcode`\~=\active\def~{{\char32}}\def-{{\char1}}%
\hbox{\amgr
{}~~~~~~~~~~~~~+~0~+~0~+~0~+~0~~~~~~~~+~{\char45}~+~{\char45}~+~{\char45}~+~{\char45}~~~~~~~~+~+~+~+~+~+~+~{\char45}}
\hbox{\amgr
{}~~~~~~~~~~~~~0~0~0~0~0~0~0~0~~~~~~~~{\char45}~0~{\char45}~0~{\char45}~0~{\char45}~0~~~~~~~~+~0~+~0~+~0~+~0}
\hbox{\amgr
{}~~~~~~~~~~~~~+~0~+~0~+~0~+~0~~~~~~~~+~{\char45}~+~{\char45}~+~{\char45}~+~{\char45}~~~~~~~~+~+~+~+~+~+~+~{\char45}}
\hbox{\amgr
{}~~~~~~~~~~~~~0~0~0~0~0~0~0~0~~~~~~~~{\char45}~0~{\char45}~0~{\char45}~0~{\char45}~0~~~~~~~~+~0~+~0~+~0~+~0}
\hbox{\amgr
{}~~~~~~~~~~~~~+~0~+~0~+~0~+~0~~~<-~~~+~{\char45}~+~{\char45}~+~{\char45}~+~{\char45}~~~->~~~+~+~+~+~+~+~+~{\char45}}
\hbox{\amgr
{}~~~~~~~~~~~~~0~0~0~0~0~0~0~0~~~~~~~~{\char45}~0~{\char45}~0~{\char45}~0~{\char45}~0~~~~~~~~+~0~+~0~+~0~+~0}
\hbox{\amgr
{}~~~~~~~~~~~~~+~0~+~0~+~0~+~0~~~~~~~~+~{\char45}~+~{\char45}~+~{\char45}~+~{\char45}~~~~~~~~+~+~+~+~+~+~+~{\char45}}
\hbox{\amgr
{}~~~~~~~~~~~~~0~0~0~0~0~0~0~0~~~~~~~~{\char45}~0~{\char45}~0~{\char45}~0~{\char45}~0~~~~~~~~{\char45}~0~{\char45}~0~{\char45}~0~{\char45}~0}}
\vskip 0.5 truecm
\par\noindent
\centerline {\smbfb Fig.6.1}
\endinsert
\par
To construct ${\cal G}$, first of all we define a map
${\cal F}:\s\rightarrow {\hat \s}={\cal F}\s$
with $\s$ an acceptable configuration and ${\hat \s}$ a local minimum of the
energy, such that the two following properties are satisfied
$$\eqalign{
H({\hat \s})&\le H(\s)\cr
\s & \prec {\hat \s}\cr}\;\; ;\Eq (6.2)$$
that is the local minimum ${\hat \s}$ is bigger than $\s$ and at a lower
energy level. Then we define the set ${\cal G}$ as the set of configurations
$\s$ such that ${\hat \s}$ is subcritical, that is
$P_{\hat \s} (\t_{\menouno}<\t_{\piuuno})\rightarrow 1$ as
$\b\rightarrow\infty$.
\par
Now we define the map ${\cal F}:\s\rightarrow {\hat \s}$; the definition is
given in the following five steps. Let $\s$ be an acceptable configuration:
\par
$(i)$ starting from $\s$ we construct the configuration $\s_1$ by turning
into zero all the minus spins of $\s$ which have at least one plus spin
among their nearest neighbour sites. We remark that $H(\s_1)\le H(\s)$ (see
Fig.3.1)  and $\s\prec\s_1$.
\par
$(ii)$ Let us denote by $c^-_1$ the minus spins cluster in the configuration
$\s_1$ which is winding around the torus and by $c^-_i$ all the other minus
spins clusters in $\s_1$. In $\s_1$
there is no direct interface $+-$, then we can conclude that every $c^-_i$
cluster is inside a zero spins cluster (see Fig.6.2). Now we consider the
configuration $\s_2$ obtained from $\s_1$ by turning into zero all the minus
spins in all the clusters $c^-_i$. The result $\s_1\prec\s_2$ is obvious. We
have, also, that  $H(\s_2)\le H(\s_1)$; indeed in every cluster
$c^-_i$ there is at least one minus spin with two zero spins among its nearest
neighbours; this spin can be transformed into zero lowering the energy. We can
repeat this argument until all the spins of the starting cluster $c^-_i$ have
been transformed into zero.
\midinsert
\vskip 0.5 truecm
\vbox{\font\amgr=cmr10 at
10truept\baselineskip0.1466667truein\lineskiplimit-\maxdimen
\catcode`\-=\active\catcode`\~=\active\def~{{\char32}}\def-{{\char1}}%
\hbox{\amgr ~~~~~~~~~~~~~~~~~~~~~~~~~~~~~~~~~{\char2}-----{\char3}}
\hbox{\amgr
{}~~~~~~~~~~~~~~~~~~~~~~~~~~~~~~~~~{\char0}~~0~~{\char5}-------{\char3}~~~~~{\char2}----{\char3}}
\hbox{\amgr
{}~~~~~~~~~~~~~~~~~~~~~~~~~{\char2}-------{\char4}~~~~~~~~~~~~~{\char0}~~~{\char2}-{\char4}~~~~{\char5}-{\char3}%
}
\hbox{\amgr
{}~~~~~~~~~~~~~~~~~~~~~~~~~{\char0}~~~~~~~~~~~~~~~~~0~~~{\char0}~~~{\char0}~~~0~~~~{\char0}}
\hbox{\amgr
{}~~~~~~~~~~~~~~~~~~~~~~~~~{\char0}~~~0~~~{\char2}----{\char3}~~~~~~~~{\char0}~{\char2}-{\char4}~~{\char2}-{\char3}%
{}~~~{\char5}---{\char3}}
\hbox{\amgr
{}~~~~~~~~~~~~~~~~~~~{\char2}-----{\char4}~~~~{\char2}--{\char4}~~~~{\char5}---{\char3}~~~~{\char5}-{\char4}~~~~%
{\char0}~{\char5}-{\char3}~~~~~{\char0}}
\hbox{\amgr
{}~~~~~~~~~~~~~~~~~~~{\char0}~~~~~~~~~~{\char0}~~~~{\char45}~~~~~~{\char0}~~~~~~~~~{\char2}-{\char4}~~~{\char0}~~%
0~~{\char0}}
\hbox{\amgr
{}~~~~~~~~~~~~~~~~~~~{\char5}--{\char3}~~~0~~~{\char5}--{\char3}~~~~{\char45}~~~{\char0}~~~0~~~{\char2}-{\char4}~%
{}~{\char45}~~{\char5}-{\char3}~~~{\char0}}
\hbox{\amgr
{}~~~~~~~~~~~~~~~~~~~~~~{\char0}~~~~~~~~~~{\char0}~~~~~~~~{\char0}~~~~~{\char2}-{\char4}~~~~~~~~~{\char0}~~~{\char5}%
-{\char3}}
\hbox{\amgr
{}~~~~~~~~~~~~~~~~~~~~~~{\char5}----{\char3}~~0~~{\char5}--{\char3}~~~{\char2}-{\char4}~~~~~{\char0}~~~{\char45}~%
{}~~~{\char45}~~{\char0}~~~~~{\char0}}
\hbox{\amgr
{}~~~~~~~~~~~~~~~~~~~~~~~~~~~{\char0}~~~~~~~~{\char5}---{\char4}~~~~~{\char2}-{\char4}~~~~~~~~~~~{\char0}~~~{\char2}%
-{\char4}}
\hbox{\amgr
{}~~~~~~~~~~~~~~~~~~~~~~~~~~~{\char5}---{\char3}~~~~~~~~~~~~~~{\char0}~~~{\char45}~~~{\char2}-----{\char4}~~~{\char0}%
}
\hbox{\amgr
{}~~~~~~~~~~~~~~~~~~~~~~~~~~~~~~~{\char0}~~~~~~~0~~~~~~{\char5}-------{\char4}~~~~~~0~~{\char0}}
\hbox{\amgr
{}~~~~~~~~~~~~~~~~~~~~~~~~~~~~~~~{\char5}---------{\char3}~~~~~~~~~~~~~~~{\char2}------{\char4}}
\hbox{\amgr
{}~~~~~~~~~~~~~~~~~~~~~~~~~~~~~~~~~~~~~~~~~{\char0}~~~~0~~~~{\char2}-----{\char4}}
\hbox{\amgr
{}~~~~~~~~~~~~~~~~~~~~~~~~~~~~~~~~~~~~~~~~~{\char5}---------{\char4}}}
\vskip 0.5 truecm
\par\noindent
\centerline {\smbfb Fig.6.2}
\endinsert
\par
$(iii)$ In $\s_2$ there is no direct interface $+-$, then we observe that
every cluster of plus spins is inside a cluster of zero spins; it can happen
that in
some of the plus spins clusters there are one or more clusters of zero spins
(see Fig.6.3). We construct the configuration $\s_3$ by removing all these
clusters of zero spins. With arguments similar to those used in step $(ii)$ one
can prove that $H(\s_3)\le H(\s_2)$ and $\s_2\prec\s_3$.
\par
$(iv)$ The configuration $\s_3$ is made of a minus spins cluster which is
winding around the torus, the zero spins clusters denoted by
$c^{0}_{i}\;\forall i\in\{1,2,...,k^{0}\}$ and the clusters with plus spins
$c^{+}_{i,j}\;\forall i\in\{1,2,...,k^{0}\}$ and
$\forall j\in\{1,2,...,k^{+}_{i}\}$. The clusters
$c^{+}_{i,j}\;\forall j\in\{1,2,...,k^{+}_{i}\}$ are all inside the cluster
$c^{0}_{i}$. We consider, now, the rectangular envelopes
$R^{0}_{i}=R(c^{0}_{i})\;\forall i\in\{1,2,...,k^{0}\}$ and the configuration
$\s_4$ obtained by filling all these rectangles with zero spins; in this step
the plus spins are not changed. It is immediate that
$H(\s_4)\le H(\s_3)$ and $\s_3\prec\s_4$.
\midinsert
\vskip 0.5 truecm
\vbox{\font\amgr=cmr10 at
10truept\baselineskip0.1466667truein\lineskiplimit-\maxdimen
\catcode`\-=\active\catcode`\~=\active\def~{{\char32}}\def-{{\char1}}%
\hbox{\amgr ~~~~~~~~~~~~~~~~~~~~~~~~~~~~~~~~~{\char2}-----{\char3}}
\hbox{\amgr
{}~~~~~~~~~~~~~~~~~~~~~~~~~~~~~~~~~{\char0}~~0~~{\char5}-------{\char3}~~~~~{\char2}----{\char3}}
\hbox{\amgr
{}~~~~~~~~~~~~~~~~~~~~~~~~~{\char2}-------{\char4}~~~~~~~{\char2}--{\char3}~~{\char0}~~~{\char2}-{\char4}~~~~{\char5}%
-{\char3}}
\hbox{\amgr
{}~~~~~~~~~~~~~~~~~~~~~~~~~{\char0}~~~~~~~~~{\char2}-----{\char4}~~{\char0}~~{\char0}~~~{\char0}~~~0~~~~{\char0}%
}
\hbox{\amgr
{}~~~~~~~~~~~~~~~~~~~~~~~~~{\char0}~~~0~~~{\char2}-{\char4}~~~+~~~~{\char0}~~{\char0}~{\char2}-{\char4}~~{\char2}%
-{\char3}~~~{\char5}---{\char3}}
\hbox{\amgr
{}~~~~~~~~~~~~~~~~~~~{\char2}-----{\char4}~~~~{\char2}--{\char4}~~~~{\char2}---{\char3}~{\char0}~~{\char5}-{\char4}%
{}~~~~{\char0}~{\char5}-{\char3}~~~~~{\char0}}
\hbox{\amgr
{}~~~~~~~~~~~~~~~~~~~{\char0}~~~~~~~~~~{\char0}+~~~~{\char2}-{\char4}~~~{\char0}~{\char0}~~~~~~~{\char2}-{\char4}%
{}~~~{\char0}~~0~~{\char0}}
\hbox{\amgr
{}~~~~~~~~~~~~~~~~~~~{\char5}--{\char3}~~~0~~~{\char5}--{\char3}~~{\char0}~~~~~{\char0}~{\char5}-------{\char4}~%
{}~+~~{\char5}-{\char3}~~~{\char0}}
\hbox{\amgr
{}~~~~~~~~~~~~~~~~~~~~~~{\char0}~~~~~~~~~~{\char0}~~{\char0}~~0~~{\char5}-----{\char3}~~~~~{\char2}--{\char3}~~{\char0}%
{}~~~{\char5}-{\char3}}
\hbox{\amgr
{}~~~~~~~~~~~~~~~~~~~~~~{\char5}----{\char3}~~0~~{\char0}~~{\char5}-----------{\char4}~~{\char2}--{\char4}~0{\char0}%
{}~~{\char0}~~0~~{\char0}}
\hbox{\amgr
{}~~~~~~~~~~~~~~~~~~~~~~~~~~~{\char0}~~~~~{\char5}------{\char3}~{\char2}---{\char3}~~~~{\char5}-----{\char4}~~{\char0}%
{}~~~{\char2}-{\char4}}
\hbox{\amgr
{}~~~~~~~~~~~~~~~~~~~~~~~~~~~{\char5}---{\char3}~~~~~~~~{\char5}-{\char4}~~~{\char0}~~+~~~~{\char2}-----{\char4}%
{}~~~{\char0}}
\hbox{\amgr
{}~~~~~~~~~~~~~~~~~~~~~~~~~~~~~~~{\char0}~~~~~~~0~~~~~~{\char5}-------{\char4}~~~~~~0~~{\char0}}
\hbox{\amgr
{}~~~~~~~~~~~~~~~~~~~~~~~~~~~~~~~{\char5}---------{\char3}~~~~~~~~~~~~~~~{\char2}------{\char4}}
\hbox{\amgr
{}~~~~~~~~~~~~~~~~~~~~~~~~~~~~~~~~~~~~~~~~~{\char0}~~~~0~~~~{\char2}-----{\char4}}
\hbox{\amgr
{}~~~~~~~~~~~~~~~~~~~~~~~~~~~~~~~~~~~~~~~~~{\char5}---------{\char4}}}
\vskip 0.5 truecm
\par\noindent
\centerline {\smbfb Fig.6.3}
\endinsert
\par
$(v)$ Apart from the plus spins cluster, the configuration $\s_4$ is made of
zero rectangular clusters placed in the ``sea" of minus spins. We obtain
the configuration $\s_5$ by means of the {\it chain construction} used in
[KO1], applied to the rectangular clusters
$R^{0}_{i}\;\forall i\in\{1,2,...,k^{0}\}$.
\par
Let us briefly describe this construction. Given a set of rectangles
$R^0_1,\dots ,R^0_l$ we partition it into maximal connected components
${\cal C}^{(1)}_j$ with $j=1,\dots ,k^{(1)}$ called {\it chains of first
generation}
$$(R^0_1\dots R^0_l)=({\cal C}^{(1)}_1\dots {\cal C}^{(1)}_{k^{(1)}})\;\; .$$
The notion of connection is given by pairwise interaction: a set
$R^0_1,\dots ,R^0_m$ of rectangles is connected if it cannot be divided
into two non--interacting parts.
\par
Now consider the $k^{(1)}$ rectangles $R({\cal C}^{(1)}_j)$ obtained as
rectangular envelope of the union of the rectangles belonging to
${\cal C}^{(1)}_j$. Partition this set of rectangles into maximal connected
component: in this way we construct the chains of second generation
${\cal C}^{(2)}_1,\dots ,{\cal C}^{(2)}_{k^{(2)}}$.
We continue in this way up to a
finite maximal order $n$ such that the chains of the $n$--th generation are
non--interacting rectangles (see [KO1] for more details).
\par
We call $\sigma_5$ this configuration containing these non--interacting
rectangular clusters ${\bar R}^{0}_{i}\;\forall i\in\{1,2,...,k^{0,f}\}$
of zero spins placed in the minus spins ``sea". With usual arguments one can
prove that $H(\sigma_{5})\leq H(\sigma_{4})$ and $\s_4\prec\s_5$.
\par
$(vi)$ By repeating the operations described in points $(iv)$ and $(v)$ for the
plus spins clusters lying in every rectangle ${\bar R}^{0}_{i}\;\forall
i\in\{1,2,...,k^{0,f}\}$, we obtain the final configuration ${\hat \s}$.
This configuration is made of the external rectangular zero spins clusters
${\bar R}^{0}_{i}\;\forall i\in\{1,2,...,k^{0,f}\}$ and the internal
non-interacting plus spins clusters
${\bar R}^{+}_{i,j}\;\forall i\in\{1,2,...,k^{0,f}\}$ and
$\forall j\in\{1,2,...,k^{+,f}_i\}$. As usual one can prove that
$H({\hat \s})\leq H(\sigma_{5})$ and $\s_5\prec {\hat \s}$.
\par
The definition of the map ${\cal F}$ is now complete, we observe that
${\hat \s}$ is a local minimum and that the properties $\equ (6.2)$ are
satisfied. Finally we remark that the map ${\cal F}$ is {\it monotone} in the
sense that
$$\sigma\prec\eta\;\Rightarrow\; {\hat\sigma}\prec{\hat\eta}\;\; ,
\Eq (6.3)$$
for every couple of acceptable configurations $\s$ and $\eta$.
\par
Now we state the following
\vskip 0.35 truecm
\noindent
{\bf Proposition 6.1.}\par\noindent
$$U({\cal G})\subset {\hat {\cal P}}\;\; . \Eq (6.4)$$
Namely the set of minima of the energy in the boundary of ${\cal G}$ is
contained in ${\hat {\cal P}}$.
\par\noindent
{\it Proof.}
\vskip 0.5 truecm
\par
In order to prove Proposition 6.1 we consider a configuration
$\eta\in\partial {\cal G}$ and we show that there exists a special saddle
$\tilde\eta$ such that $H(\eta)\ge H(\tilde\eta)$. Then Proposition 6.1 will
follow from Lemma 5.1.
\par
Let us consider $\eta\in {\cal G}$; there exists a configuration $\sigma
=\eta ^{x,b}$ with $x\in \L$ and $b\not= \eta (x)$ such that
$\sigma\in {\cal G}$. By virtue of the monotonicity of the map ${\cal F}$ (see
equation $\equ (6.3)$) and of the fact that ${\hat \sigma}$ is a
subcritical local minimum, it follows that $b < \eta (x)$; hence we also
have that $b\not= +1$.
\par
We denote by $R^0_i({\hat\s})\;\forall i\in\{ 1,...,k^0({\hat\s})\}$ and by
$R^+_{i,j}({\hat\s})\;\forall j\in\{ 1,...k^+_i({\hat\s})\}$ and
$\forall i\in\{ 1,...,k^0({\hat\s})\}$ the rectangles respectively of zeros
and pluses which appear in the configuration ${\hat\s}$; we remark that
all the rectangles $R^+_{i,j}({\hat\s})\;\forall
j\in\{ 1,...,k^+_i({\hat\s})\}$ are inside the zero rectangle
$R^0_i({\hat\s})$. In the following, by abuse of notation, we will also denote
by $R^0_i({\hat\s})$ what we will call {\it structure} $R^0_i({\hat\s})$,
namely the complex given by the ``external"
rectangle togheter with all its ``internal" rectangles of pluses (what
before we called plurirectangle is indeed a configuration containing a
unique structure).
\vskip 0.5 truecm
\par\noindent
{\it Case 1}: $b=-1$ and $\eta (x)=0$.
\par
{}From the definition of the map ${\cal F}$ easily follows that necessarily $x$
lies outside the rectangles $R^0_i({\hat\s})$.
\par
Given the configuration ${\hat\eta}$ we denote by
$R^0_i({\hat\eta})\;\forall i\in\{ 1,...,k^0({\hat\eta})\}$ and by
$R^+_{i,j}({\hat\eta})\;\forall j\in\{ 1,...,k^+_i({\hat\eta})\}$ and
$\forall i\in\{ 1,...,k^0({\hat\eta})\}$ the rectangles respectively of zeros
and pluses which appear in it. We denote by ${\bar R}^0({\hat\eta})$ the
supercritical structure among the $R^0_i({\hat\eta})$ and by
${\bar R}^0_1,...,{\bar R}^0_s$ the rectangles of zeros such that
$\forall i\in\{ 1,...,s\}\;{\bar R}^0_i$ appears in ${\hat\sigma}$ and
$\forall i\in\{ 1,...,s\}\;{\bar R}^0_i$ is ``inside" the rectangle
${\bar R}^0({\hat\eta})$.
\par
We consider, now, the configuration $\eta_1$ defined as follows: $\eta_1(x)=0$,
all the other spins are minus except for the zeros and the pluses of the
structures  ${\bar R}^0_i\;\forall i\in\{ 1,...,s\}$. It can be easily proved
that $H(\eta)\ge H(\eta_1)$. We distinguish the two cases $1.1$ and $1.2$.
\vskip 0.3 truecm
\par\noindent
{\it Case 1.1}: all the rectangles of pluses which appear in $\eta_1$ are
subcritical.
\par
We consider the configuration $\eta_{1.1}$ obtained from $\eta_1$ by changing
into zeros all the plus spins. We remark that $H(\eta_1)\ge H(\eta_{1.1})$
because $\eta_{1.1}$ has been constructed by removing subcritical rectangles of
pluses.
\par
With an Ising--like argument (see e.g. [KO1]) one can prove that
$H(\eta_{1.1})\ge H({\cal P}_2)$. Hence in the case $1.1$ we have found a
special saddle with energy lower than the starting configuration $\eta$.
\vskip 0.3 truecm
\par\noindent
{\it Case 1.2}: in $\eta_1$ there exists at least one supercritical
rectangle of pluses.
\par
We consider the configuration $\eta_{1.2}$ obtained from $\eta_1$ by removing
in every structure ${\bar R}^0_i\;\forall i\in\{ 1,...,s\}$
all the subcritical rectangles of pluses and by filling with pluses the
rectangular envelope of the union of the supercritical rectangles of pluses.
We remark that every structure ${\bar R}^0_i\;\forall i\in\{ 1,...,s\}$ in
$\eta_{1.2}$ is either ``empty" (with no rectangle of pluses inside) or it has
just a rectangle of pluses inside and this rectangle is supercritical.
\par
We denote by $Q$ the unit square centered at the site $x\in\L$; we distinguish
the two following cases:
\vskip 0.3 truecm
\par\noindent
{\it Case 1.2.1}: one of the structures
${\bar R}^0_i\;\forall i\in\{ 1,...,s\}$ of $\eta_{1.2}$ (we denote it by
${\bar R}^0_{1.2.1}$) interacts with $Q$ and the structure obtained by filling
with zeroes the rectangular envelope of ${\bar R}^0_{1.2.1}\cup Q$ is
supercritical.
\par
Let us denote by $\eta_{1.2.1}$ the configuration obtained by removing in
$\eta_1$ all the structures ${\bar R}^0_i\;\forall i\in\{ 1,...,s\}$ except
for ${\bar R}^0_{1.2.1}$. If $Q$ is adjacent to ${\bar R}^0_{1.2.1}$ then
$\eta_{1.2.1}$ is a special saddle. Otherwise $Q$ is at distance one from one
of the sides of the rectangle ${\bar R}^0_{1.2.1}$ or $Q$ and
${\bar R}^0_{1.2.1}$ touch in a corner; in this case it can be esily found a
special saddle with energy lower than $H(\eta_{1.2})$.
\par
Hence in the case $1.2.1$ a special saddle with energy lower than the starting
configuration $\eta$ has been found.
\vskip 0.3 truecm
\par\noindent
{\it Case 1.2.2}: the condition $1.2.1$ is not fulfilled.
\par
By an argument similar to the one used in [KO1] (see pages 1136--1137 therein)
we can find two structures ${\tilde R}_1$ and ${\tilde R}_2$ such that: they
are both subcritical, their external rectangles are interacting, the structure
obtained by filling with zeroes their rectangular envelope is supercritical and
$H(\eta_{1.2})\ge H({\tilde R}_1)+H({\tilde R}_2)$ (when we say
$H({\tilde R}_i)\;$ with $i\in\{ 1,2\}$ we are referring to the energy of the
configuration obtained by plunging the structure ${\tilde R}_i$ in the
``sea" of minuses). We still have to distinguish between two possible cases.
\vskip 0.3 truecm
\par\noindent
{\it Case 1.2.2.1}: both structures
$H({\tilde R}_i)\;$ with $i\in\{ 1,2\}$ have a supercritical rectangle of
pluses inside.
\par
Now we consider a just--supercritical frame whose external rectangle is
contained in the rectangular envelope of the union of the two external
rectangles of ${\tilde R}_1$ and of ${\tilde R}_2$. Such a frame surely
exists and we denote it by ${\tilde C}$.
\par
Starting from ${\tilde R}_1$ and ${\tilde R}_2$ and recalling that these
structures are subcritical, one can construct two other
structures, ${\tilde \Re}_1$ and ${\tilde \Re}_2$ (birectangles or frames),
such that the three following conditions are satisfied:
$i)$ $H({\tilde R}_1)\ge H({\tilde \Re}_1)$ and
$H({\tilde R}_2)\ge H({\tilde \Re}_2)$; $ii)$ if the two external rectangles of
the two structures ${\tilde \Re}_1$ and ${\tilde \Re}_2$ touch by a corner then
the rectangular envelope of the union of the external rectangles of
${\tilde \Re}_1$ and of ${\tilde \Re}_2$ coincides exactly  with the external
rectangle of the frame ${\tilde C}$; $iii)$ at least one of the two internal
rectangles of pluses (the one in ${\tilde \Re}_1$ or the one in
${\tilde \Re}_2$) is supercritical.
\par
If one considers the configuration $\eta_{1.2.2.1}$ obtained by plunging the
structures in the ``sea" of minus spins such that the external rectangles of
${\tilde \Re}_1$ and of ${\tilde \Re}_2$ touch by a corner, one can easily
convince himself that $H(\eta_{1.2})\ge H(\eta_{1.2.2.1})$.
\par
Finally, starting from $\eta_{1.2.2.1}$ we construct the special saddle
${\tilde\eta}$ by performing the following steps: $i)$ we fill of zeroes the
rectangular envelope of the union of the two external rectangles of zeroes in
$\eta_{1.2.2.1}$; $ii)$ we let grow the internal supercritcal rectangle of
pluses until the frame ${\tilde C}$ is reached; $iii)$ we
transform into zeroes all the pluses, except for one, of one of the four sides
of
the internal rectangle, such that a special saddle is obtained. It can be
easily proved that $H(\eta_{1.2.2.1}) > H({\tilde \eta})$ by comparing the
energy differences involved in the three steps described above. We remark that
the energy increase of the third step is largely compensated by the energy
decrease involved in the second step.
\vskip 0.3 truecm
\par\noindent
{\it Case 1.2.2.2}: one of the structures $H({\tilde R}_i)$ is ``empty", in
the sense that it has no rectangles of pluses inside.
\par
This case can be discussed with arguments similar to those used in the
Case 1.2.2.2.
\vskip 0.5 truecm
\par\noindent
{\it Case 2}: $b=-1$ and $\eta (x)=+1$.
\par
Starting from ${\hat\s}$ one can always construct a configuration $\eta_2$ such
that: $i)$ $\eta_2\in\partial {\cal G}$; $ii)$ $\exists y\in\L$ such that
$\eta_2 (y)=0$ and $\eta_2^{y,-1}\in {\cal G}$. In this way the proof has been
reduced to the Case 1.
\vskip 0.5 truecm
\par\noindent
{\it Case 3}: $b=0$.
\par
The site $x$ is inside one of the rectangles of zeroes
$R^0_i({\hat\s})\;\forall i\in\{ 1,...,k^0({\hat\s})\}$; we denote it by
${\bar R}^0$. There are two possible cases that must be considered.
\vskip 0.3 truecm
\par\noindent
{\it Case 3.1}: $x$ is not on one of the boundary slices of ${\bar R}^0$ (the
tipical situation is depicted in Fig.6.4).
\par
In this case the rectangles of zeroes in ${\hat\eta}$ coincide with
those in ${\hat\sigma}$, but the structure ${\bar R}^0({\hat\eta})$ is
supercritical (${\bar R}^0({\hat\eta})$ is the structure of ${\hat\eta}$
such that its external rectangle of zeroes coincides with ${\bar R}^0$).
\par
We consider, now, the configuration $\eta_{3.1}$ defined as follows: $i)$
$\eta_{3.1}$ is obtained starting from ${\hat\sigma}$, by removing all the
structures $R^0_i({\hat\s})\;\forall i\in\{ 1,...,k^0({\hat\s})\}$ except for
the one whose external rectangle coincides with the external rectangle of the
structure ${\bar R}^0$ (we denote this structure by
${\bar R}^0({\hat\sigma})$); $ii)$ $\eta_{3.1}(x)=+1$. It can be easily proved
that $H(\eta)\ge H(\eta_{3.1})$.
\midinsert
\vskip 0.5 truecm
\vbox{\font\amgr=cmr10 at
10truept\baselineskip0.1466667truein\lineskiplimit-\maxdimen
\catcode`\-=\active\catcode`\~=\active\def~{{\char32}}\def-{{\char1}}%
\hbox{\amgr
{}~~~~~~~~~~~~~~~~~~~~~~~~{\char2}---------------------------------------{\char3}}
\hbox{\amgr
{}~~~~~~~~~~~~~~~~~~~~~~~~{\char0}~~~~~~~~~~~~~~~~~~~~~~~{\char2}--------{\char3}~~~~~~{\char0}}
\hbox{\amgr
{}~~~~~~~~~~~~~~~~~~~~~~~~{\char0}~{\char2}------------{\char3}~~~~~~~~{\char0}~~~~~~~~{\char0}~~~~~~{\char0}%
}
\hbox{\amgr
{}~~~~~~~~~~~~~~~~~~~~~~~~{\char0}~{\char0}~~~~~~~~~~~~{\char0}~~~~~~~~{\char5}--------{\char4}~~~~~~{\char0}%
}
\hbox{\amgr
{}~~~~~~~~~~~~~~~~~~~~~~~~{\char0}~{\char0}~~~~~~~~~~~~{\char0}x~~~~~~~~~~~~~~~~~~~~~~~{\char0}}
\hbox{\amgr
{}~~~~~~~~~~~~~~~~~~~~~~~~{\char0}~{\char5}------------{\char4}~~~{\char2}--------{\char3}~~~~~~~~~~~{\char0}%
}
\hbox{\amgr
{}~~~~~~~~~~~~~~~~~~~~~~~~{\char0}~~~~~~~~~~~~~~~~~~{\char0}~~~~~~~~{\char0}~~~~~~~~~~~{\char0}}
\hbox{\amgr
{}~~~~~~~~~~~~~~~~~~~~~~~~{\char0}~~~~~~~~~~~~~~~~~~{\char0}~~~~~~~~{\char0}~~~~~{\char2}--{\char3}~~{\char0}%
}
\hbox{\amgr
{}~~~~~~~~~~~~~~~~~~~~~~~~{\char0}~~~~~~~~~~~~~~~~~~{\char5}--------{\char4}~~~~~{\char0}~~{\char0}~~{\char0}%
}
\hbox{\amgr
{}~~~~~~~~~~~~~~~~~~~~~~~~{\char0}~~~~~~~~~~~~~~~~~~~~~~~~~~~~~~~~~{\char0}~~{\char0}~~{\char0}}
\hbox{\amgr
{}~~~~~~~~~~~~~~~~~~~~~~~~{\char0}~~~~~~~~~~~~~~~~~~~~~~~~~~~~~~~~~{\char5}--{\char4}~~{\char0}}
\hbox{\amgr
{}~~~~~~~~~~~~~~~~~~~~~~~~{\char5}---------------------------------------{\char4}}}
\vskip 0.5 truecm
\par\noindent
\centerline {\smbfb Fig.6.4}
\endinsert
\par
We denote by $R_1$ the rectangular envelope of the union of the
supercritical rectangles of pluses inside ${\bar R}^0({\hat\sigma})$ and
by $R_2$ the rectangular envelope of the union of the
supercritical rectangles of pluses inside ${\bar R}^0({\hat\eta})$. We remark
that the two sctructures ${\bar R}^0({\hat\sigma})$ and
${\bar R}^0({\hat\eta})$ have different internal rectangles of pluses, even
though their external rectangles of zeroes coincide.
\par
Now we observe that there exists a rectangle $R_3$ contained in $R_2$ and
containing $R_1$ such that the configuration with all the spins minus except
for the zeroes in the rectangle ${\bar R}^0$ and the pluses in $R_3$ is an
almost--supercritical configuration. We consider the special saddle
${\tilde\eta}$ obtained by properly putting a unit plus protuberance to one of
the four sides of the internal rectangle of pluses of the almost--supercritical
configuration found before. It can be easily shown that $H(\eta_{3.1})\ge
H({\tilde\eta})$. Hence, even in this case, we have found a special saddle with
energy lower than the energy of the starting configuration
$\eta\in\partial {\cal G}$.
\vskip 0.3 truecm
\par\noindent
{\it Case 3.2}: $x$ is on one of the boundary slices of ${\bar R}^0$ (see, for
example, Fig.6.5).
\par
We construct the configuration $\eta_{3.2}$ starting from ${\hat\sigma}$ and by
turning into zero only the spin minus at a site nearest neighbour to $x$. One
can easily convince himself that $H({\hat\eta})\ge H(\eta_{3.2})$. If
$\eta_{3.2}\in\partial {\cal G}$ then the proof is reduced to Case 1; if
$\eta_{3.2}\in {\cal G}$ the proof is reduced to Case 3.1.
\par
The proof of Proposition 6.1 is now complete. $\square$
\midinsert
\vskip 0.5 truecm
\vbox{\font\amgr=cmr10 at
10truept\baselineskip0.1466667truein\lineskiplimit-\maxdimen
\catcode`\-=\active\catcode`\~=\active\def~{{\char32}}\def-{{\char1}}%
\hbox{\amgr
{}~~~~~~~~~~~~~~~~~~~~~~~~{\char2}---------------------------------------{\char3}}
\hbox{\amgr
{}~~~~~~~~~~~~~~~~~~~~~~~~{\char0}~~~~~~~~~~~~~~~~~~~~~~~{\char2}--------{\char3}~~~~~~{\char0}}
\hbox{\amgr
{}~~~~~~~~~~~~~~~~~~~~~~~~{\char0}~{\char2}------------{\char3}~~~~~~~~{\char0}~~~~~~~~{\char0}~~~~~~{\char0}%
}
\hbox{\amgr
{}~~~~~~~~~~~~~~~~~~~~~~~~{\char0}~{\char0}~~~~~~~~~~~~{\char0}~~~~~~~~{\char5}--------{\char4}~~~~~~{\char0}%
}
\hbox{\amgr
{}~~~~~~~~~~~~~~~~~~~~~~~~{\char0}~{\char0}~~~~~~~~~~~~{\char0}~~~~~~~~~~~~~~~~~~~~~~~~{\char0}}
\hbox{\amgr
{}~~~~~~~~~~~~~~~~~~~~~~~~{\char0}~{\char5}------------{\char4}~~~{\char2}--------{\char3}~~~~~~~~~~~{\char0}%
}
\hbox{\amgr
{}~~~~~~~~~~~~~~~~~~~~~~~~{\char0}~~~~~~~~~~~~~~~~~~{\char0}~~~~~~~~{\char0}~~~~~~~~~~~{\char0}}
\hbox{\amgr
{}~~~~~~~~~~~~~~~~~~~~~~~~{\char0}~~~~~~~~~~~~~~~~~~{\char0}~~~~~~~~{\char0}~~~~~{\char2}--{\char3}~~{\char0}%
}
\hbox{\amgr
{}~~~~~~~~~~~~~~~~~~~~~~~~{\char0}~~~~~~~~~~~~~~~~~~{\char5}--------{\char4}~~~~~{\char0}~~{\char0}~x{\char0}%
}
\hbox{\amgr
{}~~~~~~~~~~~~~~~~~~~~~~~~{\char0}~~~~~~~~~~~~~~~~~~~~~~~~~~~~~~~~~{\char0}~~{\char0}~~{\char0}}
\hbox{\amgr
{}~~~~~~~~~~~~~~~~~~~~~~~~{\char0}~~~~~~~~~~~~~~~~~~~~~~~~~~~~~~~~~{\char5}--{\char4}~~{\char0}}
\hbox{\amgr
{}~~~~~~~~~~~~~~~~~~~~~~~~{\char5}---------------------------------------{\char4}}}
\vskip 0.5 truecm
\par\noindent
\centerline {\smbfb Fig.6.5}
\endinsert
\par

\vfill\eject
\numsec=7\numfor=1

{\bf Section 7. Proof of the theorems.}
\par
Let us first give some definitions extending the ones given in Section 4.\par
We recall that by $ C ( l_1, l_2)$ we denote the set of configurations
containing only a frame with internal sides $ l_1, l_2$.
We recall the notation $ l := \min \{l_1,\; l_2\} , \; m := \{l_1,\; l_2\}$.
\par
We denote by $ S(l_1, l_2)$ the set of configurations obtained from
 $ C (l_1, l_2)$ by substituting one of the smaller internal sides with a
unit square protuberance namely by substituting all but one plus spins adjacent
from
the interior to one of the internal sides of length $l$ with zeroes (see
Fig.7.1).
\par
We denote by $R(l_1, l_2)$ the set of configurations containing a unique
birectangle
obtained by erasing the internal unit square protuberance from $ S(l_1, l_2)$
(see Fig.7.1). We denote by $ G(l_1, l_2)$ the set of configurations obtained
from the frame $C (l_1, l_2)$ by adding a unit square spin $0$ protuberance
to one of the longer external sides in $C (l_1, l_2)$.
A particularly relevant case will be the one $ | l_1-l_2 |  \leq 1$ where
either  $ m = l + 1 $ or $ m  = l  $.
We remark that $ G(l - 1, l )$ is obtained from the birectangle $ R(l, l)$ by
substituting one ``free" external row (or column) of zeroes of length $l+2$
with a
unit square protuberance (see Fig.7.1); similarly $ G(l - 1, l-1 )$ is obtained
from $ R(l - 1, l )$ by substituting one free external row or column of  spin 0
of length $l+1$ with a unit square protuberance.
\par
Finally let us denote by $\bar R(l_1,l_2) := R(0,0, l_1,l_2)\cup R(0,0,
l_2,l_1)$
the set of
 configurations without plus spins where the zero spins are precisely the
ones contained inside a rectangle with  sides equal,
respectively, to $l_1,l_2$.
\midinsert
\vskip 0.5 truecm
\vbox{\font\amgr=cmr10 at
10truept\baselineskip0.1466667truein\lineskiplimit-\maxdimen
\catcode`\-=\active\catcode`\~=\active\def~{{\char32}}\def-{{\char1}}%
\hbox{\amgr
{}~~~~~~~~~~~~~~~$C(l,l)$~~~~~~~~~~~~~~~~~~$S(l,l)$~~~~~~~~~~~~~~~~~$R(l,l)%
$}
\hbox{\amgr
{}~~~~~~~~{\char2}-----------------{\char3}~~~~{\char2}-----------------{\char3}~~~~{\char2}-----------%
------{\char3}}
\hbox{\amgr
{}~~~~~~~~{\char0}~{\char2}-------------{\char3}~{\char0}~~~~{\char0}~{\char2}-----------{\char3}~~~{\char0}%
{}~~~~{\char0}~{\char2}-----------{\char3}~~~{\char0}}
\hbox{\amgr
{}~~~~~~~~{\char0}~{\char0}~~~~~~~~~~~~~{\char0}~{\char0}~~~~{\char0}~{\char0}~~~~~~~~~~~{\char0}~~~{\char0}%
{}~~~~{\char0}~{\char0}~~~~~~~~~~~{\char0}~~~{\char0}}
\hbox{\amgr
{}~~~~~~~~{\char0}~{\char0}~~~~~~~~~~~~~{\char0}~{\char0}~~~~{\char0}~{\char0}~~~~~~~~~~~{\char5}-{\char3}%
{}~{\char0}~~~~{\char0}~{\char0}~~~~~~~~~~~{\char0}~~~{\char0}}
\hbox{\amgr
{}~~~~~~~~{\char0}~{\char0}~~~~~~~~~~~~~{\char0}~{\char0}~~~~{\char0}~{\char0}~~~~~~~~~~~{\char2}-{\char4}%
{}~{\char0}~~~~{\char0}~{\char0}~~~~~~~~~~~{\char0}~~~{\char0}}
\hbox{\amgr
{}~~~~~~~~{\char0}~{\char0}~~~~~~~~~~~~~{\char0}~{\char0}~~~~{\char0}~{\char0}~~~~~~~~~~~{\char0}~~~{\char0}%
{}~~~~{\char0}~{\char0}~~~~~~~~~~~{\char0}~~~{\char0}}
\hbox{\amgr
{}~~~~~~~~{\char0}~{\char0}~~~~~~~~~~~~~{\char0}~{\char0}~~~~{\char0}~{\char0}~~~~~~~~~~~{\char0}~~~{\char0}%
{}~~~~{\char0}~{\char0}~~~~~~~~~~~{\char0}~~~{\char0}}
\hbox{\amgr
{}~~~~~~~~{\char0}~{\char0}~~~~~~~~~~~~~{\char0}~{\char0}~~~~{\char0}~{\char0}~~~~~~~~~~~{\char0}~~~{\char0}%
{}~~~~{\char0}~{\char0}~~~~~~~~~~~{\char0}~~~{\char0}}
\hbox{\amgr
{}~~~~~~~~{\char0}~{\char5}-------------{\char4}~{\char0}~~~~{\char0}~{\char5}-----------{\char4}~~~{\char0}%
{}~~~~{\char0}~{\char5}-----------{\char4}~~~{\char0}}
\hbox{\amgr
{}~~~~~~~~{\char5}-----------------{\char4}~~~~{\char5}-----------------{\char4}~~~~{\char5}-----------%
------{\char4}}
\hbox{\amgr
{}~~~~~~~~~~~~~~~~$l+2$~~~~~~~~~~~~~~~~~~~$l+2$~~~~~~~~~~~~~~~~~~~$l+2$%
}
\hbox{\amgr }
\hbox{\amgr }
\hbox{\amgr }
\hbox{\amgr
{}~~~~~~~~~~~~~$G(l{\char45}1,l)$~~~~~~~~~~~~~$C(l{\char45}1,l)$~~~~~~~~~~~~~~$%
S(l{\char45}1,l)$}
\hbox{\amgr
{}~~~~~~~~{\char2}---------------{\char3}~~~~~~{\char2}---------------{\char3}~~~~~~{\char2}-----------%
----{\char3}}
\hbox{\amgr
{}~~~~~~~~{\char0}~{\char2}-----------{\char3}~{\char0}~~~~~~{\char0}~{\char2}-----------{\char3}~{\char0}%
{}~~~~~~{\char0}~{\char2}-----------{\char3}~{\char0}}
\hbox{\amgr
{}~~~~~~~~{\char0}~{\char0}~~~~~~~~~~~{\char0}~{\char0}~~~~~~{\char0}~{\char0}~~~~~~~~~~~{\char0}~{\char0}%
{}~~~~~~{\char0}~{\char0}~~~~~~~~~~~{\char0}~{\char0}}
\hbox{\amgr
{}~~~~~~~~{\char0}~{\char0}~~~~~~~~~~~{\char0}~{\char0}~~~~~~{\char0}~{\char0}~~~~~~~~~~~{\char0}~{\char0}%
{}~~~~~~{\char0}~{\char0}~~~~~~~~~~~{\char0}~{\char0}}
\hbox{\amgr
{}~~~~~~~~{\char0}~{\char0}~~~~~~~~~~~{\char0}~{\char5}-{\char3}~~~~{\char0}~{\char0}~~~~~~~~~~~{\char0}%
{}~{\char0}~~~~~~{\char0}~{\char0}~~~~~~~~~~~{\char0}~{\char0}}
\hbox{\amgr
{}~~~~~~~~{\char0}~{\char0}~~~~~~~~~~~{\char0}~{\char2}-{\char4}~~~~{\char0}~{\char0}~~~~~~~~~~~{\char0}%
{}~{\char0}~~~~~~{\char0}~{\char0}~~~~~~~~~~~{\char0}~{\char0}}
\hbox{\amgr
{}~~~~~~~~{\char0}~{\char0}~~~~~~~~~~~{\char0}~{\char0}~~~~~~{\char0}~{\char0}~~~~~~~~~~~{\char0}~{\char0}%
{}~~~~~~{\char0}~{\char0}~~~~~~~~~~~{\char0}~{\char0}}
\hbox{\amgr
{}~~~~~~~~{\char0}~{\char0}~~~~~~~~~~~{\char0}~{\char0}~~~~~~{\char0}~{\char0}~~~~~~~~~~~{\char0}~{\char0}%
{}~~~~~~{\char0}~{\char5}---{\char3}~{\char2}-----{\char4}~{\char0}}
\hbox{\amgr
{}~~~~~~~~{\char0}~{\char5}-----------{\char4}~{\char0}~~~~~~{\char0}~{\char5}-----------{\char4}~{\char0}%
{}~~~~~~{\char0}~~~~~{\char5}-{\char4}~~~~~~~{\char0}}
\hbox{\amgr
{}~~~~~~~~{\char5}---------------{\char4}~~~~~~{\char5}---------------{\char4}~~~~~~{\char5}-----------%
----{\char4}}
\hbox{\amgr
{}~~~~~~~~~~~~~~$l+1$~~~~~~~~~~~~~~~~~~~$l+1$~~~~~~~~~~~~~~~~~~~$l+1$%
}
\hbox{\amgr }
\hbox{\amgr }
\hbox{\amgr }
\hbox{\amgr
{}~~~~~~~~~~~~~$R(l{\char45}1,l)$~~~~~~~~~~~~$G(l{\char45}1,l{\char45}1)$~~~~~~~~~~%
{}~$C(l{\char45}1,l{\char45}1)$}
\hbox{\amgr
{}~~~~~~~~{\char2}---------------{\char3}~~~~~~{\char2}---------------{\char3}~~~~~~{\char2}-----------%
----{\char3}~~~~}
\hbox{\amgr
{}~~~~~~~~{\char0}~{\char2}-----------{\char3}~{\char0}~~~~~~{\char0}~{\char2}-----------{\char3}~{\char0}%
{}~~~~~~{\char0}~{\char2}-----------{\char3}~{\char0}}
\hbox{\amgr
{}~~~~~~~~{\char0}~{\char0}~~~~~~~~~~~{\char0}~{\char0}~~~~~~{\char0}~{\char0}~~~~~~~~~~~{\char0}~{\char0}%
{}~~~~~~{\char0}~{\char0}~~~~~~~~~~~{\char0}~{\char0}}
\hbox{\amgr
{}~~~~~~~~{\char0}~{\char0}~~~~~~~~~~~{\char0}~{\char0}~~~~~~{\char0}~{\char0}~~~~~~~~~~~{\char0}~{\char0}%
{}~~~~~~{\char0}~{\char0}~~~~~~~~~~~{\char0}~{\char0}}
\hbox{\amgr
{}~~~~~~~~{\char0}~{\char0}~~~~~~~~~~~{\char0}~{\char0}~~~~~~{\char0}~{\char0}~~~~~~~~~~~{\char0}~{\char0}%
{}~~~~~~{\char0}~{\char0}~~~~~~~~~~~{\char0}~{\char0}}
\hbox{\amgr
{}~~~~~~~~{\char0}~{\char0}~~~~~~~~~~~{\char0}~{\char0}~~~~~~{\char0}~{\char0}~~~~~~~~~~~{\char0}~{\char0}%
{}~~~~~~{\char0}~{\char0}~~~~~~~~~~~{\char0}~{\char0}}
\hbox{\amgr
{}~~~~~~~~{\char0}~{\char0}~~~~~~~~~~~{\char0}~{\char0}~~~~~~{\char0}~{\char0}~~~~~~~~~~~{\char0}~{\char0}%
{}~~~~~~{\char0}~{\char0}~~~~~~~~~~~{\char0}~{\char0}}
\hbox{\amgr
{}~~~~~~~~{\char0}~{\char5}-----------{\char4}~{\char0}~~~~~~{\char0}~{\char5}-----------{\char4}~{\char0}%
{}~~~~~~{\char0}~{\char5}-----------{\char4}~{\char0}}
\hbox{\amgr
{}~~~~~~~~{\char0}~~~~~~~~~~~~~~~{\char0}~~~~~~{\char5}------{\char3}~{\char2}------{\char4}~~~~~~{\char5}%
---------------{\char4}}
\hbox{\amgr
{}~~~~~~~~{\char5}---------------{\char4}~~~~~~~~~~~~~{\char5}-{\char4}}
\hbox{\amgr
{}~~~~~~~~~~~~~~$l+1$~~~~~~~~~~~~~~~~~~~$l+1$~~~~~~~~~~~~~~~~~~~$l+1$}}
\vskip 0.8 truecm
\par\noindent
\centerline{
\vbox
{
\hsize=13truecm
\baselineskip 0.35cm
\noindent
{\bf Fig.7.2}{\rm \quad
Contraction of a squared frame. The energy differences involved in each single
step of the contraction are:
$\scriptstyle {(h+\lambda)(l-1)}$,
$\scriptstyle {-[2J-(h+\lambda)]}$,
$\scriptstyle {(h-\lambda)(l+1)}$,
$\scriptstyle {-[2J-(h-\lambda)]}$,
$\scriptstyle {(h+\lambda)(l-2)}$,
$\scriptstyle {-[2J-(h+\lambda)]}$,
$\scriptstyle {(h-\lambda)l}$,
$\scriptstyle {-[2J-(h-\lambda)]}$.}
}
}
\endinsert
\par
We want to prove now Theorem 1.
\par
Let ${\cal P}$ be the set of protocritical saddles or special minimal saddles.
\par
If $ 0 < 2 \l < h $: ${\cal P} = {\cal P}_2$ in Fig.5.1; namely ${\cal P}$ is
the set of configurations with no pluses and a
unique cluster of zeroes given by a rectangle with sides $M^*, M^*-1$ with a
unit square protuberance attached to one of its longer sides.\par
If $0 < \l < h <2\l$ and $\d < { h+\l \over 2h}$ then ${\cal P}= {\cal P}_{1,a}
:=
S(l^*,l^*)$.
If $ 0 < \l <h <2\l$ and $\d > { h+\l \over 2h}$ then ${\cal P}= {\cal P}_{1,b}
:=
S(l^*-1,l^*)$  (see Fig.5.1).
\par
Now we notice that the set ${\cal G} \subset\O _{\L}$, defined in Section 6
satisfies the
following   properties:\par\noindent
1. ${\cal G}$   connected;
$\menouno \; \in \; {\cal G},\;\;\;
\piuuno \;\not \in \; {\cal G}$.
\par\noindent
2. There exists a path  $\o : \menouno
\to {\cal P} $, contained in ${\cal G}$, with
$$
H(\s) < H({\cal P}) \;\;\; \forall \s\in \o , \;\;\s \not= {\cal P}
\Eq (7.0a)
$$
and there exists a path $\o' :{\cal P} \to \piuuno $, contained in ${\cal
G}^c$, with
$$
H(\s) < H({\cal P}) \;\;\; \forall \s\in \o' ,\;\; \s \not= {\cal P}\;\; .
\Eq (7.0b)
$$
In the case ${\cal P}={\cal P}_1$ $\equ (7.0b)$ easily
follows from the arguments of proof of Proposition 4.1: $\o$ is
constructed following a sequence of shrinking subcritical droplets whereas
$\o'$ follows a sequence of growing supercritical droplets. In the case
${\cal P}={\cal P}_2$ $\equ (7.0b)$ follows from the arguments of proof of
Proposition 4.2.
\par\noindent
3. The minimal energy in $\partial {\cal G}$ is attained only for
``protocritical'' (global saddle) configurations $\sigma \in {\cal P}$; namely,
$$
\min_{\sigma\in \partial {\cal G}}(H(\sigma)-H(-\underline 1))=
H({\cal P})-H(-\underline 1)=: \;\Gamma\;\; , \Eq (7.0)
$$
$$
\min_{\sigma\in \partial {\cal G}\setminus {\cal P}}(H(\sigma)-H({\cal P}))
\;>\; 0\;\; . \Eq (7.0')
$$
\bigskip
We notice that, starting from any $\sigma\in{\cal P}$, we can change a spin
adjacent
to the unit square protuberance always present in ${\cal P}$ (from $-1$ to $0$
in ${\cal P}_2$ if
$h>2\l$ and from $0$ to $ +1$  in ${\cal P}_{1,a}$ or ${\cal P}_{1,b}$ if
$h<2\l$)
in order to get a ``stable protuberance of length 2". This protuberance is
called
stable since its growth takes place decreasing the energy while its shrinking
requires an increase of energy. The probability of the above described single
spin
change is not smaller than $ { 1\over |\L|}$ (see, for instance, [NS1] for more
details on this point).
\par
In other words, with probability separated from zero, uniformly in $\b$,
starting from ${\cal P}$, we reach the strict basin of attraction of a
supercritical minimum.  Then,
for any $\varepsilon>0$, it follows from Proposition  4.1 that the probability
to
reach $+\underline 1$ before reaching $-\underline 1$, can be bounded from
below as:
$$ P_{{\cal P}}(\tau_{+\underline 1}<\tau_{-\underline 1})\geq
\exp (-\e \b)\;\; . \Eq (7.0'')
$$
We get from Proposition 4.1 that, for $\b$ sufficiently large, the typical
time, starting from  ${\cal P}_1$ to reach $\piuuno$ is much shorter than the
typical
time to get  to ${\cal P}$ starting from $\menouno$
$$
\lim_{\beta\to\infty}P_{{\cal P}_1}(\tau_{+\underline
1}<\exp(\G_1) \mid \tau_{+\underline 1}<\tau_{-\underline 1})
=1\;\; . \Eq (7.0''')
$$
for a suitable $\G_1 \; < \; \G$.\par
Moreover by an analysis totally analogous to the one needed for the Ising
model (see for instance [NS1]) one can get the same results starting from
${\cal P}_2$; namely
$$
\lim_{\beta\to\infty}P_{{\cal P}_2}(\tau_{+\underline
1}<\exp(\G_2) \mid \tau_{+\underline 1}<\tau_{-\underline 1})
=1\;\; . \Eq (7.0'''')
$$
for a suitable $\G_2 \; < \; \G$.\par
In Appendix A we state and prove a result concerning the sequence of passages
through ${\cal P}$ and the typical time to see an ``efficient" passage through
${\cal P}$ namely one followed by a descent to $\piuuno$.
\par
{}From Propositions 3.4, 3.7 in [OS1], Proposition A.1 of Appendix A,
\equ (7.0''), \equ (7.0''') we easily get Theorem 1. $\square$
\par\bigskip
We want now to give the definition of the tube ${\cal T}$ of trajectories
appearing in the statement of Theorem 2 below. It represents  the typical
mechanism of escape from metastability in the sense that, with probability
tending to 1 as $\b$ tends to infinity, during its first excursion from
$\menouno$ to $\piuuno$, our process will follow a path in ${\cal T}$.\par
${\cal T}$ will be optimal in the sense that it cannot be really reduced
without loosing in probability.\par
${\cal T}$ involves a sequence of  ``droplets" with suitable geometric
shapes and suitable ``resistance times" in some ``permanence sets" of
configurations
related to these droplets.
The precise statement about the typical paths during the first excursion
between $\menouno$ and $\piuuno$  will involve a certain randomness of these
resistance  times inside the different permanence sets appearing in ${\cal
T}$.\par

In ${\cal T}$ we will distinguish two parts. The ``up" part ${\cal T}_u$ namely
the
ascent to ${\cal P}$ and the ``down" part ${\cal T}_d$ from ${\cal P}$ to
$\piuuno$. This second part ${\cal T}_d$ is {\it almost downhill} in the sense
that
all the paths $\o = \s_0,\s_1, \dots ,\s_i , \dots  \in \; {\cal T}_d$ will be
such
that:
$$
\s_0 ={\cal P}, \; \exists \;\;\bar T \; : \; \s _{\bar T} \; = \; \piuuno
,\;\;\;\;\;
\max _{ \s \in \o \setminus {\cal P} } H(\s)\; < \; H({\cal P}), \;\;\;
\min_{\s \in \o} H(\s) = H(\piuuno )\;\; .
$$
Whereas ${\cal T}_u$ is {\it almost uphill} in the sense that
all the paths $\o = \s_0,\s_1, \dots ,\s_i , \dots  \in \; {\cal T}_u$ will be
such
that:
$$
\s_0 =\menouno, \; \exists \;\;\bar T \; : \; \s _{\bar T} \; = \; {\cal P}
,\;\;\;\;\;
\max _{ \s \in \o \setminus {\cal P} }H(\s) \; < \; H({\cal P}), \;\;\;
\min_{\s \in \o} H(\s) = H(\menouno )\;\; .
$$
In the following we give  the definition of
the time reversed tube $\bar {\cal T} $ of ${\cal T}_u$.\par
$\bar {\cal T}$ will also be almost downhill; it will describe the typical
first
``descent" from the protocritical saddle to $\menouno$. By general arguments
based on
reversibility (see ref. [S1]), we will deduce the desired results on the first
excursion from $\menouno$ to  ${\cal P}$ saying that with probability tending
to one
as $\b$ tends to $\infty$ it takes place in the tube ${\cal T}_u$.
Then  to conclude our construction of ${\cal T}$ we will
only have to determine  ${\cal T}_d$.
\par
Let us now recall some basic definitions of [OS1] concerning the first
descent from any configuration $\h_0$ contained in a given cycle $A$ to the
bottom
$F(A)$ valid not only for our Blume-Capel Metropolis dynamics but also for a
general
``low temperature"  Markov chain satisfying Hypothesis M in Appendix A. We
refer to Appendix A where this more general set-up is introduced.  \par
We will first define in general the set of ``standard cascades" emerging from a
configuration $\h _0$; our intention is to apply a (simplified version of a)
result
 of [OS1] telling that with high probability when $\b \to \infty$ the first
descent
from  $\h _0$ to $F(A)$ follows, in a well specified way,  a standard cascade.
Thus the main model-dependent work will be to determine, in our specific case,
the
set of standard cascades.
In particular we will reduce the problem of the determination of the tube of
typical trajectories followed by our process during
its first descent  to $\menouno$ starting from a configuration $\s_0$ in ${\cal
G}$
immediately reached starting from the global saddle ${\cal P}$, along a
downhill
path entering ${\cal G}$, to find the set (denoted by $\bar {\cal T})$
of all the standard cascades (in a suitable cycle) emerging from $\s_0$.
\par \bigskip
A {\it standard cascade} emerging from a state $\h_0$ is a
sequence: $$
 {\cal T}  ( \h_0, \o_1, \h_1,\o_2, \dots , \h_{M-1}, \o_M)\;=\;
 \o_1\cup Q_1\cup \o_2, \dots , Q_{M-1}\cup \o_M \cup Q_M, \Eq (7.1)
$$
where  for
$i=1,\dots, M$: $\o_i$ is a downhill path emerging from $\h_{i-1}$
and ending inside
the ``permanence set" $Q_i$.
Each path $\o_i$ can be downhill continued up to a stable equilibrium point
$\x_i \in Q_i$.
The  $Q_i$'s
are
special  sets being a sort of generalized cycles containing also
the minimal saddles between $\x_i$ and $F(A)$; for $i=1,\dots, M-1$
$\h_{i} \in Q_i $ are minimal saddles between $\x_i$ and $F(A)$; finally
$ \x_M \subseteq Q_M \subseteq F(A)$ (see [OS1], Section 4 for more details).
\par
Notice that $\o_i$ can just reduce to  one downhill step from $\h _{i-1}$
to $Q_i$; in this case we use the convention: $\o_i = \h _{i-1} $.
\par
We do not give here the precise definition of the $Q_i$'s
since it happens that we do not really need it.
In our particular case of Metropolis dynamics for the Blume-Capel model
with particular initial conditions (of interest for our applications) as we
will
check we have some semplifications w.r.t the general case.\par
The most important is that the  $Q_i$'s, for $i=1,\dots, M-1$ are replaced by
genuine cycles $A_i$;   $\h_i$, not contained in $A_i$, is an element of
${\cal S} (A_i)$ and $\o_i$ ends in the interior of $A_i$.\par
We will apply the general theory developed in [OS1] to two cases.
In the first one, when analyzing $\bar {\cal T}$ the cycle $A$ will be the
maximal
connected set $\bar A$ in
$\O _{\L}$ containing $\menouno$ with energy less than  $H({\cal P})$.
It follows from
Proposition 3.4 in [OS1] that $\bar A$ is contained in the set ${\cal G}$
introduced
in Section 6 and that ${\cal S}(\bar A) \equiv {\cal P}$.
Always in this  case we have: $F(\bar A)\equiv
Q_M \equiv \x_M \equiv \menouno$.\par
In the second case, in the study of ${\cal T}_d$ the cycle $A$ will be the
maximal
connected component $\tilde A$
in
$\O _{\L}$ containing $\piuuno$ with energy less than  $H({\cal P})$.
It is immediate to see that $\tilde A \subset {\cal G}^c$.
In this  case we have: $F(\tilde A)\equiv
Q_M \equiv \x_M \equiv \piuuno$.\par
In both cases, as we said before,  for suitable initial conditions we will
verify
that the $Q_i$'s  for $i=1,\dots, M-1$ are replaced by
genuine cycles $A_i$; $M$ will depend on the initial configuration $\h_0$ as
well
as on the particular choice of the parameters $J,h,\l$.
$\o_i$ ends in the interior of $A_i$; $\h_i \in {\cal S}(\x_i, F(A_{i+1}))$,
not contained in $A_i$ as we said before
are minimal saddles in the boundary $\partial A_i$.
The cycles $A_i$ are precisely the maximal connected components containing
$\x_i$
with energy less than $H(\h_i)$ ($\x_i\in A_i$  are the minima towards which
$\o_i$
can be downhill continued).
\par
We consider an initial configuration $\h_0 $  corresponding to one of the
following five cases:\par
1.  $A = \bar A$: $\h_0 \in \bar A\cap [R(l,l)\cup
R(l,l+1)]$  for some $l\ge \widetilde L$. \par
2. $A = \bar A$: $\h_0 \in \bar R(M^*,M^*-1)$.\par
3. $A = \tilde A$,
$0 < \l < h <2\l$ and $\d > { h+\l \over 2h}$: $\h_0 \in \bar B
(C(l^*-1,l^*))$.\par
4. $A = \tilde A$,
$0 < \l < h <2\l$ and $\d < { h+\l \over 2h}$: $\h_0 \in \bar B (C(l^*,l^*)
)$.\par
5. $A = \tilde A$: $\h_0 \in \bar B (\bar R(M^*,M^*))$.
\par
\vskip.5cm
{\bf Remark.}\par
We could even consider much more general initial configurations
$\h _0$. It is not true (see the definition of the set $\bar D$ in Section 4
that for {\it any } $\h _0$ the simplified
version (involving genuine cycles $A_i$ in place of the sets $Q_i$) of the
general
[OS1] results holds true.
In fact with the very particular choice $\h_0 \in \bar A\cap [R(l,l)\cup
R(l,l+1)]$ as
we will see an even simpler statement holds: the $\o_i$ will be almost all
coinciding
with $\h _i$ (in the above specified sense).\par
\vskip.5cm
{\bf Warning.}\par
We want to warn the reader of the use that we are going to make, in the
construction
of the tube ${\cal T}$, of the equivalence class of configurations as it has
been
specified in the remark in the proof of Proposition 4.1. In fact,
strictly speaking what we will construct and call standard cascades are {\it
sets}
of standard cascades obtained from equivalence classes of configurations modulo
rotations, translations, inversion and ``displacement of protuberances". \par

\bigskip

Let us now start with the definition of the set $\bar {\cal T}$ of the standard
cascades emerging from a configuration $\s_0$ in ${\cal G}$
immediately reached starting from the global saddle ${\cal P}$, along a
downhill
path entering into ${\cal G}$. \par

We consider  first the case $a = h/\l <2$ . The other case of $a >
2$ is almost identical to the corresponding one for the Ising model and will be
treated later on.\par
We have to distinguish two cases: $\d \;
 = l^* - { 2J -(h-\l)\over h} <\; { h +\l \over 2h}$, when the global
saddle   ${\cal P}$ has the form
  ${\cal P}_{1,a} = S(l^*, l^*)$
given in Fig.5.1; or $\d \; >\; { h +\l \over 2h}$ when the global saddle has
the form ${\cal P}_{1,b } = S(l^* -  1, l^*)$ also given in Fig.5.1.
\par
Let us first consider the case $\d \; < \; { h +\l \over 2h}$ (like in
Fig.7.2 for $l=l^*$).
Let $\bar {\cal P}_1 = R(l^*, l^*) $ be the configuration obtained from ${\cal
P}_1$
by erasing the unit square protuberance. $\bar {\cal P}_1$ is a subcritical
birectangle; it belongs to the set ${\cal G}$ and satisfies condition 1
above.\par

To construct the tube $\bar {\cal T}$  we have basically to solve the above
described
sequence of minimax problems.
To simplify the exposition we divide the tube $\bar {\cal T}$ into four
segments corresponding to four different mechanisms of contraction; we write:
$$
\bar {\cal T} = \bar {\cal T}_1 \cup \bar {\cal T}_2 \cup\bar {\cal T}_3 \cup
\bar {\cal T}_4\;\; .
\Eq (7.2')
$$
The most relevant ones are the first and the second part. As we will see the
third
part for $h <  2\l$  reduces just to a simple downhill path.
\par
We start from the determination of the minimal saddle
$\h_1 := {\cal S}(\bar {\cal P}_1,\menouno)$ between
$\bar {\cal P}_1$ and $\menouno$.\par
{}From the results of Section 4 we know that ${\cal S}(\bar {\cal
P}_1,\menouno)$ is not
trivial in the sense that it differs from $\bar {\cal P}_1$ and
$$ \h_1 := \; {\cal S}(\bar {\cal P}_1,\menouno) \; = S(l^*- 1, l^*)
\;\; .\Eq (7.3)$$
Thus the first ``permanence set"  $Q_1$ of our standard cascade is the cycle
$A_{l^*-1, l^*}$ defined as the maximal connected set of configurations
containing
$R(l^*, l^*)$ with energy less than  $H( S(l^*-1,l^*) )$.
We recall that the basic inequality to be checked in order to get \equ (7.3) is
$$H( S(l,l) ) - H( S(l-1,l) )> 0$$
which is verified for $L^*\leq l\leq l^*-1$.\par
For any $l$: $L^* \leq l \leq l^* -1 $
we define the cycle $A_{l,l}\; (A_{l,l+1})$ as  the maximal connected
set of configurations containing $R(l,l) \; ( R(l,l+1) )$  with energy less
than
$H( S(l-1,l) )$ $(H( S(l,l)))$ (see Fig.7.2).
By extending the previous argument we get that the first part of our standard
cascade is given by:
$$
\bar {\cal T}_1 = A_{l^*-1, l^*},
S(l^*-1,l^*),A_{l^*-1,l^*-1}, S(l^*-1,l^*-1),A_{l^*-2,l^*-1},
$$
$$\dots ,
 S(\widetilde L,\widetilde L+1), A_{\widetilde L,\widetilde L}, S(\widetilde
L,\widetilde L)
\Eq (7.4)$$
Then we observe that for $l\leq \widetilde L -1$, we have
$$\eqalign{
H( S(l,l) ) &< H( G(l,l) )\cr
H( S(l,l+1) ) &< H( G(l,l+1) )\cr}\;\; ;\Eq (7.5)
$$
\equ (7.5) are the basic inequalities to get that, for $l_0 \le l < \widetilde
L$:
$$\eqalign{
{\cal S}(R(l,l+1), \menouno) &= G(l,l)\cr
{\cal S}(R(l,l), \menouno)   &= G(l-1,l)\cr}\;\; .\Eq (7.6)
$$
It is clear from the results of Section 4 that the subsequent permanence sets
are
the cycles:
$$
A^1_{\widetilde L,\widetilde L -1},A^2_{\widetilde L,\widetilde L -1},
A^1_{\widetilde L-1,\widetilde L -1},A^2_{\widetilde L-1,\widetilde L -1},
\dots,
A^1_{l,l},A^2_{l,l}, A^1_{l,l-1},A^2_{l,l-1},\dots ,
A^2_{l_0,l_0 } \Eq (7.7)
$$
where $ l_0 =[{h \over \l} +1], \;\; l_0 \leq l$;
we notice that for our present choice of the parameters: $\l < h < 2 \l$, we
have $ l_0 =2$ but we could consider a general situation
$ l_0 > 2$ as well when analizing
the contraction of a subcritical frame in the region $h > 2\l$ (case 1 above).
Moreover, for  $l_0 \leq l\leq \widetilde L $:
\par
$A^1_{l,l-1}\; = $ maximal connected component of the
set of configurations containing $R(l,l)$ with energy less than $H( G(l-1,l))$;
namely $A^1_{l,l-1} $ is the strict basin of attraction of $R(l,l)$:
$$
A^1_{l,l-1}\; = \bar B ( R(l,l))\;\; ,
$$
with bottom
$$
F(A^1_{l,l-1}) \; =  R(l,l)
$$
and minimal saddle
$$
{\cal S} (A^1_{l,l-1})\; = G (l-1,l)\;\; ;
$$
$A^2_{l,l-1}\; = $ maximal connected component
containing $C(l-1,l)$ with energy less than $H(S(l-1,l))$. We have:
$$
A^2_{l,l-1}\; = \bar B ( C(l-1,l))
$$
$$
F(A^2_{l,l-1}) \; =  C(l-1,l)
$$
and
$$
{\cal S} (A^2_{l,l-1})\; = S(l-1,l)\;\; .
$$
For  $l_0 + 1 < l \leq \widetilde L $ we define $A^1_{l-1,l-1} \; = $
maximal connected component containing $R(l-1,l)$ with energy less than
$H( G(l-1,l-1))$. We have
$$
A^1_{l-1,l-1}\; = \bar B ( R(l-1,l))
$$
$$
F(A^1_{l-1,l-1}) \; =  R(l-1,l)
$$
and
$$
{\cal S} (A^1_{l-1,l-1})\; = G (l-1,l-1)\;\; .
$$
$A^2_{l-1,l-1}\; = $ maximal connected component
 containing $C(l-1,l-1)$ with energy less than $H(S(l-1,l-1))$. We have:
$$
A^2_{l-1,l-1}\; = \bar B ( C(l-1,l-1))
$$
$$
F(A^2_{l-1,l-1}) \; =  C(l-1,l-1)
$$
and
$$
{\cal S} (A^2_{l-1,l-1})\; = S(l-1,l-1)\;\; .
$$
Then the second segment of the standard cascade is:
$$
A^1_{\widetilde L,\widetilde L -1}, G (\widetilde L -1,\widetilde L),
A^2_{\widetilde L,\widetilde L -1}, S (\widetilde L -1,\widetilde L),
A^1_{\widetilde L-1,\widetilde L -1},G (\widetilde L-1,\widetilde L -1),
A^2_{\widetilde L-1,\widetilde L -1},
$$
$$ S (\widetilde L-1,\widetilde L -1),\dots,
S(l_0,l_0 ) ,A^1_{l_0,l_0-1}\;\; . \Eq (7.8)
$$
We notice that both the first and the second part $ \bar{\cal T} _1$,
$ \bar{\cal T} _2$ of the tube $ \bar{\cal T}$ describe a contraction following
squared or almost squared frames; but whereas in the first
part the permanence sets are cycles with many minima in their interior, in the
second part they are ``one well" in the sense that they coincide with the
strict
basin of attraction of their bottoms. The typical times spent inside these
cycles and the typical states visited before leaving them are different in the
two cases of $ \bar{\cal T} _1$ and $ \bar{\cal T} _2$.\par
The third part $ \bar{\cal T} _3$, that we are going to define, corresponds to
the
shrinking of the interior rectangle of the frame $C(l_0,l_0)$.
Indeed it follows from Section 4 (see \equ (4.14) therein) that for $ l < l_0$
the lowest minimal saddles in the boundary of the basin
of attraction $A^1_{l_0,l_0-1 }$  of the
birectangle $ R(l_0,l_0) \equiv R( l_0,l_0-1, l_0+2, l_0 +2) \cup
R( l_0 -1,l_0, l_0+2, l_0 +2)$ is not $G(l_0-1, l_0)$ corresponding to $ S_2$
in
Fig.4.6 but, rather, the saddle $S_3$ in Fig.4.6; in other words starting from
the birectangle $ R(l_0,l_0)$ it is no more convenient to continue the
contraction
along frame shapes but, on the contrary, the internal rectangle starts its
independent shrinking keeping fixed the external one. It appears clear that if
$h < 2 \l$ then the shrinking and disappearing of the internal two by two
rectangle is just a down hill path where the number of internal plus spins
decreases monotonically to zero. If we were considering a general initial
condition
corresponding to the above case  1 namely the contraction of a subcritical
frame for
$h > 2\l$, then we would have had $ l_0 >2$ and the shrinking and disappearence
of the
internal rectangle would have followed a sequence of squared or almost squared
rectangular shape exactly like in the case of the standard Ising model.
\par
In the following we will consider birectangles $R(L_1,L_2;M_1,M_2)$
(see $\equ (3.3)$) also for $L_1,L_2=0,1$.
\par
Then the third part for $h < 2 \l$ is just the downhill path:
$$
\bar {\cal T}_3 = S_3^*, R(1,2 ; 4,4),R(1,1 ; 4,4),R(0,0 ; 4,4)
\Eq (7.9)
$$
where by $S_3^*$ we denote the saddle depicted in Fig.4.6 when the external
rectangle is a $ 4 \times 4$ square and the internal cluster is a ``triangle
made by 3 sites".\par
Finally the fourth part is just an Ising-like contraction of the remaining
$ 4 \times 4$ rectangle of zeroes. We will observe first a sequence of
permanece sets (corresponding to stable rectangles) and saddles and finally
the downhill path describing the disappearence of the last $ 2 \times 2$
stable rectangle.
 \par
We have: $$
\bar {\cal T}_4 = \tilde S _1 ,  \tilde R _1,
\tilde S _2 ,  \tilde R _2,
\tilde S _3 ,  \tilde R _3,
\tilde S _4 ,  \tilde R _4,
\tilde \o
\Eq (7.10)
$$
where $ \tilde R _1 =
R(0,0 ; 4,3) \cup R(0,0 ; 3,4)$, $\tilde R _2 = R(0,0 ; 3,3)$,
$\tilde R _3 =
R(0,0 ; 3,2)\cup R(0,0 ; 2,3)$,
$\tilde R _4 = R(0,0 ; 2,2)$;
the downhill path $\tilde \o$ is given by
$$
\tilde \o := \tilde S_5, \tilde R_5, \tilde R_6,\menouno\;\; ,
\Eq (7.11)
$$
with $\tilde R_5 = R(0,0 ; 1,2)\cup R(0,0 ; 2,1)$,
$\tilde R_6 = R(0,0 ; 1,1)$;
the saddles $ \tilde S_i $, $i= 1, \dots , 5$ are obtained from the
rectangles  in $ \tilde R _i$ by adding a unit square protuberance to one of
its
longer sides.\par
This concludes the definition of $\bar {\cal T}$  for $ h < 2\l,
\d \; < \; { h +\l \over 2h}$.\par
In the case $ h<2\l$,
$\d \; > \; { h +\l \over 2h}$ the definition of $\bar {\cal T}$ is almost
identical; we only have to modify a little bit at the very beginning the
definition
of $\bar {\cal T}_1$ by eliminating its first permanence set.\par
Indeed we know from Section 5 that now $ H( S( l^*-1, l^*) ) > H( S( l^*, l^*)
)$ so
that the protocritical saddle is, in this case,
${\cal P}_{1,b } = S(l^* -  1, l^*)$.
Now the configuration $\bar {\cal P}_1$ obtained from ${\cal P}_1$ by erasing
the
unit square protuberance is the subcritical rectangle
$\bar {\cal P}_1 = R ( l^*-1, l^*)$; again this belongs to case 1.
Then the first permanence set is
now  $A_{l^*-1,l^*-1}$ and we have
$$
\bar {\cal T}_1 = A_{l^*-1,l^*-1}, S(l^*-1,l^*-1),A_{l^*-2,l^*-1}\dots
 S(\widetilde L,\widetilde L+1), A_{\widetilde L,\widetilde L}, S(\widetilde
L,\widetilde L)\;\; .
\Eq (7.12)$$
The other segments of the tube $\bar {\cal T}_i , \; i=2,3,4$ are defined
exactly
as before.\par
The last  case that we have still to analyze  to construct $\bar {\cal T}$ is
$h > 2
\l$. In this case the protocritical saddle is
${\cal P} = {\cal P}_2$ and the tube
$\bar {\cal T} $ is just an Ising-like contraction along squared or almost
squared
rectangular clusters of zeroes in a sea of minuses.
\par
Now the configurations obtained by erasing the unit square protuberance,
containing a unique subcritical rectangle of zeroes in a sea of minuses is
given
by:
$$
\bar {\cal P}_2 = \bar R(M^*-1,M^*)\;\; ;
$$
notice that $\bar {\cal P}_2$ is included in the case 2 above.
\par
We observe that the appearence of a single plus spin will induce the overcoming
of an
energy barrier greater or equal to $4J -(h + \l)$. It is very easy to see that
we
can proceed in the construction of the set of the standard cascades emerging
from
$\bar R(M^*-1,M^*)$ without be forced to overcome a barrier larger then $2J$
so that certainly  in all these standard cascades, for our choice of the
parameters, we will never see a single plus spin appearing. Indeed one easily
convince
himself that the sequence of minimax problems to be solved are the exact
analogue of
the ones arising in the analysis of a subcritical contraction for a standard
Ising
model. We refer to [S1], [KO1] for more details.
For completeness in the following we summarize the results using our notation.
\par
The first permanence set is $\bar B(R(M^*-1,M^*))$.\par
Let us define the following sequences of couples of integers:
$$
(l_1,m_1), (l_2,m_2), \dots , (l_N,m_N) \;\;\; N = 2M^* -2
$$
$$
(l_1,m_1) = (M^* - 1, M^*), \;\;\;
(l_N,m_N) = (1,1); \;\;\;\; |l_i-m_i| \leq 1 \; : \; \;
m_i=l_i \;\; \hbox {or }\;\;\;m_i=l_i+1
$$

$$ \hbox {if  }(l_i,m_i) = (l,l+1) \;\;\; \hbox {then  } (l_{i+1},m_{i+1}) =
(l,l)
$$

$$
 \hbox {if  }(l_i,m_i) = (l,l)\;\;\;
\hbox {then } (l_{i+1},m_{i+1}) = (l-1,l)\;\; .
$$
Given $(l,m)$ as before: for $|l-m| \leq 1, \; 1 \leq m
 \leq M^*-1$
we denote by $ \bar S(l,m) $ the saddle obtained from $\bar R(l,m)$ by
adding a unit square protuberance (with a zero spin inside) to one of its
longest sides.
\par
We have:
$$
\bar {\cal T} = \bar B ( \bar R(l_1,m_1)), \bar S (l_2,m_2),
 \bar B (\bar R(l_2,m_2) ), \dots , \bar S(l_N,m_N), \bar R(l_N,m_N), \menouno
\;\; .
$$
This concludes the definition of $\bar {\cal T}$.\par\bigskip
Let us now pass to the definition of the descent part ${\cal T}_d$ of
 the tube ${\cal T}$. \par
We start from the case $ h < 2\l, \d > { h+\l \over 2h}$ .\par

It is immediately seen that by adding to
${\cal P}_1 = {\cal P}_{1,b} \equiv S(l^* -1,l^*)$ a unit square protuberance
to
form a stable protuberance of length 2 we get a configuration $\h_0$ included
in
case 3.
\par

We distinguish in ${\cal T}_d$ two parts: ${\cal T}_{d,1}$ and ${\cal
T}_{d,2}$.
\par
For $l^* -1 \leq l < \widetilde M -2$ we denote by $ \bar A_{l-1,l}$ the cycle
given
by the maximal connected set of configurations containing $ C({l-1, l})$
with energy less than $ H(S(l,l))$.
We easily verify that $F(\bar A_{l-1,l}) = C({l-1, l})$,
${\cal S}(\bar A_{l-1,l}) = S(l-1, l)$. \par
For $l^*  \leq l < \widetilde M -2$ we denote by $ \bar A_{l,l}$ the cycle
given
by the maximal connected set of configurations containing $ C({l, l})$
with energy less than $ H(S(l,l+1))$.
We easily get that $F(\bar A_{l,l}) = C({l, l})$,
${\cal S}(\bar A_{l,l}) = S(l, l+1)$. \par
For  $l^* -1 \leq l < \widetilde M -2$ we denote by $ \O _{l-1,l}$ the
set of downhill paths starting from $S(l-1, l)$ and ending in
$\bar A _{l-1, l} $.
Similarly, for  $l^* \leq l < \widetilde M -2$ we denote by $ \O _{l,l}$ the
set of downhill paths starting from $S(l, l)$ and ending in
$\bar A _{l, l} $.
We set
$$
{\cal T}_{d,1} = \h_0, \bar A _{l^*-1, l^*}, S(l^*,l^*),  \O _{l^*,l^*},
\bar A _{l^*, l^*},S(l^*,l^*+1),\O _{l^*,l`^*+1}, \dots,
S({\widetilde M}-2,{\widetilde M}-2)\;\; .
$$
As it has been shown in Section 4 for $l \geq \widetilde M -2$ the growth is
typically
symmetric in the sense that the probability of growth in the directions
parallel or
orthogonal to the shortest side of our supercritical frame are logarithmically
equivalent for large $\b$.
Moreover it follows from the analysis developed in Section 4 that for
$l \geq \widetilde M -2$ the set ${\cal D}$ defined in $\equ (4.20)$ do not
play any
particular role and the permanence sets are cycles given by the strict basins
of
attraction of frames $C(l_1,l_2)$ or birectangles $R(l_1,l_2)$ .
The second part ${\cal T}_{d,2}$ of ${\cal T}_d$ will describe the
supercritical
growth starting from $l= \widetilde M -2$.
To construct ${\cal T}_{d,2}$ we need some more geometrical defininitions.\par
For a given frame $C(l_1,l_2)$, we use the notation $C(l,m)$ to make explicit
the
shorter and
longer sides $l$ and $m$, respectively.\par
 We denote by $G_>(l,m)$, $G_<(l,m)$,
respectively, the saddle configurations containing a unique
droplets
 obtained by attaching  a unit square protuberance (with a zero spin inside) to
a
longer or shorter external side  of $C(l,m)$.\par
We denote by $R_>(l,m)$, $R_<(l,m)$,
respectively, the
birectangles
 obtained from $G_>(l,m)$, $G_<(l,m)$
by extending the unit square protuberance to an entire side.\par
We denote by $S_>(l,m)$, $S_<(l,m)$,
respectively, the
saddle configurations containing a unique droplet
 obtained from $R_>(l,m)$, $R_<(l,m)$
by attaching a unit square protuberance (with a plus spin inside)  to the
internal
free side.\par
We denote by $\O_>(l,m)$, $\O_<(l,m)$,
respectively, the set of all the
downhill paths emerging from
  $S_>(l,m)$, $S_<(l,m)$
and ending in $\bar B(C(l+1,m))$, $\bar B(C(l,m+1))$;
finally we denote by $\hat\O _>(l,m)$, $\hat\O _<(l,m)$ the set of all downhill
paths
emerging from $G_>(l,m)$, $G_<(l,m)$ and
ending in $\bar B(R_>(l,m))$, $\bar B(R_<(l,m))$.\par
Given $(l,m)$, we  denote  by
$\G _>(l,m)$ the sequence: $\bar B(C(l,m))$, $G_>(l,m)$, $\hat\O _>(l,m)$,
$\bar B(R_>(l,m))$, $S_>(l,m)$, $\O_>(l,m)$, $\bar B(C(l,m+1))$.
Similarly
we denote  by
$\G _<(l,m)$ the sequence: $\bar B(C(l,m))$, $G_<(l,m)$, $\hat\O _<(l,m)$,
$\bar B(R_<(l,m))$, $S_<(l,m)$, $\O_<(l,m)$, $\bar B(C(l+1,m))$.
\par
A sequence $(l_i,m_i)_{i=1,2,\dots}$ with $l_i\leq m_i$ is called {\it
regularly
increasing} if:
\par
$l_1=m_1= \widetilde M-2$  and, for any $i=1,2, \dots$, either
 $(l_{i+1},m_{i+1}) \equiv (l_i,m_i)^> := (l_{i}+1,m_{i})$ or
 $(l_{i+1},m_{i+1}) \equiv (l_i,m_i)^< := (l_{i},m_{i}+1)$.\par
If $l_i = m_i = L \; \equiv $ the side of our torus $\L$, we set $l_{i+1} =
m_{i+1} =
L$.\par Let ${\cal L}$ be the set of all regularly increasing sequences.
For any $(l_i,m_i)_{i=1,2,\dots} \in {\cal L}$ we define:
$\d (l_i,m_i) := \; >$ if $ (l_{i+1},m_{i+1}) \equiv (l_i,m_i)^>$ and
$\d (l_i,m_i) := \; <$ if $ (l_{i+1},m_{i+1}) \equiv (l_i,m_i)^<$.
\par
{}From the arguments developed in Section 4 it easily follows that the second
part of
${\cal T}_{d}$ is given by:
$$
{\cal T}_{d,2} = \cup _{(l_i,m_i)_{i=1,2,\dots} \in {\cal L}}
\G_{\d(l_1,m_1)}(l_1,m_1)
\cup \G_{\d(l_2,m_2)}(l_2,m_2)
\cup \dots,\G_{\d(l_i,m_i)}(l_i,m_i), \dots\;\; .
$$
This concludes the construction of ${\cal T}_{d}$
for the case $ h < 2\l, \d > { h+\l \over 2h}$ .\par
The case $ h < 2\l, \d < { h+\l \over 2h}$ requires only minor changes: the
only
difference is that now we have to start a step further.
Indeed it is immediately seen that by adding to
${\cal P}_1 = {\cal P}_{1,a} \equiv S(l^* ,l^*)$ a unit square protuberance to
form a stable protuberance of length 2 we get a configuration $\h_0$ included
in
case 4. We have
$$
{\cal T}_{d,1} = \h_0,
\bar A _{l^*, l^*},S(l^*,l^*+1),\O _{l^*,l`^*+1}, \dots,
S(M^*-2,M^*-2)\;\; .
$$
The rest is identical.\par
For the case  $ h > 2\l$ we have exactly the same behaviour as in the Ising
model
namely we pass to consider an initial condition like in the case 5. Then we
have a
symmetric growth along a sequence of supercritical growing rectangles of
zeroes in a
sea of minuses up to the configuration $\zero$. Subsequently we have again an
Ising-like nucleation of a protocritical droplet of pluses in the sea of
zeroes (an $L^*\times (L^* -1)$ rectangle with a unit square protuberance
attached to one
of its longer sides) up to the configuration $\piuuno$. This last case has been
already
analyzed in detail (see, for instance [NS1], [S1]). We leave the details to the
reader.\par
One can easily convince himself that this indeed concludes the
construction of  the set
of all standard cascades emerging from any of the above specified five type of
initial
conditions for any value of the parameters (not only for the subcases that we
have
explicitely treated).\par
We can now state our main result on the tube of typical trajectories during the
first
excursion between $\menouno$ and $\piuuno$.\par
Let
${\cal T} := {\cal T}_u \cup {\cal P} \cup {\cal T}_d$
with ${\cal T}_u$ given by the time reversal of the set of standard cascades in
$\bar A \subset {\cal G} $ emerging from $ \bar {\cal P}$:
${\cal T}_u := {\cal R} \bar {\cal T}  $ (the time reversal operator acts on
paths
in this way: for $\o = (x_1, x_2, \dots x_{N-1}, x_N) :\;\;{\cal R} \o =
(x_N, x_{N-1}, \dots x_2, x_1)$);   ${\cal T}_d$ given by the set of standard
cascades in $\tilde A \subset {\cal G}^c$ emerging from  $ \tilde {\cal P}$.
Let $\bar {\cal P}_1$ be either $\bar {\cal P}_1$ or $\bar {\cal P}_2$
according
to the values of the parameters $J,h,\m$;
let $\tilde {\cal P}_1$ be either $\tilde {\cal P}_1$ or $\tilde {\cal P}_2$
according
to the values of the parameters $J,h,\m$.\par
\vskip 0.35 truecm
\noindent
{\bf Theorem 2.}\par\noindent
Consider the dynamical Blume-Capel model described by the Markov chain with
transition probabilities given in $\equ (2.5)$ of Section 2. For any choice of
the
parameters $J,h,\m$ compatible with $\equ (3.14)$.\par\noindent
$i)$
$$ \lim_{\b \to \infty}  P_{\menouno} ( \s_t \in {\cal T} \; \forall \; t \in
[\bar \t
_{\menouno}, \t _{\piuuno}] ) \; = \; 1\;\; .
$$
The history of the process in ${\cal T}$ is described in the following way:
\par
\noindent
consider an initial configuration $\h_0 $  corresponding to one of the
following five cases:\par
1.  $A = \bar A$: $\h_0 \in \bar A\cap [R(l,l)\cup
R(l,l+1)]$  for some $l\ge \widetilde L$. \par
2. $A = \bar A$: $\h_0 \in \bar R(M^*,M^*-1)$.\par
3. $A = \tilde A$,
$0 < \l < h <2\l$ and $\d > { h+\l \over 2h}$: $\h_0 \in \bar B
(C(l^*-1,l^*))$.\par
4. $A = \tilde A$,
$0 < \l < h <2\l$ and $\d < { h+\l \over 2h}$: $\h_0 \in \bar B (C(l^*,l^*)
)$.\par
5. $A = \tilde A$: $\h_0 \in \bar B (\bar R(M^*,M^*))$.
\par
Then, considering for any such $A , \h_0$ the set of all standard cascades
emerging from $\h_0$ and falling into $F(A)$ we have \par \noindent
$ii)$
$$ \exists \; \d >0 \;\;
\hbox {such that}\;\; \lim _{ \b \to \infty}
P_{\h_0} (\t _{ F(A)} <
\exp ( \b [ H(\h_1) - H(F(A)) - \d] )=1\;\; ,
$$
\par \noindent
$iii)$
$$
\lim _{ \b \to \infty}
P_{\h_0} (\forall \; t \leq \t _{ F(A)} \;:\;
x_t \; \in \;{\cal T}  ( \h_0, \o_1, \h_1,\o_2, \dots , \h_{M-1}, \o_M)
$$
$$
\hbox { for some standard cascade }\; \h_0, \o_1, \h_1,\o_2, \dots , \h_{M-1},
\o_M )
\; =\; 1\;\; ,
$$
\par \noindent
$iv)$ moreover, with  probability $\to \; 1$ as $\b \to \infty$, there exists
a sequence $\h_0, \o_1, \h_1,\o_2,$ $ \dots , \h_{M-1}, \o_M $ such that
our process
starting at $t=0$ from $\h_0$,   between
$t=0$ and $t= \t _{ F(A)}$,
after having followed the initial downhill path
$\o_1$,   visits, sequentially, the sets $A_1, A_2,\dots , A_{M-1}$
exiting from $A_j$ through $\h_j$ and then following the path $\o_{j+1}$ before
entering $A_{j+1}$.\par
For every $\e >0$
with  probability tending to one as $\b \to \infty$ the  process
spends inside each $A_j$  a time $T_j (\e)$ :
$ \exp ( \b [ H(\h_j) - H(F(A_j)) - \e] ) \; < \; T_j (\e) \;< \;
\exp ( \b [ H(\h_j) - H(F(A_j)) + \e] )
$ and before exiting from $A_j$ it visits each point in $A_j$ at least $\exp \b
\e$ times .
\par\noindent
{\it Proof.}
\vskip 0.5 truecm
\par
We easily get that
$$
\lim _{\b \to \infty} P_{ {\cal P}_1}(\s_1 \in \bar {\cal P}_1| \s_1 \in {\cal
G})
= 1 \;\; .\Eq (7.20)
$$
Indeed \equ (7.20) follows from the fact that there is only one first possible
step
in any downhill path from ${\cal P}_1$ to ${\cal G}$: it corresponds to erasing
the
unit square protuberance to get $\bar {\cal P}_1$.
\par
On the other hand we have:
$$
\lim _{\b \to \infty} P_{ {\cal P}_{1,b}}(\s_1=\h_0 | \s_1 \in {\cal G}^c)
= 1\;\; .\Eq (7.21)
$$
The proof is an immediate consequence of Theorem 1, \equ (7.20), \equ (7.21),
Theorem
1 in [OS1], and the results of [S1]. $\square$

\vfill\eject
\includegraphics{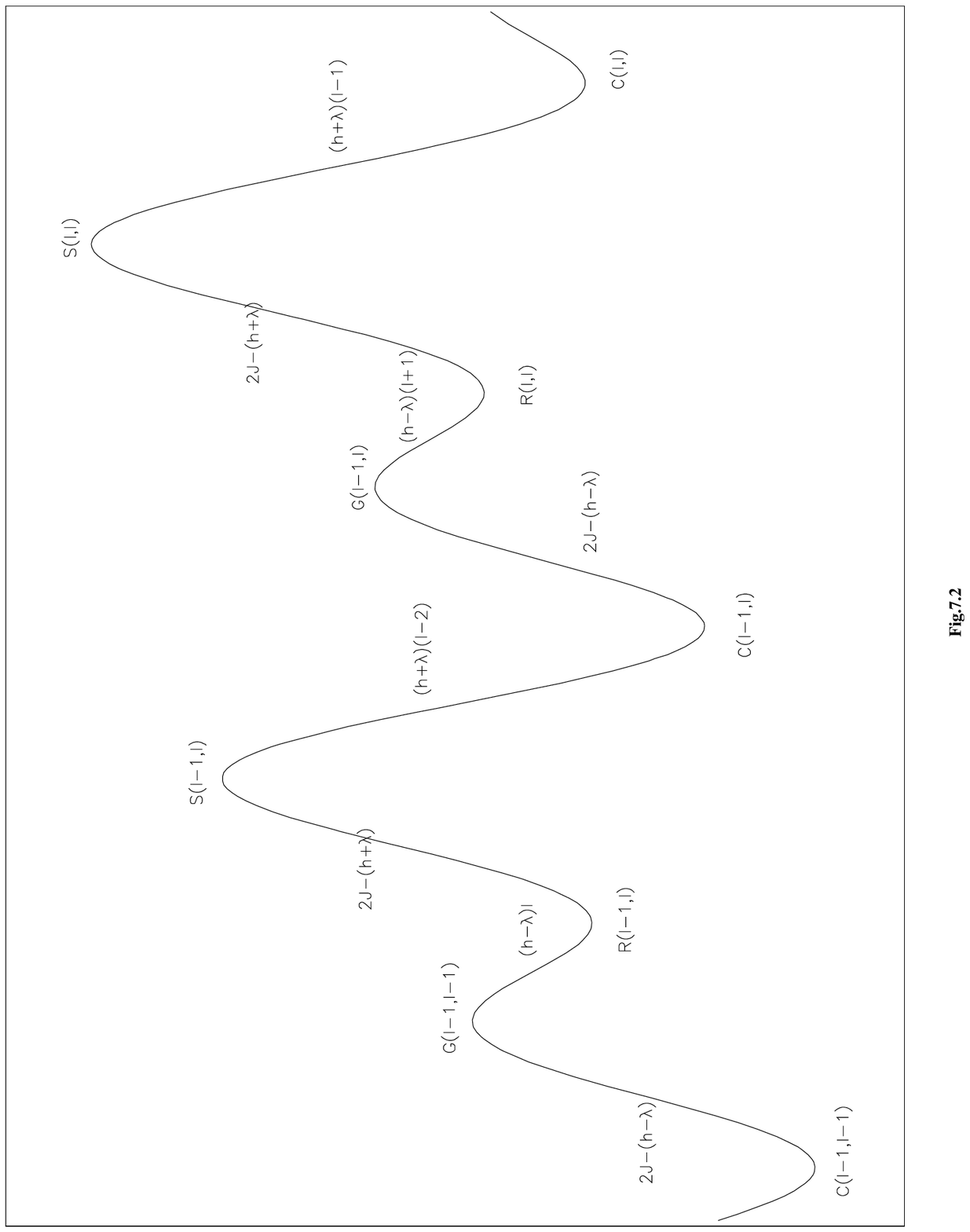}
\par
$\phantom .$

\vfill\eject
\numsec=8\numfor=1
\vskip 1cm
{\bf Section 8. Conclusions.} \par
We have described the metastable behavior of a dynamical Blume-Capel model. Our
updating rule is given by the classical Metropolis algorithm but it is clear
that
our results extend to a wide class of single-spin-flip reversible dynamics.\par
Our results refer to the asymptotic regime of  small
but fixed
magnetic field $h$ and chemical potential $\l$,
 large but fixed  volume $\L$ and very large inverse
temperature $\b$.\par
We take mainly the point of view of the so called {\it pathwise approach} to
metastability
 aiming to describe the typical behaviour of the random trajectories
of our stochastic dynamics rather than describing the evolution of the
averages.\par

 Blume-Capel model exhibits the interesting feature of the presence of
three possible phases. The equilibrium phase diagram is, consequently, very
reach
and
interesting.
The most important aspect from the  point of view of the  study of the
dynamics of metastability is the presence, near the triple point,
of two competing metastable phases. This means that, for instance, if one wants
to
describe the decay from the metastable $\menouno$ phase to the stable $\piuuno$
phase one has to take into account the presence of another metastable phase:
$\zero$.\par
We took  as initial condition the state  $\menouno$ and we analyzed the
region of parameters $0<\l <h$.
Let us subdivide it into the regions II and III defined as follows:
$$\eqalign{
{\rm II} &:= 0<\l<h<2\l\cr
{\rm III}&:= 0<2\l<h\cr}\;\; .$$
It is easily seen that, with the same arguments developed in Sections 3,4,5,6
we
could analyze the region
$$ {\rm IV} := 0<-\l<h$$
as well.
In II, III, IV the stable equilibrium phase (absolute minimum for the energy)
is
$\piuuno$ and we have:
$$
H(\menouno) > H(\zero) >  H(\piuuno)\;\; .
$$
In the region
$$ {\rm I} := 0< h< \l$$
we have
$$
H(\zero) >  H(\menouno) >  H(\piuuno)
$$
and then it is reasonable to expect and  not difficult to prove that in the
decay from $\menouno$ to $\piuuno$ the state $\zero$ does not play any role.
Indeed it is sufficient to exhibit a mechanism of transition from $\menouno$ to
$\piuuno$ involving an energy barrier smaller than $H(\zero) - H(\menouno)
$.\par
This is very easy to achieve if the volume $\L$ is sufficiently large.\par
In this paper we analyzed in detail the regions II and III which happen to be,
in a
sense, the most interesting ones. In the region IV one has the same local
minima for
the energy as in the regions II and III; they are sets of non-interacting
plurirectangles; but now the comparison between the times $t_1,t_2,t_3,t_4$
introduced in $\equ (3.11)$ changes totally. The main difference w.r.t.
the regions II, III is that now, in IV, we have:
$$
M^* < L^*\;\; ,
$$
and so we cannot even consider a possible mechanism of nucleation along a
sequence
of frames.
Indeed one has that a birectangle is supercritical if and only if the minimal
external side is not smaller than $M^*$.
Then, like in region III but in a much easier way, we can prove that the escape
from
$\menouno$ starts with an Ising-like nucleation of a protocritical droplet
${\cal
P}_2$ leading to $\zero$. But now, contrary to the region III the typical time
$T^{\menouno \to \zero}$ for going from $\menouno$ to $\zero$ is much shorter
than
the typical time
$T^{\zero \to \piuuno }$ for going from $\zero$  to $\piuuno$ so that the
asymptotics of the time  $T^{\menouno \to \piuuno }$ of the transition from
$\menouno$ to $\piuuno$ is dominated by $T^{\zero \to \piuuno }$.
\par
The situation in which a priori one could expect a  competition between the two
metastable phases would be at a first glance the union of the regions II, III,
IV. By arguing  more carefully with a heuristic analysis of the heights of the
possible
barriers between $\menouno$ and $\zero$ and between $\menouno$ and $\piuuno$
(given
by the energy of formation of suitable critical droplets) one is led to expect
that the two metastable phases  corresponding to $\menouno$ and  $\zero$ are
in a sense really
competing only around the half line $0< h= 2\l $
 separating the regions II and III. This value $h= 2\l $ depends on
the particular form of the Blume-Capel hamiltonian.\par
The main result of the present paper consists in the rigorous proof of the
above
heuristics.\par
{}From mathematical point of view we had to solve some large deviation
problems.
This kind of problems would  be extremely hard for a general non-reversible
dynamics but their treatment is very much simplified by the reversibility
property
of the dynamics.\par
In particular to get the result we  had to solve the minimax problem of the
determination of the global saddle between $\menouno$ and $\piuuno$.
This is the really hard point of the work.
We could handle the large deviation
problems \`a la Freidlin-Wentzell arising in the study of some rare events in
the
framework of our low temperature Metropolis dynamics  by taking advantage of
a general approach to the study of typical trajectories, during the first exit
from a
non-completely attracted domain, recently developed in [OS1]. Nevertheless we
still had to face the crucial model-dependent part consisting in solving some
geometrically quite involved variational problems.\par
 In particular we had to
exclude, as highly depressed in probability, any mechanism of transition based
on
coalescence and we had to single out, among many others, only very few possible
mechanisms of nucleation.
\par
\midinsert
\vskip 11 truecm\noindent
\includegraphics{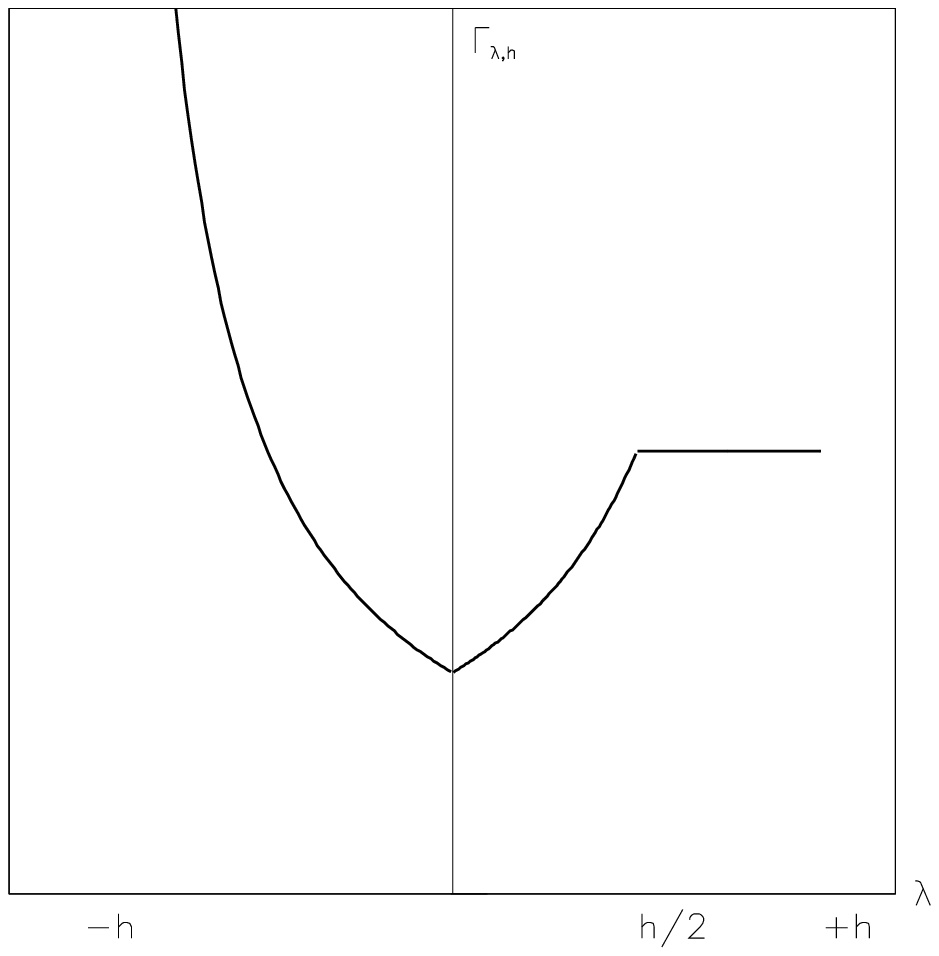}
\vskip 1 truecm
\centerline {\smbfb Fig.8.1}
\endinsert
\par
We were able to rigorously compute the lifetime of the metastable
state, namely the tipical transition times $T_{\l,h}$, for different values of
the
parameters  $\l,h$. It turns out that these transition times are given by
$$
T_{\l,h} \sim \exp (\b \G _{\l,h})$$
where the ``activation energy", for very small values of $\l,h$ has the
following
expression
$$
\G _{\l,h} \sim  {8J^2 \over h}  \;\;\;\; \hbox {for} \;\;\;\; 0< \l<h < 2\l
\Eq (8. 1)
$$
and
$$
\G _{\l,h} \sim  {4J^2 \over h-\l}  \;\;\;\; \hbox {for}  \;\;\;\;0< 2\l<h
\;\; .\Eq (8.2)
$$

The value ${4J^2 \over h-\l}$ is just the activation energy
 $\G^{\menouno \to \zero}_{\l,h}$ for the transition between
$\menouno$ and $ \zero$.
The activation energy for the transition between $ \zero$ and
$\piuuno$ is always (approximately) given by:
$$
\G^{\zero \to \piuuno}_{\l,h} \sim {4J^2 \over h+\l}\;\; .\Eq (8.2')
$$
In region III we have $\G^{\menouno \to \zero}_{\l,h} >
 \G^{\zero \to \piuuno}_{\l,h}
$ and this is the reason for \equ (8.2); but in Region IV we have the opposite
$\G^{\menouno \to \zero}_{\l,h} <
 \G^{\zero \to \piuuno}_{\l,h}$  so that we get :
$$
\G _{\l,h} \sim  {4J^2 \over h+\l}  \;\;\;\; \hbox {for}  \;\;\;\;0< -\l<h
\;\; .\Eq (8.3)
$$
This answers a question raised in [R] about the ``validity of Van't
Hoff-Arrhenius
law" which would predict, in our case a decay $\menouno \to \zero\to \piuuno$
with
 an asymptotics of the transition time determined by $T^{\menouno \to
\zero}$.\par
Our results can be interpreted by saying that this law is valid in the region
III
whereas it is violated in regions II and IV for different reasons.\par
The new phenomenon about which apparently there is no reference even in the
physics
literature is the possibility of a ``direct" transition between $\menouno $ and
$\piuuno $ and also a possible change in the mechanism of transition for
different values of the parameters.\par

Notice that if we take a fixed small value of $h$ and we vary $\l$ the analytic
expression of $\G _{\l,h}$ changes when we cross the lines $h =2\l$,
$\l=0$.\par
We draw in Fig.8.1 the graph of $\G _{\l,h}$ as a functin of $\l$ for a fixed
small
value of $h$.
\par

\vfill\eject
\numsec=9 \numfor=1

{\bf Appendix.}
\par
In this appendix we want to state and prove Proposition A1 below. It refers to
the first escape from a transient cycle $A$ (see below) and, roughly speaking,
it says that, under general hypotheses, with high probability, after many
attempts, soon or later our process will {\it really }  escape from $A$
entering into another different cycle by passing through one of the minimal
saddles of the boundary of $A$.
\par
The time for this transition has about the same asymptotics as the first
hitting time to the boundary of $A$.
\par
The result of Proposition A1 was already used before without an explicit proof
(see, e.g., [KO1] and [KO2]); it is, in fact, a simple consequence of the
strong Markov property but we think  it  useful, in order to better explain
its statement, to eventually provide an explicit proof.
\par
We will state our results in a  slightly more general set-up than the one
considered in the present work (we also use a different notation). We will
consider general Metropolis  reversible Markov chains.
\par
We suppose given an ergodic, aperiodic Markov chain $(X_t)_{t=0,1,2,\dots}$
with finite state-space $\O$ and with transition probabilities $ P(x,y)$
satisfying the following
\vskip 0.7 truecm
\noindent
{\bf Hypothesis M.}
\par\noindent
There exists a function $ H : \O \to { \bf R}^+ $  such that
$$P(x,y)= q(x,y)  \exp(-\b[ H(y) - H(x)]_+)\;\; ,\eqno (A.1)$$
where  $q(x,y) = q(y,x)$ and $(a)_+$ is the
positive part $(:= \max\{a, 0\} )$ of the real number $a$.
\vskip 0.7 truecm
\par
The above choice corresponds to a {\it  Metropolis Markov chain
} which is {\it reversible} in the sense that:
$$
\forall \; x, x' \; \in \O \;:\; \m (x) P(x,x') \;=\;  \m (x') P(x',x)
\eqno (A.2)
$$
with
$$
\m (x) \propto \exp ( - \b H(x))\;\; . \eqno (A.3)
$$
\par
One can introduce the notions of pair of communicating states, path, connected
subset of $\O$, boundary $\partial Q$ of set $Q \subset \O$
cycles, $\dots$ as the obvious generalizations of the corresponding ones given
in Section 2.
\par
For any set $Q\subset \O $ we introduce the set of all the
minima of the energy in the boundary $\partial Q$ of $Q$:
$$
U(Q) := \{ z \in \partial Q : \min _{ x \in \partial Q } H(x) = H(z) \}
\;\; .\eqno (A.4)
$$
By $F(Q)$ we denote the set of the absolute minima of the energy in the set
$Q\subset \O$:
$$
F(Q) := \{ y \in  Q : \min _{ x \in Q } H(x) = H(y) \}
\;\; .\eqno (A.5)
$$
A cycle $A$ for which there exists $y^* \in U(A)$ {\it downhill connected}
to some
point $x$ in $A^c$ (namely $\exists \;x \not \in A, $ communicating with $y^*$,
with $H(x) < H(y^*) =:H(U(A))$), is called {\it transient}; points like
$y^*$ are called {\it  (minimal) saddles}.
${\cal S}(A)$ will denote the set of all minimal saddles of $A$.
\par
Let $R=R(A)$ be the subset of $A$ to which some point in ${\cal S}(A)$ is
downhill connected:
$$
R(A) := \{ y \in A \;\; {\rm such \; that}\;\; \exists z \; \in \; {\cal S}(A)
\;\; {\rm with} \;\;  P(x,y) \; > \; 0\}
\;\; ;\eqno (A.6)
$$
let $V=V(A)$ be the analogue of $R$ outside $A$:
$$
V(A) := \{ y \not \in A \;\; {\rm such \; that}\;\; \exists z \; \in \;
{\cal S}(A)\;\; {\rm with} \;\;  P(x,y) \; > \; 0\}
\;\; .\eqno (A.7)
$$
We set:
$${\cal H} := R(A) \cup V(A)\;\; .\eqno (A.8)$$
\vskip 0.35 truecm
\noindent
{\bf Proposition A.1.}\par\noindent
Consider a transient cycle $A$. Given $\e >0$ let
$$T(\e) := \exp \b [H({\cal S}(A))-H(F(A)) + \e]\;\; .\eqno (A.9)$$
Then, for every $\e>0, \; x \in A$,
$$
\lim_{\b \to \infty} P_x( \t_{ (A\cup \partial A)^c} \; > \; T(\e) )\; = \; 0
\;\; ,\eqno (A.10)
$$
and
$$
\lim_{\b \to \infty} P_x(X_{ \t_{ (A\cup \partial A)^c}} \in \; V(A) )\; = \; 1
\;\; .\eqno (A.11) $$
\par\noindent
{\it Proof.}
\vskip 0.5 truecm
\par
{}From Hypothesis M and the definition of ${\cal S}(A)$ we know that there
exists a positive constant $c > 0$, independent of $\b$, such that
$$
\inf _{ x\in {\cal S}(A),y\in {\cal H}} P(x,y) \; > \; c,\;\;\;\;\;\;\;\;
\lim _{\b \to \infty}\sup _{ x\in {\cal S}(A),y \not \in {\cal H}} P(x,y) \; =
\; 0
\;\; .\eqno (A.12)
$$
We define, now, the sequence $\t_i$ of stopping times corresponding to
subsequent
passages of our chain $X_t$ in $\partial A$:
$$\eqalign{
\t_o      :=&\inf \{ t\geq 0    : X_t \in     \partial A\}\cr
\sigma_1  :=&\inf \{ t >   \t_o : X_t \not\in \partial A\}\cr}
\;\; ,\eqno (A.13)$$
and for $j=1,2,\dots$:
$$\eqalign{
\t_{j} :=&\inf \{ t > \s_{j}   : X_t \in      \partial A\}\cr
\s_{j} :=&\inf \{ t > \t_{j-1} : X_t \not\in  \partial A\}\cr}
\;\; .\eqno (A.14)
$$
We set, for $j=1,2,\dots$:
$$
{\cal I}_j = [\t_{j-1} +1, \t_j]\; ,\;\;\;\;\;\;
T_j:=|{\cal I}_j|= \t_j - \t_{j-1} -1 \;\; .
\eqno (A.15)
$$
Suppose that $ X_{\t_{j-1} +1} \; \in \; A$; let
$$\s^*_j := \min \{ t\; >\; \t_j \; : \; X_t \neq X_{\t_j}\}\; ;
\eqno (A.16)
$$
we say that the interval ${\cal I}_j$ is {\it good}
if the following conditions are satisfied:
$$\eqalign{
T_j &< T(\e) \cr
X_{\t_j} &\in {\cal S}(A)\cr
X_{\s^*_j} &\in{\cal H}\cr}\;\; .
$$
Let
$$ j^* := \min \{j \; : \; T_j \hbox { is not good}\}\;\; .$$
Given the integer $N$ we want to estimate, for every $x  \in A$, the
probability $P_x ( \t_{V(A)} >  N T(\e) )$.\par
We write:
$$
P_x ( \t_{V(A)}\ >  N T(\e) )= P_x ( \t_{V(A)} >
N T(\e)\; ; \; j^* > N )+P_x ( \t_{V(A)} >  N T(\e)\; ; \;
 j^* \leq N)
 \eqno (A.17)
$$
Let us consider the first event in the decomposition given in $(A.17)$:
$\{ \t_{V(A)} >
 N T(\e) ; \; j^*  > N \}$.\par
We have:
$$
P_x ( \t_{V(A)}\; >
\; N T(\e)\; ; \; j^* \; >\; N )\; \leq
$$
$$
\leq  P_x (X_{\t_1} \in {\cal S}(A);X_{\t_1+1}  \not \in  V(A),
 \dots
, X_{\t_N} \in  {\cal S}(A);X_{\t_N+1}  \not \in V(A)) \leq
$$

$$
\leq ( 1-c)^N\;\; .
\eqno (A.18)
$$
Now, from Proposition 3.7 of [OS1],
$(A.12)$  and the stationarity of our Markov chain we have
: $$
P_x ( {\cal I}_j \hbox { is not good} ) \; \leq \d (\b)
\eqno (A.19)
$$
with
$$
\lim _{\b \to \infty} \d (\b) \; = \; 0\;\; .
\eqno (A.20)
$$
{}From $(A.20)$ we get:
$$
P_x ( \t_{V(A)}\; > \; N T(\e)\; ; \;
 j^* \; \leq\; N)\; \leq \; P_x (
 j^* \; \leq\; N)\; \leq \; \d(\b) N\;\; .
\eqno (A.21)
$$
To conclude the proof it suffices to choose :
$$
N \;= \;N(\b)\; = \;1/\d(\b).
$$

\vfill\eject
\numsec=10\numfor=1
\centerline {\bf Acknowledgements.}
\vskip 1 truecm
\par\noindent
We acknowledge R. Schonmann for interesting discussions. We want to express
thanks to Istituto di Fisica Nucleare - Sezione di Bari whose financial
support made possible this collaboration. This work has been partially
supported by the grant CHRX-CT93-0411 and CIPA-CT92-4016 of the Commission
at European Communities.
\vfill\eject
\numsec=11\numfor=1
\centerline {\bf References.}
\vskip 1 truecm
\par\noindent
\item{[B]}M. Blume, Phys. Rev. {\bf 141}, 517 (1966).
\item{[BS]}J. Bricmont, J. Slawny, Journ. Stat. Phys. {\bf 54}, 89 (1989).
\item{[C]}H. W. Capel, Physica {\bf 32}, 96 (1966); {\bf 33}, 295 (1967);
{\bf 37}, 423 (1967).
\item{[CGOV]}M. Cassandro, A. Galves, E. Olivieri, M. E. Vares,
 ``Metastable behaviour of stochastic dynamics: A pathwise approach'',
Journ. Stat. Phys. {\bf 35}, 603-634 (1984).
\item{[DM]}E. I. Dinanburg, A. E. Mazel, Comm. Math. Phys. {\bf 125}, 27
(1989).
\item{[FGRN]}T. Fig, B. M. Gorman, P. A. Rikvold, M. A. Novotny,
``Numerical transfer--matrix study of a model with competing metastable
states", Phys. Rev. E {\bf 50}, 1930 (1994).
\item{[FW]} M. I. Freidlin, A. D. Wentzell, "Random Perturbations of Dynamical
Systems", Springer-Verlag (1984).
\item {[I]}S. N. Isakov, "Nonanalytic feature of the first order phase
transition in the Ising model", Comm. Math. Phys. {\bf 95}, 427-443 (1984).
\item{ [KO1]} R. Kotecky, E. Olivieri, ``Droplet dynamics for asymmetric
Ising model'', Journ. Stat. Phys. {\bf 70}, 1121-1148 (1993).
\item{ [KO2]} R. Kotecky, E. Olivieri, ``Shapes of growing droplets - a model
of
escape from a matastable phase'', Journ. Stat. Phys. {\bf 75}, 409-507 (1994).
\item{ [LR]} O. Lanford, D. Ruelle,
``Observable at infinity and states with short range correlations in
statistical mechanics", Commun. Math. Phys. {\bf 13}, 194-215 (1969).
\item{[MOS]}  F. Martinelli, E. Olivieri, E. Scoppola,
``Metastability and exponential approach to equilibrium for
low temperature stochastic Ising models'', Journ. Stat. Phys. {\bf 61},
N. 5/6 1105 (1990).
\item{[NS1]}  E. J. Neves, R. H. Schonmann,
``Critical Droplets and Metastability for a Glauber Dynamics
at Very Low Temperatures",  Comm. Math. Phys. {\bf 137}, 209 (1991).
\item{[NS2]}  E. J. Neves, R. H. Schonmann, ``Behaviour of droplets for a
class of Glauber dynamics at very low temperatures'',
Prob. Theor. Rel. Fields {\bf 91}, 331 (1992).
\item{[OS1]} E. Olivieri, E. Scoppola, ``Markov chains with exponentially
small transition probabilities: First exit problem from a general domain - I.
The reversible case''. In press on Journ. Stat. Phys..
\item{[OS2]} E. Olivieri, E. Scoppola, ``Markov chains with exponentially
small transition probabilities: First exit problem from a general domain - II.
The general case''. In preparation.
\item {[PL1]} O. Penrose, J. L. Lebowitz, "Towards a rigorous
molecular theory of metastability." In Fluctuation Phenomena (second edition).
E. W. Montroll, J. L. Lebowitz, editors. North-Holland Physics Publishing,
(1987).
\item {[PL2]} O. Penrose, J. L. Lebowitz,
"Molecular theory of metastability: an update."
Appendix to the reprinted edition of the article "Towards a rigorous
molecular theory of metastability"; by the same authors.
In Fluctuation Phenomena (second edition).
E. W. Montroll, J. L. Lebowitz, editors. North-Holland Physics Publishing,
(1987).
\item{[S1]}  R. H. Schonmann,
``The pattern of escape from metastability of a stochastic Ising model'',
Comm.  Math.  Phys. {\bf 147}, 231-240 (1992).
\item  {[S2]}R. H. Schonmann,
``Slow droplet-driven relaxation of stochastic Ising
models in the vicinity of the phase coexistence region", Comm. Math. Phys.
{\bf 161}, 1-49 (1994).
\item{[SS]} S. Shlosman, R. H. Schonmann, preprint UCLA (1994).

\vfill\eject
\end